\RequirePackage{lineno}   
\documentclass[12pt,a4paper]{article}
\usepackage{graphicx}
\usepackage{amsmath}
\usepackage{hhline}
\usepackage{amssymb}
\usepackage{times}
\usepackage{cite}
\usepackage{rotate}
\usepackage{multirow}
\usepackage{captcont}

\usepackage{subfigure}
\usepackage[pagebackref=false,colorlinks=true]{hyperref} 

\newlength{\dinwidth}
\newlength{\dinmargin}
\begin{document}  
\renewcommand{\arraystretch}{1.1}
\newcommand{\pom}{{I\!\!P}}
\newcommand{\reg}{{I\!\!R}}
\newcommand{\slowpi}{\pi_{\mathit{slow}}}
\newcommand{\fiidiii}{F_2^{D(3)}}
\newcommand{\fiidiiiarg}{\fiidiii\,(\beta,\,Q^2,\,x)}
\newcommand{\n}{1.19\pm 0.06 (stat.) \pm0.07 (syst.)}
\newcommand{\nz}{1.30\pm 0.08 (stat.)^{+0.08}_{-0.14} (syst.)}
\newcommand{\fiidiiiful}{F_2^{D(4)}\,(\beta,\,Q^2,\,x,\,t)}
\newcommand{\fiipom}{\tilde F_2^D}
\newcommand{\ALPHA}{1.10\pm0.03 (stat.) \pm0.04 (syst.)}
\newcommand{\ALPHAZ}{1.15\pm0.04 (stat.)^{+0.04}_{-0.07} (syst.)}
\newcommand{\fiipomarg}{\fiipom\,(\beta,\,Q^2)}
\newcommand{\pomflux}{f_{\pom / p}}
\newcommand{\nxpom}{1.19\pm 0.06 (stat.) \pm0.07 (syst.)}
\newcommand {\gapprox}
   {\raisebox{-0.7ex}{$\stackrel {\textstyle>}{\sim}$}}
\newcommand {\lapprox}
   {\raisebox{-0.7ex}{$\stackrel {\textstyle<}{\sim}$}}
\def\gsim{\,\lower.25ex\hbox{$\scriptstyle\sim$}\kern-1.30ex%
\raise 0.55ex\hbox{$\scriptstyle >$}\,}
\def\lsim{\,\lower.25ex\hbox{$\scriptstyle\sim$}\kern-1.30ex%
\raise 0.55ex\hbox{$\scriptstyle <$}\,}
\newcommand{\pomfluxarg}{f_{\pom / p}\,(x_\pom)}
\newcommand{\dsf}{\mbox{$F_2^{D(3)}$}}
\newcommand{\dsfva}{\mbox{$F_2^{D(3)}(\beta,Q^2,x_{I\!\!P})$}}
\newcommand{\dsfvb}{\mbox{$F_2^{D(3)}(\beta,Q^2,x)$}}
\newcommand{\dsfpom}{$F_2^{I\!\!P}$}
\newcommand{\gap}{\stackrel{>}{\sim}}
\newcommand{\lap}{\stackrel{<}{\sim}}
\newcommand{\fem}{$F_2^{em}$}
\newcommand{\tsnmp}{$\tilde{\sigma}_{NC}(e^{\mp})$}
\newcommand{\tsnm}{$\tilde{\sigma}_{NC}(e^-)$}
\newcommand{\tsnp}{$\tilde{\sigma}_{NC}(e^+)$}
\newcommand{\st}{$\star$}
\newcommand{\sst}{$\star \star$}
\newcommand{\ssst}{$\star \star \star$}
\newcommand{\sssst}{$\star \star \star \star$}
\newcommand{\tw}{\theta_W}
\newcommand{\sw}{\sin{\theta_W}}
\newcommand{\cw}{\cos{\theta_W}}
\newcommand{\sww}{\sin^2{\theta_W}}
\newcommand{\cww}{\cos^2{\theta_W}}
\newcommand{\trm}{m_{\perp}}
\newcommand{\trp}{p_{\perp}}
\newcommand{\trmm}{m_{\perp}^2}
\newcommand{\trpp}{p_{\perp}^2}
\newcommand{\alp}{\alpha_s}

\newcommand{\alps}{\alpha_s}
\newcommand{\sqrts}{$\sqrt{s}$}
\newcommand{\LO}{$O(\alpha_s^0)$}
\newcommand{\Oa}{$O(\alpha_s)$}
\newcommand{\Oaa}{$O(\alpha_s^2)$}
\newcommand{\PT}{p_{\perp}}
\newcommand{\JPSI}{J/\psi}
\newcommand{\sh}{\hat{s}}
\newcommand{\uh}{\hat{u}}
\newcommand{\MP}{m_{J/\psi}}
\newcommand{\PO}{I\!\!P}
\newcommand{\xbj}{x}
\newcommand{\xpom}{x_{\PO}}
\newcommand{\ttbs}{\char'134}
\newcommand{\xpomlo}{3\times10^{-4}}  
\newcommand{\xpomup}{0.05}  
\newcommand{\dgr}{^\circ}
\newcommand{\pbarnt}{\,\mbox{{\rm pb$^{-1}$}}}
\newcommand{\gev}{\,\mbox{GeV}}
\newcommand{\WBoson}{\mbox{$W$}}
\newcommand{\fbarn}{\,\mbox{{\rm fb}}}
\newcommand{\fbarnt}{\,\mbox{{\rm fb$^{-1}$}}}

\newcommand{\vtab}{\rule[-1mm]{0mm}{4mm}}
\newcommand{\htab}{\rule[-1mm]{0mm}{6mm}}
\newcommand{\photoproduction}{$\gamma p$}
\newcommand{\ptmiss}{$P_{T}^{\rm miss}$}
\newcommand{\epz} {$E{\rm-}p_z$}
\newcommand{\vap} {  $V_{ap}/V_p$}
\newcommand{\Zero}   {\mbox{$Z^{\circ}$}}
\newcommand{\Ftwo}   {\mbox{$\tilde{F}_2$}}
\newcommand{\Ftwoz}   {\mbox{$\tilde{F}_{2,3}$}}
\newcommand{\Fz}   {\mbox{$\tilde{F}_3$}}
\newcommand{\FL}   {\mbox{$\tilde{F}_{_{L}}$}}
\newcommand{\wtwogen} {W_2}
\newcommand{\wlgen} {W_L}
\newcommand{\xwthreegen} {xW_3}
\newcommand{\Wtwo}   {\mbox{$W_2$}}
\newcommand{\Wz}   {\mbox{$W_3$}}
\newcommand{\WL}   {\mbox{$W_{_{L}}$}}
\newcommand{\Fem}  {\mbox{$F_2$}}
\newcommand{\Fgam}  {\mbox{$F_2^{\gamma}$}}
\newcommand{\Fint} {\mbox{$F_2^{\gamma Z}$}}
\newcommand{\Fwk}  {\mbox{$F_2^{Z}$}}
\newcommand{\Ftwos} {\mbox{$F_2^{\gamma Z, Z}$}}
\newcommand{\Fzz} {\mbox{$F_3^{\gamma Z, Z}$}}
\newcommand{\Fintz} {\mbox{$F_{2,3}^{\gamma Z}$}}
\newcommand{\Fwkz}  {\mbox{$F_{2,3}^{Z}$}}
\newcommand{\Fzint} {\mbox{$F_3^{\gamma Z}$}}
\newcommand{\Fzwk}  {\mbox{$F_3^{Z}$}}
\newcommand{\Gev}  {\mbox{${\rm GeV}$}}
\newcommand{\Gevv}{\mbox{${\rm GeV}^2$}}
\newcommand{\QQ}  {\mbox{${Q^2}$}}
\newcommand{\gv}{GeV$^2\,$}
\newcommand{\bs}{\bar{s}}
\newcommand{\bc}{\bar{c}}
\newcommand{\bu}{\bar{u}}
\newcommand{\bb}{\bar{b}}
\newcommand{\bU}{\overline{U}}
\newcommand{\bD}{\overline{D}}
\newcommand{\bd}{\bar{d}}
\newcommand{\bq}{\bar{q}}    
\newcommand{\FLc}{$ F_{L}\,$} 
\newcommand{\xg}{$xg(x,Q^2)\,$}
\newcommand{\xgc}{$xg\,$}
\newcommand{\ipb}{pb$^{-1}\,$}               
\newcommand{\TOSS}{x_{{i}/{\PO}}}                                              
\newcommand{\un}[1]{\mbox{\rm #1}}
\newcommand{\pdsi}{$(\partial \sigma_r / \partial \ln y)_{Q^2}\,$}
\newcommand{\pdff}{$(\partial F_{2} / \partial \ln  Q^{2})_x\,$ }
\newcommand{\Fc}{$ F_{2}~$}
\newcommand{\amz}{$\alpha_s(M_Z^2)\,$} 

%
%
\newcommand{\qsq}{\ensuremath{Q^2} }
\newcommand{\gevsq}{\ensuremath{\mathrm{GeV}^2} }
\newcommand{\et}{\ensuremath{E_t^*} }
\newcommand{\rap}{\ensuremath{\eta^*} }
\newcommand{\gp}{\ensuremath{\gamma^*}p }
\newcommand{\dsiget}{\ensuremath{{\rm d}\sigma_{ep}/{\rm d}E_t^*} }
\newcommand{\dsigrap}{\ensuremath{{\rm d}\sigma_{ep}/{\rm d}\eta^*} }
\def\Journal#1#2#3#4{{#1} {\bf #2} (#3) #4}
\def\NCA{\em Nuovo Cimento}
\def\NIM{\em Nucl. Instrum. Methods}
\def\NIMA{{\em Nucl. Instrum. Methods} {\bf A}}
\def\NPB{{\em Nucl. Phys.}   {\bf B}}
\def\PLB{{\em Phys. Lett.}   {\bf B}}
\def\PRL{\em Phys. Rev. Lett.}
\def\PRD{{\em Phys. Rev.}    {\bf D}}
\def\ZPC{{\em Z. Phys.}      {\bf C}}
\def\EJC{{\em Eur. Phys. J.} {\bf C}}
\def\CPC{\em Comp. Phys. Commun.}

\begin{titlepage}

\noindent
\begin{flushleft}
{\tt DESY 12-107    \hfill    ISSN 0418-9833} \\
{\tt June 2012}                  \\
\end{flushleft}

\vspace{1cm}

\begin{center}
\begin{Large}

{\bf  Inclusive Deep Inelastic Scattering \\
at High \boldmath${Q^2}$ with Longitudinally \\
Polarised Lepton Beams at HERA}
\vspace{1cm}

H1 Collaboration

\end{Large}
\end{center}

\vspace{1cm}

\begin{abstract}
\noindent
Inclusive $e^{\pm}p$~single and double differential cross sections for
neutral and charged current deep inelastic scattering processes are measured with the H1
detector at HERA. The data were taken at a centre-of-mass energy of
$\sqrt{s}=319\,{\rm GeV}$ with a total integrated luminosity of 
$333.7\,{\rm pb}^{-1}$ shared between two lepton beam charges and two
longitudinal lepton polarisation modes. The differential cross
sections are measured in the range of negative four-momentum transfer squared,
$Q^2$, between $60$ and $50\,000\,{\rm GeV}^2$, and Bjorken $x$ between
$0.0008$ and $0.65$. The measurements are combined with earlier
published unpolarised H1 data to improve statistical precision and
used to determine the structure function $xF_3^{\gamma Z}$. A
measurement of the neutral current parity violating structure function
$F_2^{\gamma Z}$ is presented for the first time.  The polarisation
dependence of the charged current total cross section
is also measured. The new measurements are well described by a next-to-leading
order QCD fit based on all published H1 inclusive cross section data
which are used to extract the parton distribution functions of the proton.

\end{abstract}

\vspace{1cm}

\begin{center}
Submitted to JHEP 
\end{center}

\end{titlepage}

%
%

\begin{flushleft}

F.D.~Aaron$^{5,45}$,           
C.~Alexa$^{5}$,                
V.~Andreev$^{25}$,             
S.~Backovic$^{30}$,            
A.~Baghdasaryan$^{38}$,        
S.~Baghdasaryan$^{38}$,        
E.~Barrelet$^{29}$,            
W.~Bartel$^{11}$,              
K.~Begzsuren$^{35}$,           
A.~Belousov$^{25}$
% ...


\bigskip{\it
% ...
$ ^{5}$ National Institute for Physics and Nuclear Engineering (NIPNE) ,
          Bucharest, Romania$^{ k}$ \\
% ...
$ ^{11}$ DESY, Hamburg, Germany \\
% ...
$ ^{25}$ Lebedev Physical Institute, Moscow, Russia \\
% ...
$ ^{29}$ LPNHE, Universit\'e Pierre et Marie Curie Paris 6,
          Universit\'e Denis Diderot Paris 7, CNRS/IN2P3, Paris, France \\
 $ ^{30}$ Faculty of Science, University of Montenegro,
          Podgorica, Montenegro$^{ l}$ \\
% ...
$ ^{35}$ Institute of Physics and Technology of the Mongolian
          Academy of Sciences, Ulaanbaatar, Mongolia \\
% ...
$ ^{38}$ Yerevan Physics Institute, Yerevan, Armenia \\
% ...
$ ^{45}$ Also at Faculty of Physics, University of Bucharest,
          Bucharest, Romania \\
% ...
$ ^k$ Supported by the Romanian National Authority for Scientific Research
      under the contract PN 09370101 \\
 $ ^l$ Partially Supported by Ministry of Science of Montenegro,
      no. 05-1/3-3352 \\
}

\end{flushleft}

\newpage
\section{Introduction}
\label{sec:intro}

Precision measurements of the proton structure in neutral current (NC)
and charged current (CC) deep inelastic scattering (DIS) with
polarised lepton beams provide important information on the
understanding of parton dynamics and quantum chromodynamics (QCD). 
Previously published
measurements at the electron\footnote{In this paper ``electron'' refers generically to both
  electrons and positrons. Where distinction is required the terms
  $e^-$ and $e^+$ are used.}-proton collider  
HERA~\cite{h19497,h19899,h1hiq2,h1ccpol,h1lowestq2,h1fl,h1pdf2009,zeusbpc,zeusbpt,zeussvx,zeusnc9697,zeuscc9497,zeusnc9899,zeuscc9899,zeusnc9900,zeuscc9900}
have already provided strong constraints on the parton distribution
functions (PDFs) of the proton~\cite{h1hiq2,h1pdf2009,HERAPDF10,CTEQ6.6,MSTW2008,NNPDF2.0,GJR,ABKM}.
With access to values of four momentum transfers $\sqrt{Q^2}$
comparable to the masses of the $Z$ and $W$ bosons, precision DIS
measurements also probe the chiral structure of the electroweak
interactions.  Inclusive neutral current interactions are defined as
the process \mbox{$ep \rightarrow eX$} mediated by $\gamma/Z$
bosons, whereas inclusive charged current interactions are defined as
\mbox{$ep \rightarrow \nu X$} and are purely weak processes mediated
by $W$ bosons.

In this paper precise measurements of the inclusive neutral and 
charged current $ep$ cross sections at high $Q^2$ are presented
utilising the complete 
HERA\,II\footnote{HERA operation was split into two phases, HERA\,I
  which ran from 1992 to 2000, and HERA\,II which ran from 2003 to2007.}
data set of $333.7\,{\rm pb}^{-1}$ recorded by the H1 detector 
at a centre-of-mass energy of $\sqrt{s}= 319\,{\rm GeV}$ with
longitudinally polarised electron and positron beams.
The inclusive NC and CC polarised single differential cross sections,
${\rm d}\sigma/{\rm d}Q^2$ and the double differential reduced cross
sections ${\tilde\sigma}(x,Q^2)$ are presented for $e^+p$ and $e^-p$
scattering. The data were taken with an incident lepton beam energy
$E_e$ of $27.6\,{\rm GeV}$, whilst the energy of the unpolarised
proton beam, $E_p$, was $920\,{\rm GeV}$. The longitudinal polarisation
of the lepton beam was $\pm 35\%$ on average.

The NC data cover the $Q^2$ range from $60$ to
$50\,000\,{\rm GeV}^2$. Together with previous H1 measurements at lower
$Q^2$, down to $\sim1\,{\rm GeV}^2$~\cite{h1lowestq2,h1pdf2009}, the data
cover almost five orders of magnitude in kinematic reach. The high
$Q^2$ NC and CC data presented here give unique constraints on the
proton PDFs for Bjorken $x$ in the range $0.0008\leq x\leq 0.65$ which
is of direct relevance to all predictions for $pp$ scattering at the
LHC~\cite{LHCPrimer}. In particular, production cross sections of new
high mass states in the LHC kinematic domain are very sensitive to the
high $x$ PDFs constrained by DIS data.

The data extend to very high $Q^2$ which allows structure functions
sensitive to the interference of photon and $Z$ boson exchange
to be measured. These ``interference structure functions" access the
difference of quark and anti-quark distributions, with $xF_3^{\gamma
  Z}$, and a combination of their sums, with $F_2^{\gamma Z}$.  They
are measured by using both the charge and
polarisation dependence of the NC cross section, providing an improved
determination of $xF_3^{\gamma Z}$ and a very first measurement of
$F_2^{\gamma Z}$.

The measured inclusive cross sections are combined with previously
published unpolarised HERA\,I measurements to provide a single
coherent set of cross sections.
The sensitivity and kinematic reach of these data enable a dedicated
QCD analysis to be performed on H1 data alone. The fit procedure
takes into account all point-to-point correlated systematic
uncertainties yielding a new determination of PDFs
and their uncertainties, termed H1PDF\,2012.

This paper is organised as follows: in section~\ref{sec:theory} the
definitions of the inclusive NC and CC cross sections are given
together with their relation to the proton structure functions and
PDFs. In section~\ref{sec:det} the H1 detector and trigger system are
described as well as the HERA polarimeters. The simulation programmes
and Monte Carlo models used in the analysis are discussed in
section~\ref{sec:mc}. In section~\ref{sec:expt} the analysis procedure
is given starting with a description of the kinematic reconstruction
methods, calibration and alignment of the detector, and followed by
the event selection and assessment of the systematic uncertainties of
the measurements. The QCD analysis method is explained in
section~\ref{sec:qcdana} and the results are presented in
section~\ref{sec:results}.  The paper is summarised in
section~\ref{sec:summary}.

\section{Neutral and Charged Current Cross Sections}
\label{sec:theory}

\subsection{Neutral Currents}

The differential cross section for $e^{\pm}p$ scattering after
correction for QED radiative effects can be expressed in terms of
generalised proton structure functions $\tilde{F}$ as
\begin{equation}
\frac{\rm{d}^2\sigma^{\pm}_{\rm NC}}{{\rm d}x{\rm d}Q^2}=
\frac{2\pi\alpha^2}{xQ^4}(Y_+\tilde{F}_2^\pm \mp Y_-x\tilde{F}_3^\pm-y^2\tilde{F}_L^\pm) \cdot (1+\Delta_{\rm NC}^{\rm weak})\,,
\label{eq:ncxsec} 
\end{equation} 
where $Y_{\pm}=1\pm(1-y)^2$ and $y$ characterises the inelasticity of
the interaction. The fine structure constant is defined as
$\alpha\equiv\alpha(Q^2=0)$ and the weak radiative
corrections\footnote{The weak corrections are typically smaller than
  $1\%$ and never more than $3\%$ at the highest $Q^2$ and are not
  applied to the measured cross sections.}  $\Delta_{\rm NC}^{\rm
  weak}$ are defined as in~\cite{hector} in terms of $\alpha$ and the
$Z$ and $W$ boson masses which are taken to be $M_Z=91.187\,{\rm GeV}$ and
$M_W=80.410\,{\rm GeV}$.

The generalised structure functions, $\tilde{F}_{2,3}$, may be
written as linear combinations of the proton structure functions
$F_{2}$, $F_{2,3}^{\gamma Z}$, and $F_{2,3}^{Z}$ containing
information on QCD parton dynamics as well as on the electroweak (EW) 
couplings of the quarks to the neutral vector bosons~\cite{Klein:1983vs}. 
The structure function $F_{2}$ is
associated to pure photon exchange terms, $F_{2,3}^{\gamma Z}$
correspond to photon-$Z$ interference terms and $F_{2,3}^{Z}$ describe
the pure $Z$ exchange terms. In addition the generalised longitudinal
structure function $\tilde{F}_L$ may be similarly decomposed, however
this is an important contribution only at high $y$ and is expected to be negligible at
large $x$ and $Q^2$. The linear combinations for $\tilde{F}_2$ and
$x\tilde{F}_3$ in arbitrarily polarised
$e^{\pm}p$ scattering with lepton polarisation $P_e$ are given by
\begin{eqnarray}
 \tilde{F}^{\pm}_2 = F_2 - (v_e \pm P_e a_e) \kappa  \frac{Q^2}{Q^2+M_Z^2}    F_2^{\gamma Z} 
            + (a_e^2+v_e^2 \pm P_e 2v_e a_e) \kappa^2 \left[\frac{Q^2}{Q^2+M_Z^2}\right]^2 F_2^Z \,,
\label{eq:F2}\\
 x\tilde{F}^{\pm}_3 = -(a_e \pm P_e v_e)   \kappa    \frac{Q^2}{Q^2+M_Z^2}    xF_3^{\gamma Z} 
       + (2a_ev_e \pm P_e [v_e^2 + a_e^2] ) \kappa^2 \left[\frac{Q^2}{Q^2+M_Z^2}\right]^2 xF_3^Z\,,
\label{eq:F3}
\end{eqnarray}
with $\kappa^{-1}=4\frac{M_W^2}{M_Z^2}(1-\frac{M_W^2}{M_Z^2})$
in the on-mass-shell scheme. The quantities $v_e$ and $a_e$ are the
vector and axial-vector couplings of the electron to the $Z$ boson. 

It can be seen from equations\,\ref{eq:F2} and \ref{eq:F3} that different
combinations of structure functions may be experimentally
determined
by scattering longitudinally polarised leptons on unpolarised protons.
In particular, since $v_e$ is small, a
measurement of the polarisation asymmetry for fixed lepton charge
allows the parity violating structure function $F_2^{\gamma Z}$ to be
measured.

In the quark-parton model (QPM), the hadronic structure functions are
related to linear combinations of sums and differences of the quark
and anti-quark momentum distributions $xq(x,Q^2)$ and
$x\bar{q}(x,Q^2)$. The structure function $\tilde{F}_2$ is
determined by the sum of quarks and anti-quark momentum distributions,
whereas the structure function $x\tilde{F}_3$ is
determined by the difference of quarks and anti-quark momentum
distributions and is therefore sensitive to the valence quark
distributions:
\begin{eqnarray}
  \left[F_2,F_2^{\gamma Z},F_2^Z\right] &=& x\sum_q [e_q^2,2e_qv_q,v_q^2+a_q^2](q+\bar{q})\,, \\
  \left[xF_3^{\gamma Z},xF_3^Z \right] &=& 2x \sum_q[e_qa_q,v_qa_q](q-\bar{q})\,.
\label{eq:SF2} 
\end{eqnarray} 
Here $v_q$ and $a_q$ are the vector and axial-vector couplings of the
quarks to the $Z$ boson and $e_q$ is the charge of the quark of flavour $q$.

The reduced NC cross section is defined by
\begin{equation}
\tilde{\sigma}_{\rm NC}^{\pm}(x,Q^2)\equiv
\frac{\rm{d}^2\sigma^{\pm}_{\rm NC}}{{\rm d}x{\rm d}Q^2}\frac{xQ^4}{2\pi\alpha^2}\frac{1}{Y_+}\equiv
\left( \tilde{F}_2^\pm \mp \frac{Y_-}{Y_+}x\tilde{F}_3^\pm-\frac{y^2}{Y_+}\tilde{F}_L^\pm \right) (1+\Delta_{\rm NC}^{\rm weak})\,.
\label{eq:Rnc}
\end{equation} 


\subsection{Charged Currents}

The differential CC cross section for $e^{\pm}p$
scattering of polarised leptons with unpolarised protons, corrected
for QED radiative effects, can be expressed as
\begin{equation}
\frac{{\rm d}^2\sigma^{\pm}_{\rm CC}}{{\rm d}x {\rm d}\QQ} = (1\pm P_e)
 \frac{G_F^2}{4\pi x } \left[\frac{M_W^2}{M_W^2+Q^2} \right]^2 
\left(Y_+ W^{\pm}_2 \mp Y_{-}xW^{\pm}_3 -y^2W^{\pm}_L \right ) \cdot (1+\Delta_{\rm CC}^{\rm weak})\,,
\label{eq:ccxsec}
\end{equation}
 
where $G_F$ is the Fermi constant defined using the weak boson
masses~\cite{hector}. Here $W^{\pm}_2$, $xW^{\pm}_3$ and $W^{\pm}_L$
are the structure functions for CC $e^{\pm}p$ scattering, and
$\Delta^{\rm CC}_{\rm weak}$ represents the weak radiative corrections
for CC interactions.
From equation\,\ref{eq:ccxsec} it can be seen that the
cross section has a linear dependence on the polarisation of the
electron beam $P_e$. For a fully right handed $e^-$ beam ($P_e=1$),
or a fully left handed $e^+$ beam ($P_e=-1$) the cross section is
identically zero in the Standard Model (SM). In the QPM $W^{\pm}_L\equiv 0$, and the
structure functions $W^{\pm}_2$ and $xW^{\pm}_3$ are expressed as the
flavour dependent sum and difference of the quark and anti-quark
momentum distributions. In the CC case only the positively charged
quarks contribute to $W^-$ mediated scattering and conversely only
negatively charged quarks couple to the exchanged $W^+$ boson, thus
\begin{eqnarray}
W^-_2 \! &=&\! x(U+\overline{D})\,,\hspace{5.5mm} W^+_2 = x(\overline{U}+ D )\,,\\ 
xW^-_3 \!&=&\! x(U-\overline{D})\,,\hspace{3mm} xW^+_3 = x(D -\overline{U})\,,
\end{eqnarray}
where, below the $b$ quark mass threshold
\begin{equation}
\label{eq:ud}
 U = u + c\,,~~ 
\overline{U}= \bu + \bc\,,~~
 D = d + s\,, ~~ 
\overline{D}= \bd + \bs\,, 
\end{equation}
where $u$, $d$, $s$, $c$ represent quark densities of each flavour in
the standard notation.  Here $U$ represents the sum of up-type, and
$D$ the sum of down-type quark densities.

The reduced CC cross section is then defined as
\begin{equation}
\label{eq:Rcc}
\tilde{\sigma}_{\rm CC}(x,Q^2) \equiv  
\frac{4 \pi  x}{ G_F^2}
\left[ \frac {M_W^2+Q^2} {M_W^2} \right]^2
          \frac{{\rm d}^2 \sigma_{\rm CC}}{{\rm d}x{\rm d}Q^2}\,\,.
\end{equation}

\section{H1 Apparatus, Trigger and Data Samples}
\label{sec:det}

\subsection{The H1 Detector}

A detailed description of the H1 detector can be found
elsewhere~\cite{h1detector,h1tracker,h1lar,spacal}. The coordinate
system of H1 is defined such that the positive $z$ axis is in the
direction of the proton beam (forward direction) and the nominal
interaction point is located at $z=0$. The polar angle $\theta$ is
then defined with respect to this axis.  The detector components most
relevant to this analysis are the Liquid Argon (LAr) calorimeter,
which measures the positions and energies of particles over the range
$4^\circ<\theta<154^\circ$, the inner tracking detectors, which
measure the angles and momenta of charged particles over the range
$7^\circ<\theta<165^\circ$, and a lead-fibre calorimeter (SpaCal)
covering the range $153^\circ<\theta<177^\circ$.

The LAr calorimeter consists of an inner electromagnetic section with lead
absorbers and an outer hadronic section with steel absorbers.  The
calorimeter is divided into eight wheels along the beam axis, each
consisting of eight absorber stacks arranged in an octagonal formation
around the beam axis. The electromagnetic and the hadronic sections
are highly segmented in the transverse and the longitudinal
directions. Electromagnetic shower energies are measured with a
resolution of $\delta E/E \simeq 0.11/\sqrt{E/{\rm GeV}} \oplus 0.01$
and hadronic energies with $\delta E/E \simeq 0.46/\sqrt{E/{\rm GeV}}
\oplus 0.03$ as
determined using electron and pion test beam
data~\cite{Andrieu:1993tz,Andrieu:1994yn}.

In the central region, $25^{\circ}<\theta<155^{\circ}$, the central
tracking detector (CTD) measures the trajectories of charged particles
in two cylindrical drift chambers immersed in a uniform $1.16\,{\rm T}$
solenoidal magnetic field. The CTD also contains a further drift
chamber (COZ) between the two drift chambers to improve the $z$
coordinate reconstruction, as well as a multi-wire proportional
chamber at inner radii (CIP) mainly used for
triggering~\cite{Becker:2007ms}. The CTD measures charged particles
with a transverse momentum resolution of $\sigma(p_T)/p_T\simeq
0.2\% \, p_T/{\rm GeV} \oplus 1.5\%$. The forward
tracking detector (FTD) is used to supplement track reconstruction in
the region $7^{\circ}<\theta<30^{\circ}$~\cite{Laycock:2012xg} and improves the hadronic
final state reconstruction of forward going low momentum particles.

The CTD tracks are linked to hits in the vertex detectors: the central
silicon tracker (CST)~\cite{h1cst,h1cst2}, the forward silicon tracker
(FST), and the backward silicon tracker (BST). These detectors provide
precise spatial track reconstruction and therefore also improve the
primary vertex spatial reconstruction. The CST consists of two layers
of double-sided silicon strip detectors surrounding the beam pipe
covering an angular range of $30^\circ<\theta<150^\circ$ for tracks
passing through both layers. The FST consists of five double wheels of
single-sided strip detectors~\cite{Glushkov:2007zz} measuring the
transverse coordinates of charged particles. The BST design is very
similar to the FST and consists of six double wheels of strip
detectors~\cite{Kretzschmar:2008zz}. 

In the backward region the SpaCal provides an energy measurement for
hadronic particles, and has a hadronic energy resolution of $\delta
E/E \simeq 0.70/\sqrt{E/{\rm GeV}}\oplus 0.01$ and a resolution for
electromagnetic energy depositions of $\delta E/E \simeq
0.07/\sqrt{E/{\rm GeV}}\oplus 0.01$ measured using test beam
data~\cite{spacal_res}. It also provides a trigger used for efficiency
estimations which is based on electromagnetic energy and timing
information inside the calorimeter.

The $ep$ luminosity is determined online by measuring the event rate
for the Bethe-Heitler process of QED bremsstrahlung $ep \rightarrow
ep\gamma$. The photons are detected in the photon tagger located at
$z=-103\,{\rm m}$.  An electron tagger is placed at $z=-5.4\,{\rm m}$ adjacent to
the beam-pipe. It is used to provide information on $ep\rightarrow eX$
events at very low $Q^2$ (photoproduction) where the electron scatters
through a small angle ($\pi - \theta < 5\,{\rm mrad}$).  The overall
normalisation is determined using a precision measurement of the QED
Compton process~\cite{compton-lumi}.

At HERA transverse polarisation of the lepton beam arises naturally
through synchrotron radiation via the Sokolov-Ternov
effect~\cite{sokolov-ternov}. Spin rotators installed in the beam-line
on either side of the H1 detector allow transversely polarised leptons
to be rotated into longitudinally polarised states and back again. The
degree of polarisation is constant around the HERA ring and is
continuously measured using two independent polarimeters
LPOL~\cite{lpol} and TPOL~\cite{tpol}. The polarimeters are situated
in beam-line sections in which the beam leptons have longitudinal and
transverse polarisations, respectively. Both measurements rely on an
asymmetry in the energy spectrum of left and right handed circularly
polarised photons undergoing Compton scattering with the lepton
beam. The TPOL measurement uses in addition a spatial asymmetry. The
LPOL and TPOL measurements are averaged when both measurements are
available, otherwise only one polarimeter 
measurement is used~\cite{pubpola}.

\subsection{The Trigger}
The H1 trigger system is a three level trigger with a first level
latency of approximately $2\,\mu{\rm s}$. NC events at high $Q^2$ are
triggered mainly using information from the LAr calorimeter.  The
calorimeter has a finely segmented pointing geometry allowing the
trigger to select localised energy deposits in the electromagnetic
section of the calorimeter pointing to the nominal interaction vertex.
For electrons with energy above $11\,{\rm GeV}$ this is determined to be
$100\%$ efficient using an independently triggered sample of events.
At lower energies the triggers based on LAr information are
supplemented by using additional information from the tracking
detectors.  The LAr calorimeter electronics allow scattered leptons to
be triggered with energies as low as $5\,{\rm GeV}$, the minimum value
considered in this analysis. This gives access to the high $y$
kinematic region. For electron energies of $5\,{\rm GeV}$, the combined
trigger efficiency increases from $87\%$ to $92\%$ during the HERA\,II
run due to several incremental improvements in the trigger set-up.

The characteristic feature of CC events is a large missing transverse
momentum, $P_T^{\rm miss}$, which is identified at the trigger level
using the LAr calorimeter vector sum of energy within ``trigger towers'',
i.e. groups of trigger regions with a projective geometry pointing to
the nominal interaction vertex. At low $P_T^{\rm miss}$ the efficiency
is enhanced by use of an additional trigger requiring hadronic energy
in combination with track information
from the inner tracking chambers.  At $12\,{\rm GeV}$, the minimum $P_T^{\rm
  miss}$ considered in this analysis, the efficiency is $60\%$, rising
to $90\%$ for $P_T^{\rm miss}$ of $25\,{\rm GeV}$. This is determined from
the {\em pseudo CC} sample. This sample is constructed of NC events in
which all information from the scattered lepton is suppressed (see
section~\ref{sec:ccmeas}). The trigger energy sums are then
recalculated for the remaining hadronic final state. This sample also
provides a useful high statistics cross check of further aspects of
the CC analysis.

\subsection{Data Samples}
\label{sec:datasets}

The data sets used in this analysis are
subdivided into samples or periods of left handed and right handed polarised
lepton beams with polarisation $P_e=(N_R-N_L)/(N_R+N_L)$, where $N_R$
($N_L$) is the number of right (left) handed leptons in the beam. The
corresponding data sets are termed the $R$ and $L$ data sets
respectively. The luminosity and longitudinal lepton beam polarisation
for each data set are given in table~\ref{tab:lumi}.

\begin{table}[h]
  \begin{center}
    \begin{tabular}{r|c|c}
\hline
 & $R$ & $L$\\
\hline
\multirow{2}{*}{$e^-p$}  
& $\mathcal{L}=47.3\,{\rm pb}^{-1}$ & $\mathcal{L}=104.4\,{\rm pb}^{-1}$ \\
& $P_e=(+36.0\pm 1.0)\%$ & $P_e=(-25.8\pm 0.7)\%$ \\
\hline
\multirow{2}{*}{$e^+p$}  
& $\mathcal{L}=101.3\,{\rm pb}^{-1}$ & $\mathcal{L}=80.7\,{\rm pb}^{-1}$ \\
& $P_e=(+32.5\pm 0.7)\%$ & $P_e=(-37.0\pm 0.7)\%$ \\
\hline
\end{tabular} 
\caption{ Table of integrated luminosities, $\mathcal{L}$, and
  luminosity weighted longitudinal lepton beam polarisation, $P_e$, for
  the data sets presented here.}
\label{tab:lumi}
\end{center}
\end{table}

\section{Simulation Programs}
\label{sec:mc}

In order to determine acceptance corrections, DIS processes are
generated at leading order (LO) QCD using the {\sc
  Djangoh\,1.4}~\cite{django} Monte Carlo (MC) simulation program
which is based on {\sc Heracles\,4.6} \cite{heracles} for the
electroweak interaction and on {\sc Lepto\,6.5.1}~\cite{lepto} for the
hard matrix element calculation. The colour dipole model (CDM) as
implemented in {\sc Ariadne} \cite{cdm} is used to generate higher
order QCD dynamics.  The {\sc Jetset\,7.410} program~\cite{jetset} is
used to simulate the hadronisation process in the
`string-fragmentation' model.  Additional {\sc Djangoh} study samples
are produced in which the higher order QCD effects are simulated
using DGLAP inspired parton showers matched to the hard LO matrix
element calculation, known as MEPS.  The simulated events are produced
with PDFs from a NLO QCD fit (HERAPDF1.0) which includes combined H1
and ZEUS low $Q^2$ and high $Q^2$ NC and CC data from
HERA\,I~\cite{HERAPDF10}.
In order to improve the precision with which the acceptance corrections 
are determined, the simulated cross sections are
reweighted using the PDF set determined in this analysis, H1PDF\,2012
(see section~\ref{sec:qcdana}).

The dominant $ep$ background contribution to DIS is due to large cross
section photoproduction ($\gamma p$) processes in which energetic
$\pi^0\rightarrow\gamma\gamma$ decays or charged hadrons are
mis-identified as the scattered electron in the NC channel, or hadronic final
states produce large fake missing transverse momentum mimicking a CC
interaction.  These are simulated using the {\sc
  Pythia}\,6.224~\cite{Pythia6.2} generator with leading order parton
distribution functions for the proton and photon taken
from~\cite{Gluck:1991ee}. Additional small background contributions
arise from elastic and inelastic QED Compton processes generated with
the {\sc Wabgen} program~\cite{wabgen}; lepton pair production via two
photon interactions simulated by the {\sc Grape} code~\cite{grape};
prompt photon production in which the photon may be mis-identified as
an electron generated by {\sc Pythia}; and real $W^{\pm}/Z$
production samples produced with {\sc Epvec}~\cite{epvec}.

The detector response to events produced by the various generator
programs is simulated in detail using a program based on {\sc
  Geant3}~\cite{Geant}.  The simulation includes detailed time
dependent modelling of detector noise conditions, beam optics,
polarisation and inefficient channel maps reflecting actual running
conditions throughout the HERA\,II data taking period.  These simulated
events are then subjected to the same reconstruction, calibration,
alignment and analysis chain as the real data.

\section{Experimental Procedure}
\label{sec:expt}

\subsection{Kinematic Reconstruction}
\label{sec:kine}

Precise reconstruction of the event kinematics is crucial for the
measurement of DIS cross sections. In the NC
channel several different methods are available due to the redundancy
arising from the simultaneous reconstruction of the scattered lepton
and the hadronic final state. In contrast the CC event kinematics can
be reconstructed using only one method based on the measurement of the
hadronic final state since the neutrino escapes the detector
unobserved. Typically the quantities $Q^2$ and $y$ are reconstructed
and $x$ is obtained via the relation $Q^2=sxy$.

In NC interactions the properties of the scattered lepton are
described in terms of its energy $E^{\prime}_e$ and polar scattering
angle $\theta_e$ defined with respect to the proton direction. The
hadronic final state (HFS) is characterised using the quantities
$P_{T,h}=\sqrt{(\sum_i p_{x,i})^2+(\sum_i p_{y,i})^2}$ and
$\Sigma=\sum_i(E_i-p_{z,i})$ where the summation is performed over all
HFS particles $i$ assuming charged particles have the pion mass. Due to
the large momentum of the incident proton in the lab frame compared to
the electron, the HFS particles are often forward going (positive $z$)
and lead to losses in the forward beam-pipe. The quantities $P_{T,h}$
and $\Sigma$ are chosen due to their relative insensitivity to these
losses. The inclusive hadronic polar angle $\gamma_h$, defined by
$\tan(\gamma_h/2)=\Sigma/P_{T,h}$, is used in the calibration
procedure.

In general the scattered lepton quantities are precisely determined
whereas the hadronic quantities have moderate precision due to
particle losses and fluctuations in the hadronic shower. Isolated low
energy calorimeter deposits are classified as noise originating from
electronic sources or back-scattered low energy particles and are
excluded from the HFS.  The HFS is measured using a sophisticated
energy flow algorithm~\cite{hadroo1,hadroo2} which combines tracks
with calorimetric energy measurements in an optimum way avoiding
double counting. For each track (assumed to be a charged pion) the
measured track uncertainties are compared to the expected calorimetric
energy resolution.  The track measurement alone is used to reconstruct
the particle momentum if it has superior resolution except in cases
where an excess of energy in the calorimeter is observed originating
from neutral particles.  This is then taken into account
appropriately. If the expected calorimeter resolution is better, then
either the calorimeter information alone is used to define the hadron
momentum or, similarly to the previous case, a combination in which
the track is used and the calorimeter energy is reduced
appropriately. At high $Q^2$ and high $x$, corresponding to small
$\gamma_h$, the HFS is dominated by one or more jets thus the complete
HFS can be approximated by the sum of jet four-momenta corresponding
to localised calorimetric energy sums above threshold. This technique
allows a further suppression of noise in the hadronic reconstruction
which is important in this kinematic region.

Several reconstruction methods are used in the analysis for
determining the kinematics and for providing
systematic cross checks. The most precise method for $y\gtrsim 0.1$ is the $e$-method
which relies solely on $E^{\prime}_e$ and $\theta_e$ to reconstruct
the kinematic variables $Q^2$ and $y$ as: 
\begin{equation}
  Q^2_{e} = \frac{ (E^{\prime}_e \sin{\theta_e})^2}{ 1-y_e}\,,
 \hspace*{0.5cm} 
  y_e =1-   \frac{ E^{\prime}_e}{E_e} \sin^2\left(\frac{\theta_e}{2}\right) \,.
\label{eq:emethod}
\end{equation}
This method is used in the NC analysis region $y>0.19$.

The resolution of the $e$-method degrades at low $y$ and is also
susceptible to large QED radiative corrections at the highest and
lowest $y$. In the $\Sigma$-method \cite{sigma} $y$ is reconstructed
as \mbox{$\Sigma/(\Sigma+E^{\prime}_e(1-\cos\theta_e))$} and is
therefore less sensitive to QED radiative effects. The
$e\Sigma$-method \cite{esigma} is an optimum combination of the two
and maintains good resolution throughout the kinematic range of the NC
measurement with acceptably small QED radiative corrections. The
kinematics are determined using
\begin{equation}
  Q^2_{e\Sigma} = Q^2_{e} = \frac{ (E^{\prime}_e \sin{\theta_e})^2}{ 1-y_e}\,,
 \hspace*{0.5cm} 
  y_{e\Sigma} = 2 E_e \frac{\Sigma}{[\Sigma+E^{\prime}_e(1-\cos\theta_e) ]^2} \,.
\label{eq:esmethod}
\end{equation}
The $e\Sigma$-method is employed to reconstruct the event kinematics
for $y \leq 0.19$. In this phase space region the HFS is partially
lost in the forward beam-pipe and the influence of noise on the HFS
becomes large. In order to limit this effect the $e\Sigma$-method is
modified such that the summation in the calculation of the quantity
$\Sigma$ is performed only over hadronic jets. A
longitudinally invariant $k_T$ jet algorithm~\cite{kt-jets1,kt-jets2} is used and further
details are given in section~\ref{sec:ncmeas}.

The double angle method (DA-method)~\cite{damethod1,damethod2} provides a
useful technique for calibrating the electromagnetic (EM) and hadronic calorimeters using
$\theta_e$ and $\gamma_h$ as input. Where the HFS
is well contained within the detector ($y\gapprox 0.3$) the DA-method has good
resolution and is to first order independent of the calorimeter energy
scales. The following formulae are used to determine the kinematics:
\begin{equation}
   Q^2_{DA} = \frac{4E_e^2\sin\gamma_h(1+\cos\theta_e)}{ \sin\gamma_h+\sin\theta_e-\sin(\theta_e+\gamma_h)}\,,
  \hspace*{0.5cm} 
   y_{DA} = \frac{\sin\theta_e (1-\cos\gamma_h) }{
   \sin\gamma_h+\sin\theta_e-\sin(\theta_e+\gamma_h)} \,.
\label{eq:damethod}
\end{equation}

Finally for CC interactions the event kinematics may only be
reconstructed by the $h$-method~\cite{jbmethod} which can be
systematically studied using the NC sample.  The $h$-method kinematic
variables are reconstructed using the relations
\begin{equation}
   Q^2_{h} = \frac{P_{T,h}^2}{ 1-y_{h}}\,,
  \hspace*{1cm} 
   y_{h} = \frac{\Sigma}{ 2 \ E_e } \,.
\label{eq:hmethod}
\end{equation}

\subsection{Polar Angle Measurement and Energy Calibration}
\label{sec:calib}

In neutral current interactions the polar angle of the scattered
lepton, $\theta_e$, is determined using the position of its energy
deposit (cluster) in the LAr calorimeter, and the event vertex
reconstructed with tracks from charged particles. The relative
alignment of the calorimeter and tracking chambers is determined using
a sample of events with a well measured lepton track~\cite{tran} in
which the COZ chambers provide an accurate $z$ spatial reconstruction
of the particle trajectory. The lepton track is helically extrapolated
to an octagonal surface with inner radius $r=105\,{\rm cm}$ positioned
axially along the beam line. The shape describes the inner surface of the
LAr calorimeter. The $z$ position of the intersection of the track trajectory and this
surface defines the quantity $z_{imp}$.
The electron cluster barycentre is extrapolated to
the same surface along a straight line from the interaction
vertex. The distance between the extrapolated track and cluster is
then minimised with respect to the six alignment parameters for LAr
(three shifts and three rotations about the coordinate axes) keeping
the CTD position fixed. Four additional parameters are introduced to
allow independent shifts in $z$ for each LAr wheel. The procedure is
performed on data and checked on MC simulation to ensure that no spurious
misalignments from the method appear.  The alignment parameters are
obtained for each of four periods when the detector was moved between
data taking periods. The residual discrepancy in
$\Delta\theta=\theta_{\rm track}-\theta_{\rm clus}$ between data
and simulation determines the systematic uncertainty on the
measurement of $\theta_e$ and is shown in figure~\ref{fig:align}. The
uncertainty is taken to be $1\,\rm{mrad}$.

Cross checks are performed using the alternative $\Delta\phi$
alignment method~\cite{nikiforov} in which the difference in azimuthal
angle between track and cluster is studied as a function of $\phi_{\rm
track}$. Minimisation of this difference constrains three rotations of
the calorimeter about the tracker and two translations in $x$ and
$y$. The $z$ translation is constrained by minimising the distribution
of $\Delta\theta$. This method is found to agree well with the default
alignment procedure.

An {\em in situ} energy calibration of the electromagnetic part of the
LAr calorimeter is performed using the method described
in~\cite{nikiforov} for both data and simulation.  Briefly, a sample
of NC events in which the HFS is well contained in the detector is
used with the DA-method to predict the scattered lepton energy which
is then compared to the measured electromagnetic energy response
allowing local calibration factors to be determined in a finely
segmented grid in $z$ and $\phi$. The events used in the calibration
are required to have $E^{\prime}_e>14\,{\rm GeV}$; $44\,{\rm
  GeV}<E-P_z<66\,{\rm GeV}$ to limit radiative effects, where
$E-P_z=\Sigma+E^{\prime}_e(1-\cos\theta_e)$; $\gamma_h>{\rm
  10}^{\circ}$ to ensure good containment of the HFS and
$y_{\Sigma}<0.3$ for $z_{\rm imp}\leq 20\,{\rm cm}$ or $y_{\Sigma}<0.5$ for
$20\,{\rm cm}<z_{\rm imp}\leq 100\,{\rm cm}$ in order to obtain a good
estimate of $E_{DA}$, the predicted scattered lepton energy from the
DA-method. In each calibration region the calibration factor is taken as the
mean value of $E_{DA}/E_e^{\prime}$. The calibration is applied
octant-wise for each wheel of the LAr calorimeter. In a second step,
the calibration is applied in fine $z_{\rm imp}$ regions which become
coarser with increasing $z_{\rm imp}$ as statistical precision decreases.
The influence of non-Gaussian tails is limited by determining the
calibration factors from events where $E_{DA}/E_e^{\prime}$ deviates
by less than $\pm15\%$ of the mean. The procedure is then iterated
where the window is narrowed to $\pm10\%$. The calibration is
performed for each period of data taking separately.

The electromagnetic calibration performs well except in regions close
to $z$ and $\phi$ cracks in the detector. These local detector regions
are removed from the analysis in order to limit the size of the
corrections.  Figure~\ref{fig:em-calib-eda} shows the residual
mismatch between $E_{DA}$ and $E_e^{\prime}$ after performing the
calibration step. The residual mis-calibrations are within $\simeq
0.3\%$. In the region of large $z$, which corresponds to large $Q^2$
and large $x$, the sample size becomes small and the bin size in $z$
is increased. For the very largest $z$ the data from all run periods
are combined to provide a single calibration factor. The uncorrelated
systematic uncertainty on the electromagnetic energy scale is
estimated from the relative deviation of $E_e^{\prime}/E_{DA}$ between data and simulation and
is found to vary from $0.3\%$ in the central part of the calorimeter
to $1\%$ in the forward region where statistics are limited.

The calibration is validated with independent data samples not used in
the calibration procedure which allow checks of the calibration
linearity at low energy. These are performed using $J/\psi\rightarrow
ee$ decays and QED Compton interactions $ep\rightarrow e\gamma p$ with
$E_e^{\prime}$ of $3-8\,{\rm GeV}$ in which the lepton track momentum $P_{\rm
  track}$ is compared to the measured energy $E_{e}^{\prime}$ of the
cluster as shown in figure~\ref{fig:em-calib-eop}. The simulation on
average describes the data well in this low energy region.
Differences in the material description lead to differences in the
radiative tails of the $P_{\rm track}$ spectrum and are not of direct
relevance in this analysis.

The hadronic response of the detector is calibrated by requiring a
transverse momentum balance between the predicted $P_T$ in the
DA-method ($P_{T,DA}$) and the measured hadronic final state using a
tight selection of well reconstructed events with a single jet where
the $P_{T,DA}$ measurement is reliable to within $0.3\%$ as verified
in the simulation~\cite{kogler}. The tracks require no correction as
validated by the reconstruction of particle decays. The calorimeter calibration
constants are then determined in a minimisation procedure across the
detector acceptance separately for HFS objects inside and outside jets
and for electromagnetic and hadronic contributions to the HFS.
Calorimeter energy deposits are classified as those originating from
electromagnetic interactions and from hadronic interactions with the
help of several neural networks.  The procedure is described in detail
in~\cite{kogler} and for SpaCal in~\cite{shushkevich}.

Detailed studies of the hadronic response of the
calorimeter lead to an improved understanding of the hadronic
energy measurement. The calibration procedure is verified on a
sample of two-jet events, and on a sample in which more hadronic activity
outside of the jet is allowed. Further checks are performed by
requiring longitudinal momentum conservation of $E-P_z$ of the hadronic final
state and the scattered lepton, instead of transverse momentum
conservation. In addition the reference scale may be taken from the
scattered lepton rather than the DA-method prediction.
These studies allow the systematic uncertainty of the hadronic
scale to be reduced with respect to previous measurements
\cite{h1hiq2}. The uncorrelated part of the hadronic scale uncertainty
is reduced to $1\%$ from $1.7\%$ previously. Figure~\ref{fig:hadscale}
demonstrates the quality of the hadronic calibration showing the level
of agreement between data and simulation after the calibration
procedure. In figure~\ref{fig:hadscale-a} the mean transverse momentum
balance between the hadronic final state and the scattered lepton
versus the lepton $P_{T,e}$ is shown for the complete HERA\,II data
set. The simulation provides an accurate modelling of the data
behaviour to within $1\%$ precision. In figure~\ref{fig:hadscale-b} the
quantity $y_h$ is compared to the DA-method prediction,
$y_{DA}$, as a function of the inclusive hadronic angle, $\gamma_h$, for the full
HERA\,II data sample. Since $y_h$ is related to the longitudinal energy
flow (see equation~\ref{eq:hmethod}) this provides an alternative check of
the calibration. The simulation models the data well.

In this analysis it is the relative difference between data and
simulation that is relevant, and good agreement is found to within
$1\%$.  In addition a $0.3\%$ correlated uncertainty is considered and
accounts for a possible bias in the $P_{T}$ reconstruction in the
DA-method reference scale used in the calibration of the electron and
HFS energy. This is determined by varying $\theta_e$ and $\gamma_h$ by
the angular measurement uncertainty.

\subsection{Neutral Current Measurement Procedure}
\label{sec:ncmeas}

Inelastic $ep$ interactions are required to have a well reconstructed
interaction vertex to suppress beam induced background events. High
\qsq neutral current events are selected by requiring each event to
have a compact and isolated cluster in the
electromagnetic part of the LAr calorimeter\footnote{Small local
  detector regions are disregarded in the analysis where the cluster
  of the scattered electron is not fully contained e.g. intermediate
  space between stacks, or where the trigger is not fully efficient.}.
The scattered lepton is identified as the cluster of highest
transverse momentum. In the central detector region,
$\theta\geq30^{\circ}$, the cluster must be associated to a
CTD track.  Forward going leptons with
$\theta<30^{\circ}$ traverse the region between the FTD and CTD where
an increased amount of dead material causes electrons to shower. Since
in this kinematic region the scattered lepton has high energy and the
contribution from photoproduction background is very small, no tracker
information is required to be associated with the lepton for
$\theta<30^{\circ}$.

Energy-momentum conservation requires the variable $E-P_z$ summed over
all final state particles (including the electron) to be approximately
equal to twice the initial electron beam energy.  Restricting $E-P_z$
to be greater than $35$~GeV considerably reduces the photoproduction
background and the radiative processes in which the scattered lepton
or bremsstrahlung photons escape undetected in the lepton beam
direction.

The photoproduction background increases rapidly with decreasing
electron energy, therefore the analysis is separated into two distinct
regions: the {\em nominal} analysis ($y_e \leq 0.63$ for
$Q^2_e\leq890\,{\rm GeV}^2$ and $y_e < 0.93$ for $Q^2_e>890\,{\rm GeV}^2$) for
which the minimum electron energy is $11\,{\rm GeV}$ and the {\em high y}
analysis ($0.63<y_e<0.9$ and $56< Q^2_e<890\,{\rm GeV}^2$) for which
the minimum electron energy is $5\,{\rm GeV}$. 
The techniques employed to
contend with background in each analysis are described below.

\subsubsection{Nominal Analysis}

For the {\em nominal} analysis the small photoproduction contribution
is statistically subtracted using the background simulation. The
overall normalisation of the background simulation is checked using a
sample of data events in which the true scattered lepton is observed
in the electron tagger which, however, has limited acceptance. 

The comparison of the $e^-p$ data and the simulation is shown in
figure~\ref{fig:nc-control-ele} for the scattered lepton energy spectrum
and polar angle, and the distribution of $E-P_z$, which are all used
in the kinematic reconstruction of $x$ and $Q^2$ using the
$e\Sigma$-method. The corresponding distributions for $e^+p$ data and
simulation are shown in figure~\ref{fig:nc-control-pos}.  In the figure
the $R$ and $L$ data are combined and the simulation is normalised to
the luminosity of the data, as is also done for all later performance
figures. All distributions are described well by the simulation aside
from a small difference in normalisation which is discussed in
section~\ref{sec:fitresults} where the data are compared to the NLO QCD
fit.

For the NC analysis in the region $y<0.19$ the noise component has an
increasing influence in the transverse momentum balance
$P_{T,h}/P_{T,e}$ through its effect on $P_{T,h}$. The event
kinematics reconstructed with the $e\Sigma$-method in which the HFS
is formed from hadronic jets only, limits the noise contribution
and substantially improves the $P_{T,h}/P_{T,e}$ description.
The jets are found with the
longitudinally invariant $k_T$ jet algorithm~\cite{kt-jets1,kt-jets2}
as implemented in FastJet~\cite{fastjet1,fastjet2}
with radius parameter $R=1.0$ and are required to have transverse
momenta $P_{T,{\rm jet}}>2\,{\rm GeV}$. In figure~\ref{fig:jet-control-ele}
the quality of the simulation and its description of the $e^-p$ data
for $y_e<0.19$ can be seen for the distributions of the
$P_{T,h}/P_{T,e}$, $\gamma_h$, and $E-P_z$ where all HFS quantities
are obtained using the vector sum of 
jet four-momenta.
Distributions for the $e^+p$ sample are also shown in
figure~\ref{fig:jet-control-pos}. Overall both sets of distributions are
well described in shape by the simulation.

At low $y$, the forward going hadronic final state particles can
undergo interactions with material of the beam pipe.  In some cases
the products of these secondary interactions are incorrectly assigned
as originating from the primary vertex, producing a bias in the
determination of the primary interaction vertex position. Such cases
are recognised and corrected by considering a vertex position
calculated using a stand alone reconstruction of the track associated
with the electron cluster~\cite{nikiforov,shushkevich}.

For the nominal analysis the photoproduction contribution is low, and
this allows the electron candidate track verification in the region
$\theta\geq30^{\circ}$ to be supplemented with an alternative method
which increases efficiency. For NC events with no CTD track associated
to the electron cluster, the track verification is achieved by
searching for hits in the CIP located on the line from the interaction
vertex to the electron cluster.

This optimised treatment of the vertex determination and
verification of the electron cluster with the tracker information
improves the reliability of the vertex position determination and
increases the efficiency of the procedure to $99.5\%$.

\subsubsection{High y Analysis}

In the {\em high y} region the neutral current analysis is extended to
lower energies of the scattered electron, $E_e^{\prime}>5\,{\rm GeV}$.  At
low energies photoproduction background contributions arise due to
$\pi^0\rightarrow \gamma \gamma$ decays and charged hadrons being
mis-identified as electron candidates.  Part of this background is
suppressed by requiring a well measured track linked to the
calorimeter cluster.  The track is furthermore required to have the
same charge as the beam lepton. The remaining background in the
correctly charged sample is estimated from the number of data events
in which the detected lepton has opposite charge to the beam lepton. A
charge asymmetry can arise due to the different detector response to
particles compared to anti-particles, in particular $p$ and 
$\bar{p}$~\cite{Adloff:2000qk,h1fl2010}. By taking into account the charge asymmetry
between negative and positive background, the background estimate is
statistically subtracted from the correctly charged sample. The charge
asymmetry between fake lepton candidates in the $e^+p$ and $e^-p$ data
sets is determined by measuring the ratio of wrongly charged fake
scattered lepton candidates in $e^+p$ and $e^-p$ scattering, taking
into account the difference in luminosity. The asymmetry is found to
be $1.03\pm 0.05$.  This is cross checked using a sample of
photoproduction events in which the scattered electron is detected in
the electron tagger. Further details are given
in~\cite{shushkevich,habib}.

The $e$-method using scattered lepton variables alone has the highest
precision in this region of phase space and is used to reconstruct the
event kinematics.

Figure~\ref{fig:hiy-control-ele} shows the scattered lepton energy
spectrum, the polar angle distribution and the $E-P_z$ spectrum of
the {\em high y} sample for the $e^-p$
data before background subtraction
 and the simulation to which the background, obtained from wrongly charged lepton candidates in the data, is added.
 The corresponding distributions for the $e^+p$
data can be seen in figure~\ref{fig:hiy-control-pos}. The NC simulation
provides a good description of these distributions. The difference
between data and simulation in the $E-P_z$ spectrum is well within the
systematic uncertainty of the hadronic calibration which at high $y$
depends largely on the SpaCal (see table~\ref{tab:syserr}). 

\subsection{Charged Current Measurement Procedure}
\label{sec:ccmeas}

The selection of charged current events requires a large missing
transverse momentum, $P_T^{\rm miss}\equiv P_{T,h}\geq 12\,{\rm GeV}$,
assumed to be carried by an undetected neutrino. The event must also
have a well defined reconstructed vertex. The
kinematic variables $y$ and $Q^2$ are determined using the
$h$-method.  In order to restrict the measurement to a
region with good kinematic resolution the events are required to have
$y_h<0.85$.  In addition the measurement is confined to the region
with sufficiently high trigger efficiency by demanding
$y_h>0.03$. This criterion also restricts the measurement to the
region where any bias in the interaction vertex position due to
forward going hadronic final state particles is limited and well
modelled.

The $ep$ background is dominated by photoproduction and is suppressed
by exploiting the correlation between $P_{T,h}$ and the ratio
$V_{ap}/V_{p}$ of transverse energy flow anti-parallel and parallel to
the hadronic final state transverse momentum
vector~\cite{placakyte,tran}. This variable provides good
discrimination between the CC signal which lies at small
$V_{ap}/V_{p}$ and large $P_{T,h}$, and the background which lies at
large $V_{ap}/V_{p}$ and small $P_{T,h}$. For $P_{T,h}>25\,{\rm GeV}$,
$V_{ap}/V_{p}<0.25$ is required. For smaller transverse momenta the
maximum allowed ratio is reduced as a parabolic function of $P_{T,h}$
such that at $12\,{\rm GeV}$ values down to $V_{ap}/V_{p}=0.10-0.12$ are
accepted depending on the different data sets since the relative
photoproduction contributions differ for the four $R$/$L$ $e^{\pm}p$
samples.  The residual $ep$ background is negligible for most of the
measured kinematic domain, though it reaches $15\%$ at the lowest $Q^2$
and the highest $y$. The simulation is used to estimate this
contribution which is subtracted statistically from the CC data
sample. A systematic uncertainty of $30\%$ is attributed to 
the photoproduction background.  The non-$ep$ background is
rejected as described in~\cite{h1hiq2,placakyte} by removing 
events with topologies typical of cosmic ray and beam-gas interaction
background as well as events with timing inconsistent with the HERA bunch
crossing intervals.

All efficiencies in the CC analysis can be reliably determined from the
{\em pseudo CC} data samples of NC events, free from background
contamination, in which all information associated to the scattered
electron is suppressed.  The production of the samples involves the
identification of the scattered lepton and subsequent deletion of all
calorimetric energies associated to the cluster. All trigger related
energy sums are recalculated after removal of the trigger towers
associated with the electron. Finally all CTD hits in a road around
the electron trajectory are deleted. After removal of this information
the events are passed through the standard H1 software chain to fully
reconstruct the event including all particle trajectories and the
interaction vertex. The {\em pseudo CC} samples are then reweighted to
the CC cross section employing the original $e\Sigma$ kinematic
quantities using H1PDF\,2012. The {\em pseudo CC} samples are
produced for each data taking period to accurately describe running
conditions as closely as possible. A potential bias in the method is
studied by comparing {\em pseudo CC} data with {\em pseudo CC}
simulation using NC {\sc Djangoh} samples which are processed in
the same way. In all cases the {\em pseudo CC} data and {\em pseudo
  CC} MC are found to provide adequate descriptions of each measured
efficiency and also compare well to standard {\sc Djangoh} CC
simulation after the application of additional adjustment factors as
described below. Any remaining discrepancies are accounted for in the
systematic uncertainties as described in section~\ref{sys:sel}.

 The {\em pseudo CC} data are used to give a precise measure of the CC
 trigger efficiency for each data taking period separately. The
 efficiency is measured in each $x,Q^2$ bin and the simulation is
 reweighted to describe the observed behaviour. The efficiency is found
 to be $79\%$ at $Q^2= 300\,{\rm GeV}^2$ and reaches $98\%$ at
 $Q^2\geq3\,000\,{\rm GeV}^2$.

The $e^{+}p$ and $e^{-}p$ distributions for $P^{\rm miss}_{T}$ and
$E-P_z$ are shown for data and simulation in
figure~\ref{fig:cc-control}. The spectra are well described by the
simulation.

\subsection{Cross Section Measurement}
\label{sec:xsec}

For both the NC and CC analyses the selected event samples are
corrected for detector acceptance, efficiencies and migrations using
the simulation and converted to QED corrected cross sections.  The
quality of the simulation, in which all selection efficiency effects
are included, is shown in figures~\ref{fig:align}-\ref{fig:cc-control}
and gives a reliable determination of detector acceptance.  The
accessible kinematic ranges of the measurements depend on the
resolution of the reconstructed kinematics and are determined by
requiring the purity and stability of any measurement bin to be larger
than $30\%$ as determined from signal MC. The purity is defined as the
fraction of events generated and reconstructed in a measurement bin
from the total number of events reconstructed in the bin. The
stability is the ratio of the number of events generated and
reconstructed in a bin to the number of events generated in that
bin. The detector acceptance, $\mathcal{A}$, is obtained from the
ratio of stability divided by purity and corrects the measured signal
event yields for detector effects including resolution smearing and
selection efficiency.

The measured differential cross sections $\sigma(x,Q^2)$ are
then determined using the relation
\begin{eqnarray}
\sigma(x,Q^2) =
\frac{N-B}{\mathcal{L}\cdot\mathcal{A}}\cdot{\mathcal C}
\cdot \left(1+\Delta^{\rm QED}\right) \,,
\end{eqnarray}
where $N$ and $B$ are the selected number of data events and the
estimated number of background events respectively, $\mathcal{L}$ is
the integrated luminosity, 
$\mathcal{C}$ is the bin centre correction, and $(1+\Delta^{\rm QED})$
are the QED radiative corrections.

The bin centre correction ${\mathcal C(x_c,Q^2_c)}$ is a factor
obtained from NLO QCD expectation using H1PDF\,2012,
$\sigma^{th}$, and scales the bin integrated cross section to
a differential cross section at the kinematic point $x_c,Q^2_c$
defined as
\begin{eqnarray}
{\mathcal C}(x_c,Q^2_c) =\frac{\sigma^{th}(x_c,Q^2_c)}
{\iint_{bin}{\small {\rm d}x{\rm d}Q^2}\,\,\,\sigma^{th}(x,Q^2)}\,.
\end{eqnarray}
The single differential cross sections are obtained using the obvious
extensions to the formulae given above.

The DIS cross sections $\sigma$ can be factorised as 
\begin{eqnarray}
\sigma = \sigma^{\rm Born}(1+\Delta^{\rm
  QED})(1+\Delta^{\rm weak})\,,
\end{eqnarray}
where $\sigma^{\rm Born}$ is the Born cross section and $\Delta^{\rm
  QED}$ are the QED radiative
corrections. The measured cross sections presented in
section~\ref{sec:results}, in which the effects of QED radiation have
been corrected for, correspond to the differential cross sections
${\rm d}^2 \sigma/{\rm d}x{\rm d}Q^2$ defined in
equations\,\ref{eq:ncxsec} and \ref{eq:ccxsec}. The QED radiative
corrections are defined in~\cite{h19497,h1hiq2} and are calculated to
first order in $\alpha$ using the program {\sc
  Heracles}~\cite{heracles} as implemented in {\sc
  Djangoh}~\cite{django} and verified with the numerical analysis
programs {\sc Hector}~\cite{hector} and {\sc Eprc}~\cite{eprc}. In
order to retain sensitivity to higher order EW effects in the measured
cross sections, no $\Delta^{\rm weak}$ corrections are applied to the
measurements.

New measurements of the unpolarised cross sections are also performed.
The $L$ and $R$ data sets for $e^+p$ and $e^-p$ scattering are merged
and the cross sections are measured with a resulting small residual
polarisation of $+1.7\%$ in $e^+p$ and $-6.6\%$ in $e^-p$ data. This
remaining polarisation is corrected for using the H1PDF\,2012 fit to
yield HERA\,II cross sections with $P_e=0$.

\subsection{Systematic Uncertainties}
\label{sys:sel}
The uncertainties on the measurement lead to systematic errors on the
cross sections, which can be split into bin-to-bin correlated and
uncorrelated parts.  All the correlated systematic errors are found to
be symmetric to a good approximation and are assumed so in the
following. The total systematic error is formed by adding the
individual errors in quadrature. 

The size of each systematic uncertainty source and its region of
applicability are given in table~\ref{tab:syserr}.  Further details
can be found elsewhere~\cite{tran,shushkevich,placakyte,nikiforov,kogler}.
The influence of the systematic uncertainties on the NC and CC cross
section measurements are given in
tables~\ref{tab:ncdxdq2_eleLH}-\ref{tab:nchighy_pos}, and their origin
and method of estimation are discussed below.

\begin{table}[ht]\footnotesize
  \begin{center}
    \begin{tabular}{ l|c|r }

\hline
\multicolumn{2}{ l }{{\bf Source}} &  {\bf Uncertainty} \\
\hline
\multirow{5}{*}{Electron energy scale} 
&        $z_{\rm imp} \leq -150\,{\rm cm}$    &    $0.5\%$ unc. $\oplus$ $0.3\%$ corr. \\ 
& $-150 < z_{\rm imp} \leq  -60\,{\rm cm}$    &    $0.3\%$ unc. $\oplus$ $0.3\%$ corr. \\ 
& $-60  < z_{\rm imp} \leq  +20\,{\rm cm}$    &    $0.5\%$ unc. $\oplus$ $0.3\%$ corr. \\ 
& $+20  < z_{\rm imp} \leq +110\,{\rm cm}$    &    $0.5\%$ unc. $\oplus$ $0.3\%$ corr. \\ 
&        $z_{\rm imp} >    +110\,{\rm cm}$    &    $1.0\%$ unc. $\oplus$ $0.3\%$ corr. \\ 
\hline
Electron scale linearity &  $E_e^{\prime}<11\,{\rm GeV}$ & $0.5\%$ \\
\hline
\multirow{2}{*}{Hadronic energy scale}
& LAr \& Tracks   & $1.0\%$ unc. $\oplus$ $0.3\%$ corr. \\
& SpaCal & $5.0\%$ unc. $\oplus$ $0.3\%$ corr. \\
\hline
Polar angle
&$\theta_e$  & $1\,{\rm mrad}$ corr.\\
\hline
\multirow{3}{*}{Noise} 
& NC $y<0.19$ & $5\%$ energy not in jets , corr.\\
& NC $y>0.19$ & $20\%$ corr.\\
& CC & $20\%$ corr.\\
\hline
\multirow{5}{*}{NC trigger efficiency} 
& $e^+p$ 2003-2004 & $0.5\%$ \\
& $e^-p$ 2004-2005 & $0.6\%$ \\
& $e^-p$ 2006      & $0.5\%$ \\
& $e^+p$ 2006-2007 & $0.3\%$ \\
& NC {\em high $y$} & $0.5-1.2\%$ \\
\hline
\multirow{2}{*}{Electron track and vertex efficiency}
&$\theta_e>50^{\circ}$  & $0.2\%$ \\
&$\theta_e<50^{\circ}$, NC {\em high $y$}   &    $1.0\%$ \\
\hline
Electron charge determination  &  NC {\em high $y$}  & $0.5\%$ \\
\hline
Electron ID efficiency                       
& $z_{\rm imp}<20\,{\rm cm}\ ( \geq 20\,{\rm cm})$ & $0.2\%\ (1.0\%)$ \\
\hline
\multirow{2}{*}{Background} 
& NC, CC: $\gamma p$     & $30\%$ corr.\\
& CC: NC (others) & $10\%\ (20\%)$ corr.\\
\hline
Background $\gamma p$ charge asymmetry & NC {\em high $y$}  &  $1.03 \pm 0.05$ corr.\\

\hline
CC trigger efficiency ($\epsilon$) & & $15\%\cdot(1-\epsilon) \oplus 0.2\%$ \\
\hline
CC vertex efficiency
& $y\geq 0.15\ (<0.15)$ & $1.5\%\ (3.0\%)$ \\
\hline
CC background finder efficiency & $y\geq 0.08\ (<0.08)$ & $1.0\%\ (2.0\%)$ \\
\hline
CC $V_{ap}/V_p$ bg suppression cut  & & $\pm 0.02$ corr.\\

\hline
\multirow{3}{*}{QED radiative corrections  }
& NC $x<0.1 \, , \, 0.1 \leq x<0.3 \, , \, x\geq0.3$      & $0.3/1.0/2.0\%$ \\
& NC {\em high y}            & $1.0\%$ \\
& CC kinematics dependent    & $\sim 1.0-2.0\%$ \\
\hline
\multirow{3}{*}{Acceptance corrections}
& NC $e^{\pm}p$: $Q^2 \leq 5000 \   (> 5000)\,{\rm GeV}^2$ & $0.2\% \ (1.0\%)$ \\
& CC $e^-p$:     $Q^2 \leq 5000 \   (> 5000)\,{\rm GeV}^2$ & $0.2\% \ (1.0\%)$ \\
& CC $e^+p$:     $Q^2 \leq 5000  \  (> 5000)\,{\rm GeV}^2$ & $0.5\% \ (3.0\%)$ \\

\hline
Polarisation
& LPOL (TPOL)  & $2.0\% \ (1.9\%$) \\
\hline
Luminosity     &    &    $2.3\%$ corr. $\oplus$ $1.5\%$ unc.\\

\hline
\end{tabular} 
\caption{ Table of applied systematic uncertainties and regions of
  applicability. Uncertainties which are considered point-to-point
  correlated are labelled corr. and correspond to the sources listed
  in table~\ref{tab:correlation}. All other sources are considered
  uncorrelated. The effect of these uncertainties on the cross section
  measurements are given in the tables of
  section~\ref{sec:results}.   }
\label{tab:syserr}
\end{center}
\end{table}

\begin{description}
\item[Electron Energy Uncertainty:] 
The calibration of the
electromagnetic part of the LAr calorimeter is performed using a
subset of NC data. Uncertainties arise from the particular choice of
calibration samples, the linearity correction uncertainty, and at very
high $Q^2$ from the limited statistics due to the sharply decreasing
NC cross section. The uncertainty varies for each wheel of the LAr
calorimeter and values are listed in table~\ref{tab:syserr}.
The correlated part of the uncertainty comes
from the residual global shift between data and MC in the kinematic
peak of the $E_{e}^{\prime}$ distribution (shown in
figure~\ref{fig:nc-control}) after the calibration
procedure and is determined to be $0.3\%$. It
results in a correlated systematic error on the NC cross section which
is up to $2\%$ at low $y$ and $Q^2 \lesssim 1\,000\,{\rm GeV}^2$,
increasing to $\sim 4\%$ at larger $Q^2$.

\item[Hadronic Calibration Uncertainty:] 
An uncorrelated uncertainty of
$1\%$ is obtained for the hadronic energy measurement. The uncertainty is
determined by quantifying the agreement between data and simulation in
the mean of the $P_{T,h}/P_{T,DA}$ distribution in each $x,Q^2$
measurement bin by means of pull distributions. The pull is defined as
the difference in the mean values divided by the uncertainty which
includes the uncorrelated hadronic calibration component. An
uncertainty of $1\%$ yields a pull distribution which is Gaussian
distributed, centred on zero, and with unit standard deviation. A
$0.3\%$ correlated component to the uncertainty is considered which
originates from the calibration method due to the uncertainty of the
reference scale~\cite{kogler}.  The resulting correlated
systematic error is typically below $1\%$ for NC and CC cross
sections, and the uncorrelated component is up to $5\%$ at large $x$.

\item[Polar Angle Uncertainty:] 
A correlated $1\,{\rm mrad}$ uncertainty on the
determination of the electron polar angle is considered, accounting
for any remaining discrepancy in the measurements of $\theta_{\rm
clus}$ and $\theta_{\rm track}$ after the alignment procedure
described in section~\ref{sec:calib}. The uncertainty accommodates differences
between periods due to changes in the tracking capability of the
detector for example when the FST or the CST were not fully
operational during data taking. This leads to a typical uncertainty on
the NC reduced cross section of less than $1\%$, increasing at high
$x$. 
 
\item[Noise Subtraction Uncertainty:] 
Energy classified as noise in the
LAr calorimeter is excluded from the HFS. For $y<0.19$ in the NC analysis
the calorimetric energy not contained within
hadronic jets is classified as noise. The uncertainty on the subtracted noise is estimated to
be $5\%$ of the noise contribution. This variation encompasses all observed
differences between data and simulation in the ratio of
$P_{T,h}/P_{T,e}$ in each $x,Q^2$ measurement bin in this kinematic
region. Changing the jet algorithm to the SIScone
method~\cite{siscone}, varying the minimum jet transverse momentum
$p_{T,jet}$ by $50\%$, and varying the jet radius parameter by $25\%$
all have a minor effect on the resulting distribution of
$P_{T,h}/P_{T,e}$ and deviations are contained within the $5\%$
scaling of hadronic energy not associated with a jet. For the CC
analysis and for $y>0.19$ in the NC
analysis the noise contribution is
defined as the sum of isolated low energy calorimetric depositions.
In this kinematic region the contribution is small and a conservative
uncertainty of $20\%$ of the noise is assigned to accommodate
differences between data and simulation. As a cross check this same
noise definition is used for $y<0.19$ and it is verified that a $20\%$
noise variation also encompasses any difference between data and
simulation. The check results in larger event migrations outside the
measured region compared to
using the default jet method in this kinematic region.  This source of
systematic uncertainty gives rise to a correlated systematic error at
low $y$ comparable to or smaller than the correlated electron energy
scale uncertainty except for $x\geq 0.4$ and $Q^2 \lesssim
5\,000\,{\rm GeV}^2$ in the NC measurements where the uncertainty
due to noise rises to $5-10\%$.
 
\item[NC Trigger Efficiency Uncertainty:] 
The uncertainty on the
trigger efficiency in the NC {\em nominal} analysis is determined
separately for each data taking period to account for changing trigger
and beam conditions (using several dedicated monitor triggers). Three
trigger requirements are employed: the global timing, the event
timing and the calorimeter energy. The efficiency of global timing
criteria to suppress out of time beam related background was
continuously monitored with high precision and found to be $99\%$
initially rising to $100\%$ at the end of the HERA\,II running
period. The efficiency of the calorimeter energy trigger requirements
is determined in a fine grid in $z_{\rm imp}$ and $\phi_e$, the
azimuthal angle of the scattered lepton. Time dependent local regions
of the calorimeter with efficiencies of less than $99\%$ are rejected
in the analysis ~\cite{nikiforov}. Finally the event timing trigger
requirements were also continuously monitored in the data. After
rejection of local inefficient regions the overall trigger efficiency
is close to $100\%$ and the uncertainty is listed in
table~\ref{tab:syserr}.

\item[NC {\em high y} Trigger Efficiency Uncertainty:] 
The trigger
efficiency estimate for the {\em high y} analysis involves several
strategies due to the complex nature of the trigger designed to record
low energy electrons in a high hadronic background environment. All
efficiencies are measured individually in each data taking
period. The efficiency of the electron calorimeter energy requirement is
estimated using events triggered by the HFS in the SpaCal. This method
allows the trigger energy threshold to be accurately measured and any
potential bias is controlled by comparing this efficiency with that
determined from events triggered by the HFS in the LAr.  The same
global timing and event timing conditions as mentioned above are used
in the {\em high y} triggers. Finally the triggers place conditions on
the charged particle multiplicity. The efficiency of these track
requirements is measured with a variety of independent control samples
and the uncertainty is determined from the level of agreement between
them~\cite{shushkevich}. Taking all contributions to the trigger
condition into account leads to an error of between $0.5$ to $1.2\%$ on
the high $y$ cross sections.

\item[Electron Track-Vertex Efficiency Uncertainty:] 
In the NC analysis the
efficiency for reconstructing a track associated to the scattered
lepton and for reconstructing the interaction vertex are determined
simultaneously. The efficiency measurement is performed using a
control selection of clean NC events with $45<E-P_z<65\,{\rm GeV}$ and
additional topological algorithms are employed to remove non-$ep$ and
QED Compton backgrounds. Three algorithms are used to determine the
interaction vertex and the MC and data are compared in the efficiency
for each contributing algorithm. The kinematic dependence for the
efficiency of each algorithm is well reproduced by the simulation
after the application of a single small adjustment to the MC of
$-0.3\%$. The combined efficiency in the NC {\em nominal} analysis is
found to be $99.5\%$ in the data~\cite{shushkevich}. The
residual differences between data and simulation define the
uncorrelated systematic uncertainty which is $0.2\%$ for
$\theta_e>50^{\circ}$ and $1\%$ for $\theta_e\leq50^{\circ}$ and is
considered to be uncorrelated since a number of different vertex
reconstruction algorithms are employed. In the {\em high y} analysis
the efficiency is checked using a sample of QED Compton events which
is limited in statistical precision and a $1\%$ uncertainty is 
adopted~\cite{shushkevich}.

\item[Electron Charge Identification Efficiency Uncertainty:] 
In the NC
{\em high y} analysis the efficiency for correct charge identification
of the scattered lepton is measured in the region 
$15 <E_e^{\prime}<25\,{\rm GeV}$.  The simulation after correction
by $-0.4\%$ describes the efficiency of the data and no time
dependence is observed. Therefore all periods are combined to reduce
statistical fluctuations and a final uncertainty of $2\times0.25\%$ is
used.  The factor of two accounts for the fact that charge
mis-identification has a dual influence on the measurement by causing
a loss of signal events and also an increase of the subtracted
background~\cite{shushkevich}.

\item[Electron Identification Uncertainty:] 
A calorimetric algorithm
based on longitudinal and transverse shower shape quantities is used to identify electrons in
the NC analysis. The efficiency of this selection can be estimated
using a simple track based electron finder which searches for an
isolated high $p_T$ track associated to an electromagnetic energy
deposition. The efficiency is measured in the {\em high y} and {\em
nominal} analysis regions and is found to be well described by the
simulation and shows little time variation. Thus the complete HERA\,II
data sample is combined to estimate the efficiency at large $z_{\rm imp}$
in order to reduce statistical fluctuations. The efficiency is found
to be $98.6\%$ overall and is described by the MC to within $0.2\%$
for $z_{\rm imp}<20\,{\rm cm}$ and $1.0\%$ for 
$z_{\rm imp}>20\,{\rm cm}$~\cite{shushkevich}.

\item[Background Subtraction Uncertainty:] 
In the CC and the NC {\em
  nominal} analysis the photoproduction background is estimated from
simulation and a $30\%$ uncertainty on the subtracted photoproduction
background is assigned. A background dominated control sample is used
to determine the size of this uncertainty. For both analyses the
control samples consist of events passing the complete CC or {\em
  nominal} NC analysis selections in addition to the requirement of an
energetic electron observed upstream in the electron tagger. Such
events arise from photoproduction interactions in which the HFS
observed in the main detector gives rise to a fake electron signature
or fake missing transverse momentum in the NC and CC analyses
respectively. Due to the restricted acceptance of the electron tagger the
background samples have limited statistical precision but simulation
provides a reasonable description of the data within the estimated
uncertainty. For the CC analysis the region of $V_{ap}/V_p>0.3$ is
exclusively populated by photoproduction background. The data
distribution is well described in shape and normalisation within the
quoted uncertainty. In the {\em high y} analysis, simulation is not
used to estimate the background contribution, nevertheless the tagged
sample electron energy spectrum is well described in shape by the
simulation within an overall normalisation uncertainty of $30\%$. This
source results in a correlated systematic error of typically $\lesssim
1\%$ for the NC {\em nominal} analysis and CC cross sections. 

In the CC analysis a background contribution of NC events arises in
which the scattered lepton is poorly measured in the detector close to
uninstrumented regions. Such contributions are suppressed by a set of
topological algorithms to identify such events. The algorithms
search for single isolated CTD tracks not recognised as from the
scattered electron and opposite to the HFS~\cite{placakyte}, or
search for low energy electromagnetic clusters associated to a CTD
track with low calorimetric activity around the cluster and well
separated from the remaining HFS~\cite{tran}. These background
contributions are well simulated and subtracted using simulation. A
$10\%$ uncertainty on the amount subtracted is considered.
  
\item[NC {\em high y} Background Subtraction Uncertainty:] 
In the NC
{\em high y} analysis the photoproduction background is estimated from
wrongly charged fake lepton candidates directly from the data as
described in section~\ref{sec:ncmeas}. The asymmetry in the charge of
fake leptons is measured to be $N^-/N^+=1.03\pm0.05$ where $N^-$ is
the number of fake $e^-$ candidates and $N^+$ is the number of fake
$e^+$ candidates. The ratio is measured independently for $L$ and $R$
data samples and also for two different energy regions of the fake
lepton. All results are consistent within the large statistical
uncertainties and are combined. The resulting uncertainty on the
measured cross sections is found to be less than $1\%$~\cite{shushkevich}.

\item[CC Trigger Efficiency Uncertainty:] 
For the CC analysis the
efficiency is determined using the {\em pseudo CC} data samples and
the CC simulation is corrected in each $x$, $Q^2$ measurement bin. In
an alternative approach the efficiency is measured and parameterised
as a function of $P_{T}^{\rm miss}$ and $\gamma_h$ determined using
calorimetric information only to mimic the trigger algorithms; the
resulting differences agree within $15\%$ of the inefficiency which is
taken as the systematic uncertainty. Whilst the {\em pseudo CC} sample
benefits from the excellent kinematic resolution from the suppressed
electron and high statistics from the large NC cross section, at high
$Q^2$ approaching the EW scale (where the efficiency is close to
$100\%$) the NC and CC cross sections are of similar magnitude and
therefore the {\em pseudo CC} sample has similar statistical
uncertainty as the CC sample itself. Thus for $P_{T}^{\rm
  miss}>25\,{\rm GeV}$ the complete HERA\,II data are combined to estimate the
efficiency in this region after first checking for consistency between
the periods. An uncertainty of $0.2\%$ is included to accommodate any
remaining influence from the lack of statistical precision.

\item[CC Vertex Efficiency Uncertainty:] 
The CC vertex finding
efficiency is estimated using the {\em pseudo CC} data and MC samples
and is compared to the efficiency estimate from the CC simulation
within the range $-35<z_{\rm vtx}<+35\,{\rm cm}$ where
$z_{\rm vtx}$ is the $z$ position of the reconstructed event vertex.
The comparison is performed for each data taking
period to account for changing detector configurations. Small
adjustment factors are applied to the CC simulation so that the
efficiency agrees with the {\em pseudo CC} data samples. Residual
differences between the simulation and the {\em pseudo CC} samples are
used to determine the size of this uncertainty which is $1.5\%$ for
$y_h\geq0.15$ and $3\%$ for $y_h<0.15$.
  
\item[CC Background Finder Uncertainty:] 
The use of topological and
timing algorithms in the CC analysis to suppress non-$ep$ interactions
can lead to a signal inefficiency.  The {\em pseudo CC} data yield a
lower efficiency than the simulation by $\approx 2-3\%$~\cite{tran}. 
The simulation is therefore weighted to provide a
better description. After the adjustment all samples agree to within
$1\%$ for $y_h\geq0.08$ and $2\%$ for $y_h<0.08$.

\item[CC $\pmb{V}_{\pmb{ap}}\pmb{/}\pmb{V}_{\pmb{p}}$ Background Rejection Uncertainty:]
The correlated error due to the uncertainty of the efficiency of the
anti-photoproduction cut in the CC analysis is estimated by varying
the quantity $V_{ap}/V_p$ by $\pm 0.02$ in the simulation. The size of
the variation is determined by comparing different methods to
calculate $V_{ap}/V_p$, e.g.\ using individual calorimeter cells or
hadronic clusters, as well as using Monte Carlo samples with different
hadronisation models (CDM versus MEPS) which affect the shape of the
$V_{ap}/V_p$ distribution~\cite{tran}.  This leads to a maximum error
at low $P_{T,h}$ of up to $6\%$ in the single differential cross
section ${\rm d}\sigma_{CC}/{\rm d}x{\rm d}Q^2$.  This is the dominant
contribution to the correlated uncertainty of the CC double
differential cross section for $x\leq0.032$.

\item[QED Radiative Corrections Uncertainty:] 
An error on the NC and CC
cross sections originating from the QED radiative corrections is taken
into account. This is determined by comparing the predicted radiative
corrections from the programs {\sc Heracles} (as implemented in {\sc
Djangoh}), {\sc Hector}, and {\sc Eprc}~\cite{eprc}.  The radiative
corrections due to the exchange of two or more photons between the
lepton and the quark lines, which are not included in {\sc Djangoh},
vary with the polarity of the lepton beam.  This variation (estimated
using {\sc Eprc}) is expected to be small compared to the quoted
errors and is neglected~\cite{shushkevich}.

\item[Acceptance Correction Uncertainty:] 
The MC simulation is used to
determine acceptance corrections to the data and relies on a specific
choice of PDF. Changing the PDF used influences the acceptance which
for NC $e^{\pm}p$ and CC $e^-p$ changes by $0.2\%$ for
$Q^2<5\,000\,{\rm GeV}^2$, and by $1.0\%$ for $Q^2>5\,000\,{\rm
  GeV}^2$.  For CC $e^+p$ the changes are $0.5\%$ for
$Q^2<5\,000\,{\rm GeV}^2$ and $3.0\%$ for $Q^2>5\,000\,{\rm GeV}^2$.
  
\item[Polarisation Uncertainty:]
The independent polarisation
measurements from the TPOL and LPOL have a relative
uncertainty $\delta P/P=1.9\%$ and $2.0\%$ respectively. In general
the polarimeter measurements agree to within $1.5\%$ although
variations with time are observed and folded into the final quoted
polarisation uncertainties on the measured cross sections presented
here. In three run periods the disagreement rises to $5-10\%$
affecting approximately $16\,{\rm pb}^{-1}$ of the luminosity. In these
periods the uncertainty is enlarged~\cite{pubpola}.

\item[Luminosity Uncertainty:] 
The luminosity is measured using elastic
QED Compton events~\cite{compton-lumi} with an uncertainty of $2.3\%$,
of which $1.1\%$ is from the uncertainty in the theoretical calculation
of the elastic QED Compton process.  In addition, there is a $1.5\%$
normalisation uncertainty on each data taking period, originating from
time-dependent corrections when extrapolating the QED Compton analysis
to smaller datasets.
\end{description}

In summary the typical total systematic error is substantially
reduced compared to previous H1
publications~\cite{h19497,h19899,h1hiq2,h1ccpol} to about $1.5\%$ for
the NC double differential cross sections, and $4\%$ for the CC double
differential cross sections. For the ${\rm d} \sigma_{\rm NC(CC)} /
{\rm d} Q^2$ measurements, the error is typically $1.3\%$ (NC) and
$3\%$ (CC). This reduction is achieved through a better
understanding of the response of the detector and in particular for
the hadronic and EM calibrations, as well as the reconstruction of
polar angles. Further detailed studies also allow reductions in
the systematic uncertainties of the electron identification and the
trigger efficiency.

\section{QCD Analysis}
\label{sec:qcdana}

To assess the impact of the H1 NC and CC cross sections at high $Q^2$ 
measured with the longitudinally polarised lepton beams on the determination 
of PDFs, a new QCD analysis (H1PDF\,2012) 
is performed. 
In addition to the new HERA\,II data presented here,
the previously published unpolarised HERA\,I data at high 
$Q^2$~\cite{h19497,h19899,h1hiq2} and at low $Q^2$ \cite{h1fl2010}, 
as well as the H1 measurements at lower proton beam 
energies~\cite{h1fl2010} are used, as shown in 
table~\ref{tab:dataset}.
This analysis supersedes the previous H1PDF\,2009 fit \cite{h1pdf2009}.
\renewcommand{\arraystretch}{1.22} 
\begin{table}[htb]
 \scriptsize
 \begin{center}
 \begin{tabular}{l|l|l|r|r|c|r|c}
 \hline
 Data set          &  $x_{\rm min}$         & $x_{\rm max}$ & $Q^2_{\rm min}$ &$Q^2_{\rm max}$ & 
$\delta \mathcal{L}$ &    Ref.            & Comment   \\ 
                     &                    &           &  ($\rm{GeV}^2$)  & ($\rm{GeV}^2$)  &
  $(\%)$             &                    &           \\
 \hline
$e^+$ Combined low $Q^2$ &  $0.00004$ &  $0.20$        &  $0.5$   & $150$      
              &  $0.5$     &~\cite{h1fl2010} &  $\sqrt{s}= 301, 319 ~\rm{GeV}$\\
\hline
$e^+$ Combined low $E_p$ &  $0.00003$ &  $0.003$        &  $1.5$   & $90$      
              &  $0.5$     &~\cite{h1fl2010} &  $\sqrt{s}= 225, 252 $~$\rm{GeV}$\\
\hline
$e^+$ NC  $94$-$97$ &  $0.0032 $ &  $0.65$        &  $150$   & $30\,000$
              &  \multirow{2}{*}{$0.5\oplus 1.4$}     & \multirow{2}{*}{\cite{h19497}}  &  \multirow{2}{*}{$\sqrt{s}=301$~$\rm{GeV}$}\\
$e^+$ CC  $94$-$97$ &  $0.013  $ &  $0.40$        &  $300$   & $15\,000$  
              &       &  & \\

$e^-$ NC  $98$-$99$ &  $0.0032$  &  $0.65$        &  $150$   & $30\,000$ 
              &  \multirow{3}{*}{$0.5\oplus 1.7$}     & \multirow{2}{*}{\cite{h19899}}  &  \multirow{2}{*}{ $\sqrt{s}=319$~$\rm{GeV}$}\\
$e^-$ CC  $98$-$99$ &  $0.013  $ &  $0.40$        &  $300$   & $15\,000$
              &       &  &  \\
$e^-$ NC  $98$-$99$  {\em high $y$}&  $0.00131$ &  $0.0105$      &  $100$   & $800$ 
              &       & \multirow{3}{*}{\cite{h1hiq2}}      &  $\sqrt{s}= 319$~$\rm{GeV}$\\
$e^-$ NC  $99$-$00$ &  $0.0032$  &  $0.65$        &  $150$   & $30\,000$
              &  \multirow{2}{*}{$0.5\oplus 1.4$}     &     & $\sqrt{s}= 319$~$\rm{GeV}$; incl. {\em high $y$}\\
$e^+$ CC  $99$-$00$ &  $0.013  $ &  $0.40$        &  $300$   & $15\,000$ 
              &     &      &  $\sqrt{s}= 319$~$\rm{GeV}$\\
\hline
$e^+$ NC {\em high $y$} &  $0.0008$  &  $0.0105$        &  $60$   & $800$
              &  $2.3\oplus 1.0 \oplus 1.1$     &  &   $\sqrt{s}= 319$~$\rm{GeV}$\\
$e^-$ NC {\em high $y$} &  $0.0008$  &  $0.0105$        &  $60$   & $800$
              &  $2.3\oplus 1.2\oplus 0.8$     &       &   $\sqrt{s}= 319$~$\rm{GeV}$\\
$e^+$ NC $L$ &  $0.002 $ &  $0.65$        &  $120$   & $30\,000$
              &  \multirow{2}{*}{$2.3\oplus 1.5$}     &     &  \multirow{4}{*}{$\sqrt{s}=319$~$\rm{GeV}$}\\
$e^+$ CC $L$ &  $0.008$ &  $0.40$        &  $300$   & $15\,000$
              &       &   &  \\
$e^+$ NC $R$ &  $0.002  $ &  $0.65$        &  $120$   & $30\,000$  
              &  \multirow{2}{*}{$2.3\oplus 1.5$}     &   & \\
$e^+$ CC $R$ &  $0.008$ &  $0.40$        &  $300$   & $15\,000$ 
              &       &       &  \\
$e^-$ NC $L$ &  $0.002$  &  $0.65$        &  $120$   & $50\,000$ 
              &  \multirow{2}{*}{$2.3\oplus 1.5$}     &  &  \multirow{4}{*}{$\sqrt{s}=319$~$\rm{GeV}$}\\
$e^-$ CC $L$ &  $0.008$ &  $0.40$        &  $300$   & $30\,000$
              &      &   &  \\
$e^-$ NC $R$ &  $0.002$  &  $0.65$        &  $120$   & $30\,000$ 
              &  \multirow{2}{*}{$2.3\oplus 1.5$}     &   &  \\
$e^-$ CC $R$ &  $0.008$ &  $0.40$      &  $300$   & $15\,000$ 
              &      &       &  \\
 \hline
 \end{tabular}
 \end{center}
 \caption[RESULT] { \label{tab:dataset} Table of data sets used in
 the QCD fit.  The normalisation uncertainties of each data set
 ($\delta \mathcal{L}$) are given as well as the kinematic ranges in
 $x$ and $Q^2$. When there are two uncertainties shown, the first one
 corresponds to the correlated error across the data sets and the
 second one is the uncertainty of the relevant data sets.
 The second and third uncertainties of the NC {\em high $y$} analyses
 represent the corresponding uncertainties of the $L$ and $R$ data sets,
 respectively.}
 \end{table}
\renewcommand{\arraystretch}{1.00} 

\subsection{Analysis Framework and Settings}
\label{sec:ansatz}

The present QCD analysis uses the {\sc HERAFitter} framework~\cite{HERAPDF10, 
h1pdf2009}, an open source software package
based on the QCD evolution code {\sc QCDNUM} (v17.04)~\cite{qcdnum}.

The fit strategy follows closely the one adopted for the determination of 
the HERAPDF1.0 sets~\cite{HERAPDF10}. 
The QCD predictions for the differential cross sections
are obtained by solving the DGLAP evolution
equations~\cite{gb72a,gb72b,l75,d77,ap77} at NLO in the $\overline{\rm
MS}$ scheme with the renormalisation and factorisation scales chosen
to be $Q$.  
The heavy quark coefficient
functions are calculated in the RT general-mass variable-flavour-number
scheme~\cite{rt97}. 
The result is cross checked against the ACOT scheme variant~\cite{acot}
that takes full account of quark masses.
The heavy quark masses for charm, $m_c=1.4\,{\rm GeV}$ and beauty, $m_b=4.75\,{\rm GeV}$ are
chosen following~\cite{rt10}.
The strong coupling constant is
fixed to $\alpha_s(M^2_Z)=0.1176$~\cite{pdg2008}, as used for the HERAPDF1.0 NLO
sets.

The $\chi^2$ function which is minimised using the MINUIT 
package~\cite{MINUIT} is defined similarly to \cite{HERAPDF10} as 
\begin{equation}
\chi^2=\sum_i\frac{\left[ \mu_i - m_i\left(1-\sum_j\gamma_j^i b_j\right) \right]^2}{\delta_{i,{\rm unc}}^2 m_i^2+\delta_{i,{\rm stat}}^2\mu_i m_i\left(1-\sum_j\gamma_j^i b_j\right)} +\sum_j b_j^2 + \sum_i \ln \frac{\delta_{i,{\rm unc}}^2 m_i^2+\delta_{i,{\rm stat}}^2\mu_i m_i}{\delta_{i,{\rm unc}}^2 \mu_i^2+\delta_{i,{\rm stat}}^2\mu_i^2},
\label{eq:chi2}
\end{equation}
where $m_i$ is the theoretical prediction and $\mu_i$ is the measured cross 
section at point $i$, $(Q^2,x,s)$ with the relative statistical and 
uncorrelated systematic uncertainty $\delta_{i,{\rm stat}}$, 
$\delta_{i,{\rm unc}}$, respectively. 
The above $\chi^2$ definition takes into account that the quoted 
uncertainties are based on measured cross sections, which are subject to 
statistical fluctuations. Therefore one needs to correct for possible biases 
by using the expected instead of the observed number of events with the
corresponding errors scaled accordingly.
The correlations between data points caused by systematic uncertainties are 
also taken into account in the fit via the $\chi^2$ definition, with 
$\gamma_j^i$ denoting the relative correlated systematic uncertainties and 
$b_j$ their shifts with a penalty term $\sum_j\!b^2_j$ added.
A $\ln$ term is introduced in addition which 
arises from the likelihood transition to $\chi^2$ when the scaling 
of the errors is applied.

The systematic uncertainties for the polarised measurements of the high
$Q^2$ HERA\,II NC {\it nominal} and {\it high y} and CC cross sections are 
described in detail in section~\ref{sys:sel}. 
The correlations among the uncertainty sources
across the data sets are summarised in table~\ref{tab:correlation}. 
The new measurements reported here have a common normalisation uncertainty of 
$2.3\%$ originating from the luminosity measurement based on the QED Compton 
analysis ($\delta^{{\cal L}5}$ in table~\ref{tab:correlation}).
Each data set has an additional uncorrelated normalisation uncertainty of 
$1.5\%$ ($\delta^{{\cal L}6}-\delta^{{\cal L}9}$ in table~\ref{tab:correlation}).
The uncertainty is correlated for all measurement points within 
the data set.
The uncorrelated normalisation uncertainty for the unpolarised HERA\,II NC 
{\it high y} data is a luminosity weighted average of the left and right handed polarised
periods. 
The combined low $Q^2$ data set has $47$ sources of uncertainty which are 
assumed to be uncorrelated with those of the high $Q^2$ data sets and are not 
listed in table~\ref{tab:correlation} but are described in~\cite{h1fl2010}. 
The only exception is the common normalisation uncertainty of $0.5\%$ arising 
from the theoretical uncertainty in the Bethe-Heitler cross section.
 This is considered to be correlated with all HERA\,I data sets 
($\delta^{{\cal L}1}$ in table~\ref{tab:correlation}). 
The combined data with low proton beam energies has nine sources of correlated 
systematic uncertainty that are treated independently from all other sources 
except for $\delta^{{\cal L}1}$.

For the polarised HERA\,II data there is an additional source of
uncertainty arising from the polarisation measurement as described in
section~\ref{sys:sel}.
This affects the construction of the theoretical differential cross sections 
and it is accounted for in the QCD fit procedure by allowing the polarisation 
to vary within its uncertainties as follows:
\begin{equation}
P_e^i =  P_e^i\cdot (1\pm \delta^{Pi})\hspace{2mm} {\rm with} \hspace{2mm} \delta^{Pi}=\delta_{\rm unc}^i \cdot b_{\rm unc}^i\oplus\gamma_{\rm TPOL}^i \cdot b_{\rm TPOL}\oplus\gamma_{\rm LPOL}^i \cdot b_{\rm LPOL},
\label{eq:polar}
\end{equation} 
with index $i$ representing the four different data running periods ($\delta^{P1}-\delta^{P4}$ 
in table~\ref{tab:correlation}). 
The values for $\delta_{\rm unc}$,  
$\gamma_{\rm TPOL}$, and $\gamma_{\rm LPOL}$ are listed in 
table~\ref{tab:syserr_voica}.
They correspond to the 
uncorrelated uncertainties and to the 
two uncertainties for the polarisation 
determination method (LPOL, TPOL) which are correlated 
across different data sets. 
Note that the uncorrelated uncertainties
$\delta_{\rm unc}$ are still correlated for 
measurements within a data set. 
The free parameters
$b^i_{\rm unc}$, $b_{\rm TPOL}$ and $b_{\rm LPOL}$ 
are free parameters of the QCD fit.

\renewcommand{\arraystretch}{1.22} 
\begin{table}[htp]
\begin{center}
\begin{tabular}{l|cccccccccc}
\hline
 Data set & \multicolumn{2}{c}{$\delta^{\cal L}$} &   $\delta^E$ & $\delta^{\theta}$ &  $\delta^h$
 &$\delta^N$&
$\delta^B$& $\delta^V$& $\delta^S$ & $\delta^{\rm pol}$ \\ \hline 
$e^+$ Combined low $Q^2$ & $\delta^{{\cal L}1} $ &&&&&&&&& \\ \hline 
$e^+$ Combined low $E_p$ & $\delta^{{\cal L}1} $ &&&&&&&&& \\ \hline
$e^+$ NC $94$-$97$ & $\delta^{{\cal L}1}$ & $\delta^{{\cal L}2}$ & $\delta^{E1}$ &$\delta^{\theta1}$ & $\delta^{h1}$ & $\delta^{N1}$ & $\delta^{B1}$ & $-$ & $-$ & $-$ \\
$e^+$ CC $94$-$97$ & $\delta^{{\cal L}1}$ & $\delta^{{\cal L}2}$ & $-$ &$-$ & $\delta^{h1}$ & $\delta^{N1}$ & $\delta^{B1}$ & $\delta^{V1}$ & $-$ & $-$\\
$e^-$ NC $98$-$99$ & $\delta^{{\cal L}1}$ & $\delta^{{\cal L}3}$ & $\delta^{E1}$ &$\delta^{\theta2}$ & $\delta^{h1}$ & $\delta^{N1}$ & $\delta^{B1}$ & $-$ & $-$ & $-$\\
$e^-$ NC $98$-$99$ {\em high $y$} &  $\delta^{{\cal L}1}$ & $\delta^{{\cal L}3}$ & $\delta^{E1}$  & $\delta^{\theta 2}$ & $\delta^{h1}$ & $\delta^{N1}$ & $-$ & $-$ & $\delta^{S1}$ &$-$\\
$e^-$ CC $98$-$99$ & $\delta^{{\cal L}1}$ & $\delta^{{\cal L}3}$ & $-$ &$-$ & $\delta^{h1}$ & $\delta^{N1}$ & $\delta^{B1}$ & $\delta^{V2}$ & $-$ & $-$\\
$e^+$ NC $99$-$00$ &  $\delta^{{\cal L}1}$ & $\delta^{{\cal L}4}$ & $\delta^{E1}$  & $\delta^{\theta 2}$ & $\delta^{h1}$ & $\delta^{N1}$ & $\delta^{B1}$ & $-$ & $\delta^{S1}$ & $-$\\
$e^+$ CC $99$-$00$ & $\delta^{{\cal L}1}$ & $\delta^{{\cal L}4}$ & $-$ &$-$ & $\delta^{h1}$ & $\delta^{N1}$ & $\delta^{B1}$ & $\delta^{V2}$ & $-$ & $-$\\ \hline
$e^+$ NC {\em high $y$}   & $\delta^{{\cal L}5}$ & $\delta^{{\cal L}6}, \delta^{{\cal L}7}$ & $\delta^{E2}$ &$\delta^{\theta3}$ & $\delta^{h2}$ & $\delta^{N2}$ &$-$ & $-$ & $\delta^{S2}$ & $-$\\
$e^-$ NC {\em high $y$}   & $\delta^{{\cal L}5}$ & $\delta^{{\cal L}8}, \delta^{{\cal L}9}$ & $\delta^{E2}$ &$\delta^{\theta3}$ & $\delta^{h2}$ & $\delta^{N2}$ & $-$ & $-$ & $\delta^{S2}$ & $-$ \\
$e^+$ NC $L$  & $\delta^{{\cal L}5}$ & $\delta^{{\cal L}6}$ & $\delta^{E2}$ &$\delta^{\theta3}$ & $\delta^{h2}$ & $\delta^{N2}$ & $\delta^{B1}$ & $-$ & $-$ & $\delta^{P1}$\\
$e^+$ CC $L$  & $\delta^{{\cal L}5}$ & $\delta^{{\cal L}6}$ & $-$ &$-$ & $\delta^{h2}$ & $\delta^{N3}$ & $\delta^{B1}$ & $\delta^{V3}$ & $-$ & $\delta^{P1}$\\ 
$e^+$ NC $R$  & $\delta^{{\cal L}5}$ & $\delta^{{\cal L}7}$ & $\delta^{E2}$ &$\delta^{\theta3}$ & $\delta^{h2}$ & $\delta^{N2}$ & $\delta^{B1}$ & $-$ & $-$ & $\delta^{P2}$ \\
$e^+$ CC $R$  & $\delta^{{\cal L}5}$ & $\delta^{{\cal L}7}$ & $-$ &$-$ & $\delta^{h2}$ & $\delta^{N3}$ & $\delta^{B1}$ & $\delta^{V3}$ & $-$ & $\delta^{P2}$\\ 
$e^-$ NC $L$  & $\delta^{{\cal L}5}$ & $\delta^{{\cal L}8}$ & $\delta^{E2}$ &$\delta^{\theta3}$ & $\delta^{h2}$ & $\delta^{N2}$ & $\delta^{B1}$ & $-$ & $-$ & $\delta^{P3}$ \\
$e^-$ CC $L$  & $\delta^{{\cal L}5}$ & $\delta^{{\cal L}8}$ & $-$ &$-$ & $\delta^{h2}$ & $\delta^{N3}$ & $\delta^{B1}$ & $\delta^{V3}$ & $-$ & $\delta^{P3}$ \\ 
$e^-$ NC $R$  & $\delta^{{\cal L}5}$ & $\delta^{{\cal L}9}$ & $\delta^{E2}$ &$\delta^{\theta3}$ & $\delta^{h2}$ & $\delta^{N2}$ & $\delta^{B1}$ & $-$ & $-$ & $\delta^{P4}$\\
$e^-$ CC $R$  & $\delta^{{\cal L}5}$ & $\delta^{{\cal L}9}$ & $-$ &$-$ & $\delta^{h2}$ & $\delta^{N3}$ & $\delta^{B1}$ & $\delta^{V3}$ & $-$ & $\delta^{P4}$ \\ 
\hline                                                       
\end{tabular}        
\end{center}
\caption 
 {  Correlation of systematic error sources across different
    data sets. For each of the nine
    correlated systematic error sources one or more parameters are
    included in the fit procedure. The sources considered are due to
    the luminosity uncertainty ($\delta^{\mathcal{L}}$), the electron energy
    uncertainty ($\delta^E$), the electron polar
    angle measurement ($\delta^{\theta}$), the hadronic energy
   uncertainty ($\delta^{h}$), the uncertainty due to noise subtraction
   ($\delta^{N}$), the background subtraction error ($\delta^{B}$), the
    uncertainty in measurement of the ratio $V_{ap}/V_{p}$ ($\delta^{V}$),
    the error of the background charge asymmetry
    ($\delta^{S}$), and the error of the polarisation measurement ($\delta^{\rm pol}$). 
    The table entries indicate the correlation of the
    error sources across the data sets. For example, the uncertainty
    due to the noise subtraction is the same for all data sets in HERA\,I 
    leading
    to one common parameter in the fit ($\delta^{N1}$), whereas the $V_{ap}/V_p$ 
    uncertainty has two independently varying parameters ($\delta^{V1}$
    and $\delta^{V2}$) for the CC HERA\,I data sets.}
\label{tab:correlation}
\end{table}
\renewcommand{\arraystretch}{1.00} 

\begin{table}[htp]
\begin{center}
\begin{tabular}{l|c|cc}
\hline
$\delta^{Pi}$ (Period) & $\delta_{\rm unc}$ (\%)& $\gamma_{\rm LPOL}$ (\%)& $\gamma_{\rm TPOL}$ (\%) \\
\hline
$\delta^{P1}$ ($e^+ L$)  & $1.7$& $0.34$& $0.36$ \\
$\delta^{P2}$ ($e^+ R$)  & $2.0$& $0.48$& $0.37$ \\
$\delta^{P3}$ ($e^- L$)  & $2.6$& $0.59$& $0.53$ \\
$\delta^{P4}$ ($e^- R$)  & $2.7$& $0.55$& $0.58$ \\

\hline
\end{tabular}        
\end{center}
\caption{Uncorrelated and correlated uncertainties of the polarisation measurement for each HERA\,II running period.}
\label{tab:syserr_voica}
\end{table}

The HERA data have a minimum invariant mass of the hadronic system,
$W$, of $15$\,$\rm{GeV}$ and a maximum $x$ of $0.65$, such that they are in a
kinematic region where there is no sensitivity to target mass effects and
large-$x$ higher-twist contributions.
A minimum $Q^2$ cut of $Q^2_{\rm min}=3.5\,{\rm GeV}^2$ is imposed to remain 
in the kinematic region where perturbative QCD should be applicable.

\subsection{Parameterisations}
\label{sec:qcdparam}
Fits to determine PDFs require an ansatz for the parametrisation 
as a function of $x$ at the starting scale $Q^2_0$ of the evolution, 
here chosen to 
be $1.9$ $\rm{GeV}^2$, below the charm threshold. 
The parametrised PDFs are chosen to be
the valence quark distributions $xu_v$, $xd_v$, the $u$-type and $d$-type
anti-quark distributions $x\overline{U}$ and $x\overline{D}$ 
and the gluon distribution $xg(x)$,
according to the sensitivity of the H1 data to the PDFs.
The following functional forms are considered:
\begin{eqnarray}
&& xf(x)=A_fx^{B_f}(1-x)^{C_f}(1+D_fx+E_fx^2)\,,\\
&& xg(x)=A_g x^{B_g}(1-x)^{C_g} (1+D_g x+E_g x^2) - A^{\prime}_{g}x^{B^{\prime}_{g}}(1-x)^{C^{\prime}_{g}}\,,
\end{eqnarray}
where the $A$ to $E$ are the parameters of the fit specified below.
The parametric form for the gluon
allows extra flexibility in the low $x$ region, and 
$C^{\prime}_{g}$ is set to $25$ to suppress the negative contribution 
at high $x$.  
Relaxing the parameter $C^{\prime}_g$ does not cause significant changes to the fit results.
The normalisation parameters, $A_{u_v}$ and $A_{d_v}$, are
constrained by the quark number sum rules and $A_g$ by the momentum sum rule.
Since the H1 data have little sensitivity to the light sea flavour decomposition, 
additional assumptions are imposed.
The strange quark distribution is expressed as an $x$-independent
fraction, $f_s$, of the $d$-type sea, $f_s=x\overline{s}/x\overline{D}$,
at the starting scale, with $f_s=0.31$ as preferred by
neutrino-induced di-muon production~\cite{neutrino}.
The $B$ parameters $B_{\overline{U}}$ and $B_{\overline{D}}$, responsible for 
the shape at low $x$, are set equal, such that there is a single $B$ parameter 
for the sea distributions.
The constraint $ A_{\overline{U}} =A_{\overline{D}}(1-f_s)$ is applied to 
ensure that $x\overline{u} \to x\overline{d}$ as $x \to 0$.

The optimal parametrisation is found through a scanning
procedure which iteratively adds parameters according to the data
precision and PDF sensitivity.
Starting with a basic parametric form with $9$ parameters and all $D$ and $E$
parameters set to zero and without the negative gluon term, 
a series of $10$ parameter fits are performed with
all combinations of one extra parameter except for the negative gluon term
where two extra parameters are added. 
The fit resulting in the lowest
$\chi^2$ is then chosen as the best $10$ parameter fit. 
The process is continued adding one extra parameter till no significant 
improvement in $\chi^2$ is obtained. 
This results in a best fit with $13$ parameters which is taken as
the central fit. No further significant $\chi^2$ reduction is
achieved with $14$ parameters.

 Due to more precise data from HERA\,II an enhanced flexibility is
 allowed for the valence quark parameterisations compared to the
 H1PDF\,2009 fit, with $E_{u_v} \neq 0$ and independent parameters $B$
 for the up and down valence quark distributions.
The resulting parameterisations
at the starting scale $Q^2_0$ are
\begin{eqnarray}
xg(x) &=& A_gx^{B_g}(1-x)^{C_g}- A^{\prime}_g  x^{B^{\prime}_g}(1-x)^{25}\,,\\
xu_v(x) &=& A_{u_v}x^{B_{u_v}} (1-x)^{C_{u_v}}\left(1+E_{u_v}x^2\right)\,,\\
xd_v(x) &=& A_{d_v}x^{B_{d_v}}(1-x)^{C_{d_v}}\,,\\
x\overline{U}(x) &=& A_{\overline{U}}x^{B_{\overline{U}}}(1-x)^{C_{\overline{U}}}\,,\\
x\overline{D}(x) &=& A_{\overline{D}} x^{B_{\overline{D}}}(1-x)^{C_{\overline{D}}}\,.
\end{eqnarray}

The uncertainties in the PDF determinations arise from experimental
uncertainties as well as from assumptions in the QCD analysis. 
The PDF experimental uncertainties are estimated using a Monte Carlo 
technique~\cite{Jung:2009eq}.
The method consists of preparing $N$ replica data sets in which the central 
values of the cross sections fluctuate within their statistical and systematic 
uncertainties taking into account all point-to-point correlations.
The preparation of the data is repeated $N\simeq 400$ times and for all
these replicas complete NLO QCD fits are performed to 
extract $400$ different PDF sets.
The one standard deviation band of the experimental PDF uncertainties is
 estimated using the root-mean-squared 
of the PDF sets obtained for the replicas.
The band is then attributed to the central fit
resulting in an asymmetric uncertainty, 
as the central fit does not necessarily coincide with the mean
of the $N$ replicas. 

Parametrisation uncertainties correspond to the set of 
$14$ parameter fits considered in the $\chi^2$ optimisation (compared
to the $13$ parameter central value fit) and to the
variations of the starting scale $Q_0^2$. 
%
The uncertainties are
constructed as an envelope built from the maximal deviation at each
$x$ value from the central fit.
The variations of $Q^2_0$ mostly increase the PDF uncertainties of the sea and 
gluon at small $x$.  

Model uncertainties are evaluated by varying the input assumptions
and follow the variations adopted in HERAPDF1.0~\cite{HERAPDF10}.
The variation of input values chosen for the central fit is specified
in table~\ref{tab:fitvariation}. 
The strange quark fraction is varied between $0.23$ and $0.38$~\cite{MSTW2008}.
However, recent results from the ATLAS collaboration~\cite{atlasstrange} hint
at an unsuppressed strange quark sea distribution with $f_s=0.5$ that 
exceeds the variation range for $f_s$, as given above.
This value of $f_s$ is also studied.

The difference between the central fit and the fits corresponding to
model variations of $f_s$, $Q^2_{\rm min}$, the charm quark mass $m_c$ and 
the beauty quark mass $m_b$ are added in
quadrature, separately for positive and negative deviations, and
represent the model uncertainty of the H1PDF\,2012 fit.

\renewcommand{\arraystretch}{1.22} 
\begin{table}[htb]
  \begin{center}
    \begin{tabular}{l|c|c|c}
\hline
   Parameter &  Central Value &  Lower Limit & Upper Limit \\
\hline
 $f_s$  &    $ 0.31  $  &  $0.23$ & $0.38$  \\ 
 $ m_c$\,($\rm{GeV}$) & $1.4$  &  $1.35$ (for $Q^2_0=1.8$\,$\rm{GeV}$) & $1.65$   \\ 
 $ m_b$\,($\rm{GeV}$) & $4.75$  &  $4.3$ & $5.0$ \\
 $ Q^2_{\rm min}$\,($\rm{GeV}^2$) &  $3.5$  &  $2.5$ & $5.0$  \\
\hline
 $ Q^2_{0}$\,($\rm{GeV}^2$) &    $1.9$  &  $1.5$ ($f_s=0.29$) & $2.5$ ($m_c=1.6$, $f_s=0.34$)  \\
\hline
    \end{tabular}
    \caption{Central values of input parameters to the QCD fit and their variations.}
    \label{tab:fitvariation}
  \end{center}
\end{table}
\renewcommand{\arraystretch}{1.00} 

The total PDF uncertainty is obtained by
adding in quadrature the experimental, model and parametrisation
uncertainties.

\section{Results}
\label{sec:results}

\subsection{NC and CC Double Differential Cross Sections}
\label{sec:dxdq2}

\subsubsection{Measurements with Polarised Lepton Beams}

The reduced cross sections $\tilde{\sigma}_{\rm NC,CC}(x, Q^2)$ measured in the
kinematic range $120 \leq Q^2\leq 50\,000\,{\rm GeV}^2$ and 
$0.002\leq x\leq 0.65$
for NC, and $300 \leq Q^2\leq 30\,000\,{\rm GeV}^2$ and $0.008\leq x\leq
0.4$ for CC are shown in
figures~\ref{fig:ncdxdq2_ele}-\ref{fig:ccdxdq2_pos} and given in
tables~\ref{tab:ncdxdq2_eleLH}-\ref{tab:ccdxdq2_posRH}.
The NC cross sections corresponding to the left and right handed polarised 
lepton beams $e^\pm$ (figures~\ref{fig:ncdxdq2_ele} and \ref{fig:ncdxdq2_pos}) 
are found to agree at low $Q^2~(\lesssim1\,000\,{\rm GeV}^2)$.
At higher $Q^2$ and at high $y$, deviations are observed between the
measured cross sections of the $L$ and $R$ data sets as expected from the parity 
violation of $Z$ boson exchange at high $Q^2$.
The CC reduced cross sections for the $L$ and $R$ data sets are very
different for all $Q^2$
(figures~\ref{fig:ccdxdq2_ele} and \ref{fig:ccdxdq2_pos}) as
parity violation is maximal with $W$ boson exchange.
These cross sections
agree well with the H1PDF\,2012 fit, 
which is also shown.
Both the statistical and systematic precision have substantially
improved with respect to the corresponding measurements from HERA\,I with 
the unpolarised lepton beams.

The NC reduced cross sections for $e^\pm p$ collisions in the
phase-space of $0.19<y<0.63$ and $90\leq Q^2\leq 800\,{\rm GeV}^2$ are also
measured in $y$ and $Q^2$ bins for $P_e=0$ by combining the left and
right handed polarised data sets and correcting for small residual
polarisation effects. These cross sections are presented in
tables~\ref{tab:ncdydq2_ele} and \ref{tab:ncdydq2_pos}. However, 
these cross sections are redundant with
those presented in
tables~\ref{tab:ncdxdq2_eleLH}-\ref{tab:ncdxdq2_posRH} and therefore
they should not be used together in a fit.

The {\em high y} measurement is restricted to the $Q^2$ range $60\leq
Q^2\leq 800\,{\rm GeV}^2$ where the sensitivity to the beam polarisation
is small. Therefore the left and right handed polarised data sets are
combined for the measurements shown in figure~\ref{fig:nchighy} and
given in tables~\ref{tab:nchighy_ele} and \ref{tab:nchighy_pos}.
Within the experimental uncertainties, the two sets of measurements
are in agreement. The {\em high y} data are also well described by
H1PDF\,2012.  The error bands correspond to the total uncertainty of
the fit.  The asymmetry of the uncertainty is due to the effect of the
assumptions and the experimental uncertainty of the QCD analysis, as
described in section~\ref{sec:ansatz}.

The $L$ and $R$ data sets are combined accounting for the small
residual polarisation to provide unpolarised ($P_e=0$) cross section
measurements presented in
tables~\ref{tab:ncdxdq2_eleP0}-\ref{tab:ccdxdq2_posP0}. These are then
used in the combination with HERA\,I measurements. It should be noted
that these tables are given for completeness and they should not be
used in any fit together with the corresponding polarised cross
sections, as they are redundant.

\subsubsection{Combination with Previous H1 Measurements}
\label{sec:comb}

The new unpolarised HERA\,II cross section measurements are combined
with previously published unpolarised H1 measurements from
HERA\,I~\cite{h19497,h19899,h1hiq2}.  The combination is performed
taking into account correlated systematic uncertainties represented as
nuisance parameters~\cite{sasha- comb,h1lowestq2}. The correlation of
uncertainties across different data sets is given in
table~\ref{tab:correlation} and follows the prescription given
in~\cite{h1hiq2}. The HERA\,II systematic uncertainties are in general
considered uncorrelated with those from HERA\,I apart from the
photoproduction background uncertainty. This assumption is motivated
by improvements in the calibration procedures which lead to better
determined central values for the HERA\,II result. This approach leads
to a conservative estimate of the uncertainties for the combined
sample. In the years $1994-1997$ the data were taken at the lower
centre of mass energy of $\sqrt{s}=301$\,$\rm{GeV}$ whilst the other data
samples are taken at $\sqrt{s}=319$\,$\rm{GeV}$. To take this into account
the data at $\sqrt{s}=301$\,$\rm{GeV}$ are corrected to $\sqrt{s}=319$\,$\rm{GeV}$
using the H1PDF\,2012 parametrisation.  This correction and the
combination are only performed for data points at $y<0.35$ as at
larger $y$ the contribution of the longitudinal structure $F_L$ is
sizable, and therefore the uncertainty of this correction is
minimised. The correction is typically
$0.5-2.5$\% for $y<0.35$ and never more than $3.8$\%.

\def\totalout{   413}
\def\chicomb{ 412.1}
\def\dofcomb{   441}
\def\totalin{   854}
 
\begin{table}[htbp]
\begin{center}
\begin{tabular}{l|rr}
\hline 
Source &  Shift in units of standard deviation & Shift in \% of cross section \\
\hline 
$\delta^{\mathcal{L}1}$ (BH Theory)     & $  -0.39 $ & $  -0.19 $ \\ 
$\delta^{\mathcal{L}2}$ ($e^+$ $94$-$97$)   & $  -0.46 $ & $  -0.66 $ \\ 
$\delta^{\mathcal{L}3}$ ($e^-$ $98$-$99$)   & $  -0.69 $ & $  -1.20 $ \\ 
$\delta^{\mathcal{L}4}$ ($e^+$ $99$-$00$)   & $  -0.07 $ & $  -0.10 $ \\ \hline
$\delta^{\mathcal{L}5}$ (QEDC)          & $   0.81 $ & $   1.70 $ \\ 
$\delta^{\mathcal{L}6},\delta^{{\cal L}7}$ ($e^+ L+R$)   & $   0.84 $ & $   0.80 $ \\ 
$\delta^{{\cal L}8},\delta^{{\cal L}9}$ ($e^- L+R$)   & $   0.84 $ & $   0.89 $ \\ 

\hline
\end{tabular}
\end{center}
\caption{ Shifts of the normalisation parameters $\delta^{\cal L}$ (see
  table~\ref{tab:correlation}) both for
the luminosity measurements of HERA\,I (BH Theory) and HERA\,II (QEDC) and
for the individual normalisation of each data set after 
combination of HERA\,I and
HERA\,II measurements. The shifts are expressed in units of 
standard deviations of the parameters as well as the fractional change in the
cross sections.}
\label{tab:normshift}
\end{table}

A total of $\totalin$ data points are averaged to $\totalout$ cross section
measurements. The data show good consistency with a total $\chi^2$
per degree of freedom (ndf) of $\chi^2/{\rm ndf} = \chicomb/\dofcomb$.
Out of $22$ nuisance parameters corresponding to the correlated systematic 
error sources none develop a significant deviation from zero. The values
of the nuisance parameters for the global normalisations are given in
table~\ref{tab:normshift} which represents the values as
fractions of the normalisation uncertainty and as absolute shifts in
per cent. The adjustments of the relative normalisations are small.
The normalisation of the data collected in the years $1999-2000$
stays constant and the other HERA\,I data samples shift down by 
maximally $1.2\%$, while the HERA\,II samples shift up by maximally $1.7\%$.

The combined HERA\,I+II NC and CC cross sections are shown in
figures~\ref{fig:nccombined}-
\ref{fig:cccombined_pos} and given in
tables~\ref{tab:nccombined_ele}-
\ref{tab:cccombined_pos}. The H1PDF\,2012 fit is found to give a good
description of the $x, Q^2$ behaviour of the data.  The NC data
exhibits a strong rise with decreasing $x$ which can be interpreted as
being due to the high density of low $x$ quarks in the proton.  The
$e^-p$ data are in good agreement with the $e^+p$ measurements for
$Q^2\lesssim 1\,000\,{\rm GeV}^2$. At larger values of $Q^2$ the
$e^-p$ data are generally higher than the $e^+p$ data, as is expected
from the effects of $Z$~boson exchange. The difference is used to extract
the $xF_3^{\gamma Z}$ structure function as described in
section~\ref{sec:xf3}.

In figures~\ref{fig:cccombined_ele} and \ref{fig:cccombined_pos}, the
quark contributions from $x(u+c)$ and $(1-y)^2x(d+s)$ are
indicated for $e^-p$ and $e^+p$ data, respectively, illustrating that
the CC data can be used to separate the up- and down-type quark
distributions in the proton.

\subsection{Fit Results}
\label{sec:fitresults}

The data in the full phase space are well described by the fit
with a $\chi^2$ per degree of freedom $1569.6/1461$. 
The central fit satisfies the criteria that structure
functions are positive and $xd_v>x\overline{d}$ at large $x$.
The PDF parameters obtained from the QCD analysis 
are presented in table~\ref{tab:fitparam}.
Since the measured polarisation values are allowed to vary in the fit 
procedure within their uncertainties,
the corresponding shift parameters are minimised together 
with the PDF parameters and are shown in table~\ref{tab:polparam}.
They are found to have little correlation with the PDF parameters.
\begin{table}[htb]
  \begin{center}
    \begin{tabular}{l|rc}
\hline
Parameter & \multicolumn{2}{c}{Central value}  \\
\hline
$B_g$  & $ 0.020$&  \\
$C_g$  & $ 6.4$&  \\
$A^{\prime}_g$   & $ 0.30$&  \\
$B^{\prime}_g$   & $-0.27$& \\
$B_{u_v}$   &$ 0.70$&  \\
$C_{u_v}$   &$ 4.9$&  \\
$E_{u_v}$   &$ 12$&   \\
$B_{d_v}$   &$ 0.97$&  \\
$C_{d_v}$   &$ 5.2$&  \\
$C_{\bar{U}}$  &   $ 3.4$&  \\
$A_{\bar{D}}$  &   $ 0.17$&  \\
$B_{\bar{D}}$  &   $-0.15$&  \\
$C_{\bar{D}}$  &   $ 9.7$&   \\
\hline
\end{tabular}
\end{center}
\caption{Parameters of the central fit.}
\label{tab:fitparam} 
\end{table}

\begin{table}[htb]
  \begin{center}
    \begin{tabular}{l|r}
\hline
Parameter & Central value  \\
\hline
$b^{1(e^+pL)}_{\rm unc}$ & $0.16$  \\
$b^{2(e^+pR)}_{\rm unc}$ &  $0.19$  \\
$b^{3(e^-pL)}_{\rm unc}$ & $-0.32$  \\
$b^{4(e^-pR)}_{\rm unc}$ &  $0.50$ \\
$b_{\rm TPOL}$ & $0.11$  \\
$b_{\rm LPOL}$ & $0.11$ \\
\hline
\end{tabular}
\end{center}
\caption{Parameters corresponding to the polarisation shifts.}
\label{tab:polparam} 
\end{table}

Table \ref{tab:fitresult} summarises the partial $\chi^2$ values corresponding
to both the statistical and uncorrelated systematic uncertainties for each data 
set used in the fit.
 The total correlated $\chi^2$ value (not included in the table)
amounts to $67.4$ units.
 The systematic shifts allowed by the Hessian 
method to account for the correlations are generally small (less than one standard deviation).
Table \ref{tab:fitresult_shifts} presents the optimised 
normalisation shifts obtained by the fit.
The values are shown separately for each data period.
%
The fit results in shifting the global normalisation of the HERA\,I data 
points by $-0.7\%$ and that of the HERA\,II data points by 
$2.9\%$ corresponding to $1.3$ standard devations.
 The shift values are consistent with those from the combination 
obtained in section~\ref{sec:comb} although the numerical values are 
different due to the additional low $Q^2$ and low $E_p$ data sets 
used in the fit.

It has been observed previously that the heavy flavour scheme used here
results in a rather large $\chi^2$ value in fits for the low 
$Q^2$ data~\cite{h1fl2010}. 
In this analysis, the corresponding partial
$\chi^2$ contributions are also large.
The overall quality of the fit is improved if the ACOT scheme is used,
due to a considerably better agreement with the low $Q^2$ data 
($\simeq 30$ units improvement in $\chi^2$).
For the high $Q^2$ measurements, however,
which are the focus of this paper, using the ACOT scheme is 
slightly worse than using the RT prescription~\cite{rt97}.

\begin{table}[htb]
\begin{center}
\begin{tabular}{l|r|r}
\hline
Data Set &  Number of   &  $\chi^2$ (unc. err.) \\
 &  data points   & \\
\hline
 $e^+$ Combined low $Q^2$ &  $171$ & $ 196$\\\hline
 $e^+$ Combined low $E_p$&  $124 $ & $132 $ \\\hline
 $e^+$ NC $94$-$97$  &  $130$ & $ 92$ \\
 $e^+$ CC $94$-$97$  &  $ 25$ & $ 22$  \\
 $e^-$ NC $98$-$99$  &  $126$ & $113$ \\
 $e^-$ NC $98$-$99$ {\em high $y$} &  $13$  & $5.4$\\
 $e^-$ CC $98$-$99$  &  $ 28$ & $ 19$   \\
 $e^+$ NC $99$-$00$  &  $147$ & $144$  \\
 $e^+$ CC $99$-$00$  &  $ 28$ & $ 29$  \\ \hline 
 $e^+$ NC {\em high $y$}  &  $11$ & $5.6$  \\
 $e^-$ NC {\em high $y$}  &  $11$ & $7.7$  \\
 $e^+$ NC $L$   &  $137$ & $ 124$  \\
 $e^+$ CC $L$   &  $ 28$ & $ 46$   \\
 $e^+$ NC $R$   &  $138$ & $ 138$  \\
 $e^+$ CC $R$   &  $ 29$ & $ 40$  \\
 $e^-$ NC $L$   &  $139$ & $ 174$  \\
 $e^-$ CC $L$   &  $ 29$ & $ 27$  \\
 $e^-$ NC $R$   &  $138$ & $ 142$  \\
 $e^-$ CC $R$   &  $ 28$ & $ 16$  \\
\hline
\end{tabular}
\end{center}
\caption{Results of the H1PDF\,2012 fit. For each
data set the number of data points are given, along with the $\chi^2$
contribution determined using uncorrelated errors (unc.\ err.)\ of the
data points. }
\label{tab:fitresult} 
\end{table}

\begin{table}[htb]
\begin{center}
\begin{tabular}{l|r|r|r}
\hline
Data Period & Global & Per Period & Total\\
            &  Normalisation &  Normalisation & Normalisation\\
\hline
$e^+$ Combined low $Q^2$ & $0.993$& $-$ &$0.993$ \\\hline
$e^+$ Combined low $E_p$ & $0.993$ & $-$ &$0.993$ \\\hline
HERA\,I $e^+$ $94$-$97$ & $0.993$& $0.999$ & $0.992$ \\
HERA\,I $e^-$ $98$-$99$ & $0.993$& $1.003$ & $0.996$ \\
HERA\,I $e^+$ $99$-$00$ & $0.993$& $1.005$ &$0.998$\\\hline
HERA\,II $e^+$ $L$   & $1.029$& $0.991$ & $1.020$ \\
HERA\,II $e^+$ $R$   & $1.029$& $1.013$ &$1.042$ \\
HERA\,II $e^-$ $L$   & $1.029$& $1.010$ &$1.039$ \\
HERA\,II $e^-$ $R$   & $1.029$& $1.014$ &$1.043$ \\
\hline
\end{tabular}
\end{center}
\caption{Factors corresponding to the global luminosity normalisations (${\cal L}1$, ${\cal L}5$), the normalisation for each data period (${\cal L}2$, ${\cal L}3$, ${\cal L}4$ for HERA\,I and ${\cal L}6$, ${\cal L}7$, ${\cal L}8$, ${\cal L}9$ for HERA\,II), and the overall combined normalisation of the data sets as determined by the QCD fit.}
\label{tab:fitresult_shifts} 
\end{table}

The H1PDF\,2012 fit results are summarised in figures~\ref{fig:pdfa}-\ref{fig:pdfc}, 
shown at the starting scale $Q^2_0=1.9$\,$\rm{GeV}$$^2$, evolved to 
$Q^2=10$\,$\rm{GeV}$$^2$ and to $Q^2=M_W^2$. 
The fit result when using $f_s=0.5$ 
lies well within the uncertainty band of the H1PDF\,2012 set.

The consistency of results is checked by comparing with 
PDF fits determined from the combined unpolarised HERA\,I and
 II measurements presented in section~\ref{sec:comb}.
The resulting PDFs and shifts of the correlated sources are 
in good agreement when using the separate data sets or when using 
the combined unpolarised data.

In order to assess the impact of the new HERA\,II data, the QCD fit is
repeated under the same conditions with the new measurements excluded.
For this comparison replica data sets are generated
from the expected cross sections by using the corresponding 
experimental uncertainties, thereby resulting in symmetrical error bands.
As shown in figure~\ref{fig:hera1vs2},
the new high $Q^2$ data have a visible impact on all distributions,
especially in the $xD$ distribution.

\subsection{NC and CC Cross Sections ${ \pmb{\rm d}\pmb{\sigma}/\pmb{\rm d}\pmb{Q^2}}$ } 
\label{sec:dq2}

The single differential NC cross sections ${\rm d}\sigma_{\rm NC}/{\rm
d}Q^2$ measured for $y<0.9$ with both $e^-p$ and $e^+p$ data are shown
in figure~\ref{fig:ncdq2} (upper plots) and given in
tables~\ref{tab:ncdq2_eleLH}-\ref{tab:ncdq2_posRH}.  The data are
measured in the range $Q^2\geq 200\,{\rm GeV}^2$ up to $50\,000\,{\rm
GeV}^2$ over which the cross sections fall by more than six orders of
magnitude with increasing $Q^2$.  The cross sections are well
described by the SM expectations based on the H1PDF\,2012 fit.
The lower panel of figure~\ref{fig:ncdq2} shows the ratios of the
measurements to the corresponding SM values determined from the
H1PDF\,2012 fit. The asymmetric uncertainty represents the effect of the
assumptions and the experimental uncertainty of the QCD analysis and
is explained in section~\ref{sec:qcdparam}. Note that in this lower figure the
H1 data are scaled by the normalisation shifts
imposed by the QCD fit given in table~\ref{tab:fitresult_shifts}.

The $Q^2$ dependence of the CC cross sections 
${\rm d}\sigma_{\rm CC}/{\rm d}Q^2$ for $y<0.9$ is shown in
figure~\ref{fig:ccdq2}. In the upper figure, the strong polarisation dependence
is clearly visible. In the lower figure, the same normalisation shifts 
as for the NC data are applied. 
The CC cross sections together with the kinematic correction
factors are given in tables~\ref{tab:ccdq2_eleLH}-\ref{tab:ccdq2_posRH}.

Combining the left and right handed polarisation data sets and correcting for
the residual polarisation effects, the resulting unpolarised NC and CC
cross section measurements from HERA\,II are listed in 
tables~\ref{tab:ncdq2_eleP0}-\ref{tab:ccdq2_posP0}. 
These cross sections are combined with the corresponding
measurements from HERA\,I using the same procedure as for the combination of
the double differential sections, described in section~\ref{sec:comb}.
The results are shown in tables~\ref{tab:dq2_comb_em}-\ref{tab:dq2cc_comb_ep}.

The $Q^2$ dependence of the combined HERA\,I+II NC and CC cross
sections for $P_e=0$ is shown in
figure~\ref{fig:ncccdq2}. The NC cross sections exceed the CC cross
sections at 
$Q^2\simeq 200~\rm{GeV}^2$ by more than two orders of magnitude.  The steep
decrease of the NC cross section with increasing $Q^2$ is due to the
dominating photon exchange cross section which is proportional to $1/Q^4$. 
In contrast
the CC cross section is proportional to $\left[M_W^2/(Q^2+M_W^2)\right]^2$ and
approaches a constant value at $Q^2\simeq 300$~$\rm{GeV}$$^2$. The NC and CC cross sections are
of comparable size at $Q^2\sim 10^4$~$\rm{GeV}$$^2$, where the photon and
$Z$ exchange contributions to the NC process are of similar size to
those of $W^{\pm}$ exchange to the CC process.  
These measurements thus illustrate
the unified behaviour of the electromagnetic and the weak interactions in
DIS.

\subsection{NC Polarisation Asymmetry and $\pmb{F}_{\pmb{2}}^{\pmb{\gamma Z}}$}
\label{sec:f2gZ}

The SM predicts a difference in the NC cross section for leptons with
different helicity states arising from the chiral structure of the
neutral electroweak exchange. With longitudinally polarised lepton
beams in HERA\,II such polarisation effects can be tested, providing a
direct measure of electroweak effects in the NC cross sections. The
polarisation asymmetry, $A^{\pm}$, is defined as
\begin{equation}
A^{\pm}=\frac{2}{P^{\pm}_L-P^{\pm}_R}\cdot\frac{\sigma^\pm(P^{\pm}_L)-\sigma^\pm(P^{\pm}_R)}{\sigma^\pm(P^{\pm}_L)+\sigma^\pm(P^{\pm}_R)}\,\,\,,
\label{eq:asym}
\end{equation}
where $P^{\pm}_L$ and $P^{\pm}_R$ are the 
longitudinal lepton beam
polarisation in the $e^{\pm}p$ $R$ and $L$ data sets. To a very good
approximation $A^{\pm}$ measures the structure function ratio
$A^\pm\simeq\mp \kappa a_eQ^2/(Q^2+M^2_Z) F^{\gamma Z}_2/\tilde{F}_2$\,,
which is proportional to the product $a_ev_q$ and thus is a direct
measure of parity violation. In $e^+$ scattering $A^+$ is expected to
be positive and about equal to $-A^-$ in $e^-$ scattering. At large
$x$ the asymmetry measures the $d/u$ ratio of the valence quark
distributions according to
\begin{equation}
A^\pm\propto \pm \kappa \frac{1+d_v/u_v}{4+d_v/u_v}\,.
\end{equation}
 
The polarised single differential cross sections ${\rm
d}\sigma_{\rm NC}/{\rm d}Q^2$ are used to construct the asymmetry where it
is assumed that the correlated uncertainties of each measurement
cancel. The asymmetry is shown in figure~\ref{fig:pol_asy} compared to
the H1PDF\,2012 fit. The magnitude of the
asymmetry is observed to increase with increasing $Q^2$ and is
positive in $e^+p$ and negative in $e^-p$ scattering. The data are in
good agreement with the SM using H1PDF\,2012 and confirm
the parity violation effects of electroweak interactions at large
$Q^2$.


For a given lepton charge the difference in the left and right polarised NC cross sections is sensitive to 
$F_2^{\gamma Z}$ as well as $xF_3^{\gamma Z}$ and $xF_3^{Z}$ as given by
\small
\begin{equation}
\frac{\sigma^\pm(P_L^\pm)-\sigma^\pm(P_R^\pm)}{P_L^\pm-P_R^\pm}=\frac{\kappa Q^2}{Q^2+M_Z^2}\left[\mp a_e
  F_2^{\gamma Z} + \frac{Y_-}{Y_+}v_e xF_3^{\gamma Z} -
  \frac{Y_-}{Y_+}\frac{\kappa Q^2}{Q^2+M_Z^2}(v_e^2+a_e^2)xF_3^{Z}
  \right] \,.\label{eq:difsigma}
\end{equation}
\normalsize 

\begin{figure}[ht!]
\begin{center}
\includegraphics[width=\columnwidth]{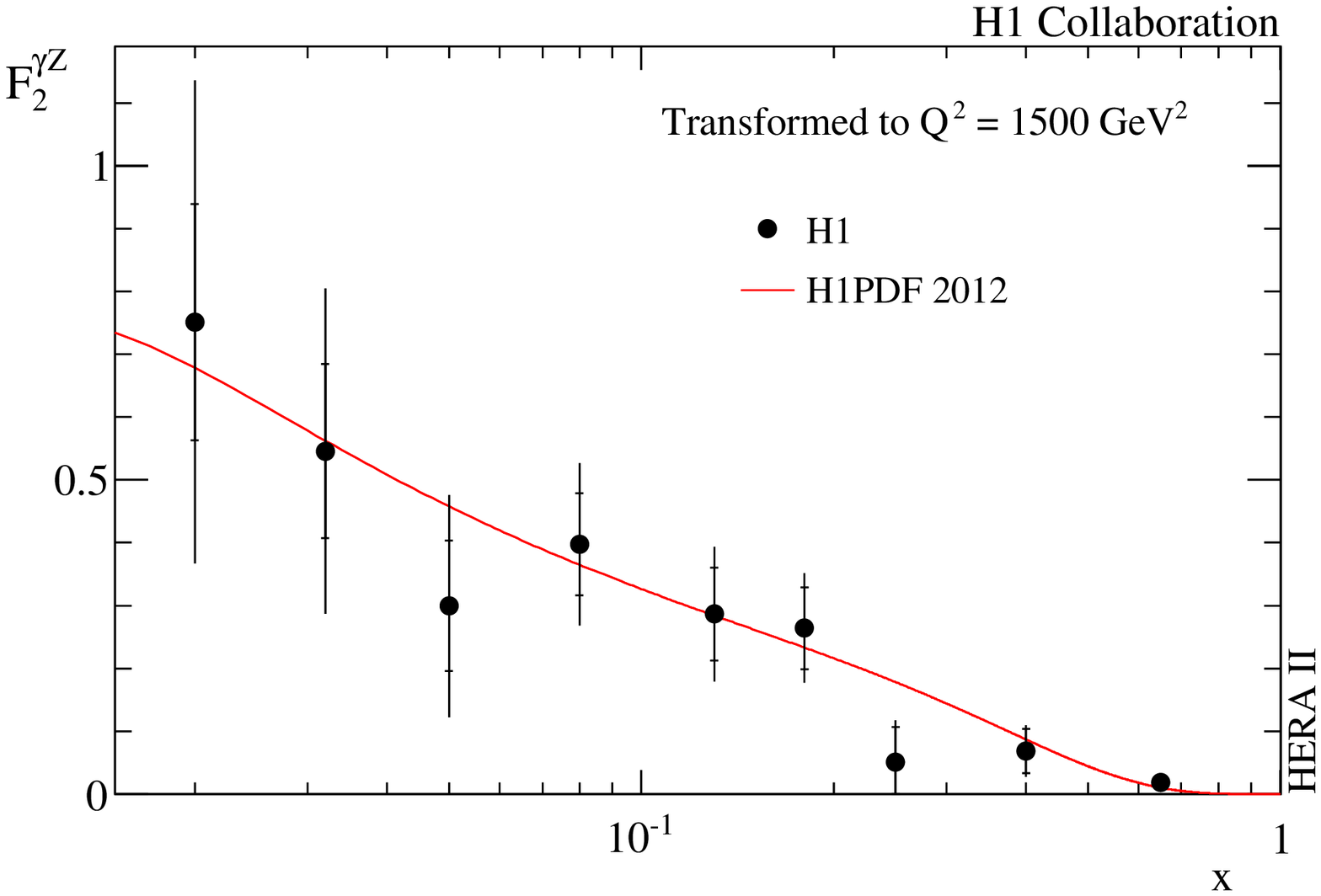}
\end{center}
\caption{ Structure function $F_2^{\gamma Z}$ transformed to 
  $Q^2=1\,500\,{\rm GeV}^2$ for data (solid points) and the expectation from
  H1PDF\,2012 (solid curve). The inner error bars represent the
  statistical uncertainties and the full error bar corresponds to the
  total measurement uncertainty.}
\label{fig:f2gZ_1500} 
\end{figure}

By taking the difference of equation~\ref{eq:difsigma} for the $e^+p$
and $e^-p$ data, the terms proportional to $xF_3^{\gamma Z}$ and
$xF_3^{Z}$ cancel and $F_2^{\gamma Z}$ can be directly extracted using
the measured cross sections. The measurement is performed for
$Q^2\geq200$~$\rm{GeV}$$^2$. It is shown in figure~\ref{fig:f2gZ} and listed in 
table~\ref{tab:f2gZ}. Only a
weak $Q^2$ dependence is expected and therefore the measurements are
transformed to a common $Q^2$ value of $1\,500$~$\rm{GeV}$$^2$ using the
H1PDF\,2012 fit and are averaged in each $x$ bin. The average is
calculated as a weighted mean using the quadratic sum of statistical
and uncorrelated systematic uncertainties. The result is displayed in
figure~\ref{fig:f2gZ_1500} in comparison to the H1PDF\,2012 fit and listed in 
table \ref{tab:f2gZ_1500}. The
correlated uncertainties of the $F_2^{\gamma Z}$ measurement consist
of contributions from the point-to-point correlated sources of
uncertainties.  The dominant contribution at low $Q^2$ and low $y$ is
the normalisation uncertainty of $1.5$\% of each data set. The global
luminosity uncertainty of 2.3\% is not included.  

\subsection{Measurement of $\pmb{x}\pmb{F}_{\pmb 3}^{\pmb{\gamma Z}}$}
\label{sec:xf3}

The new combined HERA\,I+II NC unpolarised cross section measurements for $e^+p$ 
and $e^-p$ scattering are used to update the previous measurement of 
the structure function $xF_3^{\gamma Z}$~\cite{h19899,h1hiq2}. 
Only data taken at $E_p=920$~$\rm{GeV}$ are used for this determination. 
The structure function $x\tilde{F}_3$ is obtained in a
simultaneous fit with $x\tilde{F}_3$, $\tilde{\sigma}_0^\pm \equiv 
\tilde{F}_2^\pm- y^2/Y_+ \tilde{F}_L^\pm$ and nuisance parameters for 
the systematics shifts $b_j$ being free minimisation parameters.  
The $\chi^2$ function for the minimisation is
\begin{equation} \label{eq:xf3}
\chi^2 
\left(\tilde{\sigma}_{0}^\pm, x\tilde{F}_3^\pm,b\right) = \sum_i 
\frac{\textstyle \left[ \left(\tilde{\sigma}_{0,i}^{\pm} \mp \frac{Y_-}{Y_+} x\tilde{F}_{3,i}^{\pm}\right) - 
\sum_j \Gamma_{i,j} b_j - \mu^{\pm}_i\right]^2}{\textstyle \Delta^2_i} + \sum_j b^2_j\,.
\end{equation}
Here $\mu^{\pm}_i$ is the measured central value of the reduced
$e^{\pm}p$ cross section at an $x,Q^2$ point $i$ with a combined
statistical and uncorrelated systematic uncertainty $\Delta_i =
\sqrt{\left(\Delta_{i, \rm stat}^2 + \Delta_{i,\rm syst}^2 \right)}$.
The effect of correlated error sources $j$ on the cross section
measurements is given by the systematic error matrix $\Gamma_{i,j}$.
The $\chi^2$ function depends quadratically on 
$\tilde{\sigma}_{0,i}^{\pm}$ and $x\tilde{F}_{3,i}^{\pm}$.
The minimisation of the $\chi^2$ function
with respect to these variables leads to a system of linear equations
which is solved analytically, similar to~\cite{h1fl2010}.
This procedure gives results equivalent to a determination of
$x\tilde{F}_3$ in which the systematic uncertainties are treated by
varying the measurements by each systematic error and adding the
resulting deviations in quadrature.

The dominant contribution to $x\tilde{F}_3$ arises from $\gamma Z$
interference, which allows the extraction of $xF_3^{\gamma Z}$
according to $xF_3^{\gamma Z}\simeq -x\tilde{F}_3(Q^2+M^2_Z)/(\kappa
a_e Q^2)$ where the pure $Z$ boson exchange term is neglected. This is
justified since the contribution of $xF_3^{Z}$ is suppressed by the
small coupling $v_e$ and an additional factor $\kappa Q^2/(Q^2+M_Z^2)$
(see eq.~\ref{eq:F3}). The resulting structure function for
$Q^2>1\,000$~$\rm{GeV}$$^2$ is presented in table~\ref{tab:xf3gz} and shown in
figure~\ref{fig:xf3} together with the expectations determined from
the H1PDF\,2012 fit.  Since at high $x$ and low $Q^2$ the expected
sensitivity to $x\tilde{F}_3$ is smaller than the luminosity
uncertainty, the measurement is not performed in this region.

This non-singlet structure function exhibits only a weak dependence on
$Q^2$ and therefore the measurements can be first transformed to
$Q^2=1500$~$\rm{GeV}$$^2$ using H1PDF\,2012 and then averaged for fixed $x$
values. The averaged $xF_3^{\gamma Z}$ is given in
table~\ref{tab:xf3gz_1500} and shown in figure~\ref{fig:xf3gZ_1500} in
comparison with the H1PDF\,2012 fit. The calculation from the H1PDF\,2012 fit
gives a good description of the $xF_3^{\gamma Z}$ measurement. The structure function
$xF_3^{\gamma Z}$ determines both the shape and magnitude of the
valence distribution $2u_v+d_v$ assuming the quark and anti-quark sea
distributions are the same. 
The integral of this structure function is analogous to the GLS sum
rule in neutrino scattering\cite{gls} which is in LO predicted to be $5/3$ and acquires
$\mathcal{O}(\alpha_s/\pi)$ QCD
corrections~\cite{Rizvi:2000qf}. 
The measured value using
  all HERA\,I+II data is
\begin{eqnarray}
\int_{0.016}^{0.725} {\rm d}x \,\, F_3^{\gamma Z}(x,Q^2=1\,500\,{\rm GeV}^2)=
1.22 \pm 0.09 ({\rm stat}) \pm 0.07 ({\rm syst}) \,\,,
\end{eqnarray}
which can be compared to the H1PDF\,2012 fit in the same region
$\int_{0.016}^{0.725}F_3^{\gamma Z}\,{\rm d}x = 1.16^{+0.02}_{-0.03}$
including the total estimated uncertainty. 
The extrapolation of the measurement to the full kinematic region in $x$ by applying
a scale factor determined from the H1PDF\,2012 fit, yields 
$\int_0^1 {\rm d}x \,\, F_3^{\gamma Z} = 1.69 \pm 0.12 ({\rm stat})
\pm 0.10 ({\rm syst})$. 
No additional
uncertainty due to the scale factor is considered.
This value agrees 
with the integral evaluated using the H1PDF\,2012 fit over the full $x$ range at $Q^2=1\,500$~$\rm{GeV}$$^2$ which is determined to be
$\int_{0}^{1}F_3^{\gamma Z}\,{\rm d}x = 1.595$.
The quark number sum rules are
imposed as constraints in the QCD fit and therefore this measurement
validates the sum rules.

\subsection{Total CC Cross Sections $ \pmb{\sigma}_{\pmb{\rm CC}}^{\pmb{\rm tot}}$}
\label{sec:cctot}

The total CC cross sections for $Q^2>400\,{\rm GeV}^2$ and $y<0.9$ are
listed in table~\ref{tab:cctot} for the $e^-$ and $e^+$ data and
for the different longitudinal lepton beam polarisations.  Corrections
$(k^{\pm}_{\rm cor})$ from the analysis phase space $Q^2>400\,{\rm GeV}^2$,
$p_{T,h}>12$\,$\rm{GeV}$ and $0.03<y<0.85$ are applied using the SM
expectation based on H1PDF2012 and are found to be $k^-_{\rm cor}=1.070$ for $e^-p$
and $k^+_{\rm cor}=1.063$ for $e^+p$ scattering.  
The corresponding cross sections~\cite{h1ccpol}
using the unpolarised HERA\,I data and the same kinematic corrections
are also shown in table~\ref{tab:cctot}.
\begin{table}[htb] 
\begin{center}
\begin{tabular}{ccc}
\hline 
 & $P_e~\rm{(\%)}$ & $\sigma_{\rm CC}^{\rm tot}~\rm{(pb)}$  \\ \hline
\multirow{3}{*}{$e^-p$} 
& $-25.8\pm 0.7$ & $66.5\pm 1.0_{\rm stat}\pm 1.8_{\rm syst}\pm 1.8_{\rm lumi}$ \\
& $0$            & $57.0\pm 2.2_{\rm stat}\pm 0.9_{\rm syst}\pm 1.0_{\rm lumi}$ \\
& $+36.0\pm 1.0$ & $33.7\pm 1.1_{\rm stat}\pm 0.9_{\rm syst}\pm 0.9_{\rm lumi}$ \\ \hline
\multirow{3}{*}{$e^+p$}
& $-37.0\pm 0.7$ & $17.3\pm 0.6_{\rm stat}\pm 0.6_{\rm syst}\pm 0.5_{\rm lumi}$ \\
& $0$            & $28.4\pm 0.8_{\rm stat}\pm 0.8_{\rm syst}\pm 0.4_{\rm lumi}$ \\
& $+32.5\pm 0.7$ & $36.6\pm 0.8_{\rm stat}\pm 1.2_{\rm syst}\pm 1.0_{\rm lumi}$ \\ \hline
\end{tabular} 
\end{center} 
\caption
{\label{tab:cctot} The total CC cross section $\sigma^{\rm tot}_{\rm CC}$
for $Q^2>400\,{\rm GeV}^2$ and $y<0.9$. The uncertainties correspond to the
statistical, experimental systematic and luminosity uncertainties.}
\end{table}

The cross sections are shown in figure~\ref{fig:cctot} and compared to the SM
expectations using the H1PDF\,2012 fit. 
They agree within one standard 
deviation if the normalisation factors as determined from the QCD fit are applied.
%
A linear fit to the polarisation dependence of the measured cross
sections is performed taking into account the correlated systematic
uncertainties between the measurements and is also shown in
figure~\ref{fig:cctot}. The fit is performed simultaneously to $e^-p$
and $e^+p$ data and yields a $\chi^2=2.0$ for two degrees of
freedom. The result of the fit extrapolated to the point $P_e=+1$ for $e^-p$ and 
$P_e=-1$ for $e^+p$ scattering results in
\begin{eqnarray}
&&\sigma^{\rm tot}_{\rm CC}(P_e=+1,e^-p)=-1.3\pm 2.4_{\rm exp}\pm 1.5_{\rm lumi}\pm 1.2_{\rm pol}\,{\rm pb}\,, \nonumber\\
&&\sigma^{\rm tot}_{\rm CC}(P_e=-1,e^+p)=-0.5\pm 1.3_{\rm exp}\pm 0.7_{\rm lumi}\pm 0.4_{\rm pol}\,{\rm pb}\,, \nonumber
\end{eqnarray}
where the quoted errors correspond to the experimental, luminosity and
polarisation related uncertainties. These extrapolated
cross sections are consistent with the SM prediction of a vanishing
cross section and correspond to an upper limit on $\sigma^{\rm
tot}_{\rm CC}(P_e=+1,e^-p)$ and $\sigma^{\rm tot}_{\rm CC}(P_e=-1,e^+p)$ of $4.8~\rm{pb}$ and
$2.6~\rm{pb}$ at $95\%$ confidence level (CL), 
respectively, as 
derived according
to~\cite{fc97}. This result excludes the existence of charged currents
involving right handed fermions mediated by a boson of mass $M_W^R$ 
below $214$ and $194$\,$\rm{GeV}$ at $95\%$ CL respectively, assuming SM couplings and a light right 
handed $\nu_e$. These limits are comparable with those derived earlier by
H1~\cite{h1ccpol} and ZEUS~\cite{zeuscc2010}.

\clearpage
\section{Conclusions}
\label{sec:summary}

The inclusive DIS cross section for $e^{\pm}p$ interactions at
$\sqrt{s}=319$\,GeV are measured using $333.7$\,pb$^{-1}$ of
integrated luminosity. The $e^{-}p$ data analysed here corresponds to
an almost ten-fold increase in luminosity over the HERA\,I data
set. Moreover the operation of the HERA collider with left and right
handed longitudinally polarised electron and positron beams allows
measurements in the neutral and charged current channels
 with four distinct initial states. 
The NC and CC cross
sections cover the region $Q^2\gtrsim 100$\,GeV$^2$ and Bjorken $x \gtrsim 10^{-3}$. The cross
sections are measured differentially in $Q^2$ and double
differentially in $x$ and $Q^2$. The systematic uncertainties of the
measurements are substantially reduced compared to previous
publications. In the NC channel a precision of $1.5\%$ is attained for
the  systematic uncertainty in the kinematic region
$Q^2<1\,000$\,GeV$^2$ and $y>0.1$, compared to a statistical accuracy of
about $1-3\%$.
The high inelasticity region of $0.63\leq y \leq 0.90$ for $60\leq Q^2
\leq 800$\,GeV$^2$ is measured in the NC analysis for unpolarised
$e^{\pm}p$ scattering. This phase space region is sensitive to the
$F_L$ structure function.

A NLO QCD analysis of the data is performed for $Q^2\geq 3.5$\,GeV$^2$
including all previously published H1 NC and CC cross section
measurements. The data are well described by the QCD fit over the full
phase space. The new data at high $Q^2$ provide better constraints on
the partonic structure of the proton. In particular the CC $e^+p$ data
enable an improved flavour separation at high $x$.

The NC lepton polarisation asymmetry $A^\pm$, sensitive to parity
violation, is determined separately for $e^+p$ and $e^-p$
scattering. The asymmetry is 
found to increase in magnitude with $Q^2$ in
agreement with the expectation of the Standard Model. The structure function
$F_2^{\gamma Z}$ is measured for the first time using the polarisation
dependence of the $e^{\pm}p$ NC cross section. The structure function
is reported differentially in $x,Q^2$ and the result is also averaged
at $Q^2=1\,500$~GeV$^2$.

At high $Q^2$ the structure function $xF_3^{\gamma Z}$ is determined
using unpolarised NC cross sections obtained from the complete HERA I
and HERA II data sets.  The $xF_3^{\gamma Z}$ results are averaged at
$Q^2=1\,500$\,GeV$^2$ and cover the range $0.013\leq x \leq
0.65$. The measurement integrated over $x$ validates a sum rule for
charged lepton scattering.

The polarisation dependence of the CC total cross section for $Q^2 >
400$\,GeV$^2$ and $y<0.9$ is measured and compared to the unpolarised
HERA\,I measurements. The data exhibit a linear scaling of the cross
sections with $P_e$ which is positive for $e^+p$ and negative for
$e^-p$ scattering. The data are consistent with the absence of right
handed weak currents.

The analysis reported here completes the measurements of inclusive NC
and CC cross sections with the HERA\,I and HERA\,II data samples at
$\sqrt{s}=319$\,GeV with the H1 detector.

\section*{Acknowledgements}

We are grateful to the HERA machine group whose outstanding efforts
made this experiment possible.  We thank the engineers and technicians
for their work in the construction and maintenance of the H1
detector, our funding agencies for financial support, the DESY
technical staff for continual assistance and the DESY directorate for
the hospitality which they extend to the non-DESY members of the
collaboration.

\newpage


\newpage

\newcommand{\tablepos}{htb}
\clearpage
\begin{table}[htbp] 
\begin{center} 
\tiny 
\begin{tabular}{|r|c|r|r|r|r|r|r|r|r|r|r|r|r|} 
\hline 
$Q^2$ & $x$ & $\tilde{\sigma}_{\rm NC}$ & 
$\delta_{\rm tot}$ & $\delta_{\rm stat}$ & $\delta_{\rm unc}$ & 
$\delta_{\rm unc}^{E}$ & 
$\delta_{\rm unc}^{h}$& 
$\delta_{\rm cor}$ & 
$\delta_{\rm cor}^{E^+}$ & 
$\delta_{\rm cor}^{\theta^+}$& 
$\delta_{\rm cor}^{h^+}$& 
$\delta_{\rm cor}^{N^+}$& 
$\delta_{\rm cor}^{B^+}$ \\ 
$(\rm GeV^2)$ & & & 
$(\%)$ & $(\%)$ & $(\%)$ & $(\%)$ & $(\%)$ & $(\%)$ & 
$(\%)$ & $(\%)$ & $(\%)$ & $(\%)$ & $(\%)$  
\\ \hline 
$120$ & $0.0020$ & $1.312$ & $1.73$ & $0.87$ & $1.02$ & $0.54$ & $0.09$ & $1.09$ & $-0.36$ & $-0.61$ & $0.02$ & $0.18$ & $-0.81$ \\ 
 
$120$ & $0.0032$ & $1.182$ & $1.89$ & $1.24$ & $1.21$ & $0.77$ & $0.05$ & $0.75$ & $-0.35$ & $-0.62$ & $0.02$ & $0.19$ & $-0.16$ \\ 
 
\hline 
$150$ & $0.0032$ & $1.195$ & $1.43$ & $0.73$ & $0.91$ & $0.43$ & $0.06$ & $0.81$ & $-0.32$ & $-0.59$ & $0.02$ & $0.19$ & $-0.41$ \\ 
 
$150$ & $0.0050$ & $1.071$ & $1.74$ & $0.88$ & $1.20$ & $0.86$ & $0.00$ & $0.90$ & $-0.55$ & $-0.71$ & $0.00$ & $0.04$ & $-0.04$ \\ 
 
$150$ & $0.0080$ & $0.9197$ & $2.54$ & $1.20$ & $1.85$ & $1.24$ & $1.02$ & $1.26$ & $-0.79$ & $-0.80$ & $-0.27$ & $-0.50$ & $-0.08$ \\ 
 
$150$ & $0.0130$ & $0.7984$ & $4.03$ & $1.68$ & $3.09$ & $2.78$ & $0.82$ & $1.96$ & $-1.61$ & $-0.64$ & $-0.28$ & $-0.85$ & $-0.09$ \\ 
 
\hline 
$200$ & $0.0032$ & $1.222$ & $1.81$ & $1.35$ & $0.94$ & $0.18$ & $0.07$ & $0.75$ & $0.18$ & $-0.50$ & $0.01$ & $0.15$ & $-0.51$ \\ 
 
$200$ & $0.0050$ & $1.079$ & $1.57$ & $0.96$ & $0.99$ & $0.54$ & $0.02$ & $0.75$ & $-0.44$ & $-0.58$ & $0.01$ & $0.14$ & $-0.08$ \\ 
 
$200$ & $0.0080$ & $0.9389$ & $1.93$ & $0.99$ & $1.39$ & $1.10$ & $0.00$ & $0.91$ & $-0.60$ & $-0.69$ & $0.00$ & $0.00$ & $0.00$ \\ 
 
$200$ & $0.0130$ & $0.7669$ & $1.62$ & $1.14$ & $0.89$ & $0.19$ & $0.03$ & $0.72$ & $-0.04$ & $-0.46$ & $-0.02$ & $0.55$ & $0.00$ \\ 
 
$200$ & $0.0200$ & $0.6800$ & $1.81$ & $1.23$ & $1.14$ & $0.68$ & $0.19$ & $0.70$ & $-0.47$ & $-0.39$ & $-0.08$ & $0.32$ & $-0.01$ \\ 
 
$200$ & $0.0320$ & $0.5735$ & $2.21$ & $1.38$ & $1.51$ & $1.07$ & $0.51$ & $0.85$ & $-0.60$ & $-0.56$ & $-0.19$ & $0.09$ & $0.00$ \\ 
 
$200$ & $0.0500$ & $0.5107$ & $2.97$ & $1.63$ & $1.78$ & $1.47$ & $0.02$ & $1.73$ & $-0.89$ & $-0.79$ & $-0.18$ & $1.24$ & $0.00$ \\ 
 
$200$ & $0.0800$ & $0.4341$ & $3.41$ & $1.73$ & $2.19$ & $1.90$ & $0.19$ & $1.96$ & $-1.13$ & $-0.86$ & $-0.07$ & $1.35$ & $-0.01$ \\ 
 
$200$ & $0.1300$ & $0.3521$ & $3.54$ & $2.09$ & $2.21$ & $1.36$ & $1.09$ & $1.81$ & $-0.80$ & $-1.02$ & $-0.27$ & $-1.23$ & $0.00$ \\ 
 
$200$ & $0.1800$ & $0.2987$ & $4.40$ & $2.71$ & $2.82$ & $1.17$ & $1.97$ & $2.01$ & $-0.83$ & $-1.23$ & $-0.49$ & $-1.26$ & $0.00$ \\ 
 
\hline 
$250$ & $0.0050$ & $1.096$ & $1.56$ & $1.12$ & $0.89$ & $0.30$ & $0.06$ & $0.60$ & $-0.29$ & $-0.42$ & $0.02$ & $0.18$ & $-0.27$ \\ 
 
$250$ & $0.0080$ & $0.9512$ & $1.67$ & $1.10$ & $1.01$ & $0.55$ & $0.00$ & $0.76$ & $-0.50$ & $-0.57$ & $0.00$ & $0.04$ & $0.00$ \\ 
 
$250$ & $0.0130$ & $0.8042$ & $2.08$ & $1.20$ & $1.25$ & $0.89$ & $0.18$ & $1.15$ & $0.50$ & $-0.66$ & $0.03$ & $0.79$ & $-0.02$ \\ 
 
$250$ & $0.0200$ & $0.6806$ & $2.10$ & $1.23$ & $1.31$ & $0.93$ & $0.26$ & $1.10$ & $0.35$ & $-0.57$ & $0.07$ & $0.87$ & $0.00$ \\ 
 
$250$ & $0.0320$ & $0.5813$ & $2.19$ & $1.30$ & $1.42$ & $1.10$ & $0.11$ & $1.06$ & $0.52$ & $-0.63$ & $-0.08$ & $0.67$ & $0.00$ \\ 
 
$250$ & $0.0500$ & $0.4968$ & $2.44$ & $1.48$ & $1.46$ & $1.12$ & $0.16$ & $1.27$ & $0.45$ & $-0.57$ & $-0.12$ & $1.03$ & $-0.01$ \\ 
 
$250$ & $0.0800$ & $0.4166$ & $2.94$ & $1.52$ & $1.11$ & $0.41$ & $0.34$ & $2.26$ & $0.18$ & $-0.45$ & $-0.02$ & $2.21$ & $0.00$ \\ 
 
$250$ & $0.1300$ & $0.3560$ & $2.55$ & $1.54$ & $1.82$ & $1.08$ & $0.76$ & $0.92$ & $0.51$ & $-0.56$ & $-0.22$ & $-0.46$ & $0.00$ \\ 
 
$250$ & $0.1800$ & $0.3035$ & $4.11$ & $2.11$ & $2.69$ & $1.59$ & $1.59$ & $2.28$ & $1.02$ & $-0.80$ & $-0.38$ & $-1.84$ & $0.00$ \\ 
 
\hline 
$300$ & $0.0050$ & $1.110$ & $2.23$ & $1.89$ & $1.00$ & $0.17$ & $0.08$ & $0.63$ & $-0.17$ & $-0.46$ & $0.03$ & $0.15$ & $-0.38$ \\ 
 
$300$ & $0.0080$ & $0.9626$ & $1.71$ & $1.28$ & $0.92$ & $0.35$ & $0.02$ & $0.65$ & $-0.35$ & $-0.53$ & $0.01$ & $0.10$ & $-0.08$ \\ 
 
$300$ & $0.0130$ & $0.8029$ & $1.97$ & $1.28$ & $1.18$ & $0.80$ & $0.00$ & $0.93$ & $-0.79$ & $-0.50$ & $0.00$ & $0.00$ & $0.00$ \\ 
 
$300$ & $0.0200$ & $0.6884$ & $2.00$ & $1.42$ & $1.07$ & $0.57$ & $0.17$ & $0.91$ & $0.43$ & $-0.53$ & $0.03$ & $0.61$ & $-0.02$ \\ 
 
$300$ & $0.0320$ & $0.5748$ & $2.15$ & $1.50$ & $1.20$ & $0.79$ & $0.03$ & $0.97$ & $0.46$ & $-0.55$ & $-0.03$ & $0.65$ & $0.00$ \\ 
 
$300$ & $0.0500$ & $0.4892$ & $2.57$ & $1.62$ & $1.45$ & $1.10$ & $0.18$ & $1.38$ & $0.60$ & $-0.66$ & $-0.10$ & $1.05$ & $0.00$ \\ 
 
$300$ & $0.0800$ & $0.4164$ & $3.12$ & $1.72$ & $1.48$ & $1.09$ & $0.18$ & $2.14$ & $0.47$ & $-0.69$ & $-0.06$ & $1.97$ & $-0.01$ \\ 
 
$300$ & $0.1300$ & $0.3547$ & $2.76$ & $1.71$ & $1.89$ & $1.35$ & $0.42$ & $1.07$ & $0.66$ & $-0.69$ & $-0.17$ & $0.45$ & $0.00$ \\ 
 
$300$ & $0.1800$ & $0.2961$ & $4.67$ & $2.26$ & $3.02$ & $2.14$ & $1.53$ & $2.76$ & $1.31$ & $-0.88$ & $-0.36$ & $-2.23$ & $0.00$ \\ 
 
$300$ & $0.4000$ & $0.1439$ & $6.67$ & $2.75$ & $3.63$ & $2.38$ & $1.99$ & $4.88$ & $1.39$ & $-1.03$ & $-0.41$ & $-4.55$ & $0.00$ \\ 
 
\hline 
$400$ & $0.0080$ & $1.025$ & $1.91$ & $1.54$ & $0.96$ & $0.38$ & $0.04$ & $0.59$ & $-0.33$ & $-0.41$ & $0.01$ & $0.16$ & $-0.22$ \\ 
 
$400$ & $0.0130$ & $0.8439$ & $1.93$ & $1.50$ & $0.97$ & $0.42$ & $0.00$ & $0.72$ & $-0.41$ & $-0.59$ & $0.00$ & $0.03$ & $-0.02$ \\ 
 
$400$ & $0.0200$ & $0.7106$ & $2.25$ & $1.54$ & $1.28$ & $0.91$ & $0.00$ & $1.03$ & $-0.90$ & $-0.50$ & $0.00$ & $0.00$ & $0.00$ \\ 
 
$400$ & $0.0320$ & $0.5984$ & $2.15$ & $1.63$ & $1.02$ & $0.46$ & $0.09$ & $0.95$ & $0.45$ & $-0.48$ & $-0.02$ & $0.69$ & $-0.01$ \\ 
 
$400$ & $0.0500$ & $0.4846$ & $2.32$ & $1.84$ & $1.12$ & $0.51$ & $0.34$ & $0.85$ & $0.50$ & $-0.36$ & $-0.13$ & $0.58$ & $-0.01$ \\ 
 
$400$ & $0.0800$ & $0.4188$ & $2.74$ & $1.91$ & $1.05$ & $0.27$ & $0.23$ & $1.67$ & $0.22$ & $-0.33$ & $0.05$ & $1.62$ & $0.00$ \\ 
 
$400$ & $0.1300$ & $0.3598$ & $2.65$ & $1.93$ & $1.42$ & $0.64$ & $0.16$ & $1.12$ & $0.55$ & $-0.49$ & $-0.10$ & $0.85$ & $0.00$ \\ 
 
$400$ & $0.1800$ & $0.2991$ & $4.47$ & $2.43$ & $2.57$ & $1.14$ & $1.78$ & $2.73$ & $1.06$ & $-0.54$ & $-0.48$ & $-2.41$ & $0.00$ \\ 
 
$400$ & $0.4000$ & $0.1438$ & $7.28$ & $3.09$ & $3.24$ & $1.27$ & $2.22$ & $5.74$ & $1.18$ & $-0.56$ & $-0.31$ & $-5.58$ & $0.00$ \\ 
 
\hline 
$500$ & $0.0080$ & $0.9879$ & $2.83$ & $2.57$ & $1.05$ & $0.16$ & $0.07$ & $0.52$ & $-0.24$ & $-0.39$ & $0.02$ & $0.16$ & $-0.20$ \\ 
 
$500$ & $0.0130$ & $0.8903$ & $2.15$ & $1.85$ & $0.96$ & $0.31$ & $0.03$ & $0.56$ & $-0.25$ & $-0.49$ & $0.00$ & $0.08$ & $-0.06$ \\ 
 
$500$ & $0.0200$ & $0.7270$ & $2.31$ & $1.83$ & $1.15$ & $0.69$ & $0.00$ & $0.80$ & $-0.69$ & $-0.40$ & $0.00$ & $0.00$ & $0.00$ \\ 
 
$500$ & $0.0320$ & $0.6232$ & $2.35$ & $1.87$ & $1.09$ & $0.50$ & $0.25$ & $0.91$ & $0.52$ & $-0.38$ & $-0.03$ & $0.65$ & $0.00$ \\ 
 
$500$ & $0.0500$ & $0.5411$ & $2.44$ & $1.99$ & $1.12$ & $0.56$ & $0.18$ & $0.84$ & $0.56$ & $-0.32$ & $-0.06$ & $0.54$ & $0.00$ \\ 
 
$500$ & $0.0800$ & $0.4169$ & $2.96$ & $2.27$ & $1.17$ & $0.60$ & $0.07$ & $1.49$ & $0.60$ & $-0.52$ & $-0.02$ & $1.26$ & $0.00$ \\ 
 
$500$ & $0.1300$ & $0.3658$ & $3.21$ & $2.54$ & $1.38$ & $0.44$ & $0.07$ & $1.41$ & $0.44$ & $-0.38$ & $-0.11$ & $1.28$ & $0.00$ \\ 
 
$500$ & $0.1800$ & $0.3299$ & $3.63$ & $2.86$ & $1.95$ & $0.90$ & $0.92$ & $1.11$ & $0.90$ & $-0.53$ & $-0.30$ & $-0.18$ & $0.00$ \\ 
 
$500$ & $0.2500$ & $0.2529$ & $5.34$ & $3.32$ & $2.54$ & $0.99$ & $1.69$ & $3.32$ & $0.99$ & $-0.53$ & $-0.44$ & $-3.09$ & $0.00$ \\ 
 
\hline 
$650$ & $0.0130$ & $0.8832$ & $2.38$ & $2.08$ & $1.06$ & $0.48$ & $0.05$ & $0.49$ & $-0.32$ & $-0.31$ & $0.01$ & $0.14$ & $-0.13$ \\ 
 
$650$ & $0.0200$ & $0.7582$ & $2.48$ & $2.14$ & $1.06$ & $0.41$ & $0.00$ & $0.70$ & $-0.47$ & $-0.52$ & $0.00$ & $0.03$ & $0.00$ \\ 
 
$650$ & $0.0320$ & $0.6334$ & $2.84$ & $2.23$ & $1.43$ & $1.01$ & $0.00$ & $1.04$ & $-0.94$ & $-0.45$ & $0.00$ & $0.00$ & $0.00$ \\ 
 
$650$ & $0.0500$ & $0.5229$ & $2.82$ & $2.35$ & $1.24$ & $0.68$ & $0.15$ & $0.94$ & $0.75$ & $-0.43$ & $-0.12$ & $0.34$ & $0.00$ \\ 
 
$650$ & $0.0800$ & $0.4300$ & $3.14$ & $2.66$ & $1.16$ & $0.41$ & $0.08$ & $1.22$ & $0.47$ & $-0.30$ & $0.07$ & $1.08$ & $0.00$ \\ 
 
$650$ & $0.1300$ & $0.3599$ & $3.64$ & $2.94$ & $1.55$ & $0.70$ & $0.21$ & $1.48$ & $0.72$ & $-0.51$ & $-0.06$ & $1.19$ & $0.00$ \\ 
 
$650$ & $0.1800$ & $0.3140$ & $3.77$ & $3.18$ & $1.78$ & $0.79$ & $0.46$ & $0.93$ & $0.82$ & $-0.37$ & $-0.17$ & $0.17$ & $0.00$ \\ 
 
$650$ & $0.2500$ & $0.2471$ & $5.56$ & $4.13$ & $2.44$ & $1.09$ & $1.40$ & $2.82$ & $1.06$ & $-0.44$ & $-0.34$ & $-2.55$ & $0.00$ \\ 
 
$650$ & $0.4000$ & $0.1222$ & $8.53$ & $6.14$ & $3.78$ & $1.65$ & $2.42$ & $4.56$ & $1.67$ & $-0.71$ & $-0.63$ & $-4.13$ & $0.00$ \\ 
 
\hline 

\end{tabular} 
\end{center} 
\caption[RESULT] 
{\label{tab:ncdxdq2_eleLH} The NC 
$e^-p$ reduced cross section $\tilde{\sigma}_{\rm NC}(x,Q^2)$ 
with lepton beam polarisation $P_e=-25.8$\% 
with statistical 
$(\delta_{\rm stat})$, 
total $(\delta_{tot})$, 
total uncorrelated systematic $(\delta_{\rm unc})$ 
errors, two of its contributions from the 
 electron energy error ($\delta_{unc}^{E}$)  
and the hadronic energy error  
($\delta_{\rm unc}^{h}$). 
The effect of the other uncorrelated 
systematic errors is included in $\delta_{\rm unc}$. 
In addition the correlated systematic  
$(\delta_{\rm cor})$ and its contributions from a 
positive variation of one  
standard deviation of the 
electron energy error ($\delta_{cor}^{E^+}$), of 
the polar electron angle error 
($\delta_{\rm cor}^{\theta^+}$), of the hadronic 
energy error ($\delta_{\rm cor}^{h^+}$), of the error 
due to noise subtraction ($\delta_{\rm cor}^{N^+}$) 
and of the error due to background subtraction 
($\delta_{\rm cor}^{B^+}$) are given. 
The normalisation and polarisation uncertainties are 
not included in the errors. 
The table continues on the next page.}
\end{table} 
\begin{table}[htbp] 
\begin{center} 
\tiny 
\begin{tabular}{|r|c|r|r|r|r|r|r|r|r|r|r|r|r|} 
\hline 
$Q^2$ & $x$ & $\tilde{\sigma}_{\rm NC}$ & 
$\delta_{\rm tot}$ & $\delta_{\rm stat}$ & $\delta_{\rm unc}$ & 
$\delta_{\rm unc}^{E}$ & 
$\delta_{\rm unc}^{h}$& 
$\delta_{\rm cor}$ & 
$\delta_{\rm cor}^{E^+}$ & 
$\delta_{\rm cor}^{\theta^+}$& 
$\delta_{\rm cor}^{h^+}$& 
$\delta_{\rm cor}^{N^+}$& 
$\delta_{\rm cor}^{B^+}$ \\ 
$(\rm GeV^2)$ & & & 
$(\%)$ & $(\%)$ & $(\%)$ & $(\%)$ & $(\%)$ & $(\%)$ & 
$(\%)$ & $(\%)$ & $(\%)$ & $(\%)$ & $(\%)$  
\\ \hline 
$800$ & $0.0130$ & $0.9026$ & $3.72$ & $3.50$ & $1.21$ & $0.16$ & $0.08$ & $0.41$ & $0.11$ & $-0.33$ & $0.03$ & $0.19$ & $-0.10$ \\ 
 
$800$ & $0.0200$ & $0.7206$ & $2.78$ & $2.51$ & $1.11$ & $0.46$ & $0.01$ & $0.45$ & $-0.25$ & $-0.36$ & $0.00$ & $0.08$ & $-0.06$ \\ 
 
$800$ & $0.0320$ & $0.6195$ & $3.12$ & $2.67$ & $1.40$ & $0.90$ & $0.00$ & $0.80$ & $-0.59$ & $-0.53$ & $0.00$ & $0.00$ & $0.00$ \\ 
 
$800$ & $0.0500$ & $0.5387$ & $3.04$ & $2.74$ & $1.14$ & $0.26$ & $0.08$ & $0.63$ & $0.38$ & $-0.26$ & $-0.06$ & $0.42$ & $-0.02$ \\ 
 
$800$ & $0.0800$ & $0.4321$ & $3.48$ & $3.04$ & $1.29$ & $0.45$ & $0.28$ & $1.09$ & $0.62$ & $-0.46$ & $-0.05$ & $0.77$ & $0.00$ \\ 
 
$800$ & $0.1300$ & $0.3392$ & $4.15$ & $3.58$ & $1.55$ & $0.43$ & $0.26$ & $1.42$ & $0.63$ & $-0.37$ & $-0.07$ & $1.21$ & $0.00$ \\ 
 
$800$ & $0.1800$ & $0.3127$ & $4.27$ & $3.75$ & $1.84$ & $0.59$ & $0.64$ & $0.87$ & $0.75$ & $-0.37$ & $-0.18$ & $0.16$ & $0.00$ \\ 
 
$800$ & $0.2500$ & $0.2409$ & $5.39$ & $4.63$ & $2.17$ & $0.59$ & $1.05$ & $1.72$ & $0.77$ & $-0.30$ & $-0.23$ & $-1.48$ & $0.00$ \\ 
 
$800$ & $0.4000$ & $0.1340$ & $9.10$ & $5.93$ & $4.25$ & $1.68$ & $2.96$ & $5.44$ & $1.78$ & $-0.71$ & $-0.76$ & $-5.03$ & $0.00$ \\ 
 
\hline 
$1000$ & $0.0130$ & $0.8430$ & $4.06$ & $3.45$ & $1.68$ & $0.27$ & $0.22$ & $1.34$ & $-0.03$ & $-0.20$ & $0.05$ & $0.19$ & $-1.31$ \\ 
 
$1000$ & $0.0200$ & $0.7686$ & $3.10$ & $2.87$ & $1.07$ & $0.32$ & $0.06$ & $0.45$ & $-0.18$ & $-0.37$ & $0.01$ & $0.11$ & $-0.12$ \\ 
 
$1000$ & $0.0320$ & $0.6579$ & $3.10$ & $2.82$ & $1.17$ & $0.57$ & $0.00$ & $0.50$ & $-0.26$ & $-0.43$ & $0.00$ & $0.01$ & $0.00$ \\ 
 
$1000$ & $0.0500$ & $0.5030$ & $3.67$ & $3.15$ & $1.59$ & $1.17$ & $0.00$ & $0.99$ & $-0.71$ & $-0.68$ & $0.00$ & $0.00$ & $0.00$ \\ 
 
$1000$ & $0.0800$ & $0.4311$ & $3.71$ & $3.43$ & $1.14$ & $0.05$ & $0.18$ & $0.84$ & $0.17$ & $-0.23$ & $0.05$ & $0.78$ & $0.00$ \\ 
 
$1000$ & $0.1300$ & $0.3308$ & $4.65$ & $4.21$ & $1.52$ & $0.29$ & $0.28$ & $1.26$ & $0.40$ & $-0.33$ & $-0.17$ & $1.14$ & $0.00$ \\ 
 
$1000$ & $0.1800$ & $0.3587$ & $4.48$ & $3.98$ & $1.82$ & $0.72$ & $0.53$ & $0.96$ & $0.73$ & $-0.43$ & $-0.25$ & $0.37$ & $0.00$ \\ 
 
$1000$ & $0.2500$ & $0.2591$ & $5.40$ & $4.61$ & $2.25$ & $0.90$ & $1.12$ & $1.68$ & $1.01$ & $-0.45$ & $-0.34$ & $-1.22$ & $0.00$ \\ 
 
$1000$ & $0.4000$ & $0.1266$ & $9.18$ & $6.56$ & $4.16$ & $1.84$ & $2.83$ & $4.89$ & $1.55$ & $-0.48$ & $-0.56$ & $-4.58$ & $0.00$ \\ 
 
\hline 
$1200$ & $0.0130$ & $0.9167$ & $6.04$ & $5.43$ & $2.21$ & $0.21$ & $0.27$ & $1.48$ & $0.35$ & $0.22$ & $0.05$ & $0.22$ & $-1.41$ \\ 
 
$1200$ & $0.0200$ & $0.7663$ & $3.83$ & $3.60$ & $1.22$ & $0.29$ & $0.09$ & $0.46$ & $-0.19$ & $-0.26$ & $0.02$ & $0.12$ & $-0.31$ \\ 
 
$1200$ & $0.0320$ & $0.6765$ & $3.45$ & $3.27$ & $0.98$ & $0.27$ & $0.03$ & $0.52$ & $-0.21$ & $-0.47$ & $0.00$ & $0.06$ & $-0.01$ \\ 
 
$1200$ & $0.0500$ & $0.5310$ & $3.76$ & $3.48$ & $1.27$ & $0.82$ & $0.00$ & $0.62$ & $-0.45$ & $-0.42$ & $0.00$ & $0.00$ & $0.00$ \\ 
 
$1200$ & $0.0800$ & $0.4462$ & $4.05$ & $3.73$ & $1.38$ & $0.93$ & $0.19$ & $0.77$ & $0.58$ & $-0.27$ & $-0.15$ & $0.40$ & $0.00$ \\ 
 
$1200$ & $0.1300$ & $0.3503$ & $5.66$ & $5.38$ & $1.51$ & $0.73$ & $0.09$ & $0.90$ & $0.43$ & $-0.14$ & $-0.02$ & $0.78$ & $0.00$ \\ 
 
$1200$ & $0.1800$ & $0.3217$ & $5.03$ & $4.65$ & $1.71$ & $0.83$ & $0.34$ & $0.87$ & $0.63$ & $-0.35$ & $-0.15$ & $0.47$ & $0.00$ \\ 
 
$1200$ & $0.2500$ & $0.2147$ & $6.06$ & $5.58$ & $2.16$ & $1.27$ & $0.74$ & $0.93$ & $0.82$ & $-0.27$ & $-0.19$ & $-0.29$ & $0.00$ \\ 
 
$1200$ & $0.4000$ & $0.1230$ & $9.42$ & $7.08$ & $4.36$ & $2.61$ & $2.68$ & $4.42$ & $1.58$ & $-0.33$ & $-0.62$ & $-4.06$ & $0.00$ \\ 
 
\hline 
$1500$ & $0.0200$ & $0.8069$ & $4.76$ & $4.26$ & $1.78$ & $0.09$ & $0.16$ & $1.18$ & $0.18$ & $-0.24$ & $0.04$ & $0.15$ & $-1.13$ \\ 
 
$1500$ & $0.0320$ & $0.6724$ & $4.24$ & $4.06$ & $1.12$ & $0.40$ & $0.06$ & $0.48$ & $-0.38$ & $-0.25$ & $0.01$ & $0.13$ & $-0.06$ \\ 
 
$1500$ & $0.0500$ & $0.5470$ & $4.25$ & $4.02$ & $1.25$ & $0.74$ & $0.00$ & $0.54$ & $-0.34$ & $-0.42$ & $0.00$ & $0.01$ & $0.00$ \\ 
 
$1500$ & $0.0800$ & $0.4986$ & $4.29$ & $4.05$ & $1.23$ & $0.64$ & $0.13$ & $0.70$ & $0.42$ & $-0.17$ & $-0.01$ & $0.53$ & $0.00$ \\ 
 
$1500$ & $0.1300$ & $0.3514$ & $5.63$ & $5.25$ & $1.79$ & $1.12$ & $0.18$ & $0.92$ & $0.72$ & $-0.27$ & $-0.13$ & $0.49$ & $0.00$ \\ 
 
$1500$ & $0.1800$ & $0.3066$ & $5.83$ & $5.42$ & $1.89$ & $1.13$ & $0.23$ & $1.01$ & $0.60$ & $-0.11$ & $-0.05$ & $0.80$ & $0.00$ \\ 
 
$1500$ & $0.2500$ & $0.2297$ & $6.52$ & $6.05$ & $2.24$ & $1.29$ & $0.85$ & $0.90$ & $0.68$ & $-0.18$ & $-0.25$ & $-0.51$ & $0.00$ \\ 
 
$1500$ & $0.4000$ & $0.1353$ & $10.32$ & $8.82$ & $4.02$ & $2.35$ & $2.33$ & $3.54$ & $1.51$ & $-0.30$ & $-0.61$ & $-3.13$ & $0.00$ \\ 
 
$1500$ & $0.6500$ & $0.01469$ & $19.63$ & $14.78$ & $6.93$ & $4.41$ & $4.39$ & $10.90$ & $3.27$ & $-0.40$ & $-0.95$ & $-10.34$ & $0.00$ \\ 
 
\hline 
$2000$ & $0.0219$ & $0.8967$ & $7.26$ & $6.58$ & $2.49$ & $0.20$ & $0.18$ & $1.79$ & $0.18$ & $-0.20$ & $0.03$ & $0.19$ & $-1.76$ \\ 
 
$2000$ & $0.0320$ & $0.6323$ & $5.13$ & $4.89$ & $1.50$ & $0.20$ & $0.10$ & $0.49$ & $-0.16$ & $-0.41$ & $0.03$ & $0.15$ & $-0.15$ \\ 
 
$2000$ & $0.0500$ & $0.5475$ & $5.03$ & $4.87$ & $1.19$ & $0.43$ & $0.02$ & $0.42$ & $-0.25$ & $-0.33$ & $0.01$ & $0.04$ & $-0.02$ \\ 
 
$2000$ & $0.0800$ & $0.4363$ & $5.28$ & $5.02$ & $1.43$ & $0.84$ & $0.00$ & $0.79$ & $-0.64$ & $-0.46$ & $0.00$ & $0.00$ & $0.00$ \\ 
 
$2000$ & $0.1300$ & $0.3653$ & $6.25$ & $5.98$ & $1.64$ & $0.72$ & $0.06$ & $0.80$ & $0.48$ & $-0.24$ & $0.04$ & $0.59$ & $0.00$ \\ 
 
$2000$ & $0.1800$ & $0.2969$ & $6.95$ & $6.52$ & $2.17$ & $1.45$ & $0.23$ & $1.01$ & $0.79$ & $-0.26$ & $-0.16$ & $0.54$ & $0.00$ \\ 
 
$2000$ & $0.2500$ & $0.2483$ & $7.08$ & $6.68$ & $2.20$ & $1.28$ & $0.56$ & $0.81$ & $0.67$ & $-0.30$ & $-0.13$ & $-0.31$ & $0.00$ \\ 
 
$2000$ & $0.4000$ & $0.1249$ & $9.82$ & $8.56$ & $3.87$ & $2.33$ & $1.97$ & $2.88$ & $1.36$ & $-0.13$ & $-0.55$ & $-2.47$ & $0.00$ \\ 
 
$2000$ & $0.6500$ & $0.01062$ & $23.74$ & $19.67$ & $8.15$ & $5.31$ & $5.23$ & $10.50$ & $2.50$ & $-0.39$ & $-1.12$ & $-10.13$ & $0.00$ \\ 
 
\hline 
$3000$ & $0.0320$ & $0.7671$ & $4.82$ & $4.41$ & $1.83$ & $0.25$ & $0.11$ & $0.67$ & $-0.14$ & $-0.30$ & $0.02$ & $0.11$ & $-0.57$ \\ 
 
$3000$ & $0.0500$ & $0.5870$ & $4.27$ & $4.01$ & $1.40$ & $0.15$ & $0.05$ & $0.37$ & $-0.11$ & $-0.34$ & $0.02$ & $0.08$ & $-0.05$ \\ 
 
$3000$ & $0.0800$ & $0.4900$ & $4.59$ & $4.37$ & $1.33$ & $0.55$ & $0.00$ & $0.51$ & $-0.33$ & $-0.39$ & $0.00$ & $0.01$ & $0.00$ \\ 
 
$3000$ & $0.1300$ & $0.4137$ & $5.49$ & $5.17$ & $1.72$ & $0.85$ & $0.00$ & $0.61$ & $-0.50$ & $-0.34$ & $0.00$ & $0.00$ & $0.00$ \\ 
 
$3000$ & $0.1800$ & $0.2885$ & $6.48$ & $6.14$ & $1.87$ & $1.02$ & $0.19$ & $0.83$ & $0.55$ & $-0.11$ & $-0.06$ & $0.61$ & $0.00$ \\ 
 
$3000$ & $0.2500$ & $0.2131$ & $6.96$ & $6.55$ & $2.16$ & $1.36$ & $0.40$ & $0.96$ & $0.86$ & $-0.06$ & $-0.15$ & $-0.38$ & $0.00$ \\ 
 
$3000$ & $0.4000$ & $0.1245$ & $8.85$ & $7.49$ & $4.03$ & $2.77$ & $1.83$ & $2.44$ & $1.71$ & $-0.18$ & $-0.40$ & $-1.69$ & $0.00$ \\ 
 
$3000$ & $0.6500$ & $0.01303$ & $18.15$ & $14.62$ & $6.92$ & $4.13$ & $4.71$ & $8.23$ & $2.69$ & $-0.28$ & $-1.38$ & $-7.65$ & $0.00$ \\ 
 
\hline 
$5000$ & $0.0547$ & $0.6559$ & $6.35$ & $5.98$ & $2.02$ & $0.18$ & $0.14$ & $0.72$ & $0.14$ & $-0.38$ & $0.03$ & $0.13$ & $-0.58$ \\ 
 
$5000$ & $0.0800$ & $0.5540$ & $4.92$ & $4.65$ & $1.57$ & $0.16$ & $0.05$ & $0.35$ & $-0.19$ & $-0.27$ & $0.00$ & $0.06$ & $-0.08$ \\ 
 
$5000$ & $0.1300$ & $0.4827$ & $5.55$ & $5.23$ & $1.80$ & $0.33$ & $0.01$ & $0.46$ & $0.23$ & $-0.39$ & $0.00$ & $0.01$ & $0.00$ \\ 
 
$5000$ & $0.1800$ & $0.3772$ & $6.42$ & $6.13$ & $1.87$ & $0.26$ & $0.00$ & $0.37$ & $-0.11$ & $-0.35$ & $0.00$ & $0.00$ & $0.00$ \\ 
 
$5000$ & $0.2500$ & $0.2235$ & $8.37$ & $8.02$ & $2.25$ & $0.96$ & $0.00$ & $0.79$ & $0.77$ & $-0.19$ & $0.00$ & $0.00$ & $0.00$ \\ 
 
$5000$ & $0.4000$ & $0.1051$ & $10.63$ & $9.88$ & $3.51$ & $1.93$ & $1.48$ & $1.80$ & $1.41$ & $-0.09$ & $-0.54$ & $-0.99$ & $0.00$ \\ 
 
$5000$ & $0.6500$ & $0.01439$ & $18.84$ & $16.48$ & $7.19$ & $4.40$ & $4.59$ & $5.64$ & $2.53$ & $0.24$ & $-1.02$ & $-4.93$ & $0.00$ \\ 
 
\hline 
$8000$ & $0.0875$ & $0.6428$ & $9.35$ & $8.89$ & $2.72$ & $0.56$ & $0.06$ & $1.01$ & $0.26$ & $-0.33$ & $0.02$ & $0.09$ & $-0.92$ \\ 
 
$8000$ & $0.1300$ & $0.5263$ & $7.49$ & $7.10$ & $2.35$ & $0.27$ & $0.02$ & $0.45$ & $-0.20$ & $-0.31$ & $0.01$ & $0.05$ & $-0.26$ \\ 
 
$8000$ & $0.1800$ & $0.3753$ & $8.33$ & $8.01$ & $2.27$ & $0.17$ & $0.03$ & $0.39$ & $-0.09$ & $-0.38$ & $-0.01$ & $0.02$ & $0.00$ \\ 
 
$8000$ & $0.2500$ & $0.2636$ & $9.47$ & $9.07$ & $2.64$ & $1.14$ & $0.00$ & $0.60$ & $0.60$ & $0.06$ & $0.00$ & $0.00$ & $0.00$ \\ 
 
$8000$ & $0.4000$ & $0.1164$ & $13.53$ & $12.62$ & $4.43$ & $3.21$ & $0.00$ & $2.07$ & $1.98$ & $0.60$ & $0.00$ & $0.00$ & $0.00$ \\ 
 
$8000$ & $0.6500$ & $0.01037$ & $23.03$ & $21.89$ & $6.35$ & $3.71$ & $3.52$ & $3.31$ & $2.15$ & $0.36$ & $-0.82$ & $-2.36$ & $0.00$ \\ 
 
\hline 
$12000$ & $0.1300$ & $0.7243$ & $16.19$ & $15.45$ & $4.58$ & $0.69$ & $0.15$ & $1.48$ & $0.28$ & $-0.26$ & $0.04$ & $0.13$ & $-1.42$ \\ 
 
$12000$ & $0.1800$ & $0.5335$ & $9.86$ & $9.59$ & $2.25$ & $0.11$ & $0.07$ & $0.39$ & $-0.06$ & $-0.32$ & $0.03$ & $0.09$ & $-0.19$ \\ 
 
$12000$ & $0.2500$ & $0.3098$ & $11.43$ & $11.15$ & $2.49$ & $0.98$ & $0.02$ & $0.52$ & $0.49$ & $-0.15$ & $0.01$ & $0.03$ & $0.00$ \\ 
 
$12000$ & $0.4000$ & $0.2097$ & $13.47$ & $12.42$ & $4.94$ & $4.03$ & $0.00$ & $1.68$ & $1.58$ & $0.58$ & $0.00$ & $0.00$ & $0.00$ \\ 
 
$12000$ & $0.6500$ & $0.01439$ & $28.78$ & $27.80$ & $6.56$ & $4.81$ & $3.13$ & $3.56$ & $2.71$ & $0.51$ & $-0.95$ & $-2.05$ & $0.00$ \\ 
 
\hline 
$20000$ & $0.2500$ & $0.5939$ & $13.71$ & $13.34$ & $2.94$ & $1.16$ & $0.08$ & $1.17$ & $0.39$ & $-0.17$ & $0.03$ & $0.05$ & $-1.09$ \\ 
 
$20000$ & $0.4000$ & $0.2141$ & $17.21$ & $16.55$ & $4.59$ & $3.54$ & $0.03$ & $1.11$ & $1.09$ & $0.12$ & $0.01$ & $0.03$ & $-0.16$ \\ 
 
$20000$ & $0.6500$ & $0.01850$ & $44.37$ & $40.89$ & $16.39$ & $16.00$ & $0.00$ & $5.30$ & $4.61$ & $2.63$ & $0.00$ & $0.00$ & $0.00$ \\ 
 
\hline 
$30000$ & $0.4000$ & $0.1671$ & $36.74$ & $36.01$ & $7.03$ & $3.51$ & $0.15$ & $1.93$ & $1.04$ & $-0.15$ & $0.03$ & $0.11$ & $-1.62$ \\ 
 
$30000$ & $0.6500$ & $0.04155$ & $39.86$ & $37.83$ & $11.89$ & $11.21$ & $0.00$ & $3.97$ & $3.49$ & $1.90$ & $0.00$ & $0.01$ & $0.00$ \\ 
 
\hline 
$50000$ & $0.6500$ & $0.1133$ & $58.72$ & $57.78$ & $10.19$ & $7.43$ & $0.00$ & $2.38$ & $2.03$ & $1.25$ & $0.00$ & $0.00$ & $0.00$ \\ 
 
\hline 
\end{tabular} 
\end{center} 
\captcont{continued.} 
\end{table} 

\begin{table}[htbp] 
\begin{center} 
\tiny 
\begin{tabular}{|r|c|r|r|r|r|r|r|r|r|r|r|r|r|} 
\hline 
$Q^2$ & $x$ & $\tilde{\sigma}_{\rm NC}$ & 
$\delta_{\rm tot}$ & $\delta_{\rm stat}$ & $\delta_{\rm unc}$ & 
$\delta_{\rm unc}^{E}$ & 
$\delta_{\rm unc}^{h}$& 
$\delta_{\rm cor}$ & 
$\delta_{\rm cor}^{E^+}$ & 
$\delta_{\rm cor}^{\theta^+}$& 
$\delta_{\rm cor}^{h^+}$& 
$\delta_{\rm cor}^{N^+}$& 
$\delta_{\rm cor}^{B^+}$ \\ 
$(\rm GeV^2)$ & & & 
$(\%)$ & $(\%)$ & $(\%)$ & $(\%)$ & $(\%)$ & $(\%)$ & 
$(\%)$ & $(\%)$ & $(\%)$ & $(\%)$ & $(\%)$  
\\ \hline 
$120$ & $0.0020$ & $1.316$ & $1.92$ & $1.29$ & $0.95$ & $0.38$ & $0.09$ & $1.05$ & $-0.24$ & $-0.67$ & $0.02$ & $0.18$ & $-0.75$ \\ 
 
$120$ & $0.0032$ & $1.190$ & $2.32$ & $1.78$ & $1.24$ & $0.76$ & $0.05$ & $0.83$ & $-0.40$ & $-0.68$ & $0.02$ & $0.22$ & $-0.16$ \\ 
 
\hline 
$150$ & $0.0032$ & $1.185$ & $1.62$ & $1.09$ & $0.89$ & $0.37$ & $0.07$ & $0.80$ & $-0.30$ & $-0.58$ & $0.02$ & $0.20$ & $-0.41$ \\ 
 
$150$ & $0.0050$ & $1.083$ & $2.08$ & $1.29$ & $1.30$ & $0.99$ & $0.01$ & $0.98$ & $-0.56$ & $-0.81$ & $0.00$ & $0.05$ & $-0.03$ \\ 
 
$150$ & $0.0080$ & $0.9357$ & $2.88$ & $1.78$ & $1.85$ & $1.35$ & $0.86$ & $1.31$ & $-0.86$ & $-0.86$ & $-0.30$ & $-0.38$ & $-0.06$ \\ 
 
$150$ & $0.0130$ & $0.7694$ & $4.23$ & $2.42$ & $2.90$ & $2.48$ & $1.06$ & $1.90$ & $-1.43$ & $-0.96$ & $-0.33$ & $-0.74$ & $-0.10$ \\ 
 
\hline 
$200$ & $0.0032$ & $1.166$ & $2.38$ & $2.06$ & $0.95$ & $0.14$ & $0.07$ & $0.72$ & $-0.14$ & $-0.52$ & $0.02$ & $0.16$ & $-0.46$ \\ 
 
$200$ & $0.0050$ & $1.071$ & $1.91$ & $1.46$ & $0.96$ & $0.48$ & $0.02$ & $0.77$ & $-0.39$ & $-0.65$ & $0.00$ & $0.12$ & $-0.09$ \\ 
 
$200$ & $0.0080$ & $0.9308$ & $2.19$ & $1.44$ & $1.34$ & $1.03$ & $0.00$ & $0.97$ & $-0.65$ & $-0.72$ & $0.00$ & $0.00$ & $0.00$ \\ 
 
$200$ & $0.0130$ & $0.7788$ & $2.04$ & $1.67$ & $0.89$ & $0.08$ & $0.05$ & $0.76$ & $-0.09$ & $-0.41$ & $-0.07$ & $0.62$ & $-0.03$ \\ 
 
$200$ & $0.0200$ & $0.6780$ & $2.22$ & $1.81$ & $1.05$ & $0.52$ & $0.04$ & $0.75$ & $-0.43$ & $-0.46$ & $-0.04$ & $0.40$ & $-0.01$ \\ 
 
$200$ & $0.0320$ & $0.5524$ & $2.72$ & $2.12$ & $1.36$ & $0.79$ & $0.57$ & $1.01$ & $-0.39$ & $-0.91$ & $-0.18$ & $-0.12$ & $-0.01$ \\ 
 
$200$ & $0.0500$ & $0.5222$ & $3.39$ & $2.48$ & $1.78$ & $1.44$ & $0.08$ & $1.47$ & $-0.81$ & $-0.48$ & $-0.10$ & $1.13$ & $0.00$ \\ 
 
$200$ & $0.0800$ & $0.4227$ & $3.94$ & $2.51$ & $2.30$ & $2.01$ & $0.20$ & $1.99$ & $-1.18$ & $-0.53$ & $-0.12$ & $1.50$ & $0.00$ \\ 
 
$200$ & $0.1300$ & $0.3548$ & $3.91$ & $2.84$ & $2.21$ & $1.32$ & $1.14$ & $1.52$ & $-0.84$ & $-0.88$ & $-0.30$ & $-0.86$ & $0.00$ \\ 
 
$200$ & $0.1800$ & $0.3054$ & $5.42$ & $4.12$ & $2.83$ & $1.44$ & $1.77$ & $2.10$ & $-0.96$ & $-0.75$ & $-0.31$ & $-1.69$ & $0.00$ \\ 
 
\hline 
$250$ & $0.0050$ & $1.079$ & $2.03$ & $1.69$ & $0.94$ & $0.39$ & $0.07$ & $0.62$ & $-0.39$ & $-0.34$ & $0.01$ & $0.17$ & $-0.28$ \\ 
 
$250$ & $0.0080$ & $0.9100$ & $2.15$ & $1.64$ & $1.09$ & $0.68$ & $0.00$ & $0.86$ & $-0.66$ & $-0.55$ & $0.00$ & $0.02$ & $-0.03$ \\ 
 
$250$ & $0.0130$ & $0.7825$ & $2.51$ & $1.80$ & $1.32$ & $0.94$ & $0.26$ & $1.15$ & $0.47$ & $-0.65$ & $0.04$ & $0.83$ & $-0.03$ \\ 
 
$250$ & $0.0200$ & $0.6562$ & $2.44$ & $1.86$ & $1.25$ & $0.85$ & $0.18$ & $0.98$ & $0.30$ & $-0.46$ & $0.05$ & $0.81$ & $-0.03$ \\ 
 
$250$ & $0.0320$ & $0.5549$ & $2.57$ & $1.95$ & $1.34$ & $0.99$ & $0.08$ & $0.98$ & $0.43$ & $-0.56$ & $-0.12$ & $0.66$ & $0.00$ \\ 
 
$250$ & $0.0500$ & $0.4587$ & $2.90$ & $2.20$ & $1.25$ & $0.82$ & $0.05$ & $1.41$ & $0.29$ & $-0.44$ & $-0.02$ & $1.31$ & $0.00$ \\ 
 
$250$ & $0.0800$ & $0.4222$ & $3.32$ & $2.24$ & $1.33$ & $0.88$ & $0.09$ & $2.06$ & $0.44$ & $-0.65$ & $-0.07$ & $1.90$ & $0.00$ \\ 
 
$250$ & $0.1300$ & $0.3639$ & $3.03$ & $2.29$ & $1.74$ & $1.09$ & $0.48$ & $0.95$ & $0.53$ & $-0.65$ & $-0.23$ & $0.39$ & $0.00$ \\ 
 
$250$ & $0.1800$ & $0.2977$ & $5.03$ & $3.24$ & $2.85$ & $1.90$ & $1.51$ & $2.57$ & $1.11$ & $-0.67$ & $-0.29$ & $-2.20$ & $0.00$ \\ 
 
\hline 
$300$ & $0.0050$ & $1.140$ & $3.03$ & $2.77$ & $1.03$ & $0.24$ & $0.07$ & $0.70$ & $-0.27$ & $-0.54$ & $0.02$ & $0.18$ & $-0.30$ \\ 
 
$300$ & $0.0080$ & $0.9564$ & $2.18$ & $1.89$ & $0.88$ & $0.20$ & $0.02$ & $0.63$ & $-0.20$ & $-0.58$ & $0.01$ & $0.11$ & $-0.05$ \\ 
 
$300$ & $0.0130$ & $0.7934$ & $2.35$ & $1.92$ & $1.05$ & $0.57$ & $0.00$ & $0.87$ & $-0.56$ & $-0.67$ & $0.00$ & $0.00$ & $0.00$ \\ 
 
$300$ & $0.0200$ & $0.6795$ & $2.61$ & $2.10$ & $1.15$ & $0.69$ & $0.14$ & $1.03$ & $0.55$ & $-0.61$ & $0.01$ & $0.61$ & $0.00$ \\ 
 
$300$ & $0.0320$ & $0.5821$ & $2.74$ & $2.18$ & $1.29$ & $0.90$ & $0.07$ & $1.07$ & $0.55$ & $-0.58$ & $-0.11$ & $0.70$ & $0.00$ \\ 
 
$300$ & $0.0500$ & $0.4916$ & $3.12$ & $2.46$ & $1.52$ & $1.17$ & $0.13$ & $1.17$ & $0.61$ & $-0.67$ & $-0.07$ & $0.73$ & $0.00$ \\ 
 
$300$ & $0.0800$ & $0.4185$ & $3.69$ & $2.56$ & $1.40$ & $0.91$ & $0.37$ & $2.26$ & $0.29$ & $-0.60$ & $-0.04$ & $2.15$ & $-0.01$ \\ 
 
$300$ & $0.1300$ & $0.3461$ & $3.43$ & $2.61$ & $1.89$ & $1.33$ & $0.42$ & $1.17$ & $0.73$ & $-0.60$ & $-0.17$ & $0.67$ & $0.00$ \\ 
 
$300$ & $0.1800$ & $0.2791$ & $5.61$ & $3.47$ & $3.26$ & $2.22$ & $1.87$ & $2.96$ & $1.40$ & $-0.95$ & $-0.42$ & $-2.40$ & $0.00$ \\ 
 
$300$ & $0.4000$ & $0.1494$ & $7.25$ & $3.91$ & $3.41$ & $2.10$ & $1.91$ & $5.07$ & $1.16$ & $-0.76$ & $-0.31$ & $-4.86$ & $0.00$ \\ 
 
\hline 
$400$ & $0.0080$ & $0.9779$ & $2.67$ & $2.41$ & $0.97$ & $0.36$ & $0.05$ & $0.61$ & $-0.33$ & $-0.46$ & $0.01$ & $0.15$ & $-0.18$ \\ 
 
$400$ & $0.0130$ & $0.8148$ & $2.57$ & $2.24$ & $1.04$ & $0.55$ & $0.00$ & $0.70$ & $-0.55$ & $-0.44$ & $0.00$ & $0.02$ & $-0.01$ \\ 
 
$400$ & $0.0200$ & $0.6607$ & $2.89$ & $2.36$ & $1.31$ & $0.93$ & $0.00$ & $1.04$ & $-0.93$ & $-0.47$ & $0.00$ & $0.00$ & $0.00$ \\ 
 
$400$ & $0.0320$ & $0.5791$ & $2.80$ & $2.46$ & $1.07$ & $0.50$ & $0.18$ & $0.82$ & $0.49$ & $-0.44$ & $-0.08$ & $0.48$ & $0.00$ \\ 
 
$400$ & $0.0500$ & $0.4854$ & $3.05$ & $2.70$ & $1.20$ & $0.52$ & $0.48$ & $0.79$ & $0.51$ & $-0.44$ & $-0.11$ & $0.41$ & $0.00$ \\ 
 
$400$ & $0.0800$ & $0.3915$ & $3.85$ & $3.05$ & $1.18$ & $0.27$ & $0.55$ & $2.02$ & $0.24$ & $-0.35$ & $0.07$ & $1.98$ & $0.00$ \\ 
 
$400$ & $0.1300$ & $0.3593$ & $3.59$ & $2.96$ & $1.53$ & $0.75$ & $0.39$ & $1.33$ & $0.69$ & $-0.59$ & $-0.21$ & $0.95$ & $0.00$ \\ 
 
$400$ & $0.1800$ & $0.3012$ & $5.07$ & $3.45$ & $2.34$ & $0.96$ & $1.53$ & $2.88$ & $0.90$ & $-0.44$ & $-0.31$ & $-2.68$ & $0.00$ \\ 
 
$400$ & $0.4000$ & $0.1452$ & $8.22$ & $4.99$ & $3.43$ & $1.51$ & $2.33$ & $5.55$ & $1.43$ & $-0.89$ & $-0.44$ & $-5.27$ & $0.00$ \\ 
 
\hline 
$500$ & $0.0080$ & $0.9388$ & $4.11$ & $3.93$ & $1.11$ & $0.22$ & $0.08$ & $0.48$ & $-0.20$ & $-0.37$ & $0.02$ & $0.15$ & $-0.17$ \\ 
 
$500$ & $0.0130$ & $0.8057$ & $3.02$ & $2.80$ & $1.00$ & $0.37$ & $0.02$ & $0.55$ & $-0.30$ & $-0.45$ & $0.00$ & $0.09$ & $-0.06$ \\ 
 
$500$ & $0.0200$ & $0.6732$ & $3.18$ & $2.80$ & $1.19$ & $0.73$ & $0.00$ & $0.92$ & $-0.73$ & $-0.57$ & $0.00$ & $0.00$ & $-0.01$ \\ 
 
$500$ & $0.0320$ & $0.5728$ & $3.24$ & $2.89$ & $1.11$ & $0.49$ & $0.28$ & $0.95$ & $0.49$ & $-0.35$ & $0.16$ & $0.71$ & $0.00$ \\ 
 
$500$ & $0.0500$ & $0.5054$ & $3.43$ & $3.01$ & $1.30$ & $0.77$ & $0.35$ & $1.00$ & $0.78$ & $-0.50$ & $-0.23$ & $0.32$ & $0.00$ \\ 
 
$500$ & $0.0800$ & $0.4245$ & $3.86$ & $3.28$ & $1.15$ & $0.34$ & $0.37$ & $1.68$ & $0.34$ & $-0.35$ & $0.04$ & $1.60$ & $0.00$ \\ 
 
$500$ & $0.1300$ & $0.3611$ & $4.68$ & $4.14$ & $1.56$ & $0.80$ & $0.20$ & $1.52$ & $0.80$ & $-0.54$ & $-0.11$ & $1.17$ & $0.00$ \\ 
 
$500$ & $0.1800$ & $0.3152$ & $4.60$ & $4.06$ & $1.84$ & $0.78$ & $0.72$ & $1.12$ & $0.78$ & $-0.69$ & $-0.30$ & $-0.29$ & $0.00$ \\ 
 
$500$ & $0.2500$ & $0.2397$ & $6.88$ & $5.05$ & $2.74$ & $1.22$ & $1.82$ & $3.79$ & $1.22$ & $-0.60$ & $-0.36$ & $-3.51$ & $0.00$ \\ 
 
\hline 
$650$ & $0.0130$ & $0.8564$ & $3.31$ & $3.13$ & $1.00$ & $0.25$ & $0.06$ & $0.38$ & $-0.11$ & $-0.31$ & $0.02$ & $0.14$ & $-0.12$ \\ 
 
$650$ & $0.0200$ & $0.7176$ & $3.55$ & $3.26$ & $1.20$ & $0.65$ & $0.01$ & $0.74$ & $-0.54$ & $-0.52$ & $0.00$ & $0.02$ & $-0.01$ \\ 
 
$650$ & $0.0320$ & $0.6245$ & $3.76$ & $3.33$ & $1.37$ & $0.89$ & $0.00$ & $1.07$ & $-0.87$ & $-0.62$ & $0.00$ & $0.00$ & $0.00$ \\ 
 
$650$ & $0.0500$ & $0.5392$ & $3.74$ & $3.46$ & $1.14$ & $0.42$ & $0.10$ & $0.83$ & $0.50$ & $-0.26$ & $0.07$ & $0.61$ & $0.00$ \\ 
 
$650$ & $0.0800$ & $0.4014$ & $4.36$ & $4.01$ & $1.29$ & $0.62$ & $0.20$ & $1.12$ & $0.64$ & $-0.36$ & $-0.12$ & $0.84$ & $0.00$ \\ 
 
$650$ & $0.1300$ & $0.3282$ & $5.40$ & $4.99$ & $1.50$ & $0.47$ & $0.22$ & $1.43$ & $0.52$ & $-0.25$ & $-0.13$ & $1.30$ & $0.00$ \\ 
 
$650$ & $0.1800$ & $0.3253$ & $5.04$ & $4.67$ & $1.70$ & $0.61$ & $0.26$ & $0.83$ & $0.63$ & $-0.38$ & $-0.13$ & $-0.38$ & $0.00$ \\ 
 
$650$ & $0.2500$ & $0.2467$ & $6.96$ & $5.55$ & $2.90$ & $1.44$ & $1.83$ & $3.04$ & $1.42$ & $-0.56$ & $-0.51$ & $-2.58$ & $0.00$ \\ 
 
$650$ & $0.4000$ & $0.1106$ & $9.89$ & $8.49$ & $3.44$ & $1.47$ & $1.94$ & $3.71$ & $1.46$ & $-0.74$ & $-0.33$ & $-3.31$ & $0.00$ \\ 
 
\hline 

\end{tabular} 
\end{center} 
\caption[RESULT] 
{\label{tab:ncdxdq2_eleRH} The NC 
$e^-p$ reduced cross section $\tilde{\sigma}_{\rm NC}(x,Q^2)$ 
with lepton beam polarisation $P_e=+36.0$\% 
with statistical 
$(\delta_{\rm stat})$, 
total $(\delta_{tot})$, 
total uncorrelated systematic $(\delta_{\rm unc})$ 
errors, two of its contributions from the 
 electron energy error ($\delta_{unc}^{E}$)  
and the hadronic energy error  
($\delta_{\rm unc}^{h}$). 
The effect of the other uncorrelated 
systematic errors is included in $\delta_{\rm unc}$. 
In addition the correlated systematic  
$(\delta_{\rm cor})$ and its contributions from a 
positive variation of one  
standard deviation of the 
electron energy error ($\delta_{cor}^{E^+}$), of 
the polar electron angle error 
($\delta_{\rm cor}^{\theta^+}$), of the hadronic 
energy error ($\delta_{\rm cor}^{h^+}$), of the error 
due to noise subtraction ($\delta_{\rm cor}^{N^+}$) 
and of the error due to background subtraction 
($\delta_{\rm cor}^{B^+}$) are given. 
The normalisation and polarisation uncertainties are 
not included in the errors. 
The table continues on the next page.}
\end{table} 
\begin{table}[htbp] 
\begin{center} 
\tiny 
\begin{tabular}{|r|c|r|r|r|r|r|r|r|r|r|r|r|r|} 
\hline 
$Q^2$ & $x$ & $\tilde{\sigma}_{\rm NC}$ & 
$\delta_{\rm tot}$ & $\delta_{\rm stat}$ & $\delta_{\rm unc}$ & 
$\delta_{\rm unc}^{E}$ & 
$\delta_{\rm unc}^{h}$& 
$\delta_{\rm cor}$ & 
$\delta_{\rm cor}^{E^+}$ & 
$\delta_{\rm cor}^{\theta^+}$& 
$\delta_{\rm cor}^{h^+}$& 
$\delta_{\rm cor}^{N^+}$& 
$\delta_{\rm cor}^{B^+}$ \\ 
$(\rm GeV^2)$ & & & 
$(\%)$ & $(\%)$ & $(\%)$ & $(\%)$ & $(\%)$ & $(\%)$ & 
$(\%)$ & $(\%)$ & $(\%)$ & $(\%)$ & $(\%)$  
\\ \hline 
$800$ & $0.0130$ & $0.8156$ & $5.41$ & $5.20$ & $1.42$ & $0.68$ & $0.08$ & $0.56$ & $-0.34$ & $-0.38$ & $0.01$ & $0.16$ & $-0.16$ \\ 
 
$800$ & $0.0200$ & $0.6970$ & $3.96$ & $3.76$ & $1.16$ & $0.52$ & $0.03$ & $0.44$ & $-0.33$ & $-0.27$ & $0.00$ & $0.08$ & $-0.02$ \\ 
 
$800$ & $0.0320$ & $0.5977$ & $4.18$ & $3.87$ & $1.38$ & $0.84$ & $0.00$ & $0.74$ & $-0.67$ & $-0.32$ & $0.00$ & $0.00$ & $0.00$ \\ 
 
$800$ & $0.0500$ & $0.5348$ & $4.27$ & $4.04$ & $1.17$ & $0.24$ & $0.17$ & $0.73$ & $0.48$ & $-0.35$ & $0.06$ & $0.41$ & $-0.01$ \\ 
 
$800$ & $0.0800$ & $0.3757$ & $5.10$ & $4.83$ & $1.26$ & $0.31$ & $0.07$ & $1.06$ & $0.53$ & $-0.44$ & $-0.06$ & $0.81$ & $0.00$ \\ 
 
$800$ & $0.1300$ & $0.3512$ & $6.24$ & $5.90$ & $1.57$ & $0.19$ & $0.34$ & $1.27$ & $0.45$ & $-0.40$ & $-0.03$ & $1.12$ & $-0.02$ \\ 
 
$800$ & $0.1800$ & $0.3117$ & $6.72$ & $6.28$ & $2.02$ & $0.86$ & $0.74$ & $1.28$ & $1.02$ & $-0.47$ & $-0.28$ & $-0.55$ & $0.00$ \\ 
 
$800$ & $0.2500$ & $0.2222$ & $7.37$ & $6.66$ & $2.48$ & $0.99$ & $1.32$ & $1.95$ & $1.20$ & $-0.40$ & $-0.26$ & $-1.46$ & $0.00$ \\ 
 
$800$ & $0.4000$ & $0.1183$ & $11.91$ & $9.46$ & $4.15$ & $1.43$ & $2.88$ & $5.92$ & $1.54$ & $-0.56$ & $-0.85$ & $-5.63$ & $0.00$ \\ 
 
\hline 
$1000$ & $0.0130$ & $0.8198$ & $5.57$ & $5.19$ & $1.71$ & $0.28$ & $0.22$ & $1.08$ & $0.05$ & $-0.20$ & $0.06$ & $0.18$ & $-1.05$ \\ 
 
$1000$ & $0.0200$ & $0.6965$ & $4.63$ & $4.48$ & $1.10$ & $0.37$ & $0.05$ & $0.43$ & $-0.20$ & $-0.35$ & $0.02$ & $0.10$ & $-0.08$ \\ 
 
$1000$ & $0.0320$ & $0.6199$ & $4.84$ & $4.66$ & $1.20$ & $0.58$ & $0.00$ & $0.58$ & $-0.34$ & $-0.47$ & $0.00$ & $0.01$ & $0.00$ \\ 
 
$1000$ & $0.0500$ & $0.4908$ & $5.01$ & $4.74$ & $1.42$ & $0.90$ & $0.00$ & $0.77$ & $-0.47$ & $-0.61$ & $0.00$ & $0.00$ & $0.00$ \\ 
 
$1000$ & $0.0800$ & $0.4084$ & $5.43$ & $5.21$ & $1.30$ & $0.57$ & $0.16$ & $0.85$ & $0.41$ & $-0.32$ & $0.11$ & $0.66$ & $0.00$ \\ 
 
$1000$ & $0.1300$ & $0.3811$ & $6.19$ & $5.82$ & $1.52$ & $0.19$ & $0.32$ & $1.45$ & $0.46$ & $-0.26$ & $-0.02$ & $1.35$ & $0.00$ \\ 
 
$1000$ & $0.1800$ & $0.2933$ & $6.70$ & $6.43$ & $1.79$ & $0.49$ & $0.59$ & $0.63$ & $0.41$ & $-0.33$ & $-0.12$ & $-0.32$ & $0.00$ \\ 
 
$1000$ & $0.2500$ & $0.2710$ & $7.06$ & $6.71$ & $2.01$ & $0.64$ & $0.64$ & $0.96$ & $0.75$ & $-0.24$ & $-0.25$ & $0.48$ & $0.00$ \\ 
 
$1000$ & $0.4000$ & $0.1238$ & $12.66$ & $9.92$ & $4.75$ & $2.23$ & $3.36$ & $6.28$ & $1.75$ & $-0.65$ & $-0.72$ & $-5.95$ & $0.00$ \\ 
 
\hline 
$1200$ & $0.0130$ & $0.7581$ & $9.74$ & $9.00$ & $2.85$ & $0.18$ & $0.22$ & $2.39$ & $-0.20$ & $-0.13$ & $0.04$ & $0.15$ & $-2.37$ \\ 
 
$1200$ & $0.0200$ & $0.7496$ & $5.51$ & $5.37$ & $1.17$ & $0.10$ & $0.07$ & $0.42$ & $0.09$ & $-0.32$ & $0.02$ & $0.10$ & $-0.24$ \\ 
 
$1200$ & $0.0320$ & $0.6278$ & $5.19$ & $5.03$ & $1.13$ & $0.56$ & $0.01$ & $0.55$ & $-0.38$ & $-0.39$ & $0.00$ & $0.05$ & $-0.01$ \\ 
 
$1200$ & $0.0500$ & $0.5154$ & $5.39$ & $5.22$ & $1.24$ & $0.75$ & $0.00$ & $0.56$ & $-0.34$ & $-0.44$ & $0.00$ & $0.00$ & $0.00$ \\ 
 
$1200$ & $0.0800$ & $0.4535$ & $5.77$ & $5.53$ & $1.41$ & $0.93$ & $0.24$ & $0.83$ & $0.62$ & $-0.27$ & $-0.09$ & $0.47$ & $0.00$ \\ 
 
$1200$ & $0.1300$ & $0.3312$ & $7.25$ & $7.00$ & $1.57$ & $0.81$ & $0.04$ & $1.07$ & $0.60$ & $-0.25$ & $-0.04$ & $0.85$ & $0.00$ \\ 
 
$1200$ & $0.1800$ & $0.2648$ & $7.86$ & $7.60$ & $1.93$ & $1.15$ & $0.48$ & $0.65$ & $0.57$ & $-0.12$ & $-0.10$ & $0.27$ & $0.00$ \\ 
 
$1200$ & $0.2500$ & $0.2152$ & $8.51$ & $8.26$ & $1.92$ & $0.89$ & $0.59$ & $0.71$ & $0.58$ & $-0.29$ & $-0.18$ & $0.22$ & $0.00$ \\ 
 
$1200$ & $0.4000$ & $0.1305$ & $12.00$ & $10.01$ & $4.37$ & $2.59$ & $2.70$ & $4.96$ & $1.87$ & $-0.33$ & $-0.84$ & $-4.51$ & $0.00$ \\ 
 
\hline 
$1500$ & $0.0200$ & $0.7118$ & $6.93$ & $6.68$ & $1.77$ & $0.38$ & $0.17$ & $0.60$ & $-0.13$ & $-0.31$ & $0.02$ & $0.16$ & $-0.47$ \\ 
 
$1500$ & $0.0320$ & $0.6266$ & $6.33$ & $6.22$ & $1.10$ & $0.23$ & $0.03$ & $0.46$ & $-0.32$ & $-0.31$ & $0.01$ & $0.10$ & $-0.04$ \\ 
 
$1500$ & $0.0500$ & $0.5367$ & $6.19$ & $6.00$ & $1.37$ & $0.90$ & $0.00$ & $0.66$ & $-0.51$ & $-0.42$ & $0.00$ & $0.01$ & $0.00$ \\ 
 
$1500$ & $0.0800$ & $0.4479$ & $6.51$ & $6.35$ & $1.32$ & $0.76$ & $0.07$ & $0.52$ & $0.33$ & $-0.16$ & $0.02$ & $0.37$ & $0.00$ \\ 
 
$1500$ & $0.1300$ & $0.3257$ & $10.30$ & $10.17$ & $1.48$ & $0.41$ & $0.14$ & $0.81$ & $0.41$ & $0.12$ & $0.06$ & $0.69$ & $0.00$ \\ 
 
$1500$ & $0.1800$ & $0.2626$ & $9.05$ & $8.79$ & $1.91$ & $1.11$ & $0.21$ & $1.04$ & $0.70$ & $-0.35$ & $-0.31$ & $0.61$ & $0.00$ \\ 
 
$1500$ & $0.2500$ & $0.2490$ & $9.06$ & $8.68$ & $2.43$ & $1.48$ & $0.98$ & $0.89$ & $0.82$ & $0.11$ & $0.21$ & $-0.26$ & $0.00$ \\ 
 
$1500$ & $0.4000$ & $0.09081$ & $14.08$ & $13.16$ & $3.83$ & $2.26$ & $2.00$ & $3.25$ & $1.26$ & $-0.38$ & $-0.53$ & $-2.92$ & $0.00$ \\ 
 
$1500$ & $0.6500$ & $0.01231$ & $27.10$ & $23.63$ & $7.54$ & $4.91$ & $4.80$ & $10.93$ & $2.61$ & $-0.52$ & $-1.01$ & $-10.55$ & $0.00$ \\ 
 
\hline 
$2000$ & $0.0219$ & $0.7395$ & $10.99$ & $10.62$ & $2.48$ & $0.48$ & $0.22$ & $1.37$ & $-0.27$ & $-0.20$ & $0.06$ & $0.21$ & $-1.31$ \\ 
 
$2000$ & $0.0320$ & $0.6267$ & $7.47$ & $7.29$ & $1.58$ & $0.27$ & $0.10$ & $0.49$ & $0.06$ & $-0.33$ & $0.04$ & $0.11$ & $-0.33$ \\ 
 
$2000$ & $0.0500$ & $0.5035$ & $7.69$ & $7.57$ & $1.25$ & $0.49$ & $0.01$ & $0.49$ & $-0.40$ & $-0.28$ & $0.00$ & $0.03$ & $0.00$ \\ 
 
$2000$ & $0.0800$ & $0.4419$ & $7.52$ & $7.37$ & $1.38$ & $0.70$ & $0.00$ & $0.52$ & $-0.32$ & $-0.41$ & $0.00$ & $0.00$ & $0.00$ \\ 
 
$2000$ & $0.1300$ & $0.3075$ & $9.88$ & $9.69$ & $1.71$ & $0.73$ & $0.34$ & $0.85$ & $0.17$ & $-0.18$ & $0.15$ & $0.80$ & $0.00$ \\ 
 
$2000$ & $0.1800$ & $0.2856$ & $10.11$ & $9.83$ & $2.18$ & $1.38$ & $0.43$ & $0.94$ & $0.80$ & $-0.25$ & $-0.08$ & $0.42$ & $0.00$ \\ 
 
$2000$ & $0.2500$ & $0.2258$ & $10.72$ & $10.39$ & $2.42$ & $1.46$ & $0.84$ & $1.04$ & $0.96$ & $-0.24$ & $-0.28$ & $-0.16$ & $0.00$ \\ 
 
$2000$ & $0.4000$ & $0.1182$ & $13.61$ & $12.93$ & $3.47$ & $1.96$ & $1.54$ & $2.46$ & $1.25$ & $-0.05$ & $-0.40$ & $-2.08$ & $0.00$ \\ 
 
$2000$ & $0.6500$ & $0.007810$ & $35.78$ & $33.44$ & $7.79$ & $4.64$ & $5.28$ & $10.04$ & $2.98$ & $-0.66$ & $-0.96$ & $-9.52$ & $0.00$ \\ 
 
\hline 
$3000$ & $0.0320$ & $0.5900$ & $7.76$ & $7.46$ & $1.90$ & $0.09$ & $0.12$ & $0.93$ & $0.10$ & $-0.26$ & $0.03$ & $0.13$ & $-0.88$ \\ 
 
$3000$ & $0.0500$ & $0.5803$ & $6.14$ & $5.96$ & $1.44$ & $0.28$ & $0.06$ & $0.40$ & $-0.15$ & $-0.35$ & $0.03$ & $0.09$ & $-0.08$ \\ 
 
$3000$ & $0.0800$ & $0.4753$ & $6.78$ & $6.63$ & $1.31$ & $0.40$ & $0.00$ & $0.52$ & $-0.28$ & $-0.44$ & $0.00$ & $0.01$ & $-0.02$ \\ 
 
$3000$ & $0.1300$ & $0.3423$ & $8.65$ & $8.44$ & $1.76$ & $0.87$ & $0.00$ & $0.66$ & $-0.52$ & $-0.41$ & $0.00$ & $0.00$ & $0.00$ \\ 
 
$3000$ & $0.1800$ & $0.3003$ & $9.27$ & $9.00$ & $2.05$ & $1.27$ & $0.21$ & $0.89$ & $0.78$ & $-0.09$ & $-0.09$ & $0.41$ & $0.00$ \\ 
 
$3000$ & $0.2500$ & $0.2826$ & $8.95$ & $8.59$ & $2.32$ & $1.56$ & $0.37$ & $0.95$ & $0.86$ & $-0.09$ & $0.06$ & $0.38$ & $0.00$ \\ 
 
$3000$ & $0.4000$ & $0.08914$ & $13.50$ & $12.93$ & $3.43$ & $2.12$ & $1.39$ & $1.76$ & $1.28$ & $-0.19$ & $-0.44$ & $-1.10$ & $0.00$ \\ 
 
$3000$ & $0.6500$ & $0.005010$ & $37.74$ & $35.57$ & $8.46$ & $5.69$ & $5.45$ & $9.35$ & $3.03$ & $-0.13$ & $-1.35$ & $-8.74$ & $0.00$ \\ 
 
\hline 
$5000$ & $0.0547$ & $0.5599$ & $9.80$ & $9.54$ & $2.13$ & $0.37$ & $0.10$ & $0.68$ & $-0.18$ & $-0.32$ & $0.03$ & $0.10$ & $-0.56$ \\ 
 
$5000$ & $0.0800$ & $0.4362$ & $7.86$ & $7.68$ & $1.62$ & $0.25$ & $0.06$ & $0.42$ & $-0.22$ & $-0.29$ & $0.01$ & $0.08$ & $-0.19$ \\ 
 
$5000$ & $0.1300$ & $0.3964$ & $8.70$ & $8.49$ & $1.85$ & $0.43$ & $0.02$ & $0.44$ & $0.25$ & $-0.37$ & $0.00$ & $0.02$ & $-0.04$ \\ 
 
$5000$ & $0.1800$ & $0.3454$ & $9.67$ & $9.47$ & $1.91$ & $0.29$ & $0.00$ & $0.43$ & $-0.30$ & $-0.31$ & $0.00$ & $0.00$ & $0.00$ \\ 
 
$5000$ & $0.2500$ & $0.2430$ & $17.03$ & $16.81$ & $2.56$ & $1.47$ & $0.00$ & $0.94$ & $0.93$ & $-0.10$ & $0.00$ & $0.00$ & $0.00$ \\ 
 
$5000$ & $0.4000$ & $0.1382$ & $13.32$ & $12.82$ & $3.38$ & $1.76$ & $1.27$ & $1.30$ & $1.08$ & $-0.11$ & $-0.27$ & $-0.67$ & $0.00$ \\ 
 
$5000$ & $0.6500$ & $0.01280$ & $27.43$ & $25.88$ & $6.77$ & $3.48$ & $4.66$ & $6.05$ & $2.25$ & $0.35$ & $-1.49$ & $-5.40$ & $0.00$ \\ 
 
\hline 
$8000$ & $0.0875$ & $0.5903$ & $14.27$ & $13.94$ & $2.79$ & $0.22$ & $0.16$ & $1.32$ & $0.13$ & $-0.41$ & $0.04$ & $0.16$ & $-1.23$ \\ 
 
$8000$ & $0.1300$ & $0.4711$ & $11.39$ & $11.13$ & $2.40$ & $0.29$ & $0.08$ & $0.39$ & $-0.19$ & $-0.25$ & $0.02$ & $0.09$ & $-0.20$ \\ 
 
$8000$ & $0.1800$ & $0.3787$ & $11.97$ & $11.74$ & $2.31$ & $0.17$ & $0.00$ & $0.34$ & $0.13$ & $-0.31$ & $0.00$ & $0.02$ & $0.00$ \\ 
 
$8000$ & $0.2500$ & $0.2430$ & $14.27$ & $14.04$ & $2.56$ & $0.80$ & $0.00$ & $0.34$ & $0.31$ & $0.12$ & $0.00$ & $0.00$ & $0.00$ \\ 
 
$8000$ & $0.4000$ & $0.1130$ & $19.80$ & $18.93$ & $5.29$ & $4.26$ & $0.00$ & $2.41$ & $2.38$ & $0.36$ & $0.00$ & $0.00$ & $0.00$ \\ 
 
$8000$ & $0.6500$ & $0.01617$ & $28.34$ & $26.77$ & $8.04$ & $5.47$ & $4.37$ & $4.62$ & $3.03$ & $-0.35$ & $-0.99$ & $-3.34$ & $0.00$ \\ 
 
\hline 
$12000$ & $0.1300$ & $0.6871$ & $23.67$ & $23.42$ & $3.21$ & $1.04$ & $0.09$ & $1.28$ & $0.50$ & $-0.41$ & $0.03$ & $0.08$ & $-1.10$ \\ 
 
$12000$ & $0.1800$ & $0.3882$ & $16.90$ & $16.73$ & $2.34$ & $0.50$ & $0.09$ & $0.39$ & $-0.15$ & $-0.29$ & $0.03$ & $0.10$ & $-0.19$ \\ 
 
$12000$ & $0.2500$ & $0.3008$ & $16.92$ & $16.72$ & $2.60$ & $1.20$ & $0.01$ & $0.46$ & $0.45$ & $0.12$ & $0.01$ & $0.03$ & $0.00$ \\ 
 
$12000$ & $0.4000$ & $0.1494$ & $22.50$ & $21.86$ & $4.91$ & $3.98$ & $0.00$ & $2.06$ & $1.96$ & $0.65$ & $0.00$ & $0.00$ & $0.00$ \\ 
 
$12000$ & $0.6500$ & $0.01225$ & $45.33$ & $44.83$ & $6.13$ & $4.35$ & $2.88$ & $2.67$ & $2.19$ & $0.30$ & $-0.58$ & $-1.38$ & $0.00$ \\ 
 
\hline 
$20000$ & $0.2500$ & $0.1984$ & $34.22$ & $34.11$ & $2.68$ & $1.19$ & $0.08$ & $0.57$ & $0.53$ & $-0.18$ & $0.02$ & $0.08$ & $-0.07$ \\ 
 
$20000$ & $0.4000$ & $0.2224$ & $24.77$ & $24.30$ & $4.72$ & $3.73$ & $0.02$ & $0.82$ & $0.81$ & $0.10$ & $0.01$ & $0.02$ & $0.00$ \\ 
 
$20000$ & $0.6500$ & $0.01299$ & $72.61$ & $70.89$ & $14.87$ & $14.40$ & $0.00$ & $4.95$ & $4.11$ & $2.76$ & $0.00$ & $0.00$ & $0.00$ \\ 
 
\hline 
$30000$ & $0.4000$ & $0.2553$ & $43.91$ & $43.40$ & $6.07$ & $3.07$ & $0.11$ & $2.69$ & $1.09$ & $-0.53$ & $0.02$ & $0.06$ & $-2.40$ \\ 
 
$30000$ & $0.6500$ & $0.03854$ & $58.87$ & $57.81$ & $10.51$ & $9.70$ & $0.00$ & $3.70$ & $3.26$ & $1.74$ & $0.00$ & $0.00$ & $0.00$ \\ 
 
\hline 
\end{tabular} 
\end{center} 
\captcont{continued.} 
\end{table} 

\begin{table}[htbp] 
\begin{center} 
\tiny 
\begin{tabular}{|r|c|r|r|r|r|r|r|r|r|r|r|r|r|} 
\hline 
$Q^2$ & $x$ & $\tilde{\sigma}_{\rm NC}$ & 
$\delta_{\rm tot}$ & $\delta_{\rm stat}$ & $\delta_{\rm unc}$ & 
$\delta_{\rm unc}^{E}$ & 
$\delta_{\rm unc}^{h}$& 
$\delta_{\rm cor}$ & 
$\delta_{\rm cor}^{E^+}$ & 
$\delta_{\rm cor}^{\theta^+}$& 
$\delta_{\rm cor}^{h^+}$& 
$\delta_{\rm cor}^{N^+}$& 
$\delta_{\rm cor}^{B^+}$ \\ 
$(\rm GeV^2)$ & & & 
$(\%)$ & $(\%)$ & $(\%)$ & $(\%)$ & $(\%)$ & $(\%)$ & 
$(\%)$ & $(\%)$ & $(\%)$ & $(\%)$ & $(\%)$  
\\ \hline 
$120$ & $0.0020$ & $1.342$ & $1.71$ & $0.97$ & $0.93$ & $0.55$ & $0.09$ & $1.05$ & $-0.31$ & $-0.58$ & $0.02$ & $0.17$ & $-0.79$ \\ 
 
$120$ & $0.0032$ & $1.226$ & $1.89$ & $1.38$ & $0.98$ & $0.50$ & $0.03$ & $0.84$ & $-0.28$ & $-0.76$ & $0.00$ & $0.20$ & $-0.07$ \\ 
 
\hline 
$150$ & $0.0032$ & $1.225$ & $1.42$ & $0.82$ & $0.83$ & $0.48$ & $0.06$ & $0.80$ & $-0.34$ & $-0.54$ & $0.01$ & $0.19$ & $-0.45$ \\ 
 
$150$ & $0.0050$ & $1.076$ & $1.82$ & $1.00$ & $1.22$ & $0.98$ & $0.00$ & $0.92$ & $-0.66$ & $-0.64$ & $0.00$ & $0.04$ & $-0.02$ \\ 
 
$150$ & $0.0080$ & $0.9295$ & $2.64$ & $1.36$ & $1.89$ & $1.34$ & $1.05$ & $1.24$ & $-0.84$ & $-0.56$ & $-0.31$ & $-0.66$ & $-0.05$ \\ 
 
$150$ & $0.0130$ & $0.8072$ & $4.12$ & $1.92$ & $2.85$ & $2.47$ & $1.00$ & $2.28$ & $-1.54$ & $-1.30$ & $-0.32$ & $-1.02$ & $-0.01$ \\ 
 
\hline 
$200$ & $0.0032$ & $1.240$ & $1.97$ & $1.55$ & $0.84$ & $0.13$ & $0.08$ & $0.88$ & $-0.14$ & $-0.64$ & $0.02$ & $0.18$ & $-0.56$ \\ 
 
$200$ & $0.0050$ & $1.101$ & $1.58$ & $1.08$ & $0.87$ & $0.50$ & $0.02$ & $0.75$ & $-0.41$ & $-0.62$ & $0.00$ & $0.11$ & $-0.07$ \\ 
 
$200$ & $0.0080$ & $0.9486$ & $1.98$ & $1.09$ & $1.32$ & $1.10$ & $0.00$ & $0.98$ & $-0.59$ & $-0.78$ & $0.00$ & $0.00$ & $-0.01$ \\ 
 
$200$ & $0.0130$ & $0.7920$ & $1.63$ & $1.24$ & $0.77$ & $0.10$ & $0.03$ & $0.72$ & $0.06$ & $-0.45$ & $0.01$ & $0.55$ & $-0.01$ \\ 
 
$200$ & $0.0200$ & $0.6872$ & $1.89$ & $1.38$ & $1.05$ & $0.69$ & $0.06$ & $0.75$ & $-0.35$ & $-0.55$ & $-0.15$ & $0.33$ & $-0.02$ \\ 
 
$200$ & $0.0320$ & $0.5807$ & $2.33$ & $1.61$ & $1.47$ & $1.10$ & $0.47$ & $0.85$ & $-0.62$ & $-0.54$ & $-0.17$ & $0.10$ & $0.00$ \\ 
 
$200$ & $0.0500$ & $0.5212$ & $3.20$ & $1.79$ & $2.00$ & $1.78$ & $0.06$ & $1.73$ & $-1.22$ & $-0.69$ & $-0.05$ & $1.02$ & $0.00$ \\ 
 
$200$ & $0.0800$ & $0.4303$ & $3.50$ & $1.99$ & $2.07$ & $1.81$ & $0.08$ & $2.00$ & $-1.04$ & $-0.86$ & $-0.20$ & $1.46$ & $0.00$ \\ 
 
$200$ & $0.1300$ & $0.3652$ & $3.67$ & $2.25$ & $2.30$ & $1.56$ & $1.08$ & $1.78$ & $-1.05$ & $-0.92$ & $-0.21$ & $-1.09$ & $0.00$ \\ 
 
$200$ & $0.1800$ & $0.3205$ & $4.68$ & $2.97$ & $3.01$ & $1.73$ & $1.85$ & $1.99$ & $-0.99$ & $-0.81$ & $-0.35$ & $-1.49$ & $0.00$ \\ 
 
\hline 
$250$ & $0.0050$ & $1.106$ & $1.59$ & $1.25$ & $0.78$ & $0.29$ & $0.06$ & $0.59$ & $-0.27$ & $-0.43$ & $0.01$ & $0.16$ & $-0.26$ \\ 
 
$250$ & $0.0080$ & $0.9476$ & $1.73$ & $1.24$ & $0.93$ & $0.58$ & $0.00$ & $0.76$ & $-0.50$ & $-0.57$ & $0.00$ & $0.03$ & $-0.03$ \\ 
 
$250$ & $0.0130$ & $0.7931$ & $2.09$ & $1.38$ & $1.17$ & $0.87$ & $0.17$ & $1.04$ & $0.45$ & $-0.60$ & $-0.06$ & $0.72$ & $-0.01$ \\ 
 
$250$ & $0.0200$ & $0.6765$ & $2.23$ & $1.40$ & $1.35$ & $1.07$ & $0.28$ & $1.08$ & $0.48$ & $-0.65$ & $-0.03$ & $0.72$ & $-0.01$ \\ 
 
$250$ & $0.0320$ & $0.5679$ & $2.17$ & $1.46$ & $1.29$ & $1.01$ & $0.16$ & $0.95$ & $0.39$ & $-0.46$ & $0.04$ & $0.74$ & $0.00$ \\ 
 
$250$ & $0.0500$ & $0.4983$ & $2.56$ & $1.57$ & $1.23$ & $0.90$ & $0.14$ & $1.61$ & $0.42$ & $-0.56$ & $-0.12$ & $1.44$ & $0.00$ \\ 
 
$250$ & $0.0800$ & $0.4353$ & $2.96$ & $1.71$ & $1.07$ & $0.53$ & $0.29$ & $2.17$ & $0.30$ & $-0.43$ & $0.03$ & $2.10$ & $-0.01$ \\ 
 
$250$ & $0.1300$ & $0.3763$ & $2.79$ & $1.81$ & $1.84$ & $1.13$ & $0.84$ & $1.05$ & $0.56$ & $-0.60$ & $-0.31$ & $-0.58$ & $0.00$ \\ 
 
$250$ & $0.1800$ & $0.2954$ & $4.13$ & $2.39$ & $2.49$ & $1.51$ & $1.36$ & $2.27$ & $0.84$ & $-0.68$ & $-0.24$ & $-1.98$ & $0.00$ \\ 
 
\hline 
$300$ & $0.0050$ & $1.113$ & $2.42$ & $2.13$ & $0.94$ & $0.25$ & $0.08$ & $0.68$ & $-0.25$ & $-0.39$ & $0.02$ & $0.16$ & $-0.47$ \\ 
 
$300$ & $0.0080$ & $0.9562$ & $1.80$ & $1.45$ & $0.83$ & $0.38$ & $0.02$ & $0.68$ & $-0.38$ & $-0.55$ & $0.00$ & $0.10$ & $-0.05$ \\ 
 
$300$ & $0.0130$ & $0.8024$ & $1.95$ & $1.45$ & $0.97$ & $0.61$ & $0.00$ & $0.85$ & $-0.60$ & $-0.61$ & $0.00$ & $0.00$ & $-0.01$ \\ 
 
$300$ & $0.0200$ & $0.6950$ & $2.05$ & $1.60$ & $0.92$ & $0.45$ & $0.17$ & $0.89$ & $0.31$ & $-0.44$ & $-0.04$ & $0.71$ & $-0.02$ \\ 
 
$300$ & $0.0320$ & $0.5811$ & $2.36$ & $1.69$ & $1.25$ & $0.96$ & $0.11$ & $1.07$ & $0.53$ & $-0.64$ & $-0.03$ & $0.68$ & $-0.02$ \\ 
 
$300$ & $0.0500$ & $0.4956$ & $2.63$ & $1.82$ & $1.37$ & $1.05$ & $0.24$ & $1.32$ & $0.62$ & $-0.63$ & $-0.10$ & $0.97$ & $0.00$ \\ 
 
$300$ & $0.0800$ & $0.4375$ & $3.31$ & $1.85$ & $1.44$ & $1.06$ & $0.39$ & $2.33$ & $0.47$ & $-0.61$ & $-0.08$ & $2.20$ & $-0.01$ \\ 
 
$300$ & $0.1300$ & $0.3645$ & $2.93$ & $1.93$ & $1.84$ & $1.32$ & $0.46$ & $1.20$ & $0.66$ & $-0.73$ & $-0.21$ & $-0.66$ & $0.00$ \\ 
 
$300$ & $0.1800$ & $0.3022$ & $4.79$ & $2.52$ & $2.97$ & $2.01$ & $1.65$ & $2.80$ & $1.03$ & $-0.67$ & $-0.34$ & $-2.49$ & $0.00$ \\ 
 
$300$ & $0.4000$ & $0.1520$ & $7.17$ & $3.06$ & $3.84$ & $2.51$ & $2.25$ & $5.22$ & $1.39$ & $-0.94$ & $-0.41$ & $-4.93$ & $0.00$ \\ 
 
\hline 
$400$ & $0.0080$ & $1.005$ & $2.03$ & $1.77$ & $0.83$ & $0.28$ & $0.06$ & $0.57$ & $-0.29$ & $-0.42$ & $0.01$ & $0.14$ & $-0.20$ \\ 
 
$400$ & $0.0130$ & $0.8181$ & $2.05$ & $1.71$ & $0.91$ & $0.49$ & $0.01$ & $0.69$ & $-0.48$ & $-0.48$ & $0.00$ & $0.03$ & $-0.01$ \\ 
 
$400$ & $0.0200$ & $0.6991$ & $2.35$ & $1.77$ & $1.15$ & $0.83$ & $0.00$ & $1.03$ & $-0.83$ & $-0.61$ & $0.00$ & $0.00$ & $0.00$ \\ 
 
$400$ & $0.0320$ & $0.5960$ & $2.31$ & $1.91$ & $0.95$ & $0.48$ & $0.08$ & $0.90$ & $0.48$ & $-0.43$ & $-0.05$ & $0.62$ & $0.00$ \\ 
 
$400$ & $0.0500$ & $0.4920$ & $2.51$ & $2.05$ & $1.08$ & $0.62$ & $0.26$ & $0.97$ & $0.59$ & $-0.53$ & $-0.09$ & $0.55$ & $0.00$ \\ 
 
$400$ & $0.0800$ & $0.4180$ & $3.26$ & $2.15$ & $1.03$ & $0.28$ & $0.42$ & $2.23$ & $0.24$ & $-0.27$ & $0.14$ & $2.20$ & $0.00$ \\ 
 
$400$ & $0.1300$ & $0.3590$ & $2.88$ & $2.11$ & $1.48$ & $0.77$ & $0.37$ & $1.28$ & $0.71$ & $-0.65$ & $-0.24$ & $0.81$ & $0.00$ \\ 
 
$400$ & $0.1800$ & $0.3005$ & $4.89$ & $2.76$ & $2.57$ & $1.25$ & $1.74$ & $3.11$ & $1.18$ & $-0.64$ & $-0.41$ & $-2.77$ & $0.00$ \\ 
 
$400$ & $0.4000$ & $0.1541$ & $7.61$ & $3.56$ & $3.27$ & $1.33$ & $2.25$ & $5.88$ & $1.25$ & $-0.76$ & $-0.36$ & $-5.69$ & $0.00$ \\ 
 
\hline 
$500$ & $0.0080$ & $0.9664$ & $3.15$ & $2.93$ & $1.03$ & $0.31$ & $0.07$ & $0.55$ & $-0.16$ & $-0.37$ & $0.01$ & $0.15$ & $-0.34$ \\ 
 
$500$ & $0.0130$ & $0.8628$ & $2.37$ & $2.10$ & $0.91$ & $0.41$ & $0.02$ & $0.61$ & $-0.38$ & $-0.47$ & $0.01$ & $0.10$ & $-0.04$ \\ 
 
$500$ & $0.0200$ & $0.7297$ & $2.46$ & $2.09$ & $1.00$ & $0.57$ & $0.00$ & $0.81$ & $-0.55$ & $-0.59$ & $0.00$ & $0.00$ & $0.00$ \\ 
 
$500$ & $0.0320$ & $0.5975$ & $2.68$ & $2.28$ & $1.02$ & $0.44$ & $0.37$ & $0.98$ & $0.45$ & $-0.39$ & $-0.07$ & $0.77$ & $-0.01$ \\ 
 
$500$ & $0.0500$ & $0.5146$ & $2.70$ & $2.29$ & $1.08$ & $0.59$ & $0.23$ & $0.94$ & $0.59$ & $-0.48$ & $-0.12$ & $0.53$ & $0.00$ \\ 
 
$500$ & $0.0800$ & $0.4219$ & $3.11$ & $2.48$ & $1.01$ & $0.35$ & $0.18$ & $1.58$ & $0.35$ & $-0.28$ & $-0.03$ & $1.52$ & $0.00$ \\ 
 
$500$ & $0.1300$ & $0.3936$ & $3.62$ & $2.93$ & $1.36$ & $0.54$ & $0.06$ & $1.64$ & $0.54$ & $-0.43$ & $-0.10$ & $1.48$ & $0.00$ \\ 
 
$500$ & $0.1800$ & $0.3095$ & $3.70$ & $3.13$ & $1.78$ & $0.73$ & $0.79$ & $0.88$ & $0.73$ & $-0.41$ & $-0.24$ & $-0.13$ & $0.00$ \\ 
 
$500$ & $0.2500$ & $0.2451$ & $6.29$ & $3.83$ & $2.82$ & $1.19$ & $2.00$ & $4.12$ & $1.19$ & $-0.56$ & $-0.34$ & $-3.89$ & $0.00$ \\ 
 
\hline 
$650$ & $0.0130$ & $0.8608$ & $2.56$ & $2.35$ & $0.90$ & $0.29$ & $0.04$ & $0.47$ & $-0.26$ & $-0.38$ & $0.01$ & $0.09$ & $-0.08$ \\ 
 
$650$ & $0.0200$ & $0.7302$ & $2.82$ & $2.47$ & $1.15$ & $0.73$ & $0.00$ & $0.76$ & $-0.56$ & $-0.51$ & $0.00$ & $0.02$ & $-0.02$ \\ 
 
$650$ & $0.0320$ & $0.6111$ & $2.95$ & $2.56$ & $1.16$ & $0.70$ & $0.00$ & $0.90$ & $-0.66$ & $-0.61$ & $0.00$ & $0.00$ & $0.00$ \\ 
 
$650$ & $0.0500$ & $0.4995$ & $3.15$ & $2.74$ & $1.22$ & $0.74$ & $0.20$ & $0.99$ & $0.78$ & $-0.42$ & $-0.12$ & $0.42$ & $-0.01$ \\ 
 
$650$ & $0.0800$ & $0.3952$ & $3.58$ & $3.04$ & $1.16$ & $0.54$ & $0.18$ & $1.50$ & $0.59$ & $-0.62$ & $0.05$ & $1.24$ & $0.00$ \\ 
 
$650$ & $0.1300$ & $0.3548$ & $3.94$ & $3.30$ & $1.40$ & $0.45$ & $0.02$ & $1.65$ & $0.51$ & $-0.10$ & $-0.10$ & $1.56$ & $0.00$ \\ 
 
$650$ & $0.1800$ & $0.3200$ & $4.16$ & $3.57$ & $1.80$ & $0.80$ & $0.61$ & $1.16$ & $0.82$ & $-0.58$ & $-0.23$ & $-0.53$ & $0.00$ \\ 
 
$650$ & $0.2500$ & $0.2397$ & $6.24$ & $4.65$ & $2.49$ & $1.10$ & $1.51$ & $3.32$ & $1.14$ & $-0.48$ & $-0.35$ & $-3.06$ & $0.00$ \\ 
 
$650$ & $0.4000$ & $0.1337$ & $8.49$ & $6.70$ & $3.67$ & $1.57$ & $2.32$ & $3.70$ & $1.59$ & $-0.51$ & $-0.54$ & $-3.26$ & $0.00$ \\ 
 
\hline 

\end{tabular} 
\end{center} 
\caption[RESULT] 
{\label{tab:ncdxdq2_posLH} The NC 
$e^+p$ reduced cross section $\tilde{\sigma}_{\rm NC}(x,Q^2)$ 
with lepton beam polarisation $P_e=-37.0$\% 
with statistical 
$(\delta_{\rm stat})$, 
total $(\delta_{tot})$, 
total uncorrelated systematic $(\delta_{\rm unc})$ 
errors, two of its contributions from the 
 electron energy error ($\delta_{unc}^{E}$)  
and the hadronic energy error  
($\delta_{\rm unc}^{h}$). 
The effect of the other uncorrelated 
systematic errors is included in $\delta_{\rm unc}$. 
In addition the correlated systematic  
$(\delta_{\rm cor})$ and its contributions from a 
positive variation of one  
standard deviation of the 
electron energy error ($\delta_{cor}^{E^+}$), of 
the polar electron angle error 
($\delta_{\rm cor}^{\theta^+}$), of the hadronic 
energy error ($\delta_{\rm cor}^{h^+}$), of the error 
due to noise subtraction ($\delta_{\rm cor}^{N^+}$) 
and of the error due to background subtraction 
($\delta_{\rm cor}^{B^+}$) are given. 
The normalisation and polarisation uncertainties are 
not included in the errors. 
The table continues on the next page.}
\end{table} 
\begin{table}[htbp] 
\begin{center} 
\tiny 
\begin{tabular}{|r|c|r|r|r|r|r|r|r|r|r|r|r|r|} 
\hline 
$Q^2$ & $x$ & $\tilde{\sigma}_{\rm NC}$ & 
$\delta_{\rm tot}$ & $\delta_{\rm stat}$ & $\delta_{\rm unc}$ & 
$\delta_{\rm unc}^{E}$ & 
$\delta_{\rm unc}^{h}$& 
$\delta_{\rm cor}$ & 
$\delta_{\rm cor}^{E^+}$ & 
$\delta_{\rm cor}^{\theta^+}$& 
$\delta_{\rm cor}^{h^+}$& 
$\delta_{\rm cor}^{N^+}$& 
$\delta_{\rm cor}^{B^+}$ \\ 
$(\rm GeV^2)$ & & & 
$(\%)$ & $(\%)$ & $(\%)$ & $(\%)$ & $(\%)$ & $(\%)$ & 
$(\%)$ & $(\%)$ & $(\%)$ & $(\%)$ & $(\%)$  
\\ \hline 
$800$ & $0.0130$ & $0.7821$ & $4.13$ & $3.94$ & $1.15$ & $0.17$ & $0.08$ & $0.44$ & $-0.26$ & $-0.29$ & $0.03$ & $0.13$ & $-0.14$ \\ 
 
$800$ & $0.0200$ & $0.6885$ & $3.07$ & $2.84$ & $1.06$ & $0.50$ & $0.02$ & $0.49$ & $-0.31$ & $-0.37$ & $0.00$ & $0.05$ & $-0.06$ \\ 
 
$800$ & $0.0320$ & $0.5827$ & $3.37$ & $3.09$ & $1.20$ & $0.67$ & $0.00$ & $0.61$ & $-0.51$ & $-0.33$ & $0.00$ & $0.00$ & $0.00$ \\ 
 
$800$ & $0.0500$ & $0.5163$ & $3.40$ & $3.14$ & $1.09$ & $0.27$ & $0.22$ & $0.70$ & $0.46$ & $-0.30$ & $-0.12$ & $0.42$ & $0.00$ \\ 
 
$800$ & $0.0800$ & $0.4596$ & $3.78$ & $3.35$ & $1.28$ & $0.56$ & $0.25$ & $1.19$ & $0.74$ & $-0.55$ & $-0.17$ & $0.73$ & $-0.01$ \\ 
 
$800$ & $0.1300$ & $0.3435$ & $4.73$ & $4.04$ & $1.57$ & $0.33$ & $0.51$ & $1.90$ & $0.58$ & $-0.61$ & $-0.03$ & $1.71$ & $0.00$ \\ 
 
$800$ & $0.1800$ & $0.3167$ & $4.64$ & $4.18$ & $1.85$ & $0.60$ & $0.74$ & $0.80$ & $0.64$ & $-0.44$ & $-0.16$ & $0.09$ & $0.00$ \\ 
 
$800$ & $0.2500$ & $0.2177$ & $6.15$ & $5.17$ & $2.49$ & $1.17$ & $1.27$ & $2.24$ & $1.22$ & $-0.55$ & $-0.41$ & $-1.75$ & $0.00$ \\ 
 
$800$ & $0.4000$ & $0.1220$ & $10.23$ & $7.21$ & $4.57$ & $2.36$ & $2.94$ & $5.64$ & $2.31$ & $-0.64$ & $-0.69$ & $-5.06$ & $0.00$ \\ 
 
\hline 
$1000$ & $0.0130$ & $0.8145$ & $4.35$ & $3.89$ & $1.58$ & $0.14$ & $0.26$ & $1.13$ & $-0.13$ & $-0.35$ & $0.05$ & $0.20$ & $-1.05$ \\ 
 
$1000$ & $0.0200$ & $0.7280$ & $3.45$ & $3.28$ & $0.99$ & $0.34$ & $0.04$ & $0.43$ & $-0.21$ & $-0.35$ & $0.01$ & $0.10$ & $-0.10$ \\ 
 
$1000$ & $0.0320$ & $0.5753$ & $3.60$ & $3.38$ & $1.10$ & $0.59$ & $0.00$ & $0.60$ & $-0.34$ & $-0.50$ & $0.00$ & $0.01$ & $-0.02$ \\ 
 
$1000$ & $0.0500$ & $0.4892$ & $4.11$ & $3.58$ & $1.70$ & $1.37$ & $0.00$ & $1.09$ & $-1.02$ & $-0.40$ & $0.00$ & $0.00$ & $0.00$ \\ 
 
$1000$ & $0.0800$ & $0.4180$ & $4.17$ & $3.88$ & $1.28$ & $0.66$ & $0.34$ & $0.78$ & $0.56$ & $-0.25$ & $-0.23$ & $0.43$ & $-0.01$ \\ 
 
$1000$ & $0.1300$ & $0.3302$ & $5.14$ & $4.75$ & $1.46$ & $0.33$ & $0.25$ & $1.31$ & $0.33$ & $-0.28$ & $-0.05$ & $1.24$ & $0.00$ \\ 
 
$1000$ & $0.1800$ & $0.2940$ & $5.24$ & $4.92$ & $1.68$ & $0.43$ & $0.52$ & $0.69$ & $0.42$ & $-0.22$ & $-0.06$ & $0.49$ & $0.00$ \\ 
 
$1000$ & $0.2500$ & $0.2301$ & $6.08$ & $5.56$ & $1.98$ & $0.61$ & $0.78$ & $1.45$ & $0.86$ & $-0.35$ & $-0.22$ & $-1.09$ & $0.00$ \\ 
 
$1000$ & $0.4000$ & $0.1182$ & $9.76$ & $7.75$ & $3.92$ & $1.42$ & $2.74$ & $4.46$ & $1.27$ & $-0.28$ & $-0.62$ & $-4.22$ & $0.00$ \\ 
 
\hline 
$1200$ & $0.0130$ & $0.7796$ & $7.54$ & $6.75$ & $2.51$ & $0.05$ & $0.29$ & $2.23$ & $-0.16$ & $-0.35$ & $0.04$ & $0.22$ & $-2.19$ \\ 
 
$1200$ & $0.0200$ & $0.6598$ & $4.48$ & $4.32$ & $1.09$ & $0.15$ & $0.09$ & $0.48$ & $0.17$ & $-0.35$ & $0.02$ & $0.13$ & $-0.24$ \\ 
 
$1200$ & $0.0320$ & $0.6262$ & $3.91$ & $3.76$ & $0.96$ & $0.48$ & $0.02$ & $0.47$ & $-0.27$ & $-0.38$ & $0.01$ & $0.06$ & $0.00$ \\ 
 
$1200$ & $0.0500$ & $0.5134$ & $4.22$ & $3.95$ & $1.28$ & $0.95$ & $0.00$ & $0.74$ & $-0.52$ & $-0.53$ & $0.00$ & $0.00$ & $0.00$ \\ 
 
$1200$ & $0.0800$ & $0.4159$ & $4.54$ & $4.33$ & $1.11$ & $0.64$ & $0.03$ & $0.75$ & $0.46$ & $-0.26$ & $-0.05$ & $0.53$ & $0.00$ \\ 
 
$1200$ & $0.1300$ & $0.3167$ & $5.68$ & $5.38$ & $1.62$ & $0.99$ & $0.26$ & $0.86$ & $0.53$ & $-0.24$ & $-0.05$ & $0.63$ & $0.00$ \\ 
 
$1200$ & $0.1800$ & $0.2902$ & $5.80$ & $5.47$ & $1.54$ & $0.64$ & $0.18$ & $1.14$ & $0.55$ & $-0.32$ & $-0.19$ & $0.93$ & $0.00$ \\ 
 
$1200$ & $0.2500$ & $0.2617$ & $6.17$ & $5.62$ & $2.28$ & $1.37$ & $1.01$ & $1.09$ & $0.89$ & $-0.30$ & $-0.31$ & $-0.44$ & $0.00$ \\ 
 
$1200$ & $0.4000$ & $0.1060$ & $10.52$ & $8.59$ & $4.01$ & $2.31$ & $2.41$ & $4.55$ & $1.41$ & $-0.53$ & $-0.41$ & $-4.28$ & $0.00$ \\ 
 
\hline 
$1500$ & $0.0200$ & $0.6541$ & $5.65$ & $5.31$ & $1.75$ & $0.19$ & $0.19$ & $0.83$ & $0.24$ & $-0.20$ & $0.04$ & $0.20$ & $-0.74$ \\ 
 
$1500$ & $0.0320$ & $0.5841$ & $5.32$ & $5.22$ & $0.95$ & $0.07$ & $0.04$ & $0.35$ & $-0.14$ & $-0.30$ & $0.01$ & $0.11$ & $-0.05$ \\ 
 
$1500$ & $0.0500$ & $0.5171$ & $4.72$ & $4.55$ & $1.10$ & $0.61$ & $0.00$ & $0.59$ & $-0.21$ & $-0.55$ & $0.00$ & $0.01$ & $0.00$ \\ 
 
$1500$ & $0.0800$ & $0.4591$ & $5.38$ & $5.12$ & $1.49$ & $1.13$ & $0.11$ & $0.72$ & $0.61$ & $-0.19$ & $-0.09$ & $0.32$ & $0.00$ \\ 
 
$1500$ & $0.1300$ & $0.2984$ & $6.60$ & $6.35$ & $1.54$ & $0.76$ & $0.12$ & $0.91$ & $0.52$ & $-0.31$ & $-0.09$ & $0.68$ & $0.00$ \\ 
 
$1500$ & $0.1800$ & $0.2857$ & $6.66$ & $6.32$ & $1.75$ & $0.97$ & $0.08$ & $1.13$ & $0.60$ & $-0.17$ & $0.05$ & $0.94$ & $0.00$ \\ 
 
$1500$ & $0.2500$ & $0.2522$ & $6.93$ & $6.39$ & $2.49$ & $1.57$ & $1.13$ & $1.04$ & $0.89$ & $-0.16$ & $-0.28$ & $-0.42$ & $0.00$ \\ 
 
$1500$ & $0.4000$ & $0.1182$ & $10.27$ & $8.89$ & $3.60$ & $2.07$ & $1.88$ & $3.67$ & $1.15$ & $-0.19$ & $-0.52$ & $-3.44$ & $0.00$ \\ 
 
$1500$ & $0.6500$ & $0.01535$ & $20.35$ & $16.04$ & $7.26$ & $4.36$ & $4.95$ & $10.19$ & $3.17$ & $-0.56$ & $-1.15$ & $-9.60$ & $0.00$ \\ 
 
\hline 
$2000$ & $0.0219$ & $0.6530$ & $9.14$ & $8.55$ & $2.83$ & $0.58$ & $0.21$ & $1.52$ & $0.08$ & $-0.07$ & $0.05$ & $0.19$ & $-1.50$ \\ 
 
$2000$ & $0.0320$ & $0.5371$ & $6.19$ & $5.99$ & $1.46$ & $0.20$ & $0.07$ & $0.59$ & $-0.23$ & $-0.46$ & $0.01$ & $0.10$ & $-0.26$ \\ 
 
$2000$ & $0.0500$ & $0.5043$ & $5.78$ & $5.63$ & $1.22$ & $0.64$ & $0.01$ & $0.40$ & $-0.34$ & $-0.21$ & $0.00$ & $0.05$ & $0.00$ \\ 
 
$2000$ & $0.0800$ & $0.4257$ & $5.74$ & $5.55$ & $1.25$ & $0.63$ & $0.00$ & $0.71$ & $-0.37$ & $-0.61$ & $0.00$ & $0.00$ & $0.00$ \\ 
 
$2000$ & $0.1300$ & $0.3059$ & $7.45$ & $7.22$ & $1.72$ & $0.97$ & $0.05$ & $0.69$ & $0.37$ & $-0.18$ & $-0.09$ & $0.55$ & $0.00$ \\ 
 
$2000$ & $0.1800$ & $0.2879$ & $7.67$ & $7.37$ & $1.72$ & $0.68$ & $0.28$ & $1.27$ & $0.45$ & $-0.26$ & $0.05$ & $1.16$ & $0.00$ \\ 
 
$2000$ & $0.2500$ & $0.2094$ & $8.29$ & $7.85$ & $2.40$ & $1.55$ & $0.82$ & $1.15$ & $1.05$ & $-0.32$ & $-0.28$ & $0.18$ & $0.00$ \\ 
 
$2000$ & $0.4000$ & $0.1157$ & $11.03$ & $9.92$ & $3.69$ & $2.16$ & $1.86$ & $3.12$ & $1.33$ & $-0.20$ & $-0.36$ & $-2.79$ & $0.00$ \\ 
 
$2000$ & $0.6500$ & $0.01290$ & $23.59$ & $19.28$ & $8.15$ & $5.05$ & $5.54$ & $10.87$ & $3.09$ & $-0.42$ & $-1.40$ & $-10.32$ & $0.00$ \\ 
 
\hline 
$3000$ & $0.0320$ & $0.5739$ & $6.00$ & $5.61$ & $1.92$ & $0.21$ & $0.12$ & $0.92$ & $-0.22$ & $-0.28$ & $0.02$ & $0.12$ & $-0.85$ \\ 
 
$3000$ & $0.0500$ & $0.4657$ & $5.21$ & $5.02$ & $1.34$ & $0.13$ & $0.05$ & $0.35$ & $0.12$ & $-0.31$ & $0.02$ & $0.10$ & $-0.06$ \\ 
 
$3000$ & $0.0800$ & $0.4009$ & $5.54$ & $5.36$ & $1.29$ & $0.58$ & $0.01$ & $0.55$ & $-0.25$ & $-0.49$ & $0.00$ & $0.03$ & $-0.02$ \\ 
 
$3000$ & $0.1300$ & $0.3251$ & $6.63$ & $6.38$ & $1.64$ & $0.75$ & $0.00$ & $0.77$ & $-0.61$ & $-0.48$ & $0.00$ & $0.00$ & $0.00$ \\ 
 
$3000$ & $0.1800$ & $0.2637$ & $7.49$ & $7.11$ & $2.14$ & $1.51$ & $0.09$ & $1.02$ & $0.89$ & $-0.16$ & $-0.03$ & $0.46$ & $0.00$ \\ 
 
$3000$ & $0.2500$ & $0.2157$ & $7.43$ & $7.07$ & $2.08$ & $1.33$ & $0.17$ & $0.97$ & $0.76$ & $-0.13$ & $-0.10$ & $0.58$ & $0.00$ \\ 
 
$3000$ & $0.4000$ & $0.1236$ & $9.46$ & $8.29$ & $3.88$ & $2.60$ & $1.79$ & $2.39$ & $1.54$ & $-0.07$ & $-0.48$ & $-1.77$ & $0.00$ \\ 
 
$3000$ & $0.6500$ & $0.01266$ & $21.18$ & $16.94$ & $8.41$ & $5.43$ & $5.69$ & $9.53$ & $3.46$ & $-0.23$ & $-1.22$ & $-8.79$ & $0.00$ \\ 
 
\hline 
$5000$ & $0.0547$ & $0.4226$ & $8.31$ & $7.86$ & $2.36$ & $0.23$ & $0.08$ & $1.32$ & $0.11$ & $-0.19$ & $0.03$ & $0.09$ & $-1.30$ \\ 
 
$5000$ & $0.0800$ & $0.3438$ & $6.58$ & $6.38$ & $1.57$ & $0.04$ & $0.06$ & $0.38$ & $0.03$ & $-0.32$ & $-0.02$ & $0.07$ & $-0.19$ \\ 
 
$5000$ & $0.1300$ & $0.3067$ & $7.57$ & $7.32$ & $1.83$ & $0.35$ & $0.00$ & $0.55$ & $0.24$ & $-0.49$ & $0.00$ & $0.01$ & $0.00$ \\ 
 
$5000$ & $0.1800$ & $0.2572$ & $8.45$ & $8.19$ & $1.98$ & $0.69$ & $0.00$ & $0.71$ & $-0.49$ & $-0.51$ & $0.00$ & $0.00$ & $0.00$ \\ 
 
$5000$ & $0.2500$ & $0.2211$ & $9.24$ & $8.92$ & $2.30$ & $1.06$ & $0.00$ & $0.72$ & $0.71$ & $-0.06$ & $0.00$ & $0.00$ & $0.00$ \\ 
 
$5000$ & $0.4000$ & $0.09421$ & $12.41$ & $11.65$ & $3.85$ & $2.29$ & $1.77$ & $1.83$ & $1.42$ & $0.17$ & $-0.59$ & $-0.98$ & $0.00$ \\ 
 
$5000$ & $0.6500$ & $0.006800$ & $29.05$ & $27.87$ & $6.12$ & $3.32$ & $3.81$ & $5.48$ & $1.57$ & $0.18$ & $-0.94$ & $-5.17$ & $0.00$ \\ 
 
\hline 
$8000$ & $0.0875$ & $0.2513$ & $16.23$ & $15.58$ & $3.61$ & $0.36$ & $0.12$ & $2.80$ & $-0.07$ & $-0.23$ & $0.09$ & $0.08$ & $-2.79$ \\ 
 
$8000$ & $0.1300$ & $0.2539$ & $11.45$ & $11.14$ & $2.57$ & $0.57$ & $0.06$ & $0.62$ & $0.35$ & $-0.47$ & $0.03$ & $0.04$ & $-0.20$ \\ 
 
$8000$ & $0.1800$ & $0.2290$ & $11.56$ & $11.31$ & $2.40$ & $0.22$ & $0.01$ & $0.32$ & $0.28$ & $-0.14$ & $0.00$ & $0.02$ & $0.00$ \\ 
 
$8000$ & $0.2500$ & $0.2172$ & $11.35$ & $11.01$ & $2.71$ & $1.07$ & $0.00$ & $0.65$ & $-0.42$ & $-0.50$ & $0.00$ & $0.00$ & $0.00$ \\ 
 
$8000$ & $0.4000$ & $0.1002$ & $15.96$ & $15.10$ & $4.67$ & $3.42$ & $0.00$ & $2.21$ & $2.18$ & $0.40$ & $0.00$ & $0.00$ & $0.00$ \\ 
 
$8000$ & $0.6500$ & $0.01154$ & $26.40$ & $25.07$ & $7.14$ & $4.69$ & $3.58$ & $4.23$ & $2.92$ & $-0.16$ & $-1.15$ & $-2.84$ & $0.00$ \\ 
 
\hline 
$12000$ & $0.1300$ & $0.2029$ & $29.92$ & $28.52$ & $9.03$ & $0.10$ & $0.10$ & $0.78$ & $-0.60$ & $-0.21$ & $0.03$ & $0.09$ & $-0.44$ \\ 
 
$12000$ & $0.1800$ & $0.2060$ & $17.86$ & $17.53$ & $3.05$ & $0.76$ & $0.07$ & $1.53$ & $0.18$ & $-0.41$ & $0.02$ & $0.07$ & $-1.46$ \\ 
 
$12000$ & $0.2500$ & $0.1398$ & $18.87$ & $18.70$ & $2.49$ & $0.91$ & $0.02$ & $0.57$ & $0.48$ & $-0.30$ & $0.00$ & $0.04$ & $0.00$ \\ 
 
$12000$ & $0.4000$ & $0.07077$ & $24.82$ & $24.35$ & $4.53$ & $3.49$ & $0.00$ & $1.56$ & $1.52$ & $0.34$ & $0.00$ & $0.00$ & $0.00$ \\ 
 
$12000$ & $0.6500$ & $0.007830$ & $45.47$ & $44.93$ & $6.39$ & $4.77$ & $2.77$ & $2.84$ & $2.39$ & $0.29$ & $-0.64$ & $-1.37$ & $0.00$ \\ 
 
\hline 
$20000$ & $0.2500$ & $0.1059$ & $33.02$ & $32.86$ & $3.23$ & $1.72$ & $0.09$ & $0.70$ & $0.56$ & $-0.35$ & $0.01$ & $0.09$ & $-0.22$ \\ 
 
$20000$ & $0.4000$ & $0.07575$ & $32.17$ & $31.82$ & $4.62$ & $3.56$ & $0.03$ & $1.17$ & $1.16$ & $0.12$ & $0.01$ & $0.03$ & $0.00$ \\ 
 
$20000$ & $0.6500$ & $0.01318$ & $59.92$ & $57.87$ & $14.88$ & $14.39$ & $0.00$ & $4.56$ & $4.12$ & $1.95$ & $0.00$ & $0.00$ & $0.00$ \\ 
 
\hline 
\end{tabular} 
\end{center} 
\captcont{continued.}
\end{table} 

\begin{table}[htbp] 
\begin{center} 
\tiny 
\begin{tabular}{|r|c|r|r|r|r|r|r|r|r|r|r|r|r|} 
\hline 
$Q^2$ & $x$ & $\tilde{\sigma}_{\rm NC}$ & 
$\delta_{\rm tot}$ & $\delta_{\rm stat}$ & $\delta_{\rm unc}$ & 
$\delta_{\rm unc}^{E}$ & 
$\delta_{\rm unc}^{h}$& 
$\delta_{\rm cor}$ & 
$\delta_{\rm cor}^{E^+}$ & 
$\delta_{\rm cor}^{\theta^+}$& 
$\delta_{\rm cor}^{h^+}$& 
$\delta_{\rm cor}^{N^+}$& 
$\delta_{\rm cor}^{B^+}$ \\ 
$(\rm GeV^2)$ & & & 
$(\%)$ & $(\%)$ & $(\%)$ & $(\%)$ & $(\%)$ & $(\%)$ & 
$(\%)$ & $(\%)$ & $(\%)$ & $(\%)$ & $(\%)$  
\\ \hline 
$120$ & $0.0020$ & $1.327$ & $1.63$ & $0.87$ & $0.82$ & $0.33$ & $0.08$ & $1.11$ & $-0.22$ & $-0.70$ & $0.02$ & $0.19$ & $-0.81$ \\ 
 
$120$ & $0.0032$ & $1.169$ & $1.98$ & $1.27$ & $1.24$ & $0.91$ & $0.04$ & $0.88$ & $-0.60$ & $-0.61$ & $0.01$ & $0.16$ & $-0.12$ \\ 
 
\hline 
$150$ & $0.0032$ & $1.200$ & $1.36$ & $0.74$ & $0.84$ & $0.49$ & $0.06$ & $0.77$ & $-0.38$ & $-0.53$ & $0.01$ & $0.16$ & $-0.38$ \\ 
 
$150$ & $0.0050$ & $1.075$ & $1.59$ & $0.88$ & $1.06$ & $0.78$ & $0.00$ & $0.79$ & $-0.43$ & $-0.66$ & $0.00$ & $0.03$ & $-0.03$ \\ 
 
$150$ & $0.0080$ & $0.9349$ & $2.67$ & $1.22$ & $2.00$ & $1.68$ & $0.70$ & $1.29$ & $-1.12$ & $-0.56$ & $-0.19$ & $-0.26$ & $-0.04$ \\ 
 
$150$ & $0.0130$ & $0.7687$ & $3.92$ & $1.71$ & $2.72$ & $2.32$ & $1.01$ & $2.24$ & $-1.38$ & $-1.43$ & $-0.41$ & $-0.94$ & $-0.01$ \\ 
 
\hline 
$200$ & $0.0032$ & $1.201$ & $1.86$ & $1.40$ & $0.85$ & $0.12$ & $0.08$ & $0.88$ & $-0.11$ & $-0.62$ & $0.02$ & $0.17$ & $-0.60$ \\ 
 
$200$ & $0.0050$ & $1.073$ & $1.51$ & $0.97$ & $0.89$ & $0.55$ & $0.04$ & $0.74$ & $-0.39$ & $-0.61$ & $0.01$ & $0.12$ & $-0.07$ \\ 
 
$200$ & $0.0080$ & $0.9326$ & $1.87$ & $0.98$ & $1.25$ & $1.01$ & $0.00$ & $1.00$ & $-0.58$ & $-0.82$ & $0.00$ & $0.00$ & $0.00$ \\ 
 
$200$ & $0.0130$ & $0.7830$ & $1.54$ & $1.11$ & $0.76$ & $0.03$ & $0.05$ & $0.75$ & $-0.11$ & $-0.36$ & $-0.07$ & $0.64$ & $-0.01$ \\ 
 
$200$ & $0.0200$ & $0.6755$ & $1.81$ & $1.22$ & $1.09$ & $0.75$ & $0.04$ & $0.76$ & $-0.46$ & $-0.44$ & $-0.04$ & $0.42$ & $0.00$ \\ 
 
$200$ & $0.0320$ & $0.5719$ & $2.14$ & $1.40$ & $1.37$ & $0.91$ & $0.59$ & $0.85$ & $-0.61$ & $-0.56$ & $-0.17$ & $0.14$ & $0.00$ \\ 
 
$200$ & $0.0500$ & $0.4925$ & $2.99$ & $1.62$ & $1.81$ & $1.56$ & $0.10$ & $1.75$ & $-0.92$ & $-0.89$ & $-0.11$ & $1.18$ & $0.00$ \\ 
 
$200$ & $0.0800$ & $0.4300$ & $3.38$ & $1.77$ & $2.07$ & $1.80$ & $0.24$ & $2.00$ & $-1.30$ & $-0.64$ & $-0.12$ & $1.37$ & $0.00$ \\ 
 
$200$ & $0.1300$ & $0.3489$ & $3.50$ & $1.96$ & $2.34$ & $1.75$ & $0.87$ & $1.70$ & $-0.91$ & $-0.89$ & $-0.32$ & $-1.08$ & $0.00$ \\ 
 
$200$ & $0.1800$ & $0.2994$ & $4.43$ & $2.68$ & $2.95$ & $1.52$ & $1.94$ & $1.93$ & $-0.80$ & $-1.21$ & $-0.34$ & $-1.23$ & $0.00$ \\ 
 
\hline 
$250$ & $0.0050$ & $1.102$ & $1.53$ & $1.13$ & $0.78$ & $0.29$ & $0.06$ & $0.67$ & $-0.28$ & $-0.47$ & $0.01$ & $0.16$ & $-0.34$ \\ 
 
$250$ & $0.0080$ & $0.9413$ & $1.63$ & $1.10$ & $0.97$ & $0.64$ & $0.01$ & $0.72$ & $-0.56$ & $-0.45$ & $0.00$ & $0.03$ & $-0.01$ \\ 
 
$250$ & $0.0130$ & $0.7973$ & $2.07$ & $1.22$ & $1.24$ & $0.94$ & $0.25$ & $1.13$ & $0.50$ & $-0.60$ & $0.03$ & $0.81$ & $-0.02$ \\ 
 
$250$ & $0.0200$ & $0.6883$ & $2.11$ & $1.25$ & $1.33$ & $1.08$ & $0.12$ & $1.06$ & $0.51$ & $-0.67$ & $-0.05$ & $0.65$ & $-0.01$ \\ 
 
$250$ & $0.0320$ & $0.5715$ & $2.05$ & $1.31$ & $1.13$ & $0.81$ & $0.13$ & $1.09$ & $0.35$ & $-0.50$ & $-0.01$ & $0.91$ & $0.00$ \\ 
 
$250$ & $0.0500$ & $0.4919$ & $2.42$ & $1.45$ & $1.21$ & $0.88$ & $0.15$ & $1.51$ & $0.36$ & $-0.58$ & $-0.06$ & $1.34$ & $0.00$ \\ 
 
$250$ & $0.0800$ & $0.4249$ & $2.91$ & $1.46$ & $1.07$ & $0.58$ & $0.13$ & $2.28$ & $0.39$ & $-0.57$ & $-0.08$ & $2.17$ & $0.00$ \\ 
 
$250$ & $0.1300$ & $0.3516$ & $2.52$ & $1.60$ & $1.72$ & $1.12$ & $0.56$ & $0.93$ & $0.48$ & $-0.54$ & $-0.17$ & $-0.55$ & $0.00$ \\ 
 
$250$ & $0.1800$ & $0.2913$ & $4.40$ & $2.20$ & $2.99$ & $1.90$ & $1.81$ & $2.36$ & $1.18$ & $-0.86$ & $-0.41$ & $-1.80$ & $0.00$ \\ 
 
\hline 
$300$ & $0.0050$ & $1.117$ & $2.27$ & $1.94$ & $0.95$ & $0.31$ & $0.06$ & $0.69$ & $-0.31$ & $-0.36$ & $0.02$ & $0.14$ & $-0.48$ \\ 
 
$300$ & $0.0080$ & $0.9592$ & $1.64$ & $1.29$ & $0.77$ & $0.24$ & $0.02$ & $0.67$ & $-0.24$ & $-0.62$ & $0.00$ & $0.09$ & $-0.05$ \\ 
 
$300$ & $0.0130$ & $0.7840$ & $1.85$ & $1.31$ & $0.99$ & $0.64$ & $0.00$ & $0.84$ & $-0.64$ & $-0.54$ & $0.00$ & $0.00$ & $-0.01$ \\ 
 
$300$ & $0.0200$ & $0.7024$ & $1.94$ & $1.44$ & $0.98$ & $0.59$ & $0.08$ & $0.85$ & $0.39$ & $-0.47$ & $0.01$ & $0.58$ & $-0.01$ \\ 
 
$300$ & $0.0320$ & $0.5671$ & $2.05$ & $1.51$ & $1.09$ & $0.72$ & $0.13$ & $0.86$ & $0.35$ & $-0.49$ & $-0.02$ & $0.62$ & $0.00$ \\ 
 
$300$ & $0.0500$ & $0.4836$ & $2.42$ & $1.66$ & $1.25$ & $0.92$ & $0.16$ & $1.23$ & $0.38$ & $-0.48$ & $-0.02$ & $1.07$ & $0.00$ \\ 
 
$300$ & $0.0800$ & $0.4296$ & $3.13$ & $1.71$ & $1.44$ & $1.10$ & $0.23$ & $2.20$ & $0.54$ & $-0.75$ & $-0.12$ & $1.99$ & $-0.01$ \\ 
 
$300$ & $0.1300$ & $0.3649$ & $2.92$ & $1.75$ & $1.96$ & $1.46$ & $0.54$ & $1.26$ & $0.70$ & $-0.74$ & $-0.19$ & $0.72$ & $0.00$ \\ 
 
$300$ & $0.1800$ & $0.3024$ & $4.73$ & $2.26$ & $2.92$ & $2.00$ & $1.57$ & $2.96$ & $1.10$ & $-0.68$ & $-0.26$ & $-2.65$ & $0.00$ \\ 
 
$300$ & $0.4000$ & $0.1446$ & $6.55$ & $2.95$ & $3.64$ & $2.32$ & $2.11$ & $4.58$ & $1.31$ & $-1.09$ & $-0.49$ & $-4.22$ & $0.00$ \\ 
 
\hline 
$400$ & $0.0080$ & $0.9652$ & $1.89$ & $1.60$ & $0.81$ & $0.24$ & $0.06$ & $0.58$ & $-0.25$ & $-0.45$ & $0.01$ & $0.14$ & $-0.23$ \\ 
 
$400$ & $0.0130$ & $0.8484$ & $1.85$ & $1.49$ & $0.88$ & $0.43$ & $0.01$ & $0.65$ & $-0.42$ & $-0.50$ & $0.00$ & $0.03$ & $-0.01$ \\ 
 
$400$ & $0.0200$ & $0.6976$ & $2.23$ & $1.57$ & $1.19$ & $0.88$ & $0.00$ & $1.05$ & $-0.88$ & $-0.57$ & $0.00$ & $0.00$ & $0.00$ \\ 
 
$400$ & $0.0320$ & $0.5786$ & $2.19$ & $1.68$ & $0.98$ & $0.50$ & $0.21$ & $1.00$ & $0.49$ & $-0.52$ & $0.04$ & $0.69$ & $-0.01$ \\ 
 
$400$ & $0.0500$ & $0.4884$ & $2.40$ & $1.85$ & $1.15$ & $0.70$ & $0.36$ & $1.01$ & $0.68$ & $-0.46$ & $-0.09$ & $0.58$ & $0.00$ \\ 
 
$400$ & $0.0800$ & $0.4249$ & $2.87$ & $2.00$ & $0.99$ & $0.23$ & $0.35$ & $1.81$ & $0.20$ & $-0.32$ & $-0.08$ & $1.76$ & $0.00$ \\ 
 
$400$ & $0.1300$ & $0.3464$ & $2.61$ & $1.94$ & $1.38$ & $0.53$ & $0.42$ & $1.06$ & $0.49$ & $-0.34$ & $-0.09$ & $0.88$ & $0.00$ \\ 
 
$400$ & $0.1800$ & $0.3011$ & $4.54$ & $2.51$ & $2.37$ & $0.99$ & $1.62$ & $2.95$ & $0.93$ & $-0.50$ & $-0.36$ & $-2.73$ & $0.00$ \\ 
 
$400$ & $0.4000$ & $0.1405$ & $6.94$ & $3.13$ & $3.34$ & $1.40$ & $2.31$ & $5.22$ & $1.31$ & $-0.80$ & $-0.53$ & $-4.96$ & $0.00$ \\ 
 
\hline 
$500$ & $0.0080$ & $0.9649$ & $2.94$ & $2.69$ & $1.02$ & $0.31$ & $0.10$ & $0.59$ & $-0.15$ & $-0.47$ & $0.01$ & $0.16$ & $-0.30$ \\ 
 
$500$ & $0.0130$ & $0.8436$ & $2.08$ & $1.85$ & $0.83$ & $0.19$ & $0.01$ & $0.47$ & $-0.13$ & $-0.45$ & $0.01$ & $0.08$ & $-0.04$ \\ 
 
$500$ & $0.0200$ & $0.7286$ & $2.35$ & $1.89$ & $1.09$ & $0.71$ & $0.00$ & $0.86$ & $-0.70$ & $-0.49$ & $0.00$ & $0.00$ & $0.00$ \\ 
 
$500$ & $0.0320$ & $0.5997$ & $2.34$ & $1.95$ & $1.00$ & $0.40$ & $0.35$ & $0.84$ & $0.39$ & $-0.35$ & $0.08$ & $0.65$ & $-0.02$ \\ 
 
$500$ & $0.0500$ & $0.5235$ & $2.59$ & $2.06$ & $1.21$ & $0.75$ & $0.36$ & $1.01$ & $0.75$ & $-0.49$ & $-0.19$ & $0.41$ & $0.00$ \\ 
 
$500$ & $0.0800$ & $0.4414$ & $2.96$ & $2.21$ & $1.04$ & $0.34$ & $0.35$ & $1.67$ & $0.34$ & $-0.37$ & $0.10$ & $1.59$ & $0.00$ \\ 
 
$500$ & $0.1300$ & $0.3658$ & $3.33$ & $2.49$ & $1.46$ & $0.74$ & $0.10$ & $1.66$ & $0.74$ & $-0.61$ & $0.08$ & $1.36$ & $0.00$ \\ 
 
$500$ & $0.1800$ & $0.3056$ & $3.61$ & $2.91$ & $1.88$ & $0.82$ & $0.92$ & $1.04$ & $0.82$ & $-0.43$ & $-0.20$ & $-0.44$ & $0.00$ \\ 
 
$500$ & $0.2500$ & $0.2453$ & $6.36$ & $3.60$ & $2.87$ & $1.24$ & $2.04$ & $4.38$ & $1.24$ & $-0.54$ & $-0.48$ & $-4.14$ & $0.00$ \\ 
 
\hline 
$650$ & $0.0130$ & $0.8253$ & $2.35$ & $2.14$ & $0.86$ & $0.10$ & $0.06$ & $0.44$ & $-0.15$ & $-0.36$ & $0.01$ & $0.14$ & $-0.14$ \\ 
 
$650$ & $0.0200$ & $0.7135$ & $2.51$ & $2.21$ & $1.00$ & $0.47$ & $0.01$ & $0.62$ & $-0.30$ & $-0.55$ & $0.00$ & $0.03$ & $-0.02$ \\ 
 
$650$ & $0.0320$ & $0.6530$ & $2.88$ & $2.28$ & $1.37$ & $1.00$ & $0.00$ & $1.10$ & $-0.95$ & $-0.56$ & $0.00$ & $0.00$ & $0.00$ \\ 
 
$650$ & $0.0500$ & $0.5198$ & $2.77$ & $2.38$ & $1.09$ & $0.54$ & $0.11$ & $0.88$ & $0.61$ & $-0.31$ & $0.05$ & $0.55$ & $0.00$ \\ 
 
$650$ & $0.0800$ & $0.4273$ & $3.24$ & $2.68$ & $1.10$ & $0.43$ & $0.06$ & $1.44$ & $0.44$ & $-0.42$ & $0.06$ & $1.30$ & $0.00$ \\ 
 
$650$ & $0.1300$ & $0.3794$ & $3.85$ & $3.20$ & $1.51$ & $0.70$ & $0.08$ & $1.54$ & $0.74$ & $-0.34$ & $-0.11$ & $1.29$ & $0.00$ \\ 
 
$650$ & $0.1800$ & $0.3399$ & $3.58$ & $3.07$ & $1.63$ & $0.52$ & $0.41$ & $0.87$ & $0.55$ & $-0.38$ & $-0.26$ & $0.49$ & $0.00$ \\ 
 
$650$ & $0.2500$ & $0.2335$ & $5.75$ & $3.85$ & $2.74$ & $1.46$ & $1.63$ & $3.27$ & $1.46$ & $-0.59$ & $-0.40$ & $-2.84$ & $0.00$ \\ 
 
$650$ & $0.4000$ & $0.1321$ & $8.83$ & $5.35$ & $3.95$ & $1.34$ & $2.84$ & $5.80$ & $1.36$ & $-0.64$ & $-0.64$ & $-5.57$ & $0.00$ \\ 
 
\hline 

\end{tabular} 
\end{center} 
\caption[RESULT] 
{\label{tab:ncdxdq2_posRH} The NC 
$e^+p$ reduced cross section $\tilde{\sigma}_{\rm NC}(x,Q^2)$ 
with lepton beam polarisation $P_e=+32.5$\% 
with statistical 
$(\delta_{\rm stat})$, 
total $(\delta_{tot})$, 
total uncorrelated systematic $(\delta_{\rm unc})$ 
errors, two of its contributions from the 
 electron energy error ($\delta_{unc}^{E}$)  
and the hadronic energy error  
($\delta_{\rm unc}^{h}$). 
The effect of the other uncorrelated 
systematic errors is included in $\delta_{\rm unc}$. 
In addition the correlated systematic  
$(\delta_{\rm cor})$ and its contributions from a 
positive variation of one  
standard deviation of the 
electron energy error ($\delta_{cor}^{E^+}$), of 
the polar electron angle error 
($\delta_{\rm cor}^{\theta^+}$), of the hadronic 
energy error ($\delta_{\rm cor}^{h^+}$), of the error 
due to noise subtraction ($\delta_{\rm cor}^{N^+}$) 
and of the error due to background subtraction 
($\delta_{\rm cor}^{B^+}$) are given. 
The normalisation and polarisation uncertainties are 
not included in the errors. 
The table continues on the next page.}
\end{table} 
\begin{table}[htbp] 
\begin{center} 
\tiny 
\begin{tabular}{|r|c|r|r|r|r|r|r|r|r|r|r|r|r|} 
\hline 
$Q^2$ & $x$ & $\tilde{\sigma}_{\rm NC}$ & 
$\delta_{\rm tot}$ & $\delta_{\rm stat}$ & $\delta_{\rm unc}$ & 
$\delta_{\rm unc}^{E}$ & 
$\delta_{\rm unc}^{h}$& 
$\delta_{\rm cor}$ & 
$\delta_{\rm cor}^{E^+}$ & 
$\delta_{\rm cor}^{\theta^+}$& 
$\delta_{\rm cor}^{h^+}$& 
$\delta_{\rm cor}^{N^+}$& 
$\delta_{\rm cor}^{B^+}$ \\ 
$(\rm GeV^2)$ & & & 
$(\%)$ & $(\%)$ & $(\%)$ & $(\%)$ & $(\%)$ & $(\%)$ & 
$(\%)$ & $(\%)$ & $(\%)$ & $(\%)$ & $(\%)$  
\\ \hline 
$800$ & $0.0130$ & $0.8259$ & $3.73$ & $3.47$ & $1.24$ & $0.16$ & $0.09$ & $0.60$ & $-0.32$ & $-0.41$ & $0.03$ & $0.16$ & $-0.26$ \\ 
 
$800$ & $0.0200$ & $0.6916$ & $2.75$ & $2.52$ & $1.00$ & $0.38$ & $0.01$ & $0.45$ & $-0.28$ & $-0.34$ & $0.01$ & $0.09$ & $-0.04$ \\ 
 
$800$ & $0.0320$ & $0.6241$ & $2.99$ & $2.60$ & $1.26$ & $0.77$ & $0.00$ & $0.78$ & $-0.52$ & $-0.59$ & $0.00$ & $0.00$ & $0.00$ \\ 
 
$800$ & $0.0500$ & $0.5204$ & $3.22$ & $2.90$ & $1.11$ & $0.20$ & $0.34$ & $0.84$ & $0.30$ & $-0.28$ & $0.16$ & $0.71$ & $0.00$ \\ 
 
$800$ & $0.0800$ & $0.4397$ & $3.47$ & $3.06$ & $1.24$ & $0.43$ & $0.30$ & $1.09$ & $0.73$ & $-0.34$ & $-0.09$ & $0.73$ & $0.00$ \\ 
 
$800$ & $0.1300$ & $0.3576$ & $4.12$ & $3.54$ & $1.54$ & $0.53$ & $0.05$ & $1.43$ & $0.68$ & $-0.45$ & $-0.18$ & $1.17$ & $0.00$ \\ 
 
$800$ & $0.1800$ & $0.3237$ & $4.19$ & $3.68$ & $1.73$ & $0.56$ & $0.36$ & $1.04$ & $0.75$ & $-0.50$ & $-0.12$ & $0.51$ & $0.00$ \\ 
 
$800$ & $0.2500$ & $0.2512$ & $5.24$ & $4.23$ & $2.39$ & $0.80$ & $1.38$ & $1.96$ & $0.92$ & $-0.36$ & $-0.45$ & $-1.63$ & $0.00$ \\ 
 
$800$ & $0.4000$ & $0.1186$ & $9.66$ & $6.45$ & $3.98$ & $1.76$ & $2.44$ & $5.99$ & $1.72$ & $-0.50$ & $-0.58$ & $-5.68$ & $0.00$ \\ 
 
\hline 
$1000$ & $0.0130$ & $0.7638$ & $4.22$ & $3.64$ & $1.68$ & $0.20$ & $0.22$ & $1.33$ & $-0.09$ & $-0.25$ & $0.04$ & $0.21$ & $-1.29$ \\ 
 
$1000$ & $0.0200$ & $0.7122$ & $3.17$ & $2.97$ & $0.99$ & $0.29$ & $0.06$ & $0.52$ & $-0.22$ & $-0.43$ & $0.01$ & $0.15$ & $-0.14$ \\ 
 
$1000$ & $0.0320$ & $0.6310$ & $3.14$ & $2.86$ & $1.16$ & $0.70$ & $0.00$ & $0.56$ & $-0.47$ & $-0.32$ & $0.00$ & $0.02$ & $0.00$ \\ 
 
$1000$ & $0.0500$ & $0.5286$ & $3.51$ & $3.09$ & $1.51$ & $1.13$ & $0.00$ & $0.70$ & $-0.44$ & $-0.54$ & $0.00$ & $0.00$ & $0.00$ \\ 
 
$1000$ & $0.0800$ & $0.4441$ & $3.64$ & $3.40$ & $1.15$ & $0.38$ & $0.25$ & $0.55$ & $0.24$ & $-0.20$ & $-0.06$ & $0.44$ & $0.00$ \\ 
 
$1000$ & $0.1300$ & $0.3397$ & $4.88$ & $4.49$ & $1.44$ & $0.28$ & $0.20$ & $1.26$ & $0.28$ & $0.09$ & $0.14$ & $1.22$ & $0.00$ \\ 
 
$1000$ & $0.1800$ & $0.2968$ & $4.73$ & $4.35$ & $1.59$ & $0.25$ & $0.29$ & $0.98$ & $0.50$ & $-0.42$ & $-0.30$ & $0.67$ & $0.00$ \\ 
 
$1000$ & $0.2500$ & $0.2495$ & $5.49$ & $4.74$ & $2.35$ & $1.00$ & $1.26$ & $1.48$ & $1.06$ & $-0.48$ & $-0.21$ & $-0.89$ & $0.00$ \\ 
 
$1000$ & $0.4000$ & $0.1348$ & $10.49$ & $8.52$ & $3.77$ & $1.36$ & $2.54$ & $4.81$ & $1.50$ & $-0.50$ & $-0.65$ & $-4.49$ & $0.00$ \\ 
 
\hline 
$1200$ & $0.0130$ & $0.8534$ & $6.63$ & $5.78$ & $2.56$ & $0.20$ & $0.30$ & $1.98$ & $-0.13$ & $-0.12$ & $0.07$ & $0.24$ & $-1.96$ \\ 
 
$1200$ & $0.0200$ & $0.7304$ & $3.83$ & $3.64$ & $1.12$ & $0.23$ & $0.07$ & $0.42$ & $-0.21$ & $-0.25$ & $0.02$ & $0.12$ & $-0.23$ \\ 
 
$1200$ & $0.0320$ & $0.5777$ & $3.65$ & $3.51$ & $0.93$ & $0.40$ & $0.02$ & $0.45$ & $-0.23$ & $-0.39$ & $0.00$ & $0.07$ & $-0.01$ \\ 
 
$1200$ & $0.0500$ & $0.5028$ & $3.84$ & $3.52$ & $1.33$ & $1.00$ & $0.00$ & $0.74$ & $-0.61$ & $-0.42$ & $0.00$ & $0.00$ & $-0.01$ \\ 
 
$1200$ & $0.0800$ & $0.4340$ & $4.08$ & $3.76$ & $1.35$ & $0.91$ & $0.20$ & $0.82$ & $0.62$ & $-0.33$ & $-0.12$ & $0.40$ & $0.00$ \\ 
 
$1200$ & $0.1300$ & $0.3561$ & $4.93$ & $4.55$ & $1.63$ & $1.03$ & $0.05$ & $0.95$ & $0.63$ & $-0.33$ & $-0.04$ & $0.64$ & $0.00$ \\ 
 
$1200$ & $0.1800$ & $0.3359$ & $5.64$ & $5.25$ & $1.60$ & $0.76$ & $0.23$ & $1.28$ & $0.51$ & $-0.28$ & $-0.11$ & $1.13$ & $-0.02$ \\ 
 
$1200$ & $0.2500$ & $0.2341$ & $6.45$ & $5.97$ & $2.26$ & $1.39$ & $0.93$ & $0.94$ & $0.77$ & $-0.26$ & $-0.25$ & $-0.38$ & $0.00$ \\ 
 
$1200$ & $0.4000$ & $0.1074$ & $10.58$ & $7.57$ & $4.67$ & $2.83$ & $2.99$ & $5.72$ & $1.85$ & $-0.52$ & $-0.69$ & $-5.34$ & $0.00$ \\ 
 
\hline 
$1500$ & $0.0200$ & $0.6877$ & $5.07$ & $4.63$ & $1.84$ & $0.25$ & $0.22$ & $0.95$ & $0.22$ & $-0.32$ & $0.06$ & $0.20$ & $-0.83$ \\ 
 
$1500$ & $0.0320$ & $0.5893$ & $4.45$ & $4.29$ & $1.05$ & $0.46$ & $0.05$ & $0.54$ & $-0.34$ & $-0.40$ & $0.02$ & $0.12$ & $-0.02$ \\ 
 
$1500$ & $0.0500$ & $0.5262$ & $4.18$ & $4.00$ & $1.11$ & $0.63$ & $0.00$ & $0.49$ & $-0.35$ & $-0.33$ & $0.00$ & $0.02$ & $0.00$ \\ 
 
$1500$ & $0.0800$ & $0.4314$ & $4.55$ & $4.31$ & $1.26$ & $0.80$ & $0.14$ & $0.75$ & $0.58$ & $-0.20$ & $-0.11$ & $0.42$ & $0.00$ \\ 
 
$1500$ & $0.1300$ & $0.3534$ & $5.50$ & $5.16$ & $1.71$ & $0.94$ & $0.15$ & $0.82$ & $0.50$ & $-0.20$ & $0.11$ & $0.61$ & $0.00$ \\ 
 
$1500$ & $0.1800$ & $0.3074$ & $5.71$ & $5.38$ & $1.60$ & $0.67$ & $0.13$ & $1.04$ & $0.48$ & $-0.21$ & $0.09$ & $0.89$ & $0.00$ \\ 
 
$1500$ & $0.2500$ & $0.2088$ & $6.79$ & $6.22$ & $2.47$ & $1.56$ & $1.09$ & $1.11$ & $0.98$ & $-0.22$ & $-0.44$ & $-0.21$ & $0.00$ \\ 
 
$1500$ & $0.4000$ & $0.1242$ & $9.25$ & $7.55$ & $3.84$ & $2.20$ & $2.17$ & $3.71$ & $1.26$ & $-0.28$ & $-0.50$ & $-3.45$ & $0.00$ \\ 
 
$1500$ & $0.6500$ & $0.01439$ & $19.87$ & $14.78$ & $7.35$ & $4.62$ & $4.82$ & $11.06$ & $2.53$ & $-0.14$ & $-0.88$ & $-10.73$ & $0.00$ \\ 
 
\hline 
$2000$ & $0.0219$ & $0.7131$ & $8.25$ & $7.48$ & $3.03$ & $0.74$ & $0.22$ & $1.71$ & $0.36$ & $-0.20$ & $0.06$ & $0.22$ & $-1.64$ \\ 
 
$2000$ & $0.0320$ & $0.5443$ & $5.46$ & $5.24$ & $1.47$ & $0.26$ & $0.10$ & $0.44$ & $-0.14$ & $-0.37$ & $0.03$ & $0.12$ & $-0.16$ \\ 
 
$2000$ & $0.0500$ & $0.5434$ & $4.99$ & $4.83$ & $1.21$ & $0.25$ & $0.01$ & $0.36$ & $-0.18$ & $-0.31$ & $0.01$ & $0.03$ & $0.00$ \\ 
 
$2000$ & $0.0800$ & $0.4166$ & $5.32$ & $5.03$ & $1.54$ & $1.09$ & $0.00$ & $0.79$ & $-0.53$ & $-0.58$ & $0.00$ & $0.00$ & $0.00$ \\ 
 
$2000$ & $0.1300$ & $0.3708$ & $6.98$ & $6.71$ & $1.77$ & $1.02$ & $0.13$ & $0.78$ & $0.55$ & $-0.26$ & $-0.08$ & $0.48$ & $0.00$ \\ 
 
$2000$ & $0.1800$ & $0.3058$ & $6.68$ & $6.34$ & $1.68$ & $0.51$ & $0.36$ & $1.28$ & $0.41$ & $-0.28$ & $0.33$ & $1.12$ & $0.00$ \\ 
 
$2000$ & $0.2500$ & $0.2529$ & $7.05$ & $6.45$ & $2.60$ & $1.75$ & $0.98$ & $1.15$ & $1.09$ & $-0.27$ & $-0.19$ & $-0.16$ & $0.00$ \\ 
 
$2000$ & $0.4000$ & $0.1326$ & $9.81$ & $8.15$ & $4.34$ & $2.87$ & $2.26$ & $3.33$ & $1.81$ & $-0.31$ & $-0.80$ & $-2.66$ & $0.00$ \\ 
 
$2000$ & $0.6500$ & $0.01436$ & $21.57$ & $17.19$ & $7.30$ & $4.45$ & $4.75$ & $10.80$ & $2.64$ & $0.23$ & $-0.78$ & $-10.44$ & $0.00$ \\ 
 
\hline 
$3000$ & $0.0320$ & $0.5957$ & $5.43$ & $5.01$ & $1.86$ & $0.25$ & $0.14$ & $0.96$ & $0.15$ & $-0.26$ & $0.04$ & $0.13$ & $-0.90$ \\ 
 
$3000$ & $0.0500$ & $0.5259$ & $4.47$ & $4.22$ & $1.42$ & $0.46$ & $0.08$ & $0.41$ & $-0.21$ & $-0.34$ & $0.01$ & $0.08$ & $-0.06$ \\ 
 
$3000$ & $0.0800$ & $0.4568$ & $4.65$ & $4.45$ & $1.25$ & $0.49$ & $0.01$ & $0.57$ & $-0.38$ & $-0.42$ & $0.00$ & $0.02$ & $0.00$ \\ 
 
$3000$ & $0.1300$ & $0.3442$ & $5.80$ & $5.53$ & $1.62$ & $0.72$ & $0.00$ & $0.57$ & $-0.43$ & $-0.37$ & $0.00$ & $0.00$ & $0.00$ \\ 
 
$3000$ & $0.1800$ & $0.3252$ & $7.51$ & $7.17$ & $2.03$ & $1.34$ & $0.11$ & $0.92$ & $0.69$ & $-0.14$ & $0.05$ & $0.58$ & $0.00$ \\ 
 
$3000$ & $0.2500$ & $0.2280$ & $6.61$ & $6.20$ & $2.07$ & $1.31$ & $0.22$ & $0.96$ & $0.76$ & $-0.08$ & $-0.10$ & $0.58$ & $0.00$ \\ 
 
$3000$ & $0.4000$ & $0.1160$ & $9.03$ & $7.64$ & $4.01$ & $2.65$ & $2.00$ & $2.65$ & $1.49$ & $-0.13$ & $-0.52$ & $-2.13$ & $0.00$ \\ 
 
$3000$ & $0.6500$ & $0.01250$ & $19.52$ & $15.66$ & $7.85$ & $5.15$ & $5.08$ & $8.63$ & $3.17$ & $-0.28$ & $-1.52$ & $-7.87$ & $0.00$ \\ 
 
\hline 
$5000$ & $0.0547$ & $0.4956$ & $7.11$ & $6.66$ & $2.23$ & $0.73$ & $0.13$ & $1.11$ & $0.19$ & $-0.43$ & $0.07$ & $0.11$ & $-0.99$ \\ 
 
$5000$ & $0.0800$ & $0.4546$ & $5.24$ & $4.98$ & $1.59$ & $0.28$ & $0.05$ & $0.40$ & $-0.19$ & $-0.33$ & $0.02$ & $0.07$ & $-0.12$ \\ 
 
$5000$ & $0.1300$ & $0.3598$ & $6.29$ & $5.99$ & $1.85$ & $0.49$ & $0.00$ & $0.53$ & $0.30$ & $-0.44$ & $0.00$ & $0.02$ & $0.00$ \\ 
 
$5000$ & $0.1800$ & $0.3183$ & $6.91$ & $6.59$ & $1.99$ & $0.71$ & $0.00$ & $0.49$ & $-0.35$ & $-0.34$ & $0.00$ & $0.00$ & $0.00$ \\ 
 
$5000$ & $0.2500$ & $0.2062$ & $8.71$ & $8.44$ & $2.11$ & $0.46$ & $0.00$ & $0.44$ & $0.42$ & $-0.12$ & $0.00$ & $0.00$ & $0.00$ \\ 
 
$5000$ & $0.4000$ & $0.1106$ & $10.37$ & $9.78$ & $3.26$ & $1.61$ & $1.17$ & $1.18$ & $1.03$ & $0.04$ & $-0.31$ & $-0.48$ & $0.00$ \\ 
 
$5000$ & $0.6500$ & $0.01196$ & $21.05$ & $18.62$ & $7.25$ & $4.45$ & $4.58$ & $6.62$ & $2.40$ & $-0.25$ & $-1.37$ & $-6.01$ & $0.00$ \\ 
 
\hline 
$8000$ & $0.0875$ & $0.4227$ & $11.07$ & $10.53$ & $3.08$ & $0.73$ & $0.09$ & $1.49$ & $0.25$ & $-0.25$ & $0.03$ & $0.09$ & $-1.45$ \\ 
 
$8000$ & $0.1300$ & $0.3113$ & $9.37$ & $9.01$ & $2.53$ & $0.30$ & $0.06$ & $0.39$ & $-0.16$ & $-0.24$ & $0.01$ & $0.09$ & $-0.25$ \\ 
 
$8000$ & $0.1800$ & $0.2841$ & $9.36$ & $9.01$ & $2.42$ & $0.34$ & $0.00$ & $0.68$ & $0.22$ & $-0.64$ & $0.00$ & $0.01$ & $0.00$ \\ 
 
$8000$ & $0.2500$ & $0.2178$ & $13.40$ & $13.15$ & $2.50$ & $0.22$ & $0.00$ & $0.52$ & $0.45$ & $-0.26$ & $0.00$ & $0.00$ & $0.00$ \\ 
 
$8000$ & $0.4000$ & $0.09797$ & $14.43$ & $13.63$ & $4.43$ & $3.07$ & $0.00$ & $1.67$ & $1.61$ & $0.43$ & $0.00$ & $0.00$ & $0.00$ \\ 
 
$8000$ & $0.6500$ & $0.01497$ & $21.32$ & $19.28$ & $7.87$ & $5.03$ & $4.54$ & $4.58$ & $2.51$ & $0.39$ & $-1.13$ & $-3.64$ & $0.00$ \\ 
 
\hline 
$12000$ & $0.1300$ & $0.2111$ & $28.35$ & $27.73$ & $5.32$ & $1.04$ & $0.10$ & $2.49$ & $-1.16$ & $-0.65$ & $-0.02$ & $0.09$ & $-2.10$ \\ 
 
$12000$ & $0.1800$ & $0.2186$ & $15.26$ & $15.03$ & $2.52$ & $0.67$ & $0.07$ & $0.70$ & $0.41$ & $-0.49$ & $0.02$ & $0.08$ & $-0.28$ \\ 
 
$12000$ & $0.2500$ & $0.1659$ & $15.54$ & $15.33$ & $2.47$ & $0.86$ & $0.03$ & $0.49$ & $0.36$ & $-0.34$ & $0.00$ & $0.03$ & $0.00$ \\ 
 
$12000$ & $0.4000$ & $0.1208$ & $17.61$ & $16.94$ & $4.56$ & $3.55$ & $0.00$ & $1.56$ & $1.51$ & $0.40$ & $0.00$ & $0.00$ & $0.00$ \\ 
 
$12000$ & $0.6500$ & $0.02165$ & $25.59$ & $24.29$ & $7.16$ & $5.55$ & $3.22$ & $3.62$ & $3.05$ & $0.24$ & $-0.82$ & $-1.75$ & $0.00$ \\ 
 
\hline 
$20000$ & $0.2500$ & $0.1446$ & $25.36$ & $25.06$ & $3.73$ & $2.53$ & $0.10$ & $1.10$ & $0.98$ & $0.36$ & $0.06$ & $0.11$ & $-0.30$ \\ 
 
$20000$ & $0.4000$ & $0.1094$ & $24.06$ & $23.67$ & $4.25$ & $3.08$ & $0.02$ & $0.79$ & $0.77$ & $-0.19$ & $0.01$ & $0.03$ & $0.00$ \\ 
 
$20000$ & $0.6500$ & $0.006690$ & $72.84$ & $71.00$ & $15.42$ & $14.93$ & $0.00$ & $5.22$ & $4.59$ & $2.49$ & $0.00$ & $0.00$ & $0.00$ \\ 
 
\hline 
$30000$ & $0.4000$ & $0.08055$ & $51.57$ & $51.28$ & $5.30$ & $3.09$ & $0.08$ & $1.15$ & $0.99$ & $-0.54$ & $0.02$ & $0.06$ & $-0.23$ \\ 
 
$30000$ & $0.6500$ & $0.01363$ & $71.77$ & $70.94$ & $10.44$ & $9.27$ & $0.00$ & $3.06$ & $2.79$ & $1.26$ & $0.00$ & $0.00$ & $0.00$ \\ 
 
\hline 
\end{tabular} 
\end{center} 
\captcont{continued.}
\end{table}

\clearpage
\begin{table}[htbp] 
\begin{center} 
\tiny 
\renewcommand{\arraystretch}{1.22}
\begin{tabular}{|r|c|c|l|r|r|r|r|r|r|r|r|r|} 
\hline 
$Q^2$  &$x$ & $y$ & ${\rm d}^2\sigma_{\rm CC}/{\rm d}x{\rm d}Q^2$ & 
$\delta_{\rm tot}$ & $\delta_{\rm stat}$ & $\delta_{\rm unc}$ & 
$\delta_{\rm unc}^{h}$& 
$\delta_{\rm cor}$ & 
$\delta_{\rm cor}^{V^+}$ & 
$\delta_{\rm cor}^{h^+}$& 
$\delta_{\rm cor}^{N^+}$& 
$\delta_{\rm cor}^{B^+}$ \\ 
$(\rm GeV^2)$ & & & $(\rm pb/GeV^2)$ & 
$(\%)$ & $(\%)$ & $(\%)$ & $(\%)$ & $(\%)$ & 
$(\%)$ & $(\%)$ & $(\%)$ & $(\%)$  
\\ \hline 
$300$ & $0.008$ & $0.369$ & $2.01$ & $49.9$ & $40.7$ & $23.4$ & $1.6$ & $17.0$ & $15.5$ & $0.3$ & $-1.0$ & $-5.8$ \\ 
 
$300$ & $0.013$ & $0.227$ & $0.923$ & $20.6$ & $14.4$ & $10.8$ & $2.7$ & $9.8$ & $9.4$ & $-0.9$ & $-0.3$ & $-1.2$ \\ 
 
$300$ & $0.032$ & $0.092$ & $0.305$ & $15.9$ & $14.0$ & $5.3$ & $2.0$ & $5.5$ & $4.7$ & $-0.4$ & $0.4$ & $-2.4$ \\ 
 
$300$ & $0.080$ & $0.037$ & $0.776 \cdot 10^{-1}$ & $15.4$ & $13.5$ & $5.1$ & $2.8$ & $5.3$ & $1.4$ & $-0.4$ & $-4.3$ & $-1.9$ \\ 
 
\hline 
$500$ & $0.013$ & $0.379$ & $0.790$ & $14.9$ & $9.8$ & $7.1$ & $2.3$ & $8.7$ & $8.5$ & $-0.4$ & $-0.6$ & $-0.5$ \\ 
 
$500$ & $0.032$ & $0.154$ & $0.250$ & $9.3$ & $8.1$ & $3.2$ & $1.5$ & $3.2$ & $2.9$ & $-0.7$ & $0.1$ & $-0.2$ \\ 
 
$500$ & $0.080$ & $0.062$ & $0.621 \cdot 10^{-1}$ & $10.4$ & $9.3$ & $4.2$ & $2.4$ & $1.7$ & $0.8$ & $-0.6$ & $-0.2$ & $-0.1$ \\ 
 
$500$ & $0.130$ & $0.038$ & $0.345 \cdot 10^{-1}$ & $25.4$ & $21.4$ & $5.3$ & $0.8$ & $12.7$ & $0.1$ & $0.6$ & $-12.3$ & $0.0$ \\ 
 
\hline 
$1000$ & $0.013$ & $0.757$ & $0.476$ & $14.1$ & $10.2$ & $5.7$ & $1.7$ & $7.9$ & $7.7$ & $-0.8$ & $0.1$ & $-0.5$ \\ 
 
$1000$ & $0.032$ & $0.308$ & $0.230$ & $7.2$ & $6.2$ & $2.8$ & $1.8$ & $2.2$ & $1.9$ & $-0.4$ & $0.3$ & $-0.1$ \\ 
 
$1000$ & $0.080$ & $0.123$ & $0.710 \cdot 10^{-1}$ & $7.2$ & $6.4$ & $2.9$ & $1.1$ & $1.4$ & $0.4$ & $-0.4$ & $0.9$ & $-0.1$ \\ 
 
$1000$ & $0.130$ & $0.076$ & $0.336 \cdot 10^{-1}$ & $12.3$ & $10.9$ & $3.8$ & $1.5$ & $4.1$ & $0.0$ & $-0.1$ & $-3.7$ & $0.0$ \\ 
 
\hline 
$2000$ & $0.032$ & $0.615$ & $0.148$ & $6.8$ & $5.8$ & $2.6$ & $1.0$ & $2.2$ & $1.9$ & $-0.4$ & $-0.4$ & $0.0$ \\ 
 
$2000$ & $0.080$ & $0.246$ & $0.573 \cdot 10^{-1}$ & $5.8$ & $5.2$ & $2.1$ & $0.5$ & $1.1$ & $0.1$ & $-0.1$ & $0.7$ & $0.0$ \\ 
 
$2000$ & $0.130$ & $0.152$ & $0.290 \cdot 10^{-1}$ & $8.1$ & $7.4$ & $3.1$ & $0.9$ & $1.2$ & $-0.0$ & $-0.4$ & $-0.3$ & $0.0$ \\ 
 
$2000$ & $0.250$ & $0.079$ & $0.105 \cdot 10^{-1}$ & $17.8$ & $14.6$ & $4.0$ & $0.6$ & $9.3$ & $0.0$ & $0.4$ & $-9.1$ & $0.0$ \\ 
 
\hline 
$3000$ & $0.080$ & $0.369$ & $0.397 \cdot 10^{-1}$ & $5.7$ & $5.2$ & $2.0$ & $0.2$ & $1.2$ & $-0.1$ & $0.2$ & $0.9$ & $0.0$ \\ 
 
$3000$ & $0.130$ & $0.227$ & $0.234 \cdot 10^{-1}$ & $6.6$ & $6.1$ & $2.1$ & $0.3$ & $1.3$ & $-0.0$ & $-0.1$ & $0.8$ & $0.0$ \\ 
 
$3000$ & $0.250$ & $0.118$ & $0.858 \cdot 10^{-2}$ & $9.9$ & $9.2$ & $2.9$ & $0.5$ & $2.5$ & $0.0$ & $0.2$ & $-2.1$ & $-0.2$ \\ 
 
\hline 
$5000$ & $0.080$ & $0.615$ & $0.261 \cdot 10^{-1}$ & $7.2$ & $6.6$ & $2.4$ & $0.9$ & $1.5$ & $0.1$ & $0.3$ & $1.1$ & $0.0$ \\ 
 
$5000$ & $0.130$ & $0.379$ & $0.156 \cdot 10^{-1}$ & $6.4$ & $5.8$ & $2.1$ & $0.2$ & $1.2$ & $-0.0$ & $0.1$ & $0.7$ & $0.0$ \\ 
 
$5000$ & $0.250$ & $0.197$ & $0.603 \cdot 10^{-2}$ & $7.7$ & $7.1$ & $2.5$ & $1.5$ & $1.2$ & $-0.0$ & $0.4$ & $0.4$ & $-0.1$ \\ 
 
$5000$ & $0.400$ & $0.123$ & $0.183 \cdot 10^{-2}$ & $21.5$ & $19.2$ & $5.2$ & $4.5$ & $8.0$ & $0.0$ & $0.9$ & $-7.4$ & $0.0$ \\ 
 
\hline 
$8000$ & $0.130$ & $0.606$ & $0.105 \cdot 10^{-1}$ & $8.4$ & $7.1$ & $3.6$ & $2.7$ & $2.3$ & $-0.1$ & $1.0$ & $1.8$ & $0.0$ \\ 
 
$8000$ & $0.250$ & $0.315$ & $0.320 \cdot 10^{-2}$ & $8.1$ & $7.3$ & $3.0$ & $2.1$ & $1.7$ & $-0.0$ & $0.6$ & $1.2$ & $0.0$ \\ 
 
$8000$ & $0.400$ & $0.197$ & $0.131 \cdot 10^{-2}$ & $14.8$ & $13.4$ & $5.5$ & $5.1$ & $2.7$ & $0.0$ & $1.4$ & $-0.8$ & $0.0$ \\ 
 
\hline 
$15000$ & $0.250$ & $0.591$ & $0.192 \cdot 10^{-2}$ & $10.4$ & $8.4$ & $5.4$ & $4.7$ & $2.8$ & $0.0$ & $1.2$ & $2.1$ & $0.0$ \\ 
 
$15000$ & $0.400$ & $0.369$ & $0.481 \cdot 10^{-3}$ & $13.1$ & $11.1$ & $6.2$ & $5.7$ & $2.9$ & $0.0$ & $1.9$ & $1.5$ & $0.0$ \\ 
 
\hline 
$30000$ & $0.400$ & $0.738$ & $0.200 \cdot 10^{-3}$ & $20.9$ & $17.2$ & $10.2$ & $9.7$ & $5.8$ & $0.0$ & $3.2$ & $3.9$ & $0.0$ \\ 
 
\hline 

\end{tabular} 
\end{center} 
\caption[RESULT] 
{\label{tab:ccdxdq2_eleLH} The CC 
$e^-p$ cross section ${\rm d}^2\sigma_{\rm CC}/{\rm d}x{\rm d}Q^2$ for lepton beam polarisation $P_e=-25.8$\% 
with statistical 
$(\delta_{\rm stat})$, 
total $(\delta_{\rm tot})$, 
total uncorrelated systematic $(\delta_{\rm unc})$ 
errors and one of its contributions from 
the hadronic energy error  
($\delta_{\rm unc}^{h}$). 
The effect of the other uncorrelated 
systematic errors is included in $\delta_{\rm unc}$. 
In addition the correlated systematic  
$(\delta_{\rm cor})$ and its contributions from a 
positive variation of one  
standard deviation of the 
cuts against photoproduction ($\delta_{\rm cor}^{V^+}$), of 
the hadronic 
energy error ($\delta_{\rm cor}^{h^+}$), of the error 
due to noise subtraction ($\delta_{\rm cor}^{N^+}$) 
and of the error due to background subtraction 
($\delta_{\rm cor}^{B^+}$) are given. 
The normalisation and polarisation uncertainties are not included in the errors. 
}
\end{table} 

\begin{table}[htbp] 
\begin{center} 
\tiny 
\renewcommand{\arraystretch}{1.22}
\begin{tabular}{|r|c|c|l|r|r|r|r|r|r|r|r|r|} 
\hline 
$Q^2$  &$x$ & $y$ & ${\rm d}^2\sigma_{\rm CC}/{\rm d}x{\rm d}Q^2$ & 
$\delta_{\rm tot}$ & $\delta_{\rm stat}$ & $\delta_{\rm unc}$ & 
$\delta_{\rm unc}^{h}$& 
$\delta_{\rm cor}$ & 
$\delta_{\rm cor}^{V^+}$ & 
$\delta_{\rm cor}^{h^+}$& 
$\delta_{\rm cor}^{N^+}$& 
$\delta_{\rm cor}^{B^+}$ \\ 
$(\rm GeV^2)$ & & & $(\rm pb/GeV^2)$ & 
$(\%)$ & $(\%)$ & $(\%)$ & $(\%)$ & $(\%)$ & 
$(\%)$ & $(\%)$ & $(\%)$ & $(\%)$  
\\ \hline 
$300$ & $0.008$ & $0.369$ & $1.16$ & $56.6$ & $47.2$ & $26.2$ & $1.7$ & $17.2$ & $15.3$ & $-0.8$ & $-0.9$ & $-0.9$ \\ 
 
$300$ & $0.013$ & $0.227$ & $0.423$ & $38.3$ & $35.0$ & $11.5$ & $1.9$ & $10.6$ & $9.7$ & $-0.4$ & $0.4$ & $-2.5$ \\ 
 
$300$ & $0.032$ & $0.092$ & $0.127$ & $26.0$ & $24.9$ & $5.0$ & $1.4$ & $5.5$ & $4.8$ & $-0.4$ & $0.8$ & $-1.5$ \\ 
 
$300$ & $0.080$ & $0.037$ & $0.468 \cdot 10^{-1}$ & $26.1$ & $25.2$ & $4.9$ & $2.0$ & $4.6$ & $1.5$ & $-0.5$ & $-3.0$ & $-2.3$ \\ 
 
\hline 
$500$ & $0.013$ & $0.379$ & $0.407$ & $23.4$ & $20.5$ & $7.1$ & $2.6$ & $8.6$ & $8.3$ & $-0.8$ & $-0.9$ & $-1.1$ \\ 
 
$500$ & $0.032$ & $0.154$ & $0.141$ & $17.1$ & $16.3$ & $3.7$ & $2.3$ & $3.6$ & $3.3$ & $-0.5$ & $-0.3$ & $-0.6$ \\ 
 
$500$ & $0.080$ & $0.062$ & $0.364 \cdot 10^{-1}$ & $19.0$ & $18.4$ & $4.0$ & $1.9$ & $1.9$ & $0.7$ & $-0.4$ & $-0.1$ & $-1.0$ \\ 
 
$500$ & $0.130$ & $0.038$ & $0.132 \cdot 10^{-1}$ & $52.4$ & $50.5$ & $6.0$ & $2.1$ & $12.7$ & $0.3$ & $-0.7$ & $-12.3$ & $0.0$ \\ 
 
\hline 
$1000$ & $0.013$ & $0.757$ & $0.283$ & $22.4$ & $19.9$ & $6.2$ & $2.1$ & $8.2$ & $8.1$ & $-0.7$ & $-0.3$ & $0.0$ \\ 
 
$1000$ & $0.032$ & $0.308$ & $0.115$ & $13.3$ & $12.8$ & $2.5$ & $1.4$ & $2.2$ & $1.8$ & $-0.3$ & $0.6$ & $-0.2$ \\ 
 
$1000$ & $0.080$ & $0.123$ & $0.441 \cdot 10^{-1}$ & $12.5$ & $12.0$ & $2.8$ & $0.4$ & $1.6$ & $0.3$ & $-0.2$ & $1.1$ & $-0.3$ \\ 
 
$1000$ & $0.130$ & $0.076$ & $0.128 \cdot 10^{-1}$ & $26.8$ & $26.3$ & $3.8$ & $1.4$ & $3.4$ & $0.1$ & $0.1$ & $-2.9$ & $-0.2$ \\ 
 
\hline 
$2000$ & $0.032$ & $0.615$ & $0.709 \cdot 10^{-1}$ & $13.1$ & $12.7$ & $2.6$ & $1.3$ & $2.2$ & $1.8$ & $-0.4$ & $0.1$ & $-0.5$ \\ 
 
$2000$ & $0.080$ & $0.246$ & $0.232 \cdot 10^{-1}$ & $12.6$ & $12.4$ & $2.3$ & $1.0$ & $1.1$ & $0.1$ & $-0.3$ & $0.4$ & $0.0$ \\ 
 
$2000$ & $0.130$ & $0.152$ & $0.128 \cdot 10^{-1}$ & $16.6$ & $16.3$ & $3.0$ & $0.9$ & $1.2$ & $0.0$ & $-0.3$ & $-0.1$ & $0.0$ \\ 
 
$2000$ & $0.250$ & $0.079$ & $0.545 \cdot 10^{-2}$ & $30.3$ & $28.9$ & $3.6$ & $0.3$ & $8.1$ & $0.0$ & $0.3$ & $-7.8$ & $0.0$ \\ 
 
\hline 
$3000$ & $0.080$ & $0.369$ & $0.227 \cdot 10^{-1}$ & $10.4$ & $10.1$ & $2.0$ & $0.2$ & $1.2$ & $-0.0$ & $0.3$ & $0.8$ & $-0.1$ \\ 
 
$3000$ & $0.130$ & $0.227$ & $0.914 \cdot 10^{-2}$ & $14.6$ & $14.4$ & $2.1$ & $0.6$ & $1.3$ & $-0.0$ & $-0.4$ & $0.6$ & $0.0$ \\ 
 
$3000$ & $0.250$ & $0.118$ & $0.381 \cdot 10^{-2}$ & $20.0$ & $19.6$ & $2.9$ & $0.8$ & $2.2$ & $0.0$ & $0.1$ & $-1.6$ & $0.0$ \\ 
 
\hline 
$5000$ & $0.080$ & $0.615$ & $0.135 \cdot 10^{-1}$ & $13.9$ & $13.6$ & $2.3$ & $0.9$ & $1.5$ & $0.1$ & $0.3$ & $0.9$ & $-0.1$ \\ 
 
$5000$ & $0.130$ & $0.379$ & $0.838 \cdot 10^{-2}$ & $12.0$ & $11.7$ & $2.2$ & $0.7$ & $1.5$ & $-0.0$ & $0.3$ & $1.0$ & $0.0$ \\ 
 
$5000$ & $0.250$ & $0.197$ & $0.280 \cdot 10^{-2}$ & $15.4$ & $15.1$ & $2.6$ & $1.6$ & $1.4$ & $0.0$ & $0.4$ & $0.5$ & $0.0$ \\ 
 
$5000$ & $0.400$ & $0.123$ & $0.831 \cdot 10^{-3}$ & $42.0$ & $40.9$ & $6.3$ & $5.8$ & $7.5$ & $0.0$ & $1.3$ & $-6.7$ & $0.0$ \\ 
 
\hline 
$8000$ & $0.130$ & $0.606$ & $0.539 \cdot 10^{-2}$ & $14.9$ & $14.3$ & $3.3$ & $2.3$ & $1.8$ & $-0.1$ & $0.7$ & $1.1$ & $0.0$ \\ 
 
$8000$ & $0.250$ & $0.315$ & $0.167 \cdot 10^{-2}$ & $15.9$ & $15.4$ & $2.9$ & $1.9$ & $1.8$ & $0.0$ & $0.7$ & $1.2$ & $0.0$ \\ 
 
$8000$ & $0.400$ & $0.197$ & $0.507 \cdot 10^{-3}$ & $32.2$ & $31.7$ & $5.1$ & $4.6$ & $2.8$ & $0.0$ & $1.4$ & $-0.9$ & $0.0$ \\ 
 
\hline 
$15000$ & $0.250$ & $0.591$ & $1.00 \cdot 10^{-3}$ & $18.8$ & $17.7$ & $5.7$ & $5.0$ & $2.8$ & $-0.0$ & $1.3$ & $2.0$ & $0.0$ \\ 
 
$15000$ & $0.400$ & $0.369$ & $0.265 \cdot 10^{-3}$ & $24.3$ & $23.6$ & $5.2$ & $4.6$ & $2.5$ & $0.0$ & $1.1$ & $1.1$ & $0.0$ \\ 
 
\hline 

\end{tabular} 
\end{center} 
\caption[RESULT] 
{\label{tab:ccdxdq2_eleRH} The CC 
$e^-p$ cross section ${\rm d}^2\sigma_{\rm CC}/{\rm d}x{\rm d}Q^2$ for lepton beam polarisation $P_e=+36.0$\% 
with statistical 
$(\delta_{\rm stat})$, 
total $(\delta_{\rm tot})$, 
total uncorrelated systematic $(\delta_{\rm unc})$ 
errors and one of its contributions from 
the hadronic energy error  
($\delta_{\rm unc}^{h}$). 
The effect of the other uncorrelated 
systematic errors is included in $\delta_{\rm unc}$. 
In addition the correlated systematic  
$(\delta_{\rm cor})$ and its contributions from a 
positive variation of one  
standard deviation of the 
cuts against photoproduction ($\delta_{\rm cor}^{V^+}$), of 
the hadronic 
energy error ($\delta_{\rm cor}^{h^+}$), of the error 
due to noise subtraction ($\delta_{\rm cor}^{N^+}$) 
and of the error due to background subtraction 
($\delta_{\rm cor}^{B^+}$) are given. 
The normalisation and polarisation uncertainties are not included in the errors. 
}
\end{table} 

\begin{table}[htbp] 
\begin{center} 
\tiny 
\renewcommand{\arraystretch}{1.22}
\begin{tabular}{|r|c|c|l|r|r|r|r|r|r|r|r|r|} 
\hline 
$Q^2$  &$x$ & $y$ & ${\rm d}^2\sigma_{\rm CC}/{\rm d}x{\rm d}Q^2$ & 
$\delta_{\rm tot}$ & $\delta_{\rm stat}$ & $\delta_{\rm unc}$ & 
$\delta_{\rm unc}^{h}$& 
$\delta_{\rm cor}$ & 
$\delta_{\rm cor}^{V^+}$ & 
$\delta_{\rm cor}^{h^+}$& 
$\delta_{\rm cor}^{N^+}$& 
$\delta_{\rm cor}^{B^+}$ \\ 
$(\rm GeV^2)$ & & & $(\rm pb/GeV^2)$ & 
$(\%)$ & $(\%)$ & $(\%)$ & $(\%)$ & $(\%)$ & 
$(\%)$ & $(\%)$ & $(\%)$ & $(\%)$  
\\ \hline 
$300$ & $0.008$ & $0.369$ & $1.20$ & $50.7$ & $38.5$ & $26.5$ & $2.7$ & $19.8$ & $16.5$ & $-0.9$ & $-0.7$ & $-7.7$ \\ 
 
$300$ & $0.013$ & $0.227$ & $0.409$ & $32.7$ & $28.4$ & $11.0$ & $2.1$ & $11.9$ & $9.8$ & $-0.7$ & $-0.5$ & $-5.3$ \\ 
 
$300$ & $0.032$ & $0.092$ & $0.101$ & $25.3$ & $23.6$ & $5.4$ & $2.0$ & $7.1$ & $5.1$ & $-0.6$ & $1.3$ & $-4.5$ \\ 
 
$300$ & $0.080$ & $0.037$ & $0.256 \cdot 10^{-1}$ & $28.5$ & $27.5$ & $5.2$ & $2.4$ & $5.4$ & $1.7$ & $-0.4$ & $-4.1$ & $-2.6$ \\ 
 
\hline 
$500$ & $0.013$ & $0.379$ & $0.282$ & $23.6$ & $20.4$ & $7.1$ & $1.8$ & $9.5$ & $9.3$ & $-0.7$ & $0.3$ & $-1.5$ \\ 
 
$500$ & $0.032$ & $0.154$ & $0.104$ & $16.0$ & $15.2$ & $3.1$ & $1.0$ & $3.9$ & $3.6$ & $-0.3$ & $0.7$ & $-0.9$ \\ 
 
$500$ & $0.080$ & $0.062$ & $0.383 \cdot 10^{-1}$ & $14.8$ & $14.2$ & $3.7$ & $1.3$ & $2.0$ & $0.9$ & $-0.4$ & $0.7$ & $-1.1$ \\ 
 
$500$ & $0.130$ & $0.038$ & $0.121 \cdot 10^{-1}$ & $43.8$ & $41.5$ & $6.1$ & $0.2$ & $12.6$ & $0.1$ & $0.2$ & $-12.3$ & $0.0$ \\ 
 
\hline 
$1000$ & $0.013$ & $0.757$ & $0.238$ & $21.6$ & $18.4$ & $6.3$ & $0.9$ & $9.5$ & $9.3$ & $-0.3$ & $-0.4$ & $-1.6$ \\ 
 
$1000$ & $0.032$ & $0.308$ & $0.123$ & $10.4$ & $9.9$ & $2.3$ & $0.7$ & $2.3$ & $2.1$ & $-0.2$ & $0.6$ & $-0.2$ \\ 
 
$1000$ & $0.080$ & $0.123$ & $0.202 \cdot 10^{-1}$ & $14.2$ & $13.9$ & $2.8$ & $0.6$ & $1.3$ & $0.4$ & $-0.2$ & $0.8$ & $-0.2$ \\ 
 
$1000$ & $0.130$ & $0.076$ & $0.730 \cdot 10^{-2}$ & $26.5$ & $26.1$ & $3.8$ & $1.1$ & $2.8$ & $0.0$ & $-0.3$ & $-2.3$ & $0.0$ \\ 
 
\hline 
$2000$ & $0.032$ & $0.615$ & $0.531 \cdot 10^{-1}$ & $11.8$ & $11.3$ & $2.5$ & $0.2$ & $2.4$ & $2.1$ & $-0.1$ & $0.6$ & $-0.3$ \\ 
 
$2000$ & $0.080$ & $0.246$ & $0.156 \cdot 10^{-1}$ & $11.8$ & $11.5$ & $2.0$ & $0.2$ & $1.3$ & $0.1$ & $-0.2$ & $0.9$ & $-0.1$ \\ 
 
$2000$ & $0.130$ & $0.152$ & $0.692 \cdot 10^{-2}$ & $17.4$ & $17.1$ & $2.9$ & $0.5$ & $1.4$ & $0.0$ & $0.3$ & $0.7$ & $0.0$ \\ 
 
$2000$ & $0.250$ & $0.079$ & $0.228 \cdot 10^{-2}$ & $33.3$ & $31.8$ & $3.8$ & $1.2$ & $9.2$ & $0.0$ & $0.6$ & $-8.8$ & $0.0$ \\ 
 
\hline 
$3000$ & $0.080$ & $0.369$ & $0.118 \cdot 10^{-1}$ & $11.7$ & $11.3$ & $2.5$ & $1.5$ & $1.7$ & $-0.0$ & $0.4$ & $1.4$ & $0.0$ \\ 
 
$3000$ & $0.130$ & $0.227$ & $0.539 \cdot 10^{-2}$ & $15.2$ & $14.9$ & $2.3$ & $1.0$ & $1.6$ & $-0.0$ & $0.4$ & $1.1$ & $-0.1$ \\ 
 
$3000$ & $0.250$ & $0.118$ & $0.157 \cdot 10^{-2}$ & $23.5$ & $23.1$ & $3.7$ & $2.4$ & $2.2$ & $0.0$ & $0.6$ & $-1.4$ & $0.0$ \\ 
 
\hline 
$5000$ & $0.080$ & $0.615$ & $0.359 \cdot 10^{-2}$ & $21.5$ & $21.0$ & $3.5$ & $2.5$ & $2.7$ & $-0.2$ & $0.8$ & $2.2$ & $-0.1$ \\ 
 
$5000$ & $0.130$ & $0.379$ & $0.305 \cdot 10^{-2}$ & $16.0$ & $15.6$ & $2.8$ & $1.7$ & $1.9$ & $-0.1$ & $0.3$ & $1.4$ & $0.0$ \\ 
 
$5000$ & $0.250$ & $0.197$ & $0.808 \cdot 10^{-3}$ & $22.9$ & $22.5$ & $3.1$ & $2.4$ & $2.0$ & $0.0$ & $0.7$ & $1.0$ & $0.0$ \\ 
 
$5000$ & $0.400$ & $0.123$ & $0.525 \cdot 10^{-3}$ & $42.5$ & $40.9$ & $6.7$ & $6.2$ & $9.3$ & $0.0$ & $1.5$ & $-8.3$ & $0.0$ \\ 
 
\hline 
$8000$ & $0.130$ & $0.606$ & $0.682 \cdot 10^{-3}$ & $30.6$ & $29.7$ & $5.7$ & $5.1$ & $3.2$ & $-0.2$ & $1.0$ & $2.3$ & $0.0$ \\ 
 
$8000$ & $0.250$ & $0.315$ & $0.612 \cdot 10^{-3}$ & $21.4$ & $20.5$ & $4.7$ & $4.1$ & $2.6$ & $0.0$ & $1.0$ & $1.5$ & $0.0$ \\ 
 
$8000$ & $0.400$ & $0.197$ & $0.792 \cdot 10^{-4}$ & $59.1$ & $58.3$ & $8.1$ & $7.8$ & $4.2$ & $0.0$ & $2.1$ & $0.7$ & $0.0$ \\ 
 
\hline 
$15000$ & $0.250$ & $0.591$ & $0.721 \cdot 10^{-4}$ & $46.9$ & $46.0$ & $7.5$ & $6.9$ & $4.8$ & $-0.1$ & $2.6$ & $2.9$ & $0.0$ \\ 
 
$15000$ & $0.400$ & $0.369$ & $0.311 \cdot 10^{-4}$ & $45.9$ & $45.0$ & $7.4$ & $7.0$ & $4.2$ & $0.0$ & $2.2$ & $1.3$ & $0.0$ \\ 
 
\hline 

\end{tabular} 
\end{center} 
\caption[RESULT] 
{\label{tab:ccdxdq2_posLH} The CC 
$e^+p$ cross section ${\rm d}^2\sigma_{\rm CC}/{\rm d}x{\rm d}Q^2$ for lepton beam polarisation $P_e=-37.0$\% 
with statistical 
$(\delta_{\rm stat})$, 
total $(\delta_{\rm tot})$, 
total uncorrelated systematic $(\delta_{\rm unc})$ 
errors and one of its contributions from 
the hadronic energy error  
($\delta_{\rm unc}^{h}$). 
The effect of the other uncorrelated 
systematic errors is included in $\delta_{\rm unc}$. 
In addition the correlated systematic  
$(\delta_{\rm cor})$ and its contributions from a 
positive variation of one  
standard deviation of the 
cuts against photoproduction ($\delta_{\rm cor}^{V^+}$), of 
the hadronic 
energy error ($\delta_{\rm cor}^{h^+}$), of the error 
due to noise subtraction ($\delta_{\rm cor}^{N^+}$) 
and of the error due to background subtraction 
($\delta_{\rm cor}^{B^+}$) are given. 
The normalisation and polarisation uncertainties are not included in the errors. 
}
\end{table} 

\begin{table}[htbp] 
\begin{center} 
\tiny 
\renewcommand{\arraystretch}{1.22}
\begin{tabular}{|r|c|c|l|r|r|r|r|r|r|r|r|r|} 
\hline 
$Q^2$  &$x$ & $y$ & ${\rm d}^2\sigma_{\rm CC}/{\rm d}x{\rm d}Q^2$ & 
$\delta_{\rm tot}$ & $\delta_{\rm stat}$ & $\delta_{\rm unc}$ & 
$\delta_{\rm unc}^{h}$& 
$\delta_{\rm cor}$ & 
$\delta_{\rm cor}^{V^+}$ & 
$\delta_{\rm cor}^{h^+}$& 
$\delta_{\rm cor}^{N^+}$& 
$\delta_{\rm cor}^{B^+}$ \\ 
$(\rm GeV^2)$ & & & $(\rm pb/GeV^2)$ & 
$(\%)$ & $(\%)$ & $(\%)$ & $(\%)$ & $(\%)$ & 
$(\%)$ & $(\%)$ & $(\%)$ & $(\%)$  
\\ \hline 
$300$ & $0.008$ & $0.369$ & $0.769$ & $58.3$ & $49.3$ & $25.9$ & $2.5$ & $17.3$ & $15.8$ & $-0.7$ & $-0.6$ & $-5.2$ \\ 
 
$300$ & $0.013$ & $0.227$ & $0.586$ & $25.8$ & $20.4$ & $11.5$ & $1.7$ & $11.0$ & $10.3$ & $-0.4$ & $0.3$ & $-2.8$ \\ 
 
$300$ & $0.032$ & $0.092$ & $0.270$ & $14.5$ & $11.9$ & $5.3$ & $1.6$ & $6.2$ & $5.4$ & $-0.3$ & $1.3$ & $-2.5$ \\ 
 
$300$ & $0.080$ & $0.037$ & $0.514 \cdot 10^{-1}$ & $18.1$ & $16.8$ & $5.1$ & $2.3$ & $4.2$ & $1.9$ & $-0.3$ & $-2.9$ & $-1.6$ \\ 
 
\hline 
$500$ & $0.008$ & $0.615$ & $1.56$ & $33.4$ & $23.2$ & $16.2$ & $2.2$ & $17.7$ & $17.5$ & $-0.6$ & $-0.2$ & $-0.5$ \\ 
 
$500$ & $0.013$ & $0.379$ & $0.663$ & $16.4$ & $11.4$ & $7.1$ & $2.0$ & $9.3$ & $9.2$ & $-0.5$ & $-0.3$ & $-0.7$ \\ 
 
$500$ & $0.032$ & $0.154$ & $0.250$ & $9.9$ & $8.5$ & $3.3$ & $1.6$ & $3.6$ & $3.3$ & $-0.6$ & $0.4$ & $-0.5$ \\ 
 
$500$ & $0.080$ & $0.062$ & $0.597 \cdot 10^{-1}$ & $10.6$ & $9.8$ & $3.6$ & $0.8$ & $1.5$ & $0.9$ & $-0.2$ & $0.2$ & $-0.1$ \\ 
 
$500$ & $0.130$ & $0.038$ & $0.266 \cdot 10^{-1}$ & $27.0$ & $23.7$ & $5.7$ & $1.0$ & $11.6$ & $0.2$ & $-1.0$ & $-11.3$ & $0.0$ \\ 
 
\hline 
$1000$ & $0.013$ & $0.757$ & $0.387$ & $16.9$ & $12.5$ & $6.1$ & $0.8$ & $9.5$ & $9.4$ & $-0.5$ & $0.6$ & $-0.2$ \\ 
 
$1000$ & $0.032$ & $0.308$ & $0.174$ & $8.1$ & $7.4$ & $2.4$ & $0.9$ & $2.3$ & $2.0$ & $-0.3$ & $0.7$ & $-0.1$ \\ 
 
$1000$ & $0.080$ & $0.123$ & $0.507 \cdot 10^{-1}$ & $8.4$ & $7.8$ & $2.8$ & $0.8$ & $1.3$ & $0.4$ & $-0.4$ & $0.7$ & $-0.1$ \\ 
 
$1000$ & $0.130$ & $0.076$ & $0.265 \cdot 10^{-1}$ & $12.9$ & $12.1$ & $3.7$ & $0.9$ & $2.8$ & $0.1$ & $-0.3$ & $-2.3$ & $0.0$ \\ 
 
\hline 
$2000$ & $0.032$ & $0.615$ & $0.103$ & $8.1$ & $7.3$ & $2.4$ & $0.4$ & $2.6$ & $2.2$ & $-0.2$ & $0.9$ & $0.0$ \\ 
 
$2000$ & $0.080$ & $0.246$ & $0.367 \cdot 10^{-1}$ & $7.1$ & $6.6$ & $2.0$ & $0.2$ & $1.3$ & $0.1$ & $0.1$ & $1.0$ & $-0.1$ \\ 
 
$2000$ & $0.130$ & $0.152$ & $0.164 \cdot 10^{-1}$ & $10.5$ & $9.9$ & $3.0$ & $0.8$ & $1.4$ & $0.0$ & $0.3$ & $0.6$ & $0.0$ \\ 
 
$2000$ & $0.250$ & $0.079$ & $0.470 \cdot 10^{-2}$ & $21.3$ & $19.3$ & $3.6$ & $0.6$ & $8.3$ & $0.0$ & $-0.2$ & $-8.0$ & $0.0$ \\ 
 
\hline 
$3000$ & $0.080$ & $0.369$ & $0.244 \cdot 10^{-1}$ & $7.5$ & $6.9$ & $2.4$ & $1.2$ & $1.6$ & $-0.0$ & $0.5$ & $1.2$ & $0.0$ \\ 
 
$3000$ & $0.130$ & $0.227$ & $0.152 \cdot 10^{-1}$ & $8.4$ & $8.0$ & $2.1$ & $0.5$ & $1.4$ & $-0.0$ & $0.2$ & $0.9$ & $0.0$ \\ 
 
$3000$ & $0.250$ & $0.118$ & $0.259 \cdot 10^{-2}$ & $16.4$ & $15.9$ & $3.4$ & $2.0$ & $2.1$ & $0.0$ & $0.4$ & $-1.2$ & $0.0$ \\ 
 
\hline 
$5000$ & $0.080$ & $0.615$ & $0.984 \cdot 10^{-2}$ & $12.1$ & $10.9$ & $4.0$ & $3.3$ & $3.0$ & $0.1$ & $0.7$ & $2.5$ & $-0.1$ \\ 
 
$5000$ & $0.130$ & $0.379$ & $0.627 \cdot 10^{-2}$ & $10.8$ & $9.8$ & $3.4$ & $2.7$ & $2.4$ & $-0.1$ & $0.9$ & $1.9$ & $0.0$ \\ 
 
$5000$ & $0.250$ & $0.197$ & $0.185 \cdot 10^{-2}$ & $14.0$ & $13.3$ & $3.4$ & $2.7$ & $1.8$ & $0.0$ & $0.6$ & $0.8$ & $0.0$ \\ 
 
$5000$ & $0.400$ & $0.123$ & $0.866 \cdot 10^{-3}$ & $27.7$ & $25.0$ & $7.9$ & $7.5$ & $8.6$ & $0.0$ & $1.5$ & $-7.4$ & $0.0$ \\ 
 
\hline 
$8000$ & $0.130$ & $0.606$ & $0.213 \cdot 10^{-2}$ & $17.0$ & $15.4$ & $5.6$ & $5.0$ & $3.5$ & $-0.2$ & $1.4$ & $2.6$ & $0.0$ \\ 
 
$8000$ & $0.250$ & $0.315$ & $0.988 \cdot 10^{-3}$ & $14.8$ & $13.5$ & $4.3$ & $3.6$ & $2.8$ & $0.0$ & $1.2$ & $1.8$ & $0.0$ \\ 
 
$8000$ & $0.400$ & $0.197$ & $0.295 \cdot 10^{-3}$ & $29.4$ & $27.8$ & $7.9$ & $7.7$ & $4.5$ & $0.0$ & $2.6$ & $-1.5$ & $0.0$ \\ 
 
\hline 
$15000$ & $0.250$ & $0.591$ & $0.306 \cdot 10^{-3}$ & $21.4$ & $19.0$ & $8.1$ & $7.6$ & $4.5$ & $0.1$ & $2.0$ & $2.9$ & $0.0$ \\ 
 
$15000$ & $0.400$ & $0.369$ & $0.361 \cdot 10^{-4}$ & $39.0$ & $38.0$ & $6.8$ & $6.3$ & $4.2$ & $0.0$ & $1.8$ & $1.7$ & $0.0$ \\ 
 
\hline 

\end{tabular} 
\end{center} 
\caption[RESULT] 
{\label{tab:ccdxdq2_posRH} The CC 
$e^+p$ cross section ${\rm d}^2\sigma_{\rm CC}/{\rm d}x{\rm d}Q^2$ for lepton beam polarisation $P_e=+32.5$\% 
with statistical 
$(\delta_{\rm stat})$, 
total $(\delta_{\rm tot})$, 
total uncorrelated systematic $(\delta_{\rm unc})$ 
errors and one of its contributions from 
the hadronic energy error  
($\delta_{\rm unc}^{h}$). 
The effect of the other uncorrelated 
systematic errors is included in $\delta_{\rm unc}$. 
In addition the correlated systematic  
$(\delta_{\rm cor})$ and its contributions from a 
positive variation of one  
standard deviation of the 
cuts against photoproduction ($\delta_{\rm cor}^{V^+}$), of 
the hadronic 
energy error ($\delta_{\rm cor}^{h^+}$), of the error 
due to noise subtraction ($\delta_{\rm cor}^{N^+}$) 
and of the error due to background subtraction 
($\delta_{\rm cor}^{B^+}$) are given. 
The normalisation and polarisation uncertainties are not included in the errors. 
}
\end{table}

\clearpage
\begin{table}[htbp] 
\begin{center} 
\tiny 
\begin{tabular}{|r|c|c|r|r|r|r|r|r|r|r|r|r|r|r|} 
\hline 
$Q^2$ & $x$ & $y$ & $\tilde{\sigma}_{\rm NC}$ & 
$\delta_{\rm tot}$ & $\delta_{\rm stat}$ & $\delta_{\rm unc}$ & 
$\delta_{\rm unc}^{E}$ & 
$\delta_{\rm unc}^{h}$& 
$\delta_{\rm cor}$ & 
$\delta_{\rm cor}^{E^+}$ & 
$\delta_{\rm cor}^{\theta^+}$& 
$\delta_{\rm cor}^{h^+}$& 
$\delta_{\rm cor}^{N^+}$& 
$\delta_{\rm cor}^{B^+}$ \\ 
$(\rm GeV^2)$ & & & & 
$(\%)$ & $(\%)$ & $(\%)$ & $(\%)$ & $(\%)$ & $(\%)$ & 
$(\%)$ & $(\%)$ & $(\%)$ & $(\%)$ & $(\%)$  
\\ \hline 
$90$ & $0.0015$ & $0.595$ & $1.370$ & $2.21$ & $1.21$ & $1.10$ & $0.57$ & $0.12$ & $1.48$ & $-0.06$ & $-0.69$ & $0.02$ & $0.22$ & $-1.29$ \\ 
 
\hline 
$120$ & $0.0020$ & $0.595$ & $1.293$ & $2.08$ & $1.13$ & $1.05$ & $0.49$ & $0.12$ & $1.40$ & $-0.30$ & $-0.54$ & $0.02$ & $0.20$ & $-1.24$ \\ 
 
$120$ & $0.0022$ & $0.530$ & $1.313$ & $1.78$ & $1.14$ & $0.98$ & $0.37$ & $0.07$ & $0.95$ & $-0.26$ & $-0.67$ & $0.03$ & $0.16$ & $-0.59$ \\ 
 
$120$ & $0.0025$ & $0.475$ & $1.228$ & $1.98$ & $1.32$ & $1.21$ & $0.76$ & $0.06$ & $0.84$ & $-0.40$ & $-0.64$ & $0.02$ & $0.19$ & $-0.32$ \\ 
 
$120$ & $0.0028$ & $0.425$ & $1.254$ & $2.08$ & $1.53$ & $1.13$ & $0.55$ & $0.05$ & $0.85$ & $-0.44$ & $-0.69$ & $0.02$ & $0.20$ & $-0.11$ \\ 
 
\hline 
$150$ & $0.0025$ & $0.595$ & $1.273$ & $1.93$ & $1.31$ & $0.96$ & $0.24$ & $0.10$ & $1.05$ & $-0.21$ & $-0.43$ & $0.03$ & $0.19$ & $-0.91$ \\ 
 
$150$ & $0.0028$ & $0.530$ & $1.215$ & $1.86$ & $1.30$ & $1.02$ & $0.45$ & $0.07$ & $0.86$ & $-0.39$ & $-0.55$ & $0.01$ & $0.17$ & $-0.52$ \\ 
 
$150$ & $0.0031$ & $0.475$ & $1.210$ & $1.77$ & $1.31$ & $0.96$ & $0.31$ & $0.06$ & $0.71$ & $-0.21$ & $-0.58$ & $0.02$ & $0.20$ & $-0.27$ \\ 
 
$150$ & $0.0035$ & $0.425$ & $1.164$ & $1.79$ & $1.23$ & $1.05$ & $0.56$ & $0.05$ & $0.78$ & $-0.33$ & $-0.65$ & $0.01$ & $0.20$ & $-0.16$ \\ 
 
$150$ & $0.0039$ & $0.375$ & $1.140$ & $1.85$ & $1.14$ & $1.19$ & $0.82$ & $0.01$ & $0.83$ & $-0.50$ & $-0.63$ & $0.00$ & $0.19$ & $-0.06$ \\ 
 
$150$ & $0.0045$ & $0.325$ & $1.107$ & $1.83$ & $1.15$ & $1.13$ & $0.72$ & $0.00$ & $0.86$ & $-0.52$ & $-0.69$ & $0.00$ & $0.00$ & $-0.02$ \\ 
 
$150$ & $0.0060$ & $0.245$ & $1.015$ & $2.01$ & $0.88$ & $1.44$ & $1.18$ & $0.00$ & $1.09$ & $-0.68$ & $-0.85$ & $0.00$ & $0.00$ & $-0.01$ \\ 
 
\hline 
$200$ & $0.0033$ & $0.595$ & $1.175$ & $2.17$ & $1.69$ & $0.99$ & $0.11$ & $0.08$ & $0.93$ & $0.11$ & $-0.51$ & $0.02$ & $0.14$ & $-0.76$ \\ 
 
$200$ & $0.0037$ & $0.530$ & $1.194$ & $2.03$ & $1.63$ & $0.99$ & $0.28$ & $0.07$ & $0.70$ & $-0.25$ & $-0.53$ & $0.02$ & $0.15$ & $-0.35$ \\ 
 
$200$ & $0.0041$ & $0.475$ & $1.103$ & $2.06$ & $1.71$ & $0.96$ & $0.14$ & $0.05$ & $0.64$ & $-0.17$ & $-0.58$ & $0.01$ & $0.14$ & $-0.15$ \\ 
 
$200$ & $0.0046$ & $0.425$ & $1.093$ & $2.05$ & $1.63$ & $1.02$ & $0.43$ & $0.04$ & $0.71$ & $-0.44$ & $-0.51$ & $0.01$ & $0.21$ & $-0.08$ \\ 
 
$200$ & $0.0052$ & $0.375$ & $1.083$ & $1.98$ & $1.49$ & $1.03$ & $0.48$ & $0.01$ & $0.82$ & $-0.38$ & $-0.70$ & $0.00$ & $0.18$ & $-0.04$ \\ 
 
$200$ & $0.0061$ & $0.325$ & $1.014$ & $2.08$ & $1.38$ & $1.26$ & $0.89$ & $0.00$ & $0.91$ & $-0.64$ & $-0.64$ & $0.00$ & $0.00$ & $-0.03$ \\ 
 
$200$ & $0.0080$ & $0.245$ & $0.9317$ & $1.84$ & $0.81$ & $1.37$ & $1.11$ & $0.00$ & $0.93$ & $-0.65$ & $-0.66$ & $0.00$ & $0.00$ & $0.00$ \\ 
 
\hline 
$250$ & $0.0041$ & $0.595$ & $1.136$ & $2.42$ & $2.01$ & $1.06$ & $0.25$ & $0.08$ & $0.86$ & $-0.25$ & $-0.39$ & $0.02$ & $0.20$ & $-0.70$ \\ 
 
$250$ & $0.0046$ & $0.530$ & $1.161$ & $2.26$ & $1.92$ & $1.02$ & $0.31$ & $0.07$ & $0.59$ & $-0.31$ & $-0.39$ & $0.01$ & $0.17$ & $-0.26$ \\ 
 
$250$ & $0.0052$ & $0.475$ & $1.048$ & $2.34$ & $2.00$ & $1.05$ & $0.39$ & $0.06$ & $0.59$ & $-0.39$ & $-0.39$ & $0.01$ & $0.16$ & $-0.17$ \\ 
 
$250$ & $0.0058$ & $0.425$ & $1.045$ & $2.15$ & $1.79$ & $1.03$ & $0.41$ & $0.04$ & $0.62$ & $-0.40$ & $-0.44$ & $0.02$ & $0.16$ & $-0.06$ \\ 
 
$250$ & $0.0066$ & $0.375$ & $0.9816$ & $2.09$ & $1.68$ & $1.02$ & $0.45$ & $0.01$ & $0.73$ & $-0.43$ & $-0.56$ & $0.01$ & $0.16$ & $-0.03$ \\ 
 
$250$ & $0.0076$ & $0.325$ & $0.9542$ & $1.98$ & $1.56$ & $1.04$ & $0.52$ & $0.00$ & $0.64$ & $-0.48$ & $-0.43$ & $0.00$ & $0.00$ & $-0.01$ \\ 
 
$250$ & $0.0100$ & $0.245$ & $0.8717$ & $1.75$ & $0.94$ & $1.12$ & $0.77$ & $0.00$ & $0.95$ & $-0.65$ & $-0.69$ & $0.00$ & $0.00$ & $-0.01$ \\ 
 
\hline 
$300$ & $0.0050$ & $0.595$ & $1.102$ & $2.73$ & $2.37$ & $1.10$ & $0.28$ & $0.07$ & $0.78$ & $-0.27$ & $-0.42$ & $0.02$ & $0.15$ & $-0.58$ \\ 
 
$300$ & $0.0056$ & $0.530$ & $1.110$ & $2.58$ & $2.31$ & $1.02$ & $0.12$ & $0.07$ & $0.53$ & $0.14$ & $-0.43$ & $0.03$ & $0.15$ & $-0.23$ \\ 
 
$300$ & $0.0062$ & $0.475$ & $1.049$ & $2.63$ & $2.28$ & $1.10$ & $0.40$ & $0.06$ & $0.72$ & $-0.40$ & $-0.55$ & $0.02$ & $0.15$ & $-0.19$ \\ 
 
$300$ & $0.0069$ & $0.425$ & $1.007$ & $2.43$ & $2.11$ & $1.02$ & $0.30$ & $0.04$ & $0.65$ & $-0.29$ & $-0.55$ & $0.01$ & $0.18$ & $-0.08$ \\ 
 
$300$ & $0.0079$ & $0.375$ & $0.9664$ & $2.28$ & $1.92$ & $1.03$ & $0.41$ & $0.01$ & $0.69$ & $-0.41$ & $-0.55$ & $0.00$ & $0.12$ & $-0.01$ \\ 
 
$300$ & $0.0091$ & $0.325$ & $0.9142$ & $2.17$ & $1.79$ & $1.01$ & $0.42$ & $0.00$ & $0.67$ & $-0.41$ & $-0.53$ & $0.00$ & $0.00$ & $-0.01$ \\ 
 
$300$ & $0.0121$ & $0.245$ & $0.8206$ & $1.71$ & $1.07$ & $1.03$ & $0.63$ & $0.00$ & $0.84$ & $-0.61$ & $-0.58$ & $0.00$ & $0.00$ & $0.00$ \\ 
 
\hline 
$400$ & $0.0066$ & $0.595$ & $1.049$ & $3.10$ & $2.77$ & $1.20$ & $0.43$ & $0.08$ & $0.73$ & $-0.33$ & $-0.31$ & $0.02$ & $0.18$ & $-0.54$ \\ 
 
$400$ & $0.0074$ & $0.530$ & $1.049$ & $3.02$ & $2.78$ & $1.08$ & $0.19$ & $0.05$ & $0.47$ & $-0.09$ & $-0.40$ & $0.02$ & $0.12$ & $-0.19$ \\ 
 
$400$ & $0.0083$ & $0.475$ & $0.9833$ & $3.02$ & $2.75$ & $1.11$ & $0.31$ & $0.03$ & $0.57$ & $-0.27$ & $-0.48$ & $0.02$ & $0.14$ & $-0.10$ \\ 
 
$400$ & $0.0093$ & $0.425$ & $0.9605$ & $2.85$ & $2.51$ & $1.11$ & $0.42$ & $0.04$ & $0.77$ & $-0.46$ & $-0.59$ & $0.01$ & $0.18$ & $-0.04$ \\ 
 
$400$ & $0.0105$ & $0.375$ & $0.9145$ & $2.57$ & $2.29$ & $1.03$ & $0.33$ & $0.01$ & $0.54$ & $-0.32$ & $-0.42$ & $0.00$ & $0.13$ & $-0.04$ \\ 
 
$400$ & $0.0121$ & $0.325$ & $0.8656$ & $2.45$ & $2.16$ & $1.01$ & $0.32$ & $0.00$ & $0.58$ & $-0.32$ & $-0.48$ & $0.00$ & $0.00$ & $-0.02$ \\ 
 
$400$ & $0.0161$ & $0.245$ & $0.7548$ & $1.84$ & $1.27$ & $1.03$ & $0.61$ & $0.00$ & $0.83$ & $-0.61$ & $-0.57$ & $0.00$ & $0.00$ & $-0.01$ \\ 
 
\hline 
$500$ & $0.0083$ & $0.595$ & $0.9427$ & $3.39$ & $3.16$ & $1.12$ & $0.13$ & $0.09$ & $0.52$ & $0.13$ & $-0.38$ & $0.02$ & $0.17$ & $-0.29$ \\ 
 
$500$ & $0.0093$ & $0.530$ & $0.9525$ & $3.40$ & $3.18$ & $1.11$ & $0.11$ & $0.07$ & $0.51$ & $-0.25$ & $-0.42$ & $0.02$ & $0.13$ & $-0.09$ \\ 
 
$500$ & $0.0104$ & $0.475$ & $0.8895$ & $3.65$ & $3.33$ & $1.30$ & $0.62$ & $0.06$ & $0.73$ & $-0.46$ & $-0.51$ & $0.01$ & $0.15$ & $-0.19$ \\ 
 
$500$ & $0.0116$ & $0.425$ & $0.9040$ & $3.20$ & $2.98$ & $1.09$ & $0.16$ & $0.05$ & $0.33$ & $-0.16$ & $-0.25$ & $0.00$ & $0.13$ & $-0.05$ \\ 
 
$500$ & $0.0131$ & $0.375$ & $0.8794$ & $3.12$ & $2.87$ & $1.08$ & $0.26$ & $0.01$ & $0.56$ & $-0.15$ & $-0.53$ & $0.00$ & $0.09$ & $-0.02$ \\ 
 
$500$ & $0.0152$ & $0.325$ & $0.8044$ & $3.04$ & $2.65$ & $1.22$ & $0.69$ & $0.00$ & $0.84$ & $-0.73$ & $-0.42$ & $0.00$ & $0.00$ & $-0.02$ \\ 
 
$500$ & $0.0201$ & $0.245$ & $0.7092$ & $2.02$ & $1.54$ & $1.04$ & $0.60$ & $0.00$ & $0.78$ & $-0.60$ & $-0.50$ & $0.00$ & $0.00$ & $0.00$ \\ 
 
\hline 
$650$ & $0.0108$ & $0.595$ & $0.9391$ & $3.76$ & $3.52$ & $1.21$ & $0.21$ & $0.06$ & $0.53$ & $-0.14$ & $-0.40$ & $0.02$ & $0.13$ & $-0.30$ \\ 
 
$650$ & $0.0121$ & $0.530$ & $0.8624$ & $3.84$ & $3.60$ & $1.27$ & $0.44$ & $0.08$ & $0.40$ & $-0.23$ & $-0.26$ & $0.03$ & $0.16$ & $-0.12$ \\ 
 
$650$ & $0.0135$ & $0.475$ & $0.8786$ & $3.86$ & $3.64$ & $1.24$ & $0.31$ & $0.06$ & $0.39$ & $-0.30$ & $-0.17$ & $0.02$ & $0.13$ & $-0.11$ \\ 
 
$650$ & $0.0151$ & $0.425$ & $0.8226$ & $3.81$ & $3.49$ & $1.30$ & $0.52$ & $0.03$ & $0.81$ & $-0.54$ & $-0.59$ & $0.01$ & $0.11$ & $0.00$ \\ 
 
$650$ & $0.0171$ & $0.375$ & $0.7808$ & $3.59$ & $3.30$ & $1.26$ & $0.57$ & $0.02$ & $0.63$ & $-0.45$ & $-0.42$ & $0.00$ & $0.14$ & $-0.01$ \\ 
 
$650$ & $0.0197$ & $0.325$ & $0.7520$ & $3.41$ & $3.08$ & $1.27$ & $0.62$ & $0.00$ & $0.74$ & $-0.53$ & $-0.51$ & $0.00$ & $0.00$ & $-0.01$ \\ 
 
$650$ & $0.0261$ & $0.245$ & $0.6866$ & $2.27$ & $1.82$ & $1.13$ & $0.68$ & $0.00$ & $0.78$ & $-0.62$ & $-0.47$ & $0.00$ & $0.00$ & $0.00$ \\ 
 
\hline 
$800$ & $0.0132$ & $0.595$ & $0.8143$ & $4.52$ & $4.30$ & $1.31$ & $0.21$ & $0.06$ & $0.50$ & $-0.14$ & $-0.42$ & $0.02$ & $0.16$ & $-0.18$ \\ 
 
$800$ & $0.0149$ & $0.530$ & $0.9105$ & $4.49$ & $4.25$ & $1.39$ & $0.48$ & $0.08$ & $0.47$ & $0.35$ & $0.20$ & $0.03$ & $0.18$ & $-0.15$ \\ 
 
$800$ & $0.0166$ & $0.475$ & $0.7637$ & $4.56$ & $4.34$ & $1.32$ & $0.10$ & $0.09$ & $0.53$ & $0.04$ & $-0.49$ & $0.02$ & $0.18$ & $-0.06$ \\ 
 
$800$ & $0.0185$ & $0.425$ & $0.7535$ & $4.31$ & $4.02$ & $1.49$ & $0.80$ & $0.01$ & $0.45$ & $-0.24$ & $-0.35$ & $0.00$ & $0.13$ & $-0.05$ \\ 
 
$800$ & $0.0210$ & $0.375$ & $0.6743$ & $4.22$ & $3.90$ & $1.41$ & $0.71$ & $0.01$ & $0.78$ & $-0.76$ & $-0.14$ & $0.00$ & $0.05$ & $-0.01$ \\ 
 
$800$ & $0.0242$ & $0.325$ & $0.6736$ & $3.86$ & $3.60$ & $1.31$ & $0.57$ & $0.00$ & $0.49$ & $-0.16$ & $-0.46$ & $0.00$ & $0.00$ & $-0.01$ \\ 
 
$800$ & $0.0322$ & $0.245$ & $0.6068$ & $2.68$ & $2.23$ & $1.28$ & $0.85$ & $0.00$ & $0.77$ & $-0.62$ & $-0.44$ & $0.00$ & $0.00$ & $0.00$ \\ 
 
\hline 

\end{tabular} 
\end{center} 
\caption[RESULT] 
{\label{tab:ncdydq2_ele} The NC 
$e^-p$ reduced cross section $\tilde{\sigma}_{\rm NC}(y,Q^2)$ 
for $P_e=0$ 
with statistical 
$(\delta_{\rm stat})$, 
total $(\delta_{tot})$, 
total uncorrelated systematic $(\delta_{\rm unc})$ 
errors, two of its contributions from the 
 electron energy error ($\delta_{unc}^{E}$)  
and the hadronic energy error  
($\delta_{\rm unc}^{h}$). 
The effect of the other uncorrelated 
systematic errors is included in $\delta_{\rm unc}$. 
In addition the correlated systematic  
$(\delta_{\rm cor})$ and its contributions from a 
positive variation of one  
standard deviation of the 
electron energy error ($\delta_{cor}^{E^+}$), of 
the polar electron angle error 
($\delta_{\rm cor}^{\theta^+}$), of the hadronic 
energy error ($\delta_{\rm cor}^{h^+}$), of the error 
due to noise subtraction ($\delta_{\rm cor}^{N^+}$) 
and of the error due to background subtraction 
($\delta_{\rm cor}^{B^+}$) are given. 
The normalisation and polarisation uncertainties are 
not included in the errors. 
}
\end{table} 

\begin{table}[htbp] 
\begin{center} 
\tiny 
\begin{tabular}{|r|c|c|r|r|r|r|r|r|r|r|r|r|r|r|} 
\hline 
$Q^2$ & $x$ & $y$ & $\tilde{\sigma}_{\rm NC}$ & 
$\delta_{\rm tot}$ & $\delta_{\rm stat}$ & $\delta_{\rm unc}$ & 
$\delta_{\rm unc}^{E}$ & 
$\delta_{\rm unc}^{h}$& 
$\delta_{\rm cor}$ & 
$\delta_{\rm cor}^{E^+}$ & 
$\delta_{\rm cor}^{\theta^+}$& 
$\delta_{\rm cor}^{h^+}$& 
$\delta_{\rm cor}^{N^+}$& 
$\delta_{\rm cor}^{B^+}$ \\ 
$(\rm GeV^2)$ & & & & 
$(\%)$ & $(\%)$ & $(\%)$ & $(\%)$ & $(\%)$ & $(\%)$ & 
$(\%)$ & $(\%)$ & $(\%)$ & $(\%)$ & $(\%)$  
\\ \hline 
$90$ & $0.0015$ & $0.595$ & $1.382$ & $2.20$ & $1.12$ & $1.03$ & $0.62$ & $0.09$ & $1.59$ & $-0.29$ & $-0.64$ & $0.03$ & $0.15$ & $-1.41$ \\ 
 
\hline 
$120$ & $0.0020$ & $0.595$ & $1.340$ & $1.98$ & $1.02$ & $0.90$ & $0.40$ & $0.10$ & $1.44$ & $-0.29$ & $-0.58$ & $0.02$ & $0.19$ & $-1.28$ \\ 
 
$120$ & $0.0022$ & $0.530$ & $1.312$ & $1.69$ & $1.02$ & $0.96$ & $0.56$ & $0.08$ & $0.95$ & $-0.29$ & $-0.68$ & $0.02$ & $0.18$ & $-0.58$ \\ 
 
$120$ & $0.0025$ & $0.475$ & $1.270$ & $1.79$ & $1.18$ & $1.02$ & $0.62$ & $0.06$ & $0.89$ & $-0.53$ & $-0.66$ & $0.01$ & $0.16$ & $-0.20$ \\ 
 
$120$ & $0.0028$ & $0.425$ & $1.223$ & $1.89$ & $1.42$ & $1.02$ & $0.51$ & $0.04$ & $0.74$ & $-0.29$ & $-0.64$ & $0.01$ & $0.18$ & $-0.12$ \\ 
 
\hline 
$150$ & $0.0025$ & $0.595$ & $1.240$ & $1.88$ & $1.21$ & $0.83$ & $0.13$ & $0.11$ & $1.17$ & $-0.12$ & $-0.44$ & $0.02$ & $0.18$ & $-1.07$ \\ 
 
$150$ & $0.0028$ & $0.530$ & $1.263$ & $1.59$ & $1.17$ & $0.84$ & $0.30$ & $0.07$ & $0.68$ & $-0.27$ & $-0.47$ & $0.02$ & $0.17$ & $-0.38$ \\ 
 
$150$ & $0.0031$ & $0.475$ & $1.213$ & $1.73$ & $1.19$ & $0.94$ & $0.52$ & $0.04$ & $0.83$ & $-0.48$ & $-0.61$ & $0.01$ & $0.16$ & $-0.26$ \\ 
 
$150$ & $0.0035$ & $0.425$ & $1.209$ & $1.70$ & $1.10$ & $1.06$ & $0.75$ & $0.03$ & $0.74$ & $-0.41$ & $-0.58$ & $0.01$ & $0.16$ & $-0.15$ \\ 
 
$150$ & $0.0039$ & $0.375$ & $1.160$ & $1.62$ & $1.02$ & $0.96$ & $0.62$ & $0.02$ & $0.80$ & $-0.34$ & $-0.70$ & $0.00$ & $0.18$ & $-0.03$ \\ 
 
$150$ & $0.0045$ & $0.325$ & $1.112$ & $1.85$ & $1.04$ & $1.19$ & $0.93$ & $0.00$ & $0.96$ & $-0.66$ & $-0.70$ & $0.00$ & $0.00$ & $-0.03$ \\ 
 
$150$ & $0.0060$ & $0.245$ & $1.015$ & $1.89$ & $0.80$ & $1.37$ & $1.18$ & $0.00$ & $1.03$ & $-0.70$ & $-0.75$ & $0.00$ & $0.00$ & $0.00$ \\ 
 
\hline 
$200$ & $0.0033$ & $0.595$ & $1.220$ & $2.04$ & $1.55$ & $0.90$ & $0.23$ & $0.09$ & $0.97$ & $-0.22$ & $-0.46$ & $0.02$ & $0.17$ & $-0.81$ \\ 
 
$200$ & $0.0037$ & $0.530$ & $1.177$ & $1.87$ & $1.50$ & $0.86$ & $0.17$ & $0.07$ & $0.72$ & $-0.18$ & $-0.53$ & $0.02$ & $0.17$ & $-0.41$ \\ 
 
$200$ & $0.0041$ & $0.475$ & $1.125$ & $1.90$ & $1.55$ & $0.85$ & $0.12$ & $0.06$ & $0.70$ & $-0.09$ & $-0.64$ & $0.01$ & $0.16$ & $-0.20$ \\ 
 
$200$ & $0.0046$ & $0.425$ & $1.109$ & $1.81$ & $1.44$ & $0.89$ & $0.37$ & $0.05$ & $0.64$ & $-0.29$ & $-0.54$ & $0.01$ & $0.16$ & $-0.08$ \\ 
 
$200$ & $0.0052$ & $0.375$ & $1.059$ & $1.84$ & $1.35$ & $0.94$ & $0.53$ & $0.02$ & $0.82$ & $-0.44$ & $-0.67$ & $0.01$ & $0.17$ & $-0.03$ \\ 
 
$200$ & $0.0061$ & $0.325$ & $1.038$ & $1.81$ & $1.22$ & $1.04$ & $0.71$ & $0.00$ & $0.84$ & $-0.50$ & $-0.68$ & $0.00$ & $0.00$ & $-0.03$ \\ 
 
$200$ & $0.0080$ & $0.245$ & $0.9387$ & $1.77$ & $0.73$ & $1.26$ & $1.07$ & $0.00$ & $1.01$ & $-0.64$ & $-0.79$ & $0.00$ & $0.00$ & $0.00$ \\ 
 
\hline 
$250$ & $0.0041$ & $0.595$ & $1.139$ & $2.25$ & $1.83$ & $0.94$ & $0.26$ & $0.10$ & $0.90$ & $-0.26$ & $-0.42$ & $0.03$ & $0.17$ & $-0.74$ \\ 
 
$250$ & $0.0046$ & $0.530$ & $1.130$ & $2.04$ & $1.74$ & $0.88$ & $0.16$ & $0.05$ & $0.60$ & $-0.16$ & $-0.46$ & $0.01$ & $0.12$ & $-0.33$ \\ 
 
$250$ & $0.0052$ & $0.475$ & $1.093$ & $2.07$ & $1.75$ & $0.92$ & $0.32$ & $0.06$ & $0.60$ & $-0.32$ & $-0.46$ & $0.01$ & $0.15$ & $-0.14$ \\ 
 
$250$ & $0.0058$ & $0.425$ & $1.089$ & $1.97$ & $1.60$ & $0.93$ & $0.43$ & $0.06$ & $0.68$ & $-0.41$ & $-0.49$ & $0.02$ & $0.18$ & $-0.13$ \\ 
 
$250$ & $0.0066$ & $0.375$ & $1.021$ & $1.83$ & $1.50$ & $0.87$ & $0.36$ & $0.02$ & $0.61$ & $-0.33$ & $-0.49$ & $0.00$ & $0.14$ & $-0.02$ \\ 
 
$250$ & $0.0076$ & $0.325$ & $0.9427$ & $1.96$ & $1.43$ & $1.03$ & $0.68$ & $0.00$ & $0.85$ & $-0.65$ & $-0.54$ & $0.00$ & $0.00$ & $-0.02$ \\ 
 
$250$ & $0.0100$ & $0.245$ & $0.8633$ & $1.60$ & $0.85$ & $1.05$ & $0.80$ & $0.00$ & $0.86$ & $-0.59$ & $-0.62$ & $0.00$ & $0.00$ & $-0.01$ \\ 
 
\hline 
$300$ & $0.0050$ & $0.595$ & $1.115$ & $2.49$ & $2.15$ & $0.96$ & $0.09$ & $0.08$ & $0.81$ & $-0.08$ & $-0.38$ & $0.01$ & $0.16$ & $-0.70$ \\ 
 
$300$ & $0.0056$ & $0.530$ & $1.093$ & $2.43$ & $2.12$ & $0.95$ & $0.30$ & $0.07$ & $0.71$ & $-0.33$ & $-0.51$ & $0.03$ & $0.15$ & $-0.34$ \\ 
 
$300$ & $0.0062$ & $0.475$ & $1.043$ & $2.32$ & $2.07$ & $0.90$ & $0.07$ & $0.04$ & $0.53$ & $-0.07$ & $-0.50$ & $0.01$ & $0.13$ & $-0.10$ \\ 
 
$300$ & $0.0069$ & $0.425$ & $1.016$ & $2.31$ & $1.93$ & $1.01$ & $0.54$ & $0.04$ & $0.78$ & $-0.54$ & $-0.53$ & $0.01$ & $0.17$ & $-0.10$ \\ 
 
$300$ & $0.0079$ & $0.375$ & $0.9685$ & $2.04$ & $1.74$ & $0.86$ & $0.28$ & $0.00$ & $0.62$ & $-0.28$ & $-0.54$ & $0.00$ & $0.10$ & $-0.02$ \\ 
 
$300$ & $0.0091$ & $0.325$ & $0.8966$ & $2.02$ & $1.65$ & $0.90$ & $0.42$ & $0.00$ & $0.76$ & $-0.42$ & $-0.63$ & $0.00$ & $0.00$ & $-0.01$ \\ 
 
$300$ & $0.0121$ & $0.245$ & $0.8111$ & $1.68$ & $0.98$ & $1.00$ & $0.72$ & $0.00$ & $0.93$ & $-0.72$ & $-0.58$ & $0.00$ & $0.00$ & $-0.01$ \\ 
 
\hline 
$400$ & $0.0066$ & $0.595$ & $1.021$ & $2.87$ & $2.55$ & $1.03$ & $0.22$ & $0.10$ & $0.82$ & $0.19$ & $-0.59$ & $0.02$ & $0.14$ & $-0.52$ \\ 
 
$400$ & $0.0074$ & $0.530$ & $1.048$ & $2.73$ & $2.46$ & $1.00$ & $0.28$ & $0.07$ & $0.64$ & $-0.29$ & $-0.49$ & $0.01$ & $0.14$ & $-0.25$ \\ 
 
$400$ & $0.0083$ & $0.475$ & $0.9818$ & $2.82$ & $2.50$ & $1.13$ & $0.61$ & $0.05$ & $0.67$ & $-0.61$ & $-0.20$ & $0.00$ & $0.11$ & $-0.14$ \\ 
 
$400$ & $0.0093$ & $0.425$ & $0.9174$ & $2.65$ & $2.33$ & $1.04$ & $0.52$ & $0.04$ & $0.70$ & $-0.52$ & $-0.43$ & $0.01$ & $0.17$ & $-0.08$ \\ 
 
$400$ & $0.0105$ & $0.375$ & $0.8979$ & $2.32$ & $2.09$ & $0.88$ & $0.20$ & $0.01$ & $0.52$ & $-0.20$ & $-0.47$ & $0.00$ & $0.13$ & $-0.02$ \\ 
 
$400$ & $0.0121$ & $0.325$ & $0.8714$ & $2.34$ & $1.91$ & $1.06$ & $0.66$ & $0.00$ & $0.85$ & $-0.66$ & $-0.54$ & $0.00$ & $0.00$ & $-0.01$ \\ 
 
$400$ & $0.0161$ & $0.245$ & $0.7565$ & $1.65$ & $1.15$ & $0.91$ & $0.59$ & $0.00$ & $0.75$ & $-0.58$ & $-0.47$ & $0.00$ & $0.00$ & $0.00$ \\ 
 
\hline 
$500$ & $0.0083$ & $0.595$ & $0.9509$ & $3.20$ & $2.94$ & $1.10$ & $0.41$ & $0.10$ & $0.60$ & $-0.22$ & $-0.17$ & $0.02$ & $0.17$ & $-0.50$ \\ 
 
$500$ & $0.0093$ & $0.530$ & $0.9256$ & $3.09$ & $2.88$ & $1.02$ & $0.24$ & $0.07$ & $0.40$ & $-0.08$ & $-0.35$ & $0.01$ & $0.11$ & $-0.15$ \\ 
 
$500$ & $0.0104$ & $0.475$ & $0.9175$ & $3.22$ & $2.96$ & $1.10$ & $0.42$ & $0.06$ & $0.63$ & $-0.27$ & $-0.54$ & $0.01$ & $0.17$ & $-0.07$ \\ 
 
$500$ & $0.0116$ & $0.425$ & $0.8560$ & $3.07$ & $2.82$ & $1.01$ & $0.22$ & $0.03$ & $0.67$ & $-0.15$ & $-0.63$ & $0.01$ & $0.16$ & $-0.06$ \\ 
 
$500$ & $0.0131$ & $0.375$ & $0.8663$ & $2.82$ & $2.51$ & $1.04$ & $0.47$ & $0.01$ & $0.75$ & $-0.51$ & $-0.55$ & $0.00$ & $0.09$ & $-0.02$ \\ 
 
$500$ & $0.0152$ & $0.325$ & $0.7935$ & $2.62$ & $2.42$ & $0.91$ & $0.16$ & $0.00$ & $0.44$ & $0.12$ & $-0.42$ & $0.00$ & $0.00$ & $-0.01$ \\ 
 
$500$ & $0.0201$ & $0.245$ & $0.7274$ & $1.95$ & $1.41$ & $1.00$ & $0.68$ & $0.00$ & $0.89$ & $-0.67$ & $-0.58$ & $0.00$ & $0.00$ & $0.00$ \\ 
 
\hline 
$650$ & $0.0108$ & $0.595$ & $0.8748$ & $3.49$ & $3.25$ & $1.16$ & $0.40$ & $0.06$ & $0.53$ & $-0.39$ & $-0.20$ & $0.01$ & $0.09$ & $-0.27$ \\ 
 
$650$ & $0.0121$ & $0.530$ & $0.8859$ & $3.43$ & $3.20$ & $1.08$ & $0.16$ & $0.06$ & $0.58$ & $-0.20$ & $-0.52$ & $0.02$ & $0.11$ & $-0.13$ \\ 
 
$650$ & $0.0135$ & $0.475$ & $0.7983$ & $3.66$ & $3.42$ & $1.16$ & $0.39$ & $0.04$ & $0.60$ & $-0.27$ & $-0.52$ & $0.00$ & $0.11$ & $-0.06$ \\ 
 
$650$ & $0.0151$ & $0.425$ & $0.8432$ & $3.31$ & $3.12$ & $1.07$ & $0.17$ & $0.05$ & $0.27$ & $-0.11$ & $-0.15$ & $0.02$ & $0.20$ & $-0.01$ \\ 
 
$650$ & $0.0171$ & $0.375$ & $0.7310$ & $3.39$ & $3.09$ & $1.22$ & $0.67$ & $0.01$ & $0.69$ & $-0.48$ & $-0.48$ & $0.00$ & $0.08$ & $-0.05$ \\ 
 
$650$ & $0.0197$ & $0.325$ & $0.7176$ & $3.12$ & $2.86$ & $1.08$ & $0.42$ & $0.00$ & $0.63$ & $-0.46$ & $-0.44$ & $0.00$ & $0.00$ & $-0.01$ \\ 
 
$650$ & $0.0261$ & $0.245$ & $0.6801$ & $2.07$ & $1.68$ & $0.95$ & $0.54$ & $0.00$ & $0.76$ & $-0.47$ & $-0.59$ & $0.00$ & $0.00$ & $0.00$ \\ 
 
\hline 
$800$ & $0.0132$ & $0.595$ & $0.8173$ & $4.08$ & $3.81$ & $1.32$ & $0.28$ & $0.11$ & $0.66$ & $0.37$ & $-0.44$ & $0.03$ & $0.16$ & $-0.29$ \\ 
 
$800$ & $0.0149$ & $0.530$ & $0.7529$ & $4.10$ & $3.88$ & $1.22$ & $0.29$ & $0.06$ & $0.49$ & $-0.27$ & $-0.35$ & $0.04$ & $0.12$ & $-0.15$ \\ 
 
$800$ & $0.0166$ & $0.475$ & $0.7075$ & $4.27$ & $4.04$ & $1.30$ & $0.47$ & $0.05$ & $0.50$ & $-0.35$ & $-0.30$ & $0.02$ & $0.16$ & $-0.11$ \\ 
 
$800$ & $0.0185$ & $0.425$ & $0.6978$ & $3.97$ & $3.72$ & $1.24$ & $0.49$ & $0.02$ & $0.62$ & $-0.54$ & $-0.27$ & $0.00$ & $0.13$ & $-0.05$ \\ 
 
$800$ & $0.0210$ & $0.375$ & $0.7056$ & $3.81$ & $3.41$ & $1.52$ & $1.05$ & $0.00$ & $0.72$ & $-0.51$ & $-0.50$ & $0.00$ & $0.05$ & $-0.03$ \\ 
 
$800$ & $0.0242$ & $0.325$ & $0.6539$ & $3.63$ & $3.42$ & $1.16$ & $0.39$ & $0.00$ & $0.37$ & $-0.19$ & $-0.32$ & $0.00$ & $0.00$ & $-0.01$ \\ 
 
$800$ & $0.0322$ & $0.245$ & $0.6030$ & $2.37$ & $1.98$ & $1.12$ & $0.74$ & $0.00$ & $0.67$ & $-0.47$ & $-0.48$ & $0.00$ & $0.00$ & $0.00$ \\ 
 
\hline 

\end{tabular} 
\end{center} 
\caption[RESULT] 
{\label{tab:ncdydq2_pos} The NC 
$e^+p$ reduced cross section $\tilde{\sigma}_{\rm NC}(y,Q^2)$ 
for $P_e=0$ 
with statistical 
$(\delta_{\rm stat})$, 
total $(\delta_{tot})$, 
total uncorrelated systematic $(\delta_{\rm unc})$ 
errors, two of its contributions from the 
 electron energy error ($\delta_{unc}^{E}$)  
and the hadronic energy error  
($\delta_{\rm unc}^{h}$). 
The effect of the other uncorrelated 
systematic errors is included in $\delta_{\rm unc}$. 
In addition the correlated systematic  
$(\delta_{\rm cor})$ and its contributions from a 
positive variation of one  
standard deviation of the 
electron energy error ($\delta_{cor}^{E^+}$), of 
the polar electron angle error 
($\delta_{\rm cor}^{\theta^+}$), of the hadronic 
energy error ($\delta_{\rm cor}^{h^+}$), of the error 
due to noise subtraction ($\delta_{\rm cor}^{N^+}$) 
and of the error due to background subtraction 
($\delta_{\rm cor}^{B^+}$) are given. 
The normalisation and polarisation uncertainties are 
not included in the errors. 
}
\end{table}

\clearpage
\begin{table}[htbp] 
\begin{center} 
\tiny 
\begin{tabular}{|r|c|c|r|r|r|r|r|r|r|r|r|r|r|r|} 
\hline 
$Q^2$ & $x$ & $y$ & $\tilde{\sigma}_{\rm NC}$ & 
$\delta_{\rm tot}$ & $\delta_{\rm stat}$ & $\delta_{\rm unc}$ & 
$\delta_{\rm unc}^{E}$ & 
$\delta_{\rm unc}^{h}$& 
$\delta_{\rm cor}$ & 
$\delta_{\rm cor}^{E^+}$ & 
$\delta_{\rm cor}^{\theta^+}$& 
$\delta_{\rm cor}^{h^+}$& 
$\delta_{\rm cor}^{N^+}$& 
$\delta_{\rm cor}^{S^+}$ \\ 
$(\rm GeV^2)$ & & & & 
$(\%)$ & $(\%)$ & $(\%)$ & $(\%)$ & $(\%)$ & $(\%)$ & 
$(\%)$ & $(\%)$ & $(\%)$ & $(\%)$ & $(\%)$  
\\ \hline 
$60$ & $0.0008$ & $0.75$ & $1.436$ & $2.29$ & $1.02$ & $1.84$ & $0.52$ & $0.28$ & $0.92$ & $-0.19$ & $-0.62$ & $0.05$ & $0.19$ & $-0.62$ \\ 
 
$90$ & $0.0012$ & $0.75$ & $1.437$ & $2.05$ & $0.83$ & $1.71$ & $0.31$ & $0.26$ & $0.77$ & $-0.28$ & $-0.45$ & $0.07$ & $0.20$ & $-0.52$ \\ 
 
$120$ & $0.0016$ & $0.75$ & $1.400$ & $2.15$ & $0.93$ & $1.82$ & $0.48$ & $0.28$ & $0.69$ & $-0.09$ & $-0.40$ & $0.05$ & $0.22$ & $-0.50$ \\ 
 
$150$ & $0.0020$ & $0.75$ & $1.313$ & $2.25$ & $1.05$ & $1.89$ & $0.38$ & $0.27$ & $0.61$ & $-0.19$ & $-0.34$ & $0.06$ & $0.22$ & $-0.41$ \\ 
 
$200$ & $0.0026$ & $0.75$ & $1.252$ & $2.48$ & $1.35$ & $2.00$ & $0.49$ & $0.30$ & $0.56$ & $-0.07$ & $-0.35$ & $0.06$ & $0.24$ & $-0.35$ \\ 
 
$250$ & $0.0033$ & $0.75$ & $1.244$ & $2.56$ & $1.54$ & $1.98$ & $0.31$ & $0.32$ & $0.53$ & $-0.13$ & $-0.29$ & $0.08$ & $0.26$ & $-0.32$ \\ 
 
$300$ & $0.0039$ & $0.75$ & $1.192$ & $2.70$ & $1.79$ & $1.96$ & $0.14$ & $0.31$ & $0.51$ & $-0.12$ & $-0.26$ & $0.06$ & $0.28$ & $-0.31$ \\ 
 
$400$ & $0.0052$ & $0.75$ & $1.122$ & $2.81$ & $1.95$ & $1.96$ & $0.26$ & $0.30$ & $0.50$ & $-0.06$ & $-0.35$ & $0.08$ & $0.25$ & $-0.23$ \\ 
 
$500$ & $0.0066$ & $0.75$ & $1.018$ & $2.90$ & $2.16$ & $1.90$ & $0.13$ & $0.25$ & $0.35$ & $-0.11$ & $-0.15$ & $0.09$ & $0.25$ & $-0.13$ \\ 
 
$650$ & $0.0085$ & $0.75$ & $0.9565$ & $3.13$ & $2.45$ & $1.92$ & $0.21$ & $0.21$ & $0.38$ & $0.07$ & $-0.28$ & $0.06$ & $0.23$ & $-0.08$ \\ 
 
$800$ & $0.0105$ & $0.75$ & $0.9329$ & $3.49$ & $2.85$ & $1.97$ & $0.23$ & $0.21$ & $0.39$ & $-0.22$ & $-0.20$ & $0.06$ & $0.25$ & $-0.04$ \\ 
 
\hline 
\end{tabular} 
\end{center} 
\caption[RESULT] 
{\label{tab:nchighy_ele} The NC 
$e^-p$ reduced cross section $\tilde{\sigma}_{\rm NC}(x,Q^2)$ 
for $P_e=0$ 
with statistical 
$(\delta_{\rm stat})$, 
total $(\delta_{tot})$, 
total uncorrelated systematic $(\delta_{\rm unc})$ 
errors, two of its contributions from the 
 electron energy error ($\delta_{unc}^{E}$)  
and the hadronic energy error  
($\delta_{\rm unc}^{h}$). 
The effect of the other uncorrelated 
systematic errors is included in $\delta_{\rm unc}$. 
In addition the correlated systematic  
$(\delta_{\rm cor})$ and its contributions from a 
positive variation of one  
standard deviation of the 
electron energy error ($\delta_{cor}^{E^+}$), of 
the polar electron angle error 
($\delta_{\rm cor}^{\theta^+}$), of the hadronic 
energy error ($\delta_{\rm cor}^{h^+}$), of the error 
due to noise subtraction ($\delta_{\rm cor}^{N^+}$) 
and of the error 
due to background subtraction charge asymmetry 
 ($\delta_{\rm cor}^{S^+}$) are given. 
The normalisation and polarisation uncertainties are 
not included in the errors. 
}
\end{table} 

\begin{table}[htbp] 
\begin{center} 
\tiny 
\begin{tabular}{|r|c|c|r|r|r|r|r|r|r|r|r|r|r|r|} 
\hline 
$Q^2$ & $x$ & $y$ & $\tilde{\sigma}_{\rm NC}$ & 
$\delta_{\rm tot}$ & $\delta_{\rm stat}$ & $\delta_{\rm unc}$ & 
$\delta_{\rm unc}^{E}$ & 
$\delta_{\rm unc}^{h}$& 
$\delta_{\rm cor}$ & 
$\delta_{\rm cor}^{E^+}$ & 
$\delta_{\rm cor}^{\theta^+}$& 
$\delta_{\rm cor}^{h^+}$& 
$\delta_{\rm cor}^{N^+}$& 
$\delta_{\rm cor}^{S^+}$ \\ 
$(\rm GeV^2)$ & & & & 
$(\%)$ & $(\%)$ & $(\%)$ & $(\%)$ & $(\%)$ & $(\%)$ & 
$(\%)$ & $(\%)$ & $(\%)$ & $(\%)$ & $(\%)$  
\\ \hline 
$60$ & $0.0008$ & $0.75$ & $1.453$ & $2.21$ & $0.88$ & $1.86$ & $0.70$ & $0.26$ & $0.81$ & $-0.31$ & $-0.48$ & $0.05$ & $0.17$ & $0.54$ \\ 
 
$90$ & $0.0012$ & $0.75$ & $1.460$ & $1.93$ & $0.72$ & $1.64$ & $0.13$ & $0.26$ & $0.73$ & $-0.21$ & $-0.50$ & $0.07$ & $0.21$ & $0.44$ \\ 
 
$120$ & $0.0016$ & $0.75$ & $1.379$ & $2.04$ & $0.84$ & $1.71$ & $0.36$ & $0.28$ & $0.72$ & $0.10$ & $-0.45$ & $0.07$ & $0.23$ & $0.50$ \\ 
 
$150$ & $0.0020$ & $0.75$ & $1.346$ & $2.14$ & $0.96$ & $1.81$ & $0.42$ & $0.25$ & $0.64$ & $-0.24$ & $-0.36$ & $0.07$ & $0.23$ & $0.41$ \\ 
 
$200$ & $0.0026$ & $0.75$ & $1.289$ & $2.38$ & $1.24$ & $1.94$ & $0.59$ & $0.26$ & $0.61$ & $-0.03$ & $-0.45$ & $0.06$ & $0.21$ & $0.35$ \\ 
 
$250$ & $0.0033$ & $0.75$ & $1.263$ & $2.42$ & $1.43$ & $1.88$ & $0.20$ & $0.30$ & $0.50$ & $-0.18$ & $-0.27$ & $0.05$ & $0.23$ & $0.30$ \\ 
 
$300$ & $0.0039$ & $0.75$ & $1.203$ & $2.53$ & $1.63$ & $1.88$ & $0.05$ & $0.34$ & $0.47$ & $-0.17$ & $-0.23$ & $0.08$ & $0.26$ & $0.25$ \\ 
 
$400$ & $0.0052$ & $0.75$ & $1.162$ & $2.60$ & $1.73$ & $1.89$ & $0.35$ & $0.27$ & $0.42$ & $-0.02$ & $-0.27$ & $0.07$ & $0.28$ & $0.14$ \\ 
 
$500$ & $0.0066$ & $0.75$ & $1.038$ & $2.69$ & $1.94$ & $1.83$ & $0.07$ & $0.23$ & $0.41$ & $-0.16$ & $-0.23$ & $0.07$ & $0.25$ & $0.13$ \\ 
 
$650$ & $0.0085$ & $0.75$ & $0.9795$ & $2.93$ & $2.22$ & $1.87$ & $0.23$ & $0.23$ & $0.46$ & $-0.09$ & $-0.34$ & $0.06$ & $0.27$ & $0.09$ \\ 
 
$800$ & $0.0105$ & $0.75$ & $0.9018$ & $3.31$ & $2.65$ & $1.93$ & $0.21$ & $0.24$ & $0.43$ & $0.07$ & $-0.31$ & $0.08$ & $0.28$ & $0.05$ \\ 
 
\hline 
\end{tabular} 
\end{center} 
\caption[RESULT] 
{\label{tab:nchighy_pos} The NC 
$e^+p$ reduced cross section $\tilde{\sigma}_{\rm NC}(x,Q^2)$ 
for $P_e=0$ 
with statistical 
$(\delta_{\rm stat})$, 
total $(\delta_{tot})$, 
total uncorrelated systematic $(\delta_{\rm unc})$ 
errors, two of its contributions from the 
 electron energy error ($\delta_{unc}^{E}$)  
and the hadronic energy error  
($\delta_{\rm unc}^{h}$). 
The effect of the other uncorrelated 
systematic errors is included in $\delta_{\rm unc}$. 
In addition the correlated systematic  
$(\delta_{\rm cor})$ and its contributions from a 
positive variation of one  
standard deviation of the 
electron energy error ($\delta_{cor}^{E^+}$), of 
the polar electron angle error 
($\delta_{\rm cor}^{\theta^+}$), of the hadronic 
energy error ($\delta_{\rm cor}^{h^+}$), of the error 
due to noise subtraction ($\delta_{\rm cor}^{N^+}$) 
and of the error 
due to background subtraction charge asymmetry 
 ($\delta_{\rm cor}^{S^+}$) are given. 
The normalisation and polarisation uncertainties are 
not included in the errors. 
}
\end{table}

\clearpage
\begin{table}[htbp] 
\begin{center} 
\tiny 
\begin{tabular}{|r|c|r|r|r|r|r|r|r|r|r|r|r|r|} 
\hline 
$Q^2$ & $x$ & $\tilde{\sigma}_{\rm NC}$ & 
$\delta_{\rm tot}$ & $\delta_{\rm stat}$ & $\delta_{\rm unc}$ & 
$\delta_{\rm unc}^{E}$ & 
$\delta_{\rm unc}^{h}$& 
$\delta_{\rm cor}$ & 
$\delta_{\rm cor}^{E^+}$ & 
$\delta_{\rm cor}^{\theta^+}$& 
$\delta_{\rm cor}^{h^+}$& 
$\delta_{\rm cor}^{N^+}$& 
$\delta_{\rm cor}^{B^+}$ \\ 
$(\rm GeV^2)$ & & & 
$(\%)$ & $(\%)$ & $(\%)$ & $(\%)$ & $(\%)$ & $(\%)$ & 
$(\%)$ & $(\%)$ & $(\%)$ & $(\%)$ & $(\%)$  
\\ \hline 
$120$ & $0.0020$ & $1.313$ & $1.61$ & $0.72$ & $0.95$ & $0.49$ & $0.09$ & $1.08$ & $-0.32$ & $-0.63$ & $0.02$ & $0.18$ & $-0.79$ \\ 
 
$120$ & $0.0032$ & $1.184$ & $1.72$ & $1.02$ & $1.15$ & $0.77$ & $0.05$ & $0.78$ & $-0.36$ & $-0.64$ & $0.02$ & $0.20$ & $-0.16$ \\ 
 
\hline 
$150$ & $0.0032$ & $1.191$ & $1.34$ & $0.61$ & $0.88$ & $0.41$ & $0.06$ & $0.81$ & $-0.32$ & $-0.59$ & $0.02$ & $0.19$ & $-0.41$ \\ 
 
$150$ & $0.0050$ & $1.074$ & $1.69$ & $0.73$ & $1.21$ & $0.90$ & $0.00$ & $0.93$ & $-0.55$ & $-0.74$ & $0.00$ & $0.04$ & $-0.04$ \\ 
 
$150$ & $0.0080$ & $0.9242$ & $2.43$ & $1.00$ & $1.81$ & $1.28$ & $0.96$ & $1.28$ & $-0.82$ & $-0.82$ & $-0.28$ & $-0.46$ & $-0.07$ \\ 
 
$150$ & $0.0130$ & $0.7884$ & $3.81$ & $1.38$ & $2.98$ & $2.68$ & $0.90$ & $1.93$ & $-1.55$ & $-0.74$ & $-0.30$ & $-0.81$ & $-0.10$ \\ 
 
\hline 
$200$ & $0.0032$ & $1.204$ & $1.60$ & $1.13$ & $0.86$ & $0.14$ & $0.07$ & $0.74$ & $-0.15$ & $-0.51$ & $0.02$ & $0.15$ & $-0.49$ \\ 
 
$200$ & $0.0050$ & $1.075$ & $1.45$ & $0.80$ & $0.95$ & $0.52$ & $0.02$ & $0.75$ & $-0.43$ & $-0.60$ & $0.00$ & $0.13$ & $-0.09$ \\ 
 
$200$ & $0.0080$ & $0.9355$ & $1.82$ & $0.81$ & $1.34$ & $1.07$ & $0.00$ & $0.93$ & $-0.62$ & $-0.70$ & $0.00$ & $0.00$ & $0.00$ \\ 
 
$200$ & $0.0130$ & $0.7699$ & $1.45$ & $0.94$ & $0.83$ & $0.16$ & $0.01$ & $0.73$ & $-0.06$ & $-0.44$ & $-0.04$ & $0.57$ & $-0.01$ \\ 
 
$200$ & $0.0200$ & $0.6787$ & $1.63$ & $1.02$ & $1.05$ & $0.63$ & $0.14$ & $0.71$ & $-0.46$ & $-0.41$ & $-0.06$ & $0.34$ & $-0.01$ \\ 
 
$200$ & $0.0320$ & $0.5663$ & $2.02$ & $1.16$ & $1.41$ & $0.98$ & $0.53$ & $0.88$ & $-0.53$ & $-0.67$ & $-0.19$ & $0.07$ & $0.00$ \\ 
 
$200$ & $0.0500$ & $0.5139$ & $2.74$ & $1.36$ & $1.72$ & $1.46$ & $0.04$ & $1.64$ & $-0.86$ & $-0.69$ & $-0.16$ & $1.20$ & $0.00$ \\ 
 
$200$ & $0.0800$ & $0.4301$ & $3.25$ & $1.42$ & $2.17$ & $1.93$ & $0.19$ & $1.96$ & $-1.15$ & $-0.76$ & $-0.09$ & $1.40$ & $-0.01$ \\ 
 
$200$ & $0.1300$ & $0.3527$ & $3.22$ & $1.68$ & $2.15$ & $1.35$ & $1.10$ & $1.71$ & $-0.82$ & $-0.98$ & $-0.28$ & $-1.11$ & $0.00$ \\ 
 
$200$ & $0.1800$ & $0.3006$ & $4.08$ & $2.27$ & $2.74$ & $1.26$ & $1.90$ & $2.01$ & $-0.87$ & $-1.08$ & $-0.43$ & $-1.40$ & $0.00$ \\ 
 
\hline 
$250$ & $0.0050$ & $1.089$ & $1.41$ & $0.93$ & $0.86$ & $0.33$ & $0.06$ & $0.61$ & $-0.32$ & $-0.40$ & $0.02$ & $0.18$ & $-0.27$ \\ 
 
$250$ & $0.0080$ & $0.9371$ & $1.57$ & $0.91$ & $1.00$ & $0.59$ & $0.00$ & $0.79$ & $-0.55$ & $-0.56$ & $0.00$ & $0.03$ & $-0.01$ \\ 
 
$250$ & $0.0130$ & $0.7965$ & $1.96$ & $1.00$ & $1.23$ & $0.90$ & $0.20$ & $1.15$ & $0.49$ & $-0.66$ & $0.03$ & $0.80$ & $-0.02$ \\ 
 
$250$ & $0.0200$ & $0.6722$ & $1.93$ & $1.03$ & $1.25$ & $0.91$ & $0.23$ & $1.06$ & $0.33$ & $-0.54$ & $0.06$ & $0.85$ & $-0.01$ \\ 
 
$250$ & $0.0320$ & $0.5724$ & $2.02$ & $1.08$ & $1.35$ & $1.07$ & $0.10$ & $1.03$ & $0.49$ & $-0.61$ & $-0.09$ & $0.67$ & $0.00$ \\ 
 
$250$ & $0.0500$ & $0.4843$ & $2.24$ & $1.23$ & $1.34$ & $1.02$ & $0.11$ & $1.30$ & $0.40$ & $-0.53$ & $-0.09$ & $1.12$ & $0.00$ \\ 
 
$250$ & $0.0800$ & $0.4179$ & $2.75$ & $1.26$ & $1.09$ & $0.56$ & $0.26$ & $2.19$ & $0.27$ & $-0.51$ & $-0.03$ & $2.11$ & $0.00$ \\ 
 
$250$ & $0.1300$ & $0.3581$ & $2.35$ & $1.28$ & $1.75$ & $1.08$ & $0.68$ & $0.93$ & $0.52$ & $-0.59$ & $-0.22$ & $-0.44$ & $0.00$ \\ 
 
$250$ & $0.1800$ & $0.3014$ & $4.00$ & $1.77$ & $2.69$ & $1.69$ & $1.56$ & $2.37$ & $1.05$ & $-0.76$ & $-0.35$ & $-1.95$ & $0.00$ \\ 
 
\hline 
$300$ & $0.0050$ & $1.117$ & $1.92$ & $1.56$ & $0.91$ & $0.19$ & $0.07$ & $0.65$ & $-0.20$ & $-0.48$ & $0.03$ & $0.16$ & $-0.35$ \\ 
 
$300$ & $0.0080$ & $0.9592$ & $1.51$ & $1.06$ & $0.86$ & $0.31$ & $0.02$ & $0.64$ & $-0.31$ & $-0.55$ & $0.01$ & $0.10$ & $-0.07$ \\ 
 
$300$ & $0.0130$ & $0.7987$ & $1.77$ & $1.06$ & $1.10$ & $0.73$ & $0.00$ & $0.90$ & $-0.72$ & $-0.55$ & $0.00$ & $0.00$ & $0.00$ \\ 
 
$300$ & $0.0200$ & $0.6846$ & $1.84$ & $1.18$ & $1.04$ & $0.61$ & $0.16$ & $0.95$ & $0.47$ & $-0.56$ & $0.02$ & $0.61$ & $-0.02$ \\ 
 
$300$ & $0.0320$ & $0.5763$ & $1.98$ & $1.24$ & $1.18$ & $0.83$ & $0.04$ & $1.00$ & $0.49$ & $-0.56$ & $-0.05$ & $0.66$ & $0.00$ \\ 
 
$300$ & $0.0500$ & $0.4893$ & $2.36$ & $1.35$ & $1.42$ & $1.12$ & $0.16$ & $1.31$ & $0.60$ & $-0.66$ & $-0.08$ & $0.95$ & $0.00$ \\ 
 
$300$ & $0.0800$ & $0.4165$ & $2.95$ & $1.43$ & $1.39$ & $1.03$ & $0.24$ & $2.17$ & $0.42$ & $-0.66$ & $-0.05$ & $2.02$ & $-0.01$ \\ 
 
$300$ & $0.1300$ & $0.3516$ & $2.58$ & $1.43$ & $1.85$ & $1.34$ & $0.42$ & $1.10$ & $0.68$ & $-0.66$ & $-0.17$ & $0.52$ & $0.00$ \\ 
 
$300$ & $0.1800$ & $0.2905$ & $4.57$ & $1.90$ & $3.05$ & $2.16$ & $1.63$ & $2.82$ & $1.34$ & $-0.90$ & $-0.38$ & $-2.28$ & $0.00$ \\ 
 
$300$ & $0.4000$ & $0.1454$ & $6.46$ & $2.25$ & $3.50$ & $2.29$ & $1.97$ & $4.94$ & $1.32$ & $-0.95$ & $-0.38$ & $-4.65$ & $0.00$ \\ 
 
\hline 
$400$ & $0.0080$ & $1.009$ & $1.70$ & $1.30$ & $0.91$ & $0.38$ & $0.05$ & $0.60$ & $-0.33$ & $-0.43$ & $0.01$ & $0.15$ & $-0.20$ \\ 
 
$400$ & $0.0130$ & $0.8332$ & $1.72$ & $1.25$ & $0.94$ & $0.46$ & $0.00$ & $0.71$ & $-0.45$ & $-0.54$ & $0.00$ & $0.03$ & $-0.02$ \\ 
 
$400$ & $0.0200$ & $0.6936$ & $2.07$ & $1.29$ & $1.24$ & $0.92$ & $0.00$ & $1.03$ & $-0.91$ & $-0.49$ & $0.00$ & $0.00$ & $0.00$ \\ 
 
$400$ & $0.0320$ & $0.5913$ & $1.90$ & $1.36$ & $0.96$ & $0.47$ & $0.03$ & $0.91$ & $0.46$ & $-0.47$ & $-0.04$ & $0.62$ & $0.00$ \\ 
 
$400$ & $0.0500$ & $0.4839$ & $2.04$ & $1.52$ & $1.07$ & $0.51$ & $0.38$ & $0.83$ & $0.50$ & $-0.38$ & $-0.12$ & $0.53$ & $-0.01$ \\ 
 
$400$ & $0.0800$ & $0.4096$ & $2.60$ & $1.62$ & $1.00$ & $0.27$ & $0.33$ & $1.77$ & $0.23$ & $-0.34$ & $0.04$ & $1.73$ & $0.00$ \\ 
 
$400$ & $0.1300$ & $0.3591$ & $2.44$ & $1.62$ & $1.39$ & $0.67$ & $0.20$ & $1.19$ & $0.59$ & $-0.52$ & $-0.13$ & $0.88$ & $0.00$ \\ 
 
$400$ & $0.1800$ & $0.2993$ & $4.20$ & $1.99$ & $2.44$ & $1.08$ & $1.70$ & $2.78$ & $1.01$ & $-0.51$ & $-0.43$ & $-2.50$ & $0.00$ \\ 
 
$400$ & $0.4000$ & $0.1440$ & $7.04$ & $2.64$ & $3.23$ & $1.34$ & $2.25$ & $5.68$ & $1.26$ & $-0.66$ & $-0.35$ & $-5.49$ & $0.00$ \\ 
 
\hline 
$500$ & $0.0080$ & $0.9703$ & $2.40$ & $2.15$ & $0.95$ & $0.18$ & $0.07$ & $0.51$ & $-0.23$ & $-0.38$ & $0.02$ & $0.15$ & $-0.19$ \\ 
 
$500$ & $0.0130$ & $0.8618$ & $1.87$ & $1.54$ & $0.90$ & $0.33$ & $0.03$ & $0.55$ & $-0.26$ & $-0.48$ & $0.00$ & $0.08$ & $-0.06$ \\ 
 
$500$ & $0.0200$ & $0.7087$ & $2.07$ & $1.53$ & $1.11$ & $0.70$ & $0.00$ & $0.83$ & $-0.70$ & $-0.45$ & $0.00$ & $0.00$ & $0.00$ \\ 
 
$500$ & $0.0320$ & $0.6061$ & $2.09$ & $1.57$ & $1.03$ & $0.50$ & $0.26$ & $0.92$ & $0.51$ & $-0.37$ & $0.06$ & $0.67$ & $0.00$ \\ 
 
$500$ & $0.0500$ & $0.5288$ & $2.18$ & $1.66$ & $1.10$ & $0.62$ & $0.23$ & $0.88$ & $0.63$ & $-0.38$ & $-0.11$ & $0.47$ & $0.00$ \\ 
 
$500$ & $0.0800$ & $0.4184$ & $2.64$ & $1.87$ & $1.06$ & $0.52$ & $0.13$ & $1.53$ & $0.52$ & $-0.47$ & $0.01$ & $1.36$ & $0.00$ \\ 
 
$500$ & $0.1300$ & $0.3636$ & $2.92$ & $2.17$ & $1.33$ & $0.55$ & $0.05$ & $1.43$ & $0.55$ & $-0.43$ & $-0.07$ & $1.25$ & $0.00$ \\ 
 
$500$ & $0.1800$ & $0.3247$ & $3.17$ & $2.34$ & $1.83$ & $0.87$ & $0.86$ & $1.09$ & $0.87$ & $-0.58$ & $-0.30$ & $-0.11$ & $0.00$ \\ 
 
$500$ & $0.2500$ & $0.2483$ & $5.10$ & $2.77$ & $2.52$ & $1.06$ & $1.73$ & $3.46$ & $1.06$ & $-0.55$ & $-0.42$ & $-3.22$ & $0.00$ \\ 
 
\hline 
$650$ & $0.0130$ & $0.8722$ & $2.03$ & $1.73$ & $0.96$ & $0.41$ & $0.05$ & $0.45$ & $-0.26$ & $-0.31$ & $0.02$ & $0.14$ & $-0.13$ \\ 
 
$650$ & $0.0200$ & $0.7433$ & $2.17$ & $1.79$ & $1.01$ & $0.48$ & $0.01$ & $0.71$ & $-0.49$ & $-0.52$ & $0.00$ & $0.03$ & $-0.01$ \\ 
 
$650$ & $0.0320$ & $0.6288$ & $2.51$ & $1.85$ & $1.34$ & $0.98$ & $0.00$ & $1.04$ & $-0.92$ & $-0.50$ & $0.00$ & $0.00$ & $0.00$ \\ 
 
$650$ & $0.0500$ & $0.5264$ & $2.40$ & $1.95$ & $1.10$ & $0.60$ & $0.08$ & $0.89$ & $0.68$ & $-0.38$ & $-0.07$ & $0.42$ & $0.00$ \\ 
 
$650$ & $0.0800$ & $0.4200$ & $2.73$ & $2.22$ & $1.08$ & $0.47$ & $0.12$ & $1.18$ & $0.52$ & $-0.32$ & $-0.07$ & $1.01$ & $0.00$ \\ 
 
$650$ & $0.1300$ & $0.3491$ & $3.25$ & $2.54$ & $1.42$ & $0.63$ & $0.17$ & $1.46$ & $0.66$ & $-0.43$ & $-0.08$ & $1.23$ & $0.00$ \\ 
 
$650$ & $0.1800$ & $0.3167$ & $3.22$ & $2.63$ & $1.64$ & $0.74$ & $0.40$ & $0.87$ & $0.76$ & $-0.37$ & $-0.15$ & $0.15$ & $0.00$ \\ 
 
$650$ & $0.2500$ & $0.2464$ & $5.05$ & $3.33$ & $2.48$ & $1.20$ & $1.53$ & $2.88$ & $1.17$ & $-0.48$ & $-0.39$ & $-2.56$ & $0.00$ \\ 
 
$650$ & $0.4000$ & $0.1183$ & $7.48$ & $5.00$ & $3.52$ & $1.60$ & $2.28$ & $4.29$ & $1.61$ & $-0.72$ & $-0.54$ & $-3.88$ & $0.00$ \\ 
 
\hline 

\end{tabular} 
\end{center} 
\caption[RESULT] 
{\label{tab:ncdxdq2_eleP0} The NC 
$e^-p$ reduced cross section $\tilde{\sigma}_{\rm NC}(x,Q^2)$ 
for $P_e=0$ 
with statistical 
$(\delta_{\rm stat})$, 
total $(\delta_{tot})$, 
total uncorrelated systematic $(\delta_{\rm unc})$ 
errors, two of its contributions from the 
 electron energy error ($\delta_{unc}^{E}$)  
and the hadronic energy error  
($\delta_{\rm unc}^{h}$). 
The effect of the other uncorrelated 
systematic errors is included in $\delta_{\rm unc}$. 
In addition the correlated systematic  
$(\delta_{\rm cor})$ and its contributions from a 
positive variation of one  
standard deviation of the 
electron energy error ($\delta_{cor}^{E^+}$), of 
the polar electron angle error 
($\delta_{\rm cor}^{\theta^+}$), of the hadronic 
energy error ($\delta_{\rm cor}^{h^+}$), of the error 
due to noise subtraction ($\delta_{\rm cor}^{N^+}$) 
and of the error due to background subtraction 
($\delta_{\rm cor}^{B^+}$) are given. 
The normalisation and polarisation uncertainties are 
not included in the errors. 
The table continues on the next page.}
\end{table} 
\begin{table}[htbp] 
\begin{center} 
\tiny 
\begin{tabular}{|r|c|r|r|r|r|r|r|r|r|r|r|r|r|} 
\hline 
$Q^2$ & $x$ & $\tilde{\sigma}_{\rm NC}$ & 
$\delta_{\rm tot}$ & $\delta_{\rm stat}$ & $\delta_{\rm unc}$ & 
$\delta_{\rm unc}^{E}$ & 
$\delta_{\rm unc}^{h}$& 
$\delta_{\rm cor}$ & 
$\delta_{\rm cor}^{E^+}$ & 
$\delta_{\rm cor}^{\theta^+}$& 
$\delta_{\rm cor}^{h^+}$& 
$\delta_{\rm cor}^{N^+}$& 
$\delta_{\rm cor}^{B^+}$ \\ 
$(\rm GeV^2)$ & & & 
$(\%)$ & $(\%)$ & $(\%)$ & $(\%)$ & $(\%)$ & $(\%)$ & 
$(\%)$ & $(\%)$ & $(\%)$ & $(\%)$ & $(\%)$  
\\ \hline 
$800$ & $0.0130$ & $0.8721$ & $3.12$ & $2.91$ & $1.05$ & $0.21$ & $0.08$ & $0.42$ & $0.10$ & $-0.35$ & $0.02$ & $0.18$ & $-0.12$ \\ 
 
$800$ & $0.0200$ & $0.7107$ & $2.37$ & $2.09$ & $1.03$ & $0.48$ & $0.02$ & $0.44$ & $-0.27$ & $-0.34$ & $0.00$ & $0.08$ & $-0.05$ \\ 
 
$800$ & $0.0320$ & $0.6105$ & $2.67$ & $2.20$ & $1.30$ & $0.88$ & $0.00$ & $0.77$ & $-0.62$ & $-0.47$ & $0.00$ & $0.00$ & $0.00$ \\ 
 
$800$ & $0.0500$ & $0.5357$ & $2.57$ & $2.27$ & $1.00$ & $0.24$ & $0.11$ & $0.65$ & $0.41$ & $-0.29$ & $-0.02$ & $0.42$ & $-0.02$ \\ 
 
$800$ & $0.0800$ & $0.4131$ & $3.01$ & $2.57$ & $1.12$ & $0.41$ & $0.21$ & $1.08$ & $0.59$ & $-0.45$ & $-0.06$ & $0.78$ & $0.00$ \\ 
 
$800$ & $0.1300$ & $0.3417$ & $3.64$ & $3.08$ & $1.38$ & $0.36$ & $0.21$ & $1.37$ & $0.58$ & $-0.38$ & $-0.04$ & $1.19$ & $0.00$ \\ 
 
$800$ & $0.1800$ & $0.3115$ & $3.81$ & $3.24$ & $1.75$ & $0.67$ & $0.67$ & $0.97$ & $0.83$ & $-0.40$ & $-0.21$ & $0.19$ & $0.00$ \\ 
 
$800$ & $0.2500$ & $0.2344$ & $4.70$ & $3.81$ & $2.10$ & $0.71$ & $1.13$ & $1.78$ & $0.90$ & $-0.33$ & $-0.24$ & $-1.48$ & $0.00$ \\ 
 
$800$ & $0.4000$ & $0.1288$ & $8.54$ & $5.03$ & $4.06$ & $1.61$ & $2.94$ & $5.58$ & $1.71$ & $-0.66$ & $-0.79$ & $-5.21$ & $0.00$ \\ 
 
\hline 
$1000$ & $0.0130$ & $0.8320$ & $3.50$ & $2.87$ & $1.54$ & $0.27$ & $0.22$ & $1.26$ & $-0.03$ & $-0.20$ & $0.05$ & $0.18$ & $-1.23$ \\ 
 
$1000$ & $0.0200$ & $0.7428$ & $2.64$ & $2.42$ & $0.98$ & $0.33$ & $0.05$ & $0.44$ & $-0.19$ & $-0.37$ & $0.01$ & $0.11$ & $-0.11$ \\ 
 
$1000$ & $0.0320$ & $0.6432$ & $2.70$ & $2.42$ & $1.08$ & $0.57$ & $0.00$ & $0.53$ & $-0.29$ & $-0.44$ & $0.00$ & $0.01$ & $0.00$ \\ 
 
$1000$ & $0.0500$ & $0.4971$ & $3.14$ & $2.63$ & $1.45$ & $1.09$ & $0.00$ & $0.92$ & $-0.64$ & $-0.66$ & $0.00$ & $0.00$ & $0.00$ \\ 
 
$1000$ & $0.0800$ & $0.4223$ & $3.16$ & $2.87$ & $1.02$ & $0.21$ & $0.17$ & $0.83$ & $0.25$ & $-0.26$ & $0.07$ & $0.75$ & $0.00$ \\ 
 
$1000$ & $0.1300$ & $0.3452$ & $3.90$ & $3.41$ & $1.35$ & $0.26$ & $0.23$ & $1.32$ & $0.41$ & $-0.31$ & $-0.12$ & $1.20$ & $0.00$ \\ 
 
$1000$ & $0.1800$ & $0.3367$ & $3.87$ & $3.38$ & $1.67$ & $0.65$ & $0.55$ & $0.85$ & $0.63$ & $-0.40$ & $-0.21$ & $0.35$ & $0.00$ \\ 
 
$1000$ & $0.2500$ & $0.2619$ & $4.54$ & $3.80$ & $2.03$ & $0.82$ & $0.97$ & $1.44$ & $0.93$ & $-0.38$ & $-0.31$ & $-0.97$ & $0.00$ \\ 
 
$1000$ & $0.4000$ & $0.1253$ & $8.71$ & $5.47$ & $4.20$ & $1.96$ & $2.99$ & $5.31$ & $1.61$ & $-0.53$ & $-0.61$ & $-4.99$ & $0.00$ \\ 
 
\hline 
$1200$ & $0.0130$ & $0.8623$ & $5.36$ & $4.65$ & $2.03$ & $0.09$ & $0.25$ & $1.73$ & $0.23$ & $0.15$ & $0.05$ & $0.20$ & $-1.70$ \\ 
 
$1200$ & $0.0200$ & $0.7572$ & $3.22$ & $2.99$ & $1.11$ & $0.17$ & $0.08$ & $0.43$ & $-0.11$ & $-0.28$ & $0.02$ & $0.11$ & $-0.29$ \\ 
 
$1200$ & $0.0320$ & $0.6579$ & $2.94$ & $2.74$ & $0.94$ & $0.36$ & $0.02$ & $0.52$ & $-0.27$ & $-0.45$ & $0.00$ & $0.06$ & $-0.01$ \\ 
 
$1200$ & $0.0500$ & $0.5236$ & $3.19$ & $2.90$ & $1.19$ & $0.80$ & $0.00$ & $0.60$ & $-0.42$ & $-0.43$ & $0.00$ & $0.00$ & $0.00$ \\ 
 
$1200$ & $0.0800$ & $0.4463$ & $3.45$ & $3.10$ & $1.31$ & $0.93$ & $0.20$ & $0.79$ & $0.59$ & $-0.27$ & $-0.13$ & $0.42$ & $0.00$ \\ 
 
$1200$ & $0.1300$ & $0.3428$ & $4.64$ & $4.31$ & $1.44$ & $0.75$ & $0.06$ & $0.95$ & $0.48$ & $-0.17$ & $-0.03$ & $0.80$ & $0.00$ \\ 
 
$1200$ & $0.1800$ & $0.3026$ & $4.39$ & $3.97$ & $1.69$ & $0.93$ & $0.38$ & $0.80$ & $0.61$ & $-0.28$ & $-0.14$ & $0.41$ & $0.00$ \\ 
 
$1200$ & $0.2500$ & $0.2140$ & $5.10$ & $4.63$ & $1.99$ & $1.15$ & $0.69$ & $0.83$ & $0.75$ & $-0.28$ & $-0.19$ & $-0.17$ & $0.00$ \\ 
 
$1200$ & $0.4000$ & $0.1249$ & $8.54$ & $5.78$ & $4.29$ & $2.60$ & $2.69$ & $4.59$ & $1.67$ & $-0.33$ & $-0.69$ & $-4.20$ & $0.00$ \\ 
 
\hline 
$1500$ & $0.0200$ & $0.7724$ & $4.08$ & $3.59$ & $1.66$ & $0.10$ & $0.16$ & $0.99$ & $0.14$ & $-0.27$ & $0.03$ & $0.15$ & $-0.93$ \\ 
 
$1500$ & $0.0320$ & $0.6540$ & $3.57$ & $3.40$ & $1.01$ & $0.35$ & $0.05$ & $0.47$ & $-0.36$ & $-0.27$ & $0.01$ & $0.12$ & $-0.05$ \\ 
 
$1500$ & $0.0500$ & $0.5406$ & $3.60$ & $3.34$ & $1.21$ & $0.79$ & $0.00$ & $0.57$ & $-0.39$ & $-0.42$ & $0.00$ & $0.01$ & $0.00$ \\ 
 
$1500$ & $0.0800$ & $0.4800$ & $3.66$ & $3.42$ & $1.16$ & $0.68$ & $0.09$ & $0.65$ & $0.39$ & $-0.17$ & $0.00$ & $0.48$ & $0.00$ \\ 
 
$1500$ & $0.1300$ & $0.3415$ & $5.10$ & $4.77$ & $1.57$ & $0.91$ & $0.17$ & $0.87$ & $0.63$ & $-0.22$ & $-0.07$ & $0.55$ & $0.00$ \\ 
 
$1500$ & $0.1800$ & $0.2914$ & $5.06$ & $4.62$ & $1.80$ & $1.12$ & $0.22$ & $1.00$ & $0.63$ & $-0.18$ & $-0.13$ & $0.74$ & $0.00$ \\ 
 
$1500$ & $0.2500$ & $0.2344$ & $5.50$ & $4.96$ & $2.20$ & $1.35$ & $0.89$ & $0.88$ & $0.72$ & $-0.15$ & $-0.22$ & $-0.43$ & $0.00$ \\ 
 
$1500$ & $0.4000$ & $0.1208$ & $9.06$ & $7.42$ & $3.87$ & $2.32$ & $2.23$ & $3.45$ & $1.43$ & $-0.33$ & $-0.59$ & $-3.06$ & $0.00$ \\ 
 
$1500$ & $0.6500$ & $0.01387$ & $18.01$ & $12.53$ & $6.95$ & $4.57$ & $4.52$ & $10.90$ & $3.07$ & $-0.44$ & $-0.97$ & $-10.41$ & $0.00$ \\ 
 
\hline 
$2000$ & $0.0219$ & $0.8405$ & $6.23$ & $5.59$ & $2.19$ & $0.25$ & $0.19$ & $1.66$ & $0.16$ & $-0.20$ & $0.04$ & $0.20$ & $-1.63$ \\ 
 
$2000$ & $0.0320$ & $0.6258$ & $4.32$ & $4.06$ & $1.41$ & $0.20$ & $0.10$ & $0.47$ & $-0.10$ & $-0.39$ & $0.03$ & $0.14$ & $-0.21$ \\ 
 
$2000$ & $0.0500$ & $0.5298$ & $4.26$ & $4.10$ & $1.08$ & $0.45$ & $0.02$ & $0.43$ & $-0.30$ & $-0.31$ & $0.01$ & $0.04$ & $-0.02$ \\ 
 
$2000$ & $0.0800$ & $0.4349$ & $4.41$ & $4.15$ & $1.31$ & $0.80$ & $0.00$ & $0.71$ & $-0.55$ & $-0.45$ & $0.00$ & $0.00$ & $0.00$ \\ 
 
$2000$ & $0.1300$ & $0.3449$ & $5.37$ & $5.09$ & $1.53$ & $0.72$ & $0.11$ & $0.80$ & $0.39$ & $-0.22$ & $0.07$ & $0.65$ & $0.00$ \\ 
 
$2000$ & $0.1800$ & $0.2915$ & $5.89$ & $5.43$ & $2.06$ & $1.43$ & $0.29$ & $0.98$ & $0.80$ & $-0.26$ & $-0.13$ & $0.50$ & $0.00$ \\ 
 
$2000$ & $0.2500$ & $0.2398$ & $6.07$ & $5.62$ & $2.14$ & $1.33$ & $0.64$ & $0.86$ & $0.76$ & $-0.28$ & $-0.18$ & $-0.22$ & $0.00$ \\ 
 
$2000$ & $0.4000$ & $0.1221$ & $8.46$ & $7.14$ & $3.63$ & $2.22$ & $1.84$ & $2.75$ & $1.33$ & $-0.11$ & $-0.51$ & $-2.35$ & $0.00$ \\ 
 
$2000$ & $0.6500$ & $0.009670$ & $21.36$ & $16.95$ & $7.85$ & $5.10$ & $5.25$ & $10.35$ & $2.65$ & $-0.48$ & $-1.07$ & $-9.94$ & $0.00$ \\ 
 
\hline 
$3000$ & $0.0320$ & $0.7040$ & $4.25$ & $3.80$ & $1.74$ & $0.18$ & $0.11$ & $0.74$ & $-0.12$ & $-0.29$ & $0.02$ & $0.11$ & $-0.66$ \\ 
 
$3000$ & $0.0500$ & $0.5792$ & $3.61$ & $3.33$ & $1.34$ & $0.18$ & $0.06$ & $0.37$ & $-0.09$ & $-0.34$ & $0.02$ & $0.08$ & $-0.06$ \\ 
 
$3000$ & $0.0800$ & $0.4805$ & $3.88$ & $3.65$ & $1.23$ & $0.50$ & $0.00$ & $0.51$ & $-0.31$ & $-0.40$ & $0.00$ & $0.01$ & $0.00$ \\ 
 
$3000$ & $0.1300$ & $0.3877$ & $4.75$ & $4.41$ & $1.63$ & $0.86$ & $0.00$ & $0.62$ & $-0.50$ & $-0.36$ & $0.00$ & $0.00$ & $0.00$ \\ 
 
$3000$ & $0.1800$ & $0.2895$ & $5.45$ & $5.07$ & $1.81$ & $1.09$ & $0.19$ & $0.84$ & $0.62$ & $-0.11$ & $-0.07$ & $0.56$ & $0.00$ \\ 
 
$3000$ & $0.2500$ & $0.2321$ & $5.69$ & $5.21$ & $2.11$ & $1.42$ & $0.39$ & $0.91$ & $0.86$ & $-0.07$ & $-0.12$ & $0.26$ & $0.00$ \\ 
 
$3000$ & $0.4000$ & $0.1124$ & $7.82$ & $6.48$ & $3.76$ & $2.58$ & $1.70$ & $2.23$ & $1.58$ & $-0.18$ & $-0.42$ & $-1.51$ & $0.00$ \\ 
 
$3000$ & $0.6500$ & $0.01050$ & $17.55$ & $13.52$ & $7.22$ & $4.58$ & $4.93$ & $8.56$ & $2.79$ & $-0.21$ & $-1.37$ & $-7.97$ & $0.00$ \\ 
 
\hline 
$5000$ & $0.0547$ & $0.6174$ & $5.46$ & $5.07$ & $1.90$ & $0.12$ & $0.13$ & $0.69$ & $0.05$ & $-0.36$ & $0.03$ & $0.13$ & $-0.57$ \\ 
 
$5000$ & $0.0800$ & $0.5100$ & $4.26$ & $3.98$ & $1.50$ & $0.19$ & $0.05$ & $0.36$ & $-0.20$ & $-0.28$ & $0.01$ & $0.07$ & $-0.11$ \\ 
 
$5000$ & $0.1300$ & $0.4498$ & $4.79$ & $4.46$ & $1.70$ & $0.28$ & $0.01$ & $0.45$ & $0.24$ & $-0.38$ & $0.00$ & $0.01$ & $-0.01$ \\ 
 
$5000$ & $0.1800$ & $0.3628$ & $5.45$ & $5.15$ & $1.75$ & $0.27$ & $0.00$ & $0.38$ & $-0.17$ & $-0.34$ & $0.00$ & $0.00$ & $0.00$ \\ 
 
$5000$ & $0.2500$ & $0.2268$ & $8.07$ & $7.73$ & $2.19$ & $1.11$ & $0.00$ & $0.82$ & $0.81$ & $-0.12$ & $0.00$ & $0.00$ & $0.00$ \\ 
 
$5000$ & $0.4000$ & $0.1141$ & $8.67$ & $7.82$ & $3.34$ & $1.88$ & $1.42$ & $1.66$ & $1.31$ & $-0.04$ & $-0.46$ & $-0.90$ & $0.00$ \\ 
 
$5000$ & $0.6500$ & $0.01375$ & $16.53$ & $13.90$ & $6.85$ & $4.13$ & $4.61$ & $5.75$ & $2.45$ & $0.20$ & $-1.16$ & $-5.07$ & $0.00$ \\ 
 
\hline 
$8000$ & $0.0875$ & $0.6162$ & $7.97$ & $7.49$ & $2.49$ & $0.39$ & $0.09$ & $1.09$ & $0.22$ & $-0.35$ & $0.02$ & $0.11$ & $-1.00$ \\ 
 
$8000$ & $0.1300$ & $0.5010$ & $6.40$ & $5.98$ & $2.23$ & $0.28$ & $0.03$ & $0.43$ & $-0.19$ & $-0.29$ & $0.01$ & $0.06$ & $-0.25$ \\ 
 
$8000$ & $0.1800$ & $0.3708$ & $6.96$ & $6.61$ & $2.14$ & $0.16$ & $0.02$ & $0.36$ & $-0.05$ & $-0.36$ & $0.01$ & $0.02$ & $0.00$ \\ 
 
$8000$ & $0.2500$ & $0.2533$ & $8.02$ & $7.62$ & $2.46$ & $1.04$ & $0.00$ & $0.52$ & $0.52$ & $0.03$ & $0.00$ & $0.00$ & $0.00$ \\ 
 
$8000$ & $0.4000$ & $0.1138$ & $11.63$ & $10.50$ & $4.51$ & $3.51$ & $0.00$ & $2.16$ & $2.09$ & $0.53$ & $0.00$ & $0.00$ & $0.00$ \\ 
 
$8000$ & $0.6500$ & $0.01193$ & $18.53$ & $16.94$ & $6.54$ & $4.19$ & $3.75$ & $3.66$ & $2.39$ & $0.23$ & $-0.87$ & $-2.63$ & $0.00$ \\ 
 
\hline 
$12000$ & $0.1300$ & $0.6999$ & $13.48$ & $12.90$ & $3.66$ & $0.51$ & $0.13$ & $1.42$ & $0.34$ & $-0.30$ & $0.04$ & $0.12$ & $-1.34$ \\ 
 
$12000$ & $0.1800$ & $0.4797$ & $8.61$ & $8.32$ & $2.20$ & $0.17$ & $0.08$ & $0.38$ & $-0.03$ & $-0.31$ & $0.03$ & $0.10$ & $-0.19$ \\ 
 
$12000$ & $0.2500$ & $0.3018$ & $9.61$ & $9.27$ & $2.47$ & $1.05$ & $0.02$ & $0.50$ & $0.48$ & $-0.13$ & $0.01$ & $0.03$ & $0.00$ \\ 
 
$12000$ & $0.4000$ & $0.1878$ & $11.98$ & $10.80$ & $4.89$ & $4.02$ & $0.00$ & $1.79$ & $1.68$ & $0.60$ & $0.00$ & $0.00$ & $0.00$ \\ 
 
$12000$ & $0.6500$ & $0.01352$ & $24.68$ & $23.62$ & $6.34$ & $4.68$ & $3.06$ & $3.31$ & $2.56$ & $0.45$ & $-0.84$ & $-1.86$ & $0.00$ \\ 
 
\hline 
$20000$ & $0.2500$ & $0.4615$ & $12.77$ & $12.42$ & $2.81$ & $1.17$ & $0.08$ & $0.93$ & $0.43$ & $-0.17$ & $0.03$ & $0.06$ & $-0.81$ \\ 
 
$20000$ & $0.4000$ & $0.2125$ & $14.46$ & $13.68$ & $4.57$ & $3.59$ & $0.03$ & $1.03$ & $1.02$ & $0.12$ & $0.01$ & $0.03$ & $-0.11$ \\ 
 
$20000$ & $0.6500$ & $0.01647$ & $39.16$ & $35.42$ & $15.88$ & $15.54$ & $0.00$ & $5.20$ & $4.46$ & $2.67$ & $0.00$ & $0.00$ & $0.00$ \\ 
 
\hline 
$30000$ & $0.4000$ & $0.1903$ & $28.38$ & $27.72$ & $5.74$ & $3.39$ & $0.14$ & $2.13$ & $1.05$ & $-0.25$ & $0.03$ & $0.09$ & $-1.83$ \\ 
 
$30000$ & $0.6500$ & $0.03988$ & $33.85$ & $31.66$ & $11.35$ & $10.79$ & $0.00$ & $3.90$ & $3.42$ & $1.86$ & $0.00$ & $0.01$ & $0.00$ \\ 
 
\hline 
$50000$ & $0.6500$ & $0.07518$ & $58.83$ & $57.79$ & $10.56$ & $8.88$ & $0.00$ & $3.06$ & $2.83$ & $1.17$ & $0.00$ & $0.00$ & $0.00$ \\ 
 
\hline 
\end{tabular} 
\end{center} 
\captcont{continued.}
\end{table} 

\begin{table}[htbp] 
\begin{center} 
\tiny 
\begin{tabular}{|r|c|r|r|r|r|r|r|r|r|r|r|r|r|} 
\hline 
$Q^2$ & $x$ & $\tilde{\sigma}_{\rm NC}$ & 
$\delta_{\rm tot}$ & $\delta_{\rm stat}$ & $\delta_{\rm unc}$ & 
$\delta_{\rm unc}^{E}$ & 
$\delta_{\rm unc}^{h}$& 
$\delta_{\rm cor}$ & 
$\delta_{\rm cor}^{E^+}$ & 
$\delta_{\rm cor}^{\theta^+}$& 
$\delta_{\rm cor}^{h^+}$& 
$\delta_{\rm cor}^{N^+}$& 
$\delta_{\rm cor}^{B^+}$ \\ 
$(\rm GeV^2)$ & & & 
$(\%)$ & $(\%)$ & $(\%)$ & $(\%)$ & $(\%)$ & $(\%)$ & 
$(\%)$ & $(\%)$ & $(\%)$ & $(\%)$ & $(\%)$  
\\ \hline 
$120$ & $0.0020$ & $1.334$ & $1.49$ & $0.65$ & $0.80$ & $0.43$ & $0.09$ & $1.08$ & $-0.26$ & $-0.65$ & $0.02$ & $0.18$ & $-0.80$ \\ 
 
$120$ & $0.0032$ & $1.194$ & $1.63$ & $0.94$ & $1.03$ & $0.73$ & $0.03$ & $0.84$ & $-0.46$ & $-0.67$ & $0.01$ & $0.18$ & $-0.10$ \\ 
 
\hline 
$150$ & $0.0032$ & $1.211$ & $1.25$ & $0.55$ & $0.80$ & $0.49$ & $0.06$ & $0.78$ & $-0.36$ & $-0.53$ & $0.01$ & $0.18$ & $-0.41$ \\ 
 
$150$ & $0.0050$ & $1.075$ & $1.53$ & $0.66$ & $1.09$ & $0.87$ & $0.00$ & $0.84$ & $-0.53$ & $-0.65$ & $0.00$ & $0.04$ & $-0.03$ \\ 
 
$150$ & $0.0080$ & $0.9323$ & $2.44$ & $0.91$ & $1.89$ & $1.53$ & $0.85$ & $1.25$ & $-1.00$ & $-0.56$ & $-0.24$ & $-0.44$ & $-0.05$ \\ 
 
$150$ & $0.0130$ & $0.7854$ & $3.75$ & $1.28$ & $2.72$ & $2.39$ & $1.01$ & $2.25$ & $-1.45$ & $-1.37$ & $-0.37$ & $-0.97$ & $-0.01$ \\ 
 
\hline 
$200$ & $0.0032$ & $1.218$ & $1.55$ & $1.04$ & $0.74$ & $0.13$ & $0.08$ & $0.88$ & $-0.11$ & $-0.63$ & $0.02$ & $0.18$ & $-0.58$ \\ 
 
$200$ & $0.0050$ & $1.085$ & $1.33$ & $0.72$ & $0.84$ & $0.53$ & $0.03$ & $0.74$ & $-0.40$ & $-0.61$ & $0.01$ & $0.12$ & $-0.07$ \\ 
 
$200$ & $0.0080$ & $0.9394$ & $1.75$ & $0.73$ & $1.24$ & $1.05$ & $0.00$ & $0.99$ & $-0.59$ & $-0.80$ & $0.00$ & $0.00$ & $-0.01$ \\ 
 
$200$ & $0.0130$ & $0.7867$ & $1.30$ & $0.83$ & $0.68$ & $0.02$ & $0.02$ & $0.73$ & $-0.09$ & $-0.40$ & $-0.03$ & $0.60$ & $-0.01$ \\ 
 
$200$ & $0.0200$ & $0.6805$ & $1.55$ & $0.91$ & $1.00$ & $0.72$ & $0.02$ & $0.75$ & $-0.41$ & $-0.49$ & $-0.09$ & $0.38$ & $-0.01$ \\ 
 
$200$ & $0.0320$ & $0.5756$ & $1.91$ & $1.05$ & $1.34$ & $0.99$ & $0.54$ & $0.85$ & $-0.61$ & $-0.55$ & $-0.17$ & $0.13$ & $0.00$ \\ 
 
$200$ & $0.0500$ & $0.5051$ & $2.79$ & $1.20$ & $1.83$ & $1.66$ & $0.04$ & $1.73$ & $-1.05$ & $-0.80$ & $-0.08$ & $1.11$ & $0.00$ \\ 
 
$200$ & $0.0800$ & $0.4300$ & $3.12$ & $1.32$ & $2.00$ & $1.80$ & $0.17$ & $1.99$ & $-1.18$ & $-0.74$ & $-0.15$ & $1.41$ & $0.00$ \\ 
 
$200$ & $0.1300$ & $0.3560$ & $3.20$ & $1.48$ & $2.25$ & $1.67$ & $0.96$ & $1.73$ & $-0.97$ & $-0.90$ & $-0.27$ & $-1.08$ & $0.00$ \\ 
 
$200$ & $0.1800$ & $0.3085$ & $4.00$ & $1.99$ & $2.88$ & $1.61$ & $1.90$ & $1.94$ & $-0.88$ & $-1.04$ & $-0.34$ & $-1.34$ & $0.00$ \\ 
 
\hline 
$250$ & $0.0050$ & $1.103$ & $1.27$ & $0.84$ & $0.72$ & $0.29$ & $0.06$ & $0.63$ & $-0.28$ & $-0.45$ & $0.01$ & $0.16$ & $-0.30$ \\ 
 
$250$ & $0.0080$ & $0.9437$ & $1.43$ & $0.82$ & $0.91$ & $0.61$ & $0.00$ & $0.73$ & $-0.53$ & $-0.50$ & $0.00$ & $0.03$ & $-0.02$ \\ 
 
$250$ & $0.0130$ & $0.7952$ & $1.83$ & $0.92$ & $1.15$ & $0.91$ & $0.21$ & $1.09$ & $0.48$ & $-0.60$ & $0.03$ & $0.77$ & $-0.02$ \\ 
 
$250$ & $0.0200$ & $0.6829$ & $1.92$ & $0.93$ & $1.29$ & $1.07$ & $0.19$ & $1.07$ & $0.49$ & $-0.66$ & $-0.02$ & $0.68$ & $-0.01$ \\ 
 
$250$ & $0.0320$ & $0.5697$ & $1.82$ & $0.98$ & $1.14$ & $0.90$ & $0.04$ & $1.03$ & $0.36$ & $-0.48$ & $0.01$ & $0.83$ & $0.00$ \\ 
 
$250$ & $0.0500$ & $0.4946$ & $2.21$ & $1.07$ & $1.15$ & $0.89$ & $0.10$ & $1.55$ & $0.39$ & $-0.57$ & $-0.09$ & $1.39$ & $0.00$ \\ 
 
$250$ & $0.0800$ & $0.4294$ & $2.67$ & $1.11$ & $0.98$ & $0.56$ & $0.20$ & $2.22$ & $0.31$ & $-0.51$ & $-0.05$ & $2.14$ & $0.00$ \\ 
 
$250$ & $0.1300$ & $0.3623$ & $2.31$ & $1.20$ & $1.72$ & $1.13$ & $0.68$ & $0.98$ & $0.51$ & $-0.57$ & $-0.23$ & $-0.56$ & $0.00$ \\ 
 
$250$ & $0.1800$ & $0.2930$ & $3.91$ & $1.62$ & $2.71$ & $1.73$ & $1.61$ & $2.31$ & $1.03$ & $-0.78$ & $-0.34$ & $-1.88$ & $0.00$ \\ 
 
\hline 
$300$ & $0.0050$ & $1.115$ & $1.79$ & $1.44$ & $0.81$ & $0.28$ & $0.07$ & $0.68$ & $-0.28$ & $-0.37$ & $0.02$ & $0.15$ & $-0.48$ \\ 
 
$300$ & $0.0080$ & $0.9575$ & $1.38$ & $0.96$ & $0.73$ & $0.30$ & $0.02$ & $0.67$ & $-0.30$ & $-0.59$ & $0.00$ & $0.10$ & $-0.05$ \\ 
 
$300$ & $0.0130$ & $0.7917$ & $1.59$ & $0.97$ & $0.93$ & $0.63$ & $0.00$ & $0.84$ & $-0.62$ & $-0.57$ & $0.00$ & $0.00$ & $-0.01$ \\ 
 
$300$ & $0.0200$ & $0.6989$ & $1.63$ & $1.07$ & $0.88$ & $0.53$ & $0.12$ & $0.86$ & $0.36$ & $-0.46$ & $-0.01$ & $0.64$ & $-0.02$ \\ 
 
$300$ & $0.0320$ & $0.5730$ & $1.83$ & $1.13$ & $1.09$ & $0.83$ & $0.09$ & $0.95$ & $0.43$ & $-0.55$ & $-0.02$ & $0.64$ & $-0.01$ \\ 
 
$300$ & $0.0500$ & $0.4888$ & $2.15$ & $1.23$ & $1.23$ & $0.98$ & $0.14$ & $1.26$ & $0.49$ & $-0.54$ & $-0.05$ & $1.03$ & $0.00$ \\ 
 
$300$ & $0.0800$ & $0.4330$ & $2.92$ & $1.26$ & $1.37$ & $1.08$ & $0.30$ & $2.25$ & $0.51$ & $-0.69$ & $-0.06$ & $2.08$ & $-0.01$ \\ 
 
$300$ & $0.1300$ & $0.3646$ & $2.58$ & $1.30$ & $1.86$ & $1.40$ & $0.51$ & $1.23$ & $0.68$ & $-0.73$ & $-0.20$ & $0.69$ & $0.00$ \\ 
 
$300$ & $0.1800$ & $0.3022$ & $4.42$ & $1.68$ & $2.89$ & $2.01$ & $1.61$ & $2.89$ & $1.07$ & $-0.68$ & $-0.30$ & $-2.58$ & $0.00$ \\ 
 
$300$ & $0.4000$ & $0.1479$ & $6.45$ & $2.13$ & $3.67$ & $2.40$ & $2.17$ & $4.86$ & $1.34$ & $-1.02$ & $-0.45$ & $-4.53$ & $0.00$ \\ 
 
\hline 
$400$ & $0.0080$ & $0.9823$ & $1.51$ & $1.19$ & $0.74$ & $0.26$ & $0.06$ & $0.58$ & $-0.27$ & $-0.44$ & $0.01$ & $0.14$ & $-0.22$ \\ 
 
$400$ & $0.0130$ & $0.8346$ & $1.54$ & $1.12$ & $0.82$ & $0.45$ & $0.01$ & $0.67$ & $-0.45$ & $-0.49$ & $0.00$ & $0.03$ & $-0.01$ \\ 
 
$400$ & $0.0200$ & $0.6979$ & $1.92$ & $1.17$ & $1.11$ & $0.86$ & $0.00$ & $1.04$ & $-0.86$ & $-0.58$ & $0.00$ & $0.00$ & $0.00$ \\ 
 
$400$ & $0.0320$ & $0.5861$ & $1.81$ & $1.26$ & $0.88$ & $0.49$ & $0.15$ & $0.95$ & $0.48$ & $-0.48$ & $-0.03$ & $0.66$ & $-0.01$ \\ 
 
$400$ & $0.0500$ & $0.4898$ & $1.99$ & $1.38$ & $1.04$ & $0.67$ & $0.32$ & $0.99$ & $0.64$ & $-0.49$ & $-0.09$ & $0.57$ & $0.00$ \\ 
 
$400$ & $0.0800$ & $0.4217$ & $2.63$ & $1.46$ & $0.90$ & $0.25$ & $0.38$ & $1.99$ & $0.22$ & $-0.30$ & $-0.02$ & $1.96$ & $0.00$ \\ 
 
$400$ & $0.1300$ & $0.3518$ & $2.27$ & $1.43$ & $1.35$ & $0.64$ & $0.40$ & $1.14$ & $0.59$ & $-0.47$ & $-0.16$ & $0.85$ & $0.00$ \\ 
 
$400$ & $0.1800$ & $0.3007$ & $4.28$ & $1.86$ & $2.40$ & $1.11$ & $1.68$ & $3.02$ & $1.04$ & $-0.56$ & $-0.38$ & $-2.75$ & $0.00$ \\ 
 
$400$ & $0.4000$ & $0.1464$ & $6.80$ & $2.35$ & $3.22$ & $1.37$ & $2.28$ & $5.51$ & $1.28$ & $-0.78$ & $-0.46$ & $-5.28$ & $0.00$ \\ 
 
\hline 
$500$ & $0.0080$ & $0.9650$ & $2.24$ & $1.98$ & $0.87$ & $0.31$ & $0.09$ & $0.57$ & $-0.15$ & $-0.42$ & $0.01$ & $0.16$ & $-0.32$ \\ 
 
$500$ & $0.0130$ & $0.8515$ & $1.67$ & $1.39$ & $0.76$ & $0.29$ & $0.02$ & $0.52$ & $-0.24$ & $-0.46$ & $0.01$ & $0.09$ & $-0.04$ \\ 
 
$500$ & $0.0200$ & $0.7287$ & $1.90$ & $1.40$ & $0.97$ & $0.65$ & $0.00$ & $0.83$ & $-0.64$ & $-0.54$ & $0.00$ & $0.00$ & $0.00$ \\ 
 
$500$ & $0.0320$ & $0.5984$ & $1.96$ & $1.48$ & $0.91$ & $0.41$ & $0.36$ & $0.90$ & $0.42$ & $-0.37$ & $0.07$ & $0.70$ & $-0.01$ \\ 
 
$500$ & $0.0500$ & $0.5193$ & $2.10$ & $1.53$ & $1.06$ & $0.68$ & $0.30$ & $0.97$ & $0.68$ & $-0.49$ & $-0.16$ & $0.46$ & $0.00$ \\ 
 
$500$ & $0.0800$ & $0.4325$ & $2.49$ & $1.65$ & $0.90$ & $0.34$ & $0.28$ & $1.63$ & $0.34$ & $-0.33$ & $0.06$ & $1.56$ & $0.00$ \\ 
 
$500$ & $0.1300$ & $0.3780$ & $2.83$ & $1.90$ & $1.30$ & $0.65$ & $0.04$ & $1.65$ & $0.65$ & $-0.53$ & $-0.07$ & $1.41$ & $0.00$ \\ 
 
$500$ & $0.1800$ & $0.3072$ & $2.91$ & $2.13$ & $1.73$ & $0.78$ & $0.86$ & $0.96$ & $0.78$ & $-0.42$ & $-0.22$ & $-0.30$ & $0.00$ \\ 
 
$500$ & $0.2500$ & $0.2451$ & $5.72$ & $2.63$ & $2.76$ & $1.22$ & $2.02$ & $4.27$ & $1.22$ & $-0.55$ & $-0.42$ & $-4.03$ & $0.00$ \\ 
 
\hline 
$650$ & $0.0130$ & $0.8403$ & $1.81$ & $1.58$ & $0.76$ & $0.18$ & $0.05$ & $0.43$ & $-0.16$ & $-0.37$ & $0.01$ & $0.12$ & $-0.11$ \\ 
 
$650$ & $0.0200$ & $0.7202$ & $2.02$ & $1.65$ & $0.95$ & $0.58$ & $0.00$ & $0.67$ & $-0.41$ & $-0.53$ & $0.00$ & $0.02$ & $-0.02$ \\ 
 
$650$ & $0.0320$ & $0.6338$ & $2.30$ & $1.70$ & $1.17$ & $0.87$ & $0.00$ & $1.01$ & $-0.82$ & $-0.58$ & $0.00$ & $0.00$ & $0.00$ \\ 
 
$650$ & $0.0500$ & $0.5105$ & $2.26$ & $1.80$ & $1.02$ & $0.63$ & $0.11$ & $0.92$ & $0.69$ & $-0.35$ & $0.06$ & $0.49$ & $0.00$ \\ 
 
$650$ & $0.0800$ & $0.4128$ & $2.67$ & $2.01$ & $0.97$ & $0.48$ & $0.11$ & $1.46$ & $0.50$ & $-0.50$ & $0.06$ & $1.28$ & $0.00$ \\ 
 
$650$ & $0.1300$ & $0.3681$ & $3.08$ & $2.31$ & $1.31$ & $0.59$ & $0.05$ & $1.57$ & $0.64$ & $-0.21$ & $-0.10$ & $1.41$ & $0.00$ \\ 
 
$650$ & $0.1800$ & $0.3309$ & $2.94$ & $2.33$ & $1.56$ & $0.64$ & $0.50$ & $0.89$ & $0.66$ & $-0.46$ & $-0.25$ & $0.28$ & $0.00$ \\ 
 
$650$ & $0.2500$ & $0.2360$ & $5.10$ & $2.97$ & $2.51$ & $1.30$ & $1.58$ & $3.29$ & $1.32$ & $-0.55$ & $-0.38$ & $-2.93$ & $0.00$ \\ 
 
$650$ & $0.4000$ & $0.1328$ & $7.39$ & $4.22$ & $3.65$ & $1.44$ & $2.61$ & $4.84$ & $1.46$ & $-0.58$ & $-0.59$ & $-4.54$ & $0.00$ \\ 
 
\hline 

\end{tabular} 
\end{center} 
\caption[RESULT] 
{\label{tab:ncdxdq2_posP0} The NC 
$e^+p$ reduced cross section $\tilde{\sigma}_{\rm NC}(x,Q^2)$ 
for $P_e=0$ 
with statistical 
$(\delta_{\rm stat})$, 
total $(\delta_{tot})$, 
total uncorrelated systematic $(\delta_{\rm unc})$ 
errors, two of its contributions from the 
 electron energy error ($\delta_{unc}^{E}$)  
and the hadronic energy error  
($\delta_{\rm unc}^{h}$). 
The effect of the other uncorrelated 
systematic errors is included in $\delta_{\rm unc}$. 
In addition the correlated systematic  
$(\delta_{\rm cor})$ and its contributions from a 
positive variation of one  
standard deviation of the 
electron energy error ($\delta_{cor}^{E^+}$), of 
the polar electron angle error 
($\delta_{\rm cor}^{\theta^+}$), of the hadronic 
energy error ($\delta_{\rm cor}^{h^+}$), of the error 
due to noise subtraction ($\delta_{\rm cor}^{N^+}$) 
and of the error due to background subtraction 
($\delta_{\rm cor}^{B^+}$) are given. 
The normalisation and polarisation uncertainties are 
not included in the errors. 
The table continues on the next page.}
\end{table} 
\begin{table}[htbp] 
\begin{center} 
\tiny 
\begin{tabular}{|r|c|r|r|r|r|r|r|r|r|r|r|r|r|} 
\hline 
$Q^2$ & $x$ & $\tilde{\sigma}_{\rm NC}$ & 
$\delta_{\rm tot}$ & $\delta_{\rm stat}$ & $\delta_{\rm unc}$ & 
$\delta_{\rm unc}^{E}$ & 
$\delta_{\rm unc}^{h}$& 
$\delta_{\rm cor}$ & 
$\delta_{\rm cor}^{E^+}$ & 
$\delta_{\rm cor}^{\theta^+}$& 
$\delta_{\rm cor}^{h^+}$& 
$\delta_{\rm cor}^{N^+}$& 
$\delta_{\rm cor}^{B^+}$ \\ 
$(\rm GeV^2)$ & & & 
$(\%)$ & $(\%)$ & $(\%)$ & $(\%)$ & $(\%)$ & $(\%)$ & 
$(\%)$ & $(\%)$ & $(\%)$ & $(\%)$ & $(\%)$  
\\ \hline 
$800$ & $0.0130$ & $0.8056$ & $2.83$ & $2.60$ & $0.96$ & $0.17$ & $0.08$ & $0.53$ & $-0.29$ & $-0.36$ & $0.03$ & $0.14$ & $-0.21$ \\ 
 
$800$ & $0.0200$ & $0.6896$ & $2.14$ & $1.89$ & $0.89$ & $0.43$ & $0.02$ & $0.46$ & $-0.29$ & $-0.35$ & $0.00$ & $0.08$ & $-0.05$ \\ 
 
$800$ & $0.0320$ & $0.6051$ & $2.38$ & $1.99$ & $1.11$ & $0.73$ & $0.00$ & $0.70$ & $-0.51$ & $-0.48$ & $0.00$ & $0.00$ & $0.00$ \\ 
 
$800$ & $0.0500$ & $0.5182$ & $2.44$ & $2.13$ & $0.91$ & $0.23$ & $0.22$ & $0.76$ & $0.37$ & $-0.29$ & $0.13$ & $0.58$ & $0.00$ \\ 
 
$800$ & $0.0800$ & $0.4481$ & $2.74$ & $2.26$ & $1.07$ & $0.48$ & $0.28$ & $1.13$ & $0.73$ & $-0.43$ & $-0.12$ & $0.73$ & $0.00$ \\ 
 
$800$ & $0.1300$ & $0.3511$ & $3.39$ & $2.66$ & $1.33$ & $0.44$ & $0.19$ & $1.63$ & $0.63$ & $-0.52$ & $-0.09$ & $1.40$ & $0.00$ \\ 
 
$800$ & $0.1800$ & $0.3203$ & $3.32$ & $2.76$ & $1.59$ & $0.58$ & $0.53$ & $0.92$ & $0.70$ & $-0.47$ & $-0.14$ & $0.33$ & $0.00$ \\ 
 
$800$ & $0.2500$ & $0.2364$ & $4.48$ & $3.27$ & $2.26$ & $0.96$ & $1.33$ & $2.07$ & $1.04$ & $-0.44$ & $-0.43$ & $-1.68$ & $0.00$ \\ 
 
$800$ & $0.4000$ & $0.1200$ & $8.56$ & $4.81$ & $4.02$ & $2.02$ & $2.66$ & $5.83$ & $1.97$ & $-0.56$ & $-0.63$ & $-5.42$ & $0.00$ \\ 
 
\hline 
$1000$ & $0.0130$ & $0.7857$ & $3.27$ & $2.66$ & $1.45$ & $0.14$ & $0.24$ & $1.24$ & $-0.10$ & $-0.29$ & $0.04$ & $0.21$ & $-1.18$ \\ 
 
$1000$ & $0.0200$ & $0.7184$ & $2.41$ & $2.20$ & $0.85$ & $0.31$ & $0.05$ & $0.48$ & $-0.22$ & $-0.39$ & $0.01$ & $0.13$ & $-0.12$ \\ 
 
$1000$ & $0.0320$ & $0.6057$ & $2.48$ & $2.18$ & $1.02$ & $0.65$ & $0.00$ & $0.57$ & $-0.41$ & $-0.39$ & $0.00$ & $0.01$ & $-0.01$ \\ 
 
$1000$ & $0.0500$ & $0.5105$ & $2.90$ & $2.34$ & $1.49$ & $1.24$ & $0.00$ & $0.84$ & $-0.69$ & $-0.48$ & $0.00$ & $0.00$ & $0.00$ \\ 
 
$1000$ & $0.0800$ & $0.4320$ & $2.84$ & $2.56$ & $1.04$ & $0.50$ & $0.29$ & $0.64$ & $0.38$ & $-0.22$ & $-0.13$ & $0.44$ & $-0.01$ \\ 
 
$1000$ & $0.1300$ & $0.3352$ & $3.73$ & $3.27$ & $1.26$ & $0.30$ & $0.22$ & $1.27$ & $0.30$ & $0.14$ & $0.09$ & $1.23$ & $0.00$ \\ 
 
$1000$ & $0.1800$ & $0.2953$ & $3.66$ & $3.26$ & $1.44$ & $0.33$ & $0.39$ & $0.84$ & $0.46$ & $-0.33$ & $-0.20$ & $0.59$ & $0.00$ \\ 
 
$1000$ & $0.2500$ & $0.2407$ & $4.38$ & $3.61$ & $2.01$ & $0.83$ & $1.05$ & $1.46$ & $0.98$ & $-0.43$ & $-0.21$ & $-0.98$ & $0.00$ \\ 
 
$1000$ & $0.4000$ & $0.1273$ & $8.40$ & $5.95$ & $3.66$ & $1.38$ & $2.63$ & $4.66$ & $1.40$ & $-0.41$ & $-0.63$ & $-4.37$ & $0.00$ \\ 
 
\hline 
$1200$ & $0.0130$ & $0.8194$ & $5.30$ & $4.39$ & $2.11$ & $0.13$ & $0.30$ & $2.09$ & $-0.14$ & $-0.22$ & $0.06$ & $0.23$ & $-2.06$ \\ 
 
$1200$ & $0.0200$ & $0.6982$ & $2.98$ & $2.78$ & $0.98$ & $0.13$ & $0.08$ & $0.40$ & $0.08$ & $-0.29$ & $0.02$ & $0.12$ & $-0.23$ \\ 
 
$1200$ & $0.0320$ & $0.5986$ & $2.74$ & $2.56$ & $0.85$ & $0.43$ & $0.02$ & $0.46$ & $-0.24$ & $-0.39$ & $0.00$ & $0.06$ & $-0.01$ \\ 
 
$1200$ & $0.0500$ & $0.5067$ & $3.00$ & $2.63$ & $1.24$ & $0.98$ & $0.00$ & $0.74$ & $-0.57$ & $-0.47$ & $0.00$ & $0.00$ & $-0.01$ \\ 
 
$1200$ & $0.0800$ & $0.4254$ & $3.16$ & $2.84$ & $1.14$ & $0.80$ & $0.11$ & $0.78$ & $0.56$ & $-0.30$ & $-0.09$ & $0.45$ & $0.00$ \\ 
 
$1200$ & $0.1300$ & $0.3382$ & $3.90$ & $3.47$ & $1.52$ & $1.01$ & $0.13$ & $0.91$ & $0.59$ & $-0.29$ & $-0.05$ & $0.63$ & $0.00$ \\ 
 
$1200$ & $0.1800$ & $0.3154$ & $4.28$ & $3.83$ & $1.45$ & $0.71$ & $0.05$ & $1.22$ & $0.53$ & $-0.30$ & $-0.14$ & $1.04$ & $-0.01$ \\ 
 
$1200$ & $0.2500$ & $0.2462$ & $4.77$ & $4.12$ & $2.18$ & $1.38$ & $0.97$ & $1.00$ & $0.83$ & $-0.28$ & $-0.28$ & $-0.41$ & $0.00$ \\ 
 
$1200$ & $0.4000$ & $0.1067$ & $8.83$ & $5.68$ & $4.30$ & $2.61$ & $2.74$ & $5.22$ & $1.66$ & $-0.52$ & $-0.57$ & $-4.88$ & $0.00$ \\ 
 
\hline 
$1500$ & $0.0200$ & $0.6717$ & $3.95$ & $3.49$ & $1.62$ & $0.13$ & $0.21$ & $0.89$ & $0.23$ & $-0.27$ & $0.05$ & $0.20$ & $-0.79$ \\ 
 
$1500$ & $0.0320$ & $0.5863$ & $3.46$ & $3.33$ & $0.85$ & $0.25$ & $0.05$ & $0.45$ & $-0.25$ & $-0.36$ & $0.02$ & $0.11$ & $-0.03$ \\ 
 
$1500$ & $0.0500$ & $0.5213$ & $3.21$ & $3.00$ & $1.00$ & $0.62$ & $0.00$ & $0.51$ & $-0.29$ & $-0.42$ & $0.00$ & $0.02$ & $0.00$ \\ 
 
$1500$ & $0.0800$ & $0.4430$ & $3.61$ & $3.31$ & $1.25$ & $0.94$ & $0.04$ & $0.74$ & $0.59$ & $-0.19$ & $-0.10$ & $0.38$ & $0.00$ \\ 
 
$1500$ & $0.1300$ & $0.3287$ & $4.36$ & $4.01$ & $1.49$ & $0.86$ & $0.06$ & $0.85$ & $0.51$ & $-0.25$ & $0.06$ & $0.64$ & $0.00$ \\ 
 
$1500$ & $0.1800$ & $0.2974$ & $4.50$ & $4.10$ & $1.53$ & $0.80$ & $0.04$ & $1.07$ & $0.53$ & $-0.20$ & $0.05$ & $0.91$ & $0.00$ \\ 
 
$1500$ & $0.2500$ & $0.2276$ & $5.16$ & $4.46$ & $2.38$ & $1.56$ & $1.11$ & $1.06$ & $0.94$ & $-0.19$ & $-0.37$ & $-0.26$ & $0.00$ \\ 
 
$1500$ & $0.4000$ & $0.1214$ & $7.74$ & $5.75$ & $3.63$ & $2.15$ & $2.05$ & $3.70$ & $1.21$ & $-0.24$ & $-0.51$ & $-3.45$ & $0.00$ \\ 
 
$1500$ & $0.6500$ & $0.01480$ & $16.82$ & $10.87$ & $7.13$ & $4.51$ & $4.88$ & $10.67$ & $2.81$ & $-0.32$ & $-1.00$ & $-10.24$ & $0.00$ \\ 
 
\hline 
$2000$ & $0.0219$ & $0.6850$ & $6.33$ & $5.63$ & $2.40$ & $0.55$ & $0.21$ & $1.62$ & $0.24$ & $-0.14$ & $0.05$ & $0.20$ & $-1.58$ \\ 
 
$2000$ & $0.0320$ & $0.5398$ & $4.19$ & $3.94$ & $1.33$ & $0.22$ & $0.09$ & $0.50$ & $-0.17$ & $-0.41$ & $0.02$ & $0.11$ & $-0.20$ \\ 
 
$2000$ & $0.0500$ & $0.5250$ & $3.82$ & $3.67$ & $1.01$ & $0.42$ & $0.01$ & $0.37$ & $-0.25$ & $-0.26$ & $0.00$ & $0.04$ & $0.00$ \\ 
 
$2000$ & $0.0800$ & $0.4199$ & $4.01$ & $3.73$ & $1.28$ & $0.89$ & $0.00$ & $0.75$ & $-0.46$ & $-0.60$ & $0.00$ & $0.00$ & $0.00$ \\ 
 
$2000$ & $0.1300$ & $0.3413$ & $5.26$ & $4.96$ & $1.60$ & $1.00$ & $0.10$ & $0.73$ & $0.47$ & $-0.21$ & $-0.09$ & $0.51$ & $0.00$ \\ 
 
$2000$ & $0.1800$ & $0.2973$ & $5.19$ & $4.80$ & $1.50$ & $0.58$ & $0.25$ & $1.26$ & $0.43$ & $-0.28$ & $0.19$ & $1.14$ & $0.00$ \\ 
 
$2000$ & $0.2500$ & $0.2331$ & $5.64$ & $4.98$ & $2.39$ & $1.66$ & $0.91$ & $1.14$ & $1.07$ & $-0.29$ & $-0.23$ & $-0.01$ & $0.00$ \\ 
 
$2000$ & $0.4000$ & $0.1250$ & $8.10$ & $6.30$ & $3.94$ & $2.57$ & $2.09$ & $3.23$ & $1.61$ & $-0.27$ & $-0.62$ & $-2.71$ & $0.00$ \\ 
 
$2000$ & $0.6500$ & $0.01368$ & $18.37$ & $12.83$ & $7.46$ & $4.72$ & $5.11$ & $10.82$ & $2.84$ & $-0.28$ & $-1.06$ & $-10.38$ & $0.00$ \\ 
 
\hline 
$3000$ & $0.0320$ & $0.5847$ & $4.22$ & $3.74$ & $1.72$ & $0.20$ & $0.13$ & $0.93$ & $0.05$ & $-0.27$ & $0.03$ & $0.13$ & $-0.88$ \\ 
 
$3000$ & $0.0500$ & $0.4979$ & $3.48$ & $3.23$ & $1.26$ & $0.22$ & $0.07$ & $0.35$ & $-0.07$ & $-0.33$ & $0.02$ & $0.09$ & $-0.06$ \\ 
 
$3000$ & $0.0800$ & $0.4309$ & $3.65$ & $3.42$ & $1.15$ & $0.52$ & $0.01$ & $0.56$ & $-0.33$ & $-0.45$ & $0.00$ & $0.02$ & $-0.01$ \\ 
 
$3000$ & $0.1300$ & $0.3349$ & $4.49$ & $4.18$ & $1.50$ & $0.73$ & $0.00$ & $0.65$ & $-0.50$ & $-0.42$ & $0.00$ & $0.00$ & $0.00$ \\ 
 
$3000$ & $0.1800$ & $0.2973$ & $5.61$ & $5.17$ & $1.95$ & $1.41$ & $0.05$ & $0.95$ & $0.78$ & $-0.15$ & $-0.04$ & $0.53$ & $0.00$ \\ 
 
$3000$ & $0.2500$ & $0.2221$ & $5.14$ & $4.66$ & $1.95$ & $1.32$ & $0.20$ & $0.97$ & $0.76$ & $-0.10$ & $-0.10$ & $0.58$ & $0.00$ \\ 
 
$3000$ & $0.4000$ & $0.1191$ & $7.27$ & $5.62$ & $3.86$ & $2.62$ & $1.91$ & $2.54$ & $1.51$ & $-0.10$ & $-0.50$ & $-1.98$ & $0.00$ \\ 
 
$3000$ & $0.6500$ & $0.01256$ & $16.63$ & $11.50$ & $7.93$ & $5.27$ & $5.35$ & $9.02$ & $3.30$ & $-0.26$ & $-1.39$ & $-8.28$ & $0.00$ \\ 
 
\hline 
$5000$ & $0.0547$ & $0.4615$ & $5.59$ & $5.08$ & $2.01$ & $0.52$ & $0.11$ & $1.18$ & $0.16$ & $-0.33$ & $0.05$ & $0.10$ & $-1.12$ \\ 
 
$5000$ & $0.0800$ & $0.4038$ & $4.20$ & $3.93$ & $1.44$ & $0.15$ & $0.06$ & $0.38$ & $-0.11$ & $-0.33$ & $0.02$ & $0.07$ & $-0.15$ \\ 
 
$5000$ & $0.1300$ & $0.3351$ & $4.96$ & $4.63$ & $1.68$ & $0.36$ & $0.00$ & $0.54$ & $0.27$ & $-0.46$ & $0.00$ & $0.02$ & $0.00$ \\ 
 
$5000$ & $0.1800$ & $0.2902$ & $5.45$ & $5.14$ & $1.75$ & $0.49$ & $0.00$ & $0.50$ & $-0.31$ & $-0.39$ & $0.00$ & $0.00$ & $0.00$ \\ 
 
$5000$ & $0.2500$ & $0.2126$ & $6.47$ & $6.13$ & $1.97$ & $0.71$ & $0.00$ & $0.55$ & $0.54$ & $-0.08$ & $0.00$ & $0.00$ & $0.00$ \\ 
 
$5000$ & $0.4000$ & $0.1030$ & $8.32$ & $7.49$ & $3.32$ & $1.90$ & $1.42$ & $1.45$ & $1.19$ & $0.05$ & $-0.43$ & $-0.69$ & $0.00$ \\ 
 
$5000$ & $0.6500$ & $0.009650$ & $17.88$ & $15.48$ & $6.50$ & $3.98$ & $4.26$ & $6.14$ & $2.06$ & $-0.07$ & $-1.19$ & $-5.66$ & $0.00$ \\ 
 
\hline 
$8000$ & $0.0875$ & $0.3440$ & $9.35$ & $8.73$ & $2.69$ & $0.33$ & $0.11$ & $2.01$ & $0.13$ & $-0.24$ & $0.05$ & $0.09$ & $-1.99$ \\ 
 
$8000$ & $0.1300$ & $0.2846$ & $7.38$ & $7.00$ & $2.29$ & $0.34$ & $0.06$ & $0.45$ & $-0.20$ & $-0.33$ & $0.02$ & $0.07$ & $-0.23$ \\ 
 
$8000$ & $0.1800$ & $0.2586$ & $7.39$ & $7.05$ & $2.17$ & $0.29$ & $0.01$ & $0.48$ & $0.18$ & $-0.44$ & $0.00$ & $0.02$ & $0.00$ \\ 
 
$8000$ & $0.2500$ & $0.2164$ & $9.14$ & $8.83$ & $2.30$ & $0.56$ & $0.00$ & $0.49$ & $0.43$ & $-0.25$ & $0.00$ & $0.00$ & $0.00$ \\ 
 
$8000$ & $0.4000$ & $0.09858$ & $11.16$ & $10.12$ & $4.30$ & $3.21$ & $0.00$ & $1.89$ & $1.84$ & $0.42$ & $0.00$ & $0.00$ & $0.00$ \\ 
 
$8000$ & $0.6500$ & $0.01341$ & $17.48$ & $15.28$ & $7.24$ & $4.89$ & $4.15$ & $4.42$ & $2.67$ & $0.25$ & $-1.14$ & $-3.32$ & $0.00$ \\ 
 
\hline 
$12000$ & $0.1300$ & $0.2077$ & $20.59$ & $19.89$ & $5.03$ & $0.64$ & $0.10$ & $1.66$ & $-0.76$ & $-0.46$ & $0.03$ & $0.09$ & $-1.40$ \\ 
 
$12000$ & $0.1800$ & $0.2120$ & $11.71$ & $11.41$ & $2.45$ & $0.63$ & $0.07$ & $0.93$ & $0.31$ & $-0.46$ & $0.02$ & $0.08$ & $-0.75$ \\ 
 
$12000$ & $0.2500$ & $0.1536$ & $12.11$ & $11.86$ & $2.40$ & $0.88$ & $0.02$ & $0.52$ & $0.41$ & $-0.32$ & $0.00$ & $0.04$ & $0.00$ \\ 
 
$12000$ & $0.4000$ & $0.09789$ & $14.69$ & $13.91$ & $4.47$ & $3.53$ & $0.00$ & $1.56$ & $1.51$ & $0.38$ & $0.00$ & $0.00$ & $0.00$ \\ 
 
$12000$ & $0.6500$ & $0.01541$ & $22.64$ & $21.37$ & $6.72$ & $5.23$ & $3.04$ & $3.30$ & $2.78$ & $0.26$ & $-0.75$ & $-1.59$ & $0.00$ \\ 
 
\hline 
$20000$ & $0.2500$ & $0.1267$ & $20.22$ & $19.92$ & $3.32$ & $2.22$ & $0.09$ & $0.90$ & $0.82$ & $0.22$ & $0.04$ & $0.10$ & $-0.27$ \\ 
 
$20000$ & $0.4000$ & $0.09396$ & $19.49$ & $18.99$ & $4.30$ & $3.27$ & $0.02$ & $0.93$ & $0.92$ & $-0.07$ & $0.01$ & $0.03$ & $0.00$ \\ 
 
$20000$ & $0.6500$ & $0.009430$ & $47.59$ & $44.86$ & $15.10$ & $14.72$ & $0.00$ & $4.96$ & $4.41$ & $2.28$ & $0.00$ & $0.00$ & $0.00$ \\ 
 
\hline 
$30000$ & $0.4000$ & $0.05285$ & $46.67$ & $46.39$ & $5.08$ & $3.43$ & $0.07$ & $1.04$ & $0.99$ & $-0.21$ & $0.01$ & $0.06$ & $-0.24$ \\ 
 
$30000$ & $0.6500$ & $0.007890$ & $71.87$ & $71.13$ & $9.87$ & $9.04$ & $0.00$ & $3.00$ & $2.73$ & $1.24$ & $0.00$ & $0.00$ & $0.00$ \\ 
 
\hline 
\end{tabular} 
\end{center} 
\captcont{continued.}
\end{table} 

\begin{table}[htbp] 
\begin{center} 
\tiny 
\renewcommand{\arraystretch}{1.22}
\begin{tabular}{|r|c|c|l|r|r|r|r|r|r|r|r|r|} 
\hline 
$Q^2$  &$x$ & $y$ & ${\rm d}^2\sigma_{\rm CC}/{\rm d}x{\rm d}Q^2$ & 
$\delta_{\rm tot}$ & $\delta_{\rm stat}$ & $\delta_{\rm unc}$ & 
$\delta_{\rm unc}^{h}$& 
$\delta_{\rm cor}$ & 
$\delta_{\rm cor}^{V^+}$ & 
$\delta_{\rm cor}^{h^+}$& 
$\delta_{\rm cor}^{N^+}$& 
$\delta_{\rm cor}^{B^+}$ \\ 
$(\rm GeV^2)$ & & & $(\rm pb/GeV^2)$ & 
$(\%)$ & $(\%)$ & $(\%)$ & $(\%)$ & $(\%)$ & 
$(\%)$ & $(\%)$ & $(\%)$ & $(\%)$  
\\ \hline 
$300$ & $0.008$ & $0.369$ & $1.64$ & $44.7$ & $33.8$ & $23.9$ & $1.6$ & $16.8$ & $15.5$ & $-0.3$ & $-1.0$ & $-4.9$ \\ 
 
$300$ & $0.013$ & $0.227$ & $0.720$ & $19.9$ & $13.4$ & $10.9$ & $2.6$ & $9.9$ & $9.4$ & $-0.8$ & $-0.3$ & $-1.4$ \\ 
 
$300$ & $0.032$ & $0.092$ & $0.234$ & $14.5$ & $12.4$ & $5.2$ & $1.9$ & $5.4$ & $4.7$ & $-0.4$ & $0.5$ & $-2.2$ \\ 
 
$300$ & $0.080$ & $0.037$ & $0.640 \cdot 10^{-1}$ & $13.9$ & $11.9$ & $5.1$ & $2.6$ & $5.0$ & $1.4$ & $-0.4$ & $-4.0$ & $-2.0$ \\ 
 
\hline 
$500$ & $0.013$ & $0.379$ & $0.629$ & $14.2$ & $8.8$ & $7.1$ & $2.3$ & $8.6$ & $8.5$ & $-0.5$ & $-0.7$ & $-0.6$ \\ 
 
$500$ & $0.032$ & $0.154$ & $0.202$ & $8.6$ & $7.2$ & $3.3$ & $1.7$ & $3.2$ & $3.0$ & $-0.6$ & $0.1$ & $-0.3$ \\ 
 
$500$ & $0.080$ & $0.062$ & $0.508 \cdot 10^{-1}$ & $9.4$ & $8.3$ & $4.2$ & $2.3$ & $1.5$ & $0.8$ & $-0.6$ & $-0.2$ & $-0.3$ \\ 
 
$500$ & $0.130$ & $0.038$ & $0.261 \cdot 10^{-1}$ & $24.0$ & $19.7$ & $5.4$ & $0.7$ & $12.6$ & $0.1$ & $0.6$ & $-12.3$ & $0.0$ \\ 
 
\hline 
$1000$ & $0.013$ & $0.757$ & $0.390$ & $13.3$ & $9.1$ & $5.8$ & $1.8$ & $7.9$ & $7.8$ & $-0.8$ & $0.0$ & $-0.4$ \\ 
 
$1000$ & $0.032$ & $0.308$ & $0.182$ & $6.6$ & $5.6$ & $2.8$ & $1.7$ & $2.1$ & $1.9$ & $-0.4$ & $0.3$ & $-0.1$ \\ 
 
$1000$ & $0.080$ & $0.123$ & $0.588 \cdot 10^{-1}$ & $6.5$ & $5.7$ & $2.9$ & $0.9$ & $1.4$ & $0.4$ & $-0.3$ & $0.9$ & $-0.1$ \\ 
 
$1000$ & $0.130$ & $0.076$ & $0.255 \cdot 10^{-1}$ & $11.5$ & $10.1$ & $3.8$ & $1.5$ & $3.8$ & $0.0$ & $-0.1$ & $-3.6$ & $-0.1$ \\ 
 
\hline 
$2000$ & $0.032$ & $0.615$ & $0.116$ & $6.3$ & $5.3$ & $2.6$ & $1.1$ & $2.1$ & $1.9$ & $-0.4$ & $-0.3$ & $-0.1$ \\ 
 
$2000$ & $0.080$ & $0.246$ & $0.438 \cdot 10^{-1}$ & $5.4$ & $4.8$ & $2.1$ & $0.6$ & $1.0$ & $0.1$ & $-0.1$ & $0.7$ & $0.0$ \\ 
 
$2000$ & $0.130$ & $0.152$ & $0.225 \cdot 10^{-1}$ & $7.5$ & $6.8$ & $3.0$ & $0.9$ & $1.1$ & $-0.0$ & $-0.4$ & $-0.3$ & $0.0$ \\ 
 
$2000$ & $0.250$ & $0.079$ & $0.841 \cdot 10^{-2}$ & $16.3$ & $13.0$ & $3.9$ & $0.5$ & $9.0$ & $0.0$ & $0.3$ & $-8.8$ & $0.0$ \\ 
 
\hline 
$3000$ & $0.080$ & $0.369$ & $0.323 \cdot 10^{-1}$ & $5.2$ & $4.6$ & $2.0$ & $0.2$ & $1.1$ & $-0.1$ & $0.1$ & $0.9$ & $0.0$ \\ 
 
$3000$ & $0.130$ & $0.227$ & $0.177 \cdot 10^{-1}$ & $6.1$ & $5.6$ & $2.1$ & $0.3$ & $1.1$ & $-0.0$ & $-0.1$ & $0.8$ & $0.0$ \\ 
 
$3000$ & $0.250$ & $0.118$ & $0.665 \cdot 10^{-2}$ & $9.1$ & $8.3$ & $2.9$ & $0.6$ & $2.3$ & $0.0$ & $0.1$ & $-2.0$ & $-0.2$ \\ 
 
\hline 
$5000$ & $0.080$ & $0.615$ & $0.208 \cdot 10^{-1}$ & $6.6$ & $5.9$ & $2.4$ & $0.9$ & $1.4$ & $0.1$ & $0.2$ & $1.1$ & $0.0$ \\ 
 
$5000$ & $0.130$ & $0.379$ & $0.125 \cdot 10^{-1}$ & $5.8$ & $5.2$ & $2.1$ & $0.2$ & $1.1$ & $-0.0$ & $0.1$ & $0.8$ & $0.0$ \\ 
 
$5000$ & $0.250$ & $0.197$ & $0.471 \cdot 10^{-2}$ & $7.1$ & $6.4$ & $2.6$ & $1.5$ & $1.1$ & $-0.0$ & $0.4$ & $0.3$ & $-0.1$ \\ 
 
$5000$ & $0.400$ & $0.123$ & $0.142 \cdot 10^{-2}$ & $19.8$ & $17.4$ & $5.4$ & $4.7$ & $7.7$ & $0.0$ & $1.0$ & $-7.2$ & $0.0$ \\ 
 
\hline 
$8000$ & $0.130$ & $0.606$ & $0.834 \cdot 10^{-2}$ & $7.7$ & $6.4$ & $3.5$ & $2.6$ & $2.1$ & $-0.1$ & $0.9$ & $1.6$ & $0.0$ \\ 
 
$8000$ & $0.250$ & $0.315$ & $0.256 \cdot 10^{-2}$ & $7.5$ & $6.6$ & $3.0$ & $2.1$ & $1.6$ & $-0.0$ & $0.6$ & $1.2$ & $0.0$ \\ 
 
$8000$ & $0.400$ & $0.197$ & $0.991 \cdot 10^{-3}$ & $13.7$ & $12.3$ & $5.5$ & $5.0$ & $2.4$ & $0.0$ & $1.4$ & $-0.8$ & $0.0$ \\ 
 
\hline 
$15000$ & $0.250$ & $0.591$ & $0.153 \cdot 10^{-2}$ & $9.8$ & $7.6$ & $5.5$ & $4.8$ & $2.7$ & $-0.0$ & $1.3$ & $2.0$ & $0.0$ \\ 
 
$15000$ & $0.400$ & $0.369$ & $0.388 \cdot 10^{-3}$ & $12.1$ & $10.0$ & $6.1$ & $5.6$ & $2.7$ & $0.0$ & $1.7$ & $1.4$ & $0.0$ \\ 
 
\hline 
$30000$ & $0.400$ & $0.738$ & $0.149 \cdot 10^{-3}$ & $19.5$ & $15.6$ & $10.2$ & $9.7$ & $5.4$ & $0.0$ & $3.1$ & $3.8$ & $0.0$ \\ 
 
\hline 

\end{tabular} 
\end{center} 
\caption[RESULT] 
{\label{tab:ccdxdq2_eleP0} The CC 
$e^-p$ cross section ${\rm d}^2\sigma_{\rm CC}/{\rm d}x{\rm d}Q^2$ for lepton beam polarisation $P_e=0$ 
with statistical 
$(\delta_{\rm stat})$, 
total $(\delta_{\rm tot})$, 
total uncorrelated systematic $(\delta_{\rm unc})$ 
errors and one of its contributions from 
the hadronic energy error  
($\delta_{\rm unc}^{h}$). 
The effect of the other uncorrelated 
systematic errors is included in $\delta_{\rm unc}$. 
In addition the correlated systematic  
$(\delta_{\rm cor})$ and its contributions from a 
positive variation of one  
standard deviation of the 
cuts against photoproduction ($\delta_{\rm cor}^{V^+}$), of 
the hadronic 
energy error ($\delta_{\rm cor}^{h^+}$), of the error 
due to noise subtraction ($\delta_{\rm cor}^{N^+}$) 
and of the error due to background subtraction 
($\delta_{\rm cor}^{B^+}$) are given. 
The normalisation and polarisation uncertainties are not included in the errors. 
}
\end{table} 

\begin{table}[htbp] 
\begin{center} 
\tiny 
\renewcommand{\arraystretch}{1.22}
\begin{tabular}{|r|c|c|l|r|r|r|r|r|r|r|r|r|} 
\hline 
$Q^2$  &$x$ & $y$ & ${\rm d}^2\sigma_{\rm CC}/{\rm d}x{\rm d}Q^2$ & 
$\delta_{\rm tot}$ & $\delta_{\rm stat}$ & $\delta_{\rm unc}$ & 
$\delta_{\rm unc}^{h}$& 
$\delta_{\rm cor}$ & 
$\delta_{\rm cor}^{V^+}$ & 
$\delta_{\rm cor}^{h^+}$& 
$\delta_{\rm cor}^{N^+}$& 
$\delta_{\rm cor}^{B^+}$ \\ 
$(\rm GeV^2)$ & & & $(\rm pb/GeV^2)$ & 
$(\%)$ & $(\%)$ & $(\%)$ & $(\%)$ & $(\%)$ & 
$(\%)$ & $(\%)$ & $(\%)$ & $(\%)$  
\\ \hline 
$300$ & $0.008$ & $0.369$ & $0.938$ & $44.0$ & $30.7$ & $26.0$ & $2.5$ & $17.9$ & $16.0$ & $-0.8$ & $-0.6$ & $-5.9$ \\ 
 
$300$ & $0.013$ & $0.227$ & $0.499$ & $23.0$ & $16.6$ & $11.3$ & $1.8$ & $11.1$ & $10.2$ & $-0.5$ & $-0.3$ & $-3.5$ \\ 
 
$300$ & $0.032$ & $0.092$ & $0.192$ & $13.5$ & $10.7$ & $5.3$ & $1.7$ & $6.4$ & $5.3$ & $-0.4$ & $1.3$ & $-3.1$ \\ 
 
$300$ & $0.080$ & $0.037$ & $0.393 \cdot 10^{-1}$ & $15.9$ & $14.4$ & $5.1$ & $2.3$ & $4.4$ & $1.9$ & $-0.3$ & $-3.2$ & $-1.9$ \\ 
 
\hline 
$500$ & $0.013$ & $0.379$ & $0.486$ & $15.4$ & $10.0$ & $7.1$ & $1.9$ & $9.4$ & $9.2$ & $-0.6$ & $-0.1$ & $-0.9$ \\ 
 
$500$ & $0.032$ & $0.154$ & $0.182$ & $8.9$ & $7.4$ & $3.2$ & $1.5$ & $3.6$ & $3.4$ & $-0.5$ & $0.5$ & $-0.6$ \\ 
 
$500$ & $0.080$ & $0.062$ & $0.493 \cdot 10^{-1}$ & $9.0$ & $8.1$ & $3.6$ & $0.9$ & $1.4$ & $0.9$ & $-0.2$ & $0.3$ & $-0.4$ \\ 
 
$500$ & $0.130$ & $0.038$ & $0.199 \cdot 10^{-1}$ & $24.4$ & $20.6$ & $5.7$ & $0.7$ & $11.7$ & $0.2$ & $-0.7$ & $-11.5$ & $0.0$ \\ 
 
\hline 
$1000$ & $0.013$ & $0.757$ & $0.316$ & $15.3$ & $10.3$ & $6.1$ & $0.9$ & $9.5$ & $9.4$ & $-0.5$ & $0.5$ & $-0.6$ \\ 
 
$1000$ & $0.032$ & $0.308$ & $0.149$ & $6.8$ & $5.9$ & $2.3$ & $0.9$ & $2.2$ & $2.0$ & $-0.3$ & $0.6$ & $-0.1$ \\ 
 
$1000$ & $0.080$ & $0.123$ & $0.366 \cdot 10^{-1}$ & $7.5$ & $6.8$ & $2.8$ & $0.8$ & $1.2$ & $0.4$ & $-0.3$ & $0.8$ & $-0.1$ \\ 
 
$1000$ & $0.130$ & $0.076$ & $0.177 \cdot 10^{-1}$ & $11.9$ & $10.9$ & $3.7$ & $0.9$ & $2.6$ & $0.1$ & $-0.3$ & $-2.3$ & $0.0$ \\ 
 
\hline 
$2000$ & $0.032$ & $0.615$ & $0.793 \cdot 10^{-1}$ & $7.0$ & $6.1$ & $2.4$ & $0.3$ & $2.5$ & $2.2$ & $-0.2$ & $0.8$ & $-0.1$ \\ 
 
$2000$ & $0.080$ & $0.246$ & $0.269 \cdot 10^{-1}$ & $6.2$ & $5.8$ & $2.0$ & $0.1$ & $1.2$ & $0.1$ & $-0.0$ & $0.9$ & $-0.1$ \\ 
 
$2000$ & $0.130$ & $0.152$ & $0.120 \cdot 10^{-1}$ & $9.2$ & $8.6$ & $3.0$ & $0.8$ & $1.2$ & $0.0$ & $0.3$ & $0.7$ & $0.0$ \\ 
 
$2000$ & $0.250$ & $0.079$ & $0.356 \cdot 10^{-2}$ & $18.9$ & $16.5$ & $3.7$ & $0.8$ & $8.4$ & $0.0$ & $0.2$ & $-8.2$ & $0.0$ \\ 
 
\hline 
$3000$ & $0.080$ & $0.369$ & $0.185 \cdot 10^{-1}$ & $6.6$ & $5.9$ & $2.5$ & $1.3$ & $1.5$ & $-0.0$ & $0.5$ & $1.2$ & $0.0$ \\ 
 
$3000$ & $0.130$ & $0.227$ & $0.107 \cdot 10^{-1}$ & $7.5$ & $7.1$ & $2.1$ & $0.6$ & $1.3$ & $-0.0$ & $0.2$ & $0.9$ & $0.0$ \\ 
 
$3000$ & $0.250$ & $0.118$ & $0.210 \cdot 10^{-2}$ & $13.7$ & $13.1$ & $3.5$ & $2.1$ & $1.8$ & $0.0$ & $0.4$ & $-1.3$ & $0.0$ \\ 
 
$3000$ & $0.400$ & $0.074$ & $0.236 \cdot 10^{-3}$ & $73.4$ & $71.0$ & $6.6$ & $5.6$ & $17.6$ & $0.0$ & $0.7$ & $-16.9$ & $0.0$ \\ 
 
\hline 
$5000$ & $0.080$ & $0.615$ & $0.696 \cdot 10^{-2}$ & $10.9$ & $9.7$ & $3.9$ & $3.1$ & $2.8$ & $0.1$ & $0.8$ & $2.4$ & $-0.1$ \\ 
 
$5000$ & $0.130$ & $0.379$ & $0.476 \cdot 10^{-2}$ & $9.3$ & $8.3$ & $3.2$ & $2.4$ & $2.1$ & $-0.1$ & $0.7$ & $1.7$ & $0.0$ \\ 
 
$5000$ & $0.250$ & $0.197$ & $0.136 \cdot 10^{-2}$ & $12.2$ & $11.4$ & $3.3$ & $2.6$ & $1.6$ & $0.0$ & $0.6$ & $0.8$ & $0.0$ \\ 
 
$5000$ & $0.400$ & $0.123$ & $0.695 \cdot 10^{-3}$ & $24.3$ & $21.3$ & $7.7$ & $7.2$ & $8.4$ & $0.0$ & $1.5$ & $-7.6$ & $0.0$ \\ 
 
\hline 
$8000$ & $0.130$ & $0.606$ & $0.146 \cdot 10^{-2}$ & $15.4$ & $13.7$ & $5.6$ & $5.0$ & $3.2$ & $-0.2$ & $1.3$ & $2.5$ & $0.0$ \\ 
 
$8000$ & $0.250$ & $0.315$ & $0.802 \cdot 10^{-3}$ & $12.7$ & $11.3$ & $4.4$ & $3.8$ & $2.5$ & $0.0$ & $1.2$ & $1.8$ & $0.0$ \\ 
 
$8000$ & $0.400$ & $0.197$ & $0.195 \cdot 10^{-3}$ & $26.8$ & $25.1$ & $8.0$ & $7.7$ & $3.8$ & $0.0$ & $2.5$ & $-1.2$ & $0.0$ \\ 
 
\hline 
$15000$ & $0.250$ & $0.591$ & $0.201 \cdot 10^{-3}$ & $20.0$ & $17.6$ & $7.9$ & $7.4$ & $4.2$ & $-0.1$ & $2.2$ & $2.9$ & $0.0$ \\ 
 
$15000$ & $0.400$ & $0.369$ & $0.335 \cdot 10^{-4}$ & $30.2$ & $29.0$ & $7.0$ & $6.5$ & $3.6$ & $0.0$ & $1.9$ & $1.6$ & $0.0$ \\ 
 
\hline 

\end{tabular} 
\end{center} 
\caption[RESULT] 
{\label{tab:ccdxdq2_posP0} The CC 
$e^+p$ cross section ${\rm d}^2\sigma_{\rm CC}/{\rm d}x{\rm d}Q^2$ for lepton beam polarisation $P_e=0$ 
with statistical 
$(\delta_{\rm stat})$, 
total $(\delta_{\rm tot})$, 
total uncorrelated systematic $(\delta_{\rm unc})$ 
errors and one of its contributions from 
the hadronic energy error  
($\delta_{\rm unc}^{h}$). 
The effect of the other uncorrelated 
systematic errors is included in $\delta_{\rm unc}$. 
In addition the correlated systematic  
$(\delta_{\rm cor})$ and its contributions from a 
positive variation of one  
standard deviation of the 
cuts against photoproduction ($\delta_{\rm cor}^{V^+}$), of 
the hadronic 
energy error ($\delta_{\rm cor}^{h^+}$), of the error 
due to noise subtraction ($\delta_{\rm cor}^{N^+}$) 
and of the error due to background subtraction 
($\delta_{\rm cor}^{B^+}$) are given. 
The normalisation and polarisation uncertainties are not included in the errors. 
}
\end{table}

\renewcommand{\arraystretch}{1.22}

\clearpage
\begin{table}[\tablepos] 
\begin{center} 
\footnotesize
\begin{tabular}{|r|c|c|r|r|r|r|r|r|c|} 
\hline 
$Q^2$ & $x$ & $y$ & $\tilde{\sigma}_{\rm NC}$  & $F_{2}$ & 
$\delta_{\rm stat}$  & $\delta_{\rm unc}$ &   $\delta_{\rm cor}$ & $\delta_{\rm tot}$ & $\sqrt{s}$      \\ 
$(\rm GeV^2)$ &  &   &   &         & 
$(\%)$          & $(\%)$        & $(\%)$          & $(\%)$   & $(\rm GeV)$           \\
\hline 
$   60 $ &$ 0.00080 $ & $ 0.750 $ &$ 1.465  $ & ---  & $  1.01 $ &$  1.84 $ &$  1.65 $ &$ 2.67$ & $   319$ \\
\hline
$   90 $ &$ 0.00118 $ & $ 0.750 $ &$ 1.468  $ & ---  & $  0.82 $ &$  1.71 $ &$  1.61 $ &$ 2.49$ & $   319$ \\
\hline
$  100 $ &$ 0.00131 $ & $ 0.750 $ &$ 1.355  $ & ---  & $  4.02 $ &$  5.50 $ &$  1.75 $ &$ 7.03$ & $   319$ \\
$  100 $ &$ 0.00200 $ & $ 0.494 $ &$ 1.271  $ &$ 1.316  $ &$  3.43 $ &$  3.30 $ &$  1.89 $ &$ 5.12$ & $   319$ \\
\hline
$  120 $ &$ 0.00158 $ & $ 0.750 $ &$ 1.421  $ & ---  & $  0.92 $ &$  1.71 $ &$  1.57 $ &$ 2.50$ & $   319$ \\
$  120 $ &$ 0.00200 $ & $ 0.593 $ &$ 1.340  $ & ---  & $  0.70 $ &$  0.91 $ &$  1.73 $ &$ 2.08$ & $   319$ \\
\hline
\end{tabular} 
\end{center} 
\caption[RESULT] 
{ Combined HERA\,I+II NC $e^-p$ reduced cross sections for $P_e=0$. 
Here $F_2$ is the dominant electromagnetic structure function and
$\delta_{\rm stat}$, $\delta_{\rm unc}$, $\delta_{\rm cor}$ and $\delta_{\rm tot}$ 
are the relative statistical, total uncorrelated, correlated systematic and 
total uncertainties, respectively. 
The table continues on the next pages.}
\label{tab:nccombined_ele}
\end{table}

\begin{table}[\tablepos] 
\begin{center} 
\footnotesize
\begin{tabular}{|r|c|c|r|r|r|r|r|r|c|} 
\hline 
$Q^2$ & $x$ & $y$ & $\tilde{\sigma}_{\rm NC}$  & $F_{2}$ & 
$\delta_{\rm stat}$  & $\delta_{\rm unc}$ &   $\delta_{\rm cor}$ & $\delta_{\rm tot}$  & $\sqrt{s}$     \\ 
 $(\rm GeV^2)$ &  &   &    &         & 
$(\%)$          & $(\%)$        & $(\%)$          & $(\%)$    & $(\rm GeV)$          \\
\hline 
$  300 $ &$ 0.00394 $ & $ 0.750 $ &$ 1.202  $ & ---  & $  1.71 $ &$  1.83 $ &$  1.52 $ &$ 2.93$ & $   319$ \\
$  300 $ &$ 0.00500 $ & $ 0.593 $ &$ 1.143  $ & ---  & $  1.49 $ &$  0.89 $ &$  1.56 $ &$ 2.33$ & $   319$ \\
$  300 $ &$ 0.00800 $ & $ 0.371 $ &$ 0.9817 $ &$ 0.9921 $ &$  1.01 $ &$  0.83 $ &$  1.52 $ &$ 2.01$ & $   319$ \\
$  300 $ &$ 0.01300 $ & $ 0.228 $ &$ 0.8203 $ &$ 0.8204 $ &$  1.01 $ &$  1.05 $ &$  1.56 $ &$ 2.14$ & $   319$ \\
$  300 $ &$ 0.02000 $ & $ 0.148 $ &$ 0.7002 $ &$ 0.6985 $ &$  1.12 $ &$  1.01 $ &$  1.57 $ &$ 2.18$ & $   319$ \\
$  300 $ &$ 0.03200 $ & $ 0.093 $ &$ 0.5882 $ &$ 0.5863 $ &$  1.18 $ &$  1.13 $ &$  1.57 $ &$ 2.26$ & $   319$ \\
\hline
\end{tabular} 
\end{center} 
\captcont 
{\label{tab:nccombined_ele_c}continued.} 
\end{table} 
\begin{table}[\tablepos] 
\begin{center} 
\footnotesize
\begin{tabular}{|r|c|c|r|r|r|r|r|r|c|} 
\hline 
$Q^2$ & $x$ & $y$ & $\tilde{\sigma}_{\rm NC}$  & $F_{2}$ & 
$\delta_{\rm stat}$  & $\delta_{\rm unc}$ &   $\delta_{\rm cor}$ & $\delta_{\rm tot}$  & $\sqrt{s}$     \\ 
 $(\rm GeV^2)$ &  &   &    &         & 
$(\%)$          & $(\%)$        & $(\%)$          & $(\%)$    & $(\rm GeV)$          \\
\hline 
$  650 $ &$ 0.00853 $ & $ 0.750 $ &$ 0.9854 $ & ---  & $  2.32 $ &$  1.82 $ &$  1.50 $ &$ 3.31$ & $   319$ \\
$  650 $ &$ 0.01300 $ & $ 0.494 $ &$ 0.8981 $ &$ 0.9072 $ &$  1.65 $ &$  0.93 $ &$  1.51 $ &$ 2.42$ & $   319$ \\
$  650 $ &$ 0.02000 $ & $ 0.321 $ &$ 0.7604 $ &$ 0.7562 $ &$  1.71 $ &$  0.98 $ &$  1.53 $ &$ 2.50$ & $   319$ \\
$  650 $ &$ 0.03200 $ & $ 0.201 $ &$ 0.6454 $ &$ 0.6389 $ &$  1.75 $ &$  1.27 $ &$  1.59 $ &$ 2.69$ & $   319$ \\
$  650 $ &$ 0.05000 $ & $ 0.128 $ &$ 0.5368 $ &$ 0.5316 $ &$  1.85 $ &$  1.07 $ &$  1.58 $ &$ 2.66$ & $   319$ \\
$  650 $ &$ 0.08000 $ & $ 0.080 $ &$ 0.4291 $ &$ 0.4249 $ &$  2.10 $ &$  1.07 $ &$  1.59 $ &$ 2.84$ & $   319$ \\
\hline
\end{tabular} 
\end{center} 
\captcont 
{continued.} 
\end{table} 

\begin{table}[\tablepos] 
\begin{center} 
\footnotesize 
\begin{tabular}{|r|c|c|r|r|r|r|r|r|c|} 
\hline 
$Q^2$ & $x$ & $y$ & $\tilde{\sigma}_{\rm NC}$  & $F_{2}$ & 
$\delta_{\rm stat}$  & $\delta_{\rm unc}$ &   $\delta_{\rm cor}$ & $\delta_{\rm tot}$  & $\sqrt{s}$     \\ 
$(\rm GeV^2)$ &  &   &    &         & 
$(\%)$          & $(\%)$        & $(\%)$          & $(\%)$  & $(\rm GeV)$            \\
\hline 
$ 1500 $ &$ 0.02000 $ & $ 0.750 $ &$ 0.7971 $ & ---  & $  3.41 $ &$  1.58 $ &$  1.74 $ &$ 4.14$ & $   319$ \\
$ 1500 $ &$ 0.03200 $ & $ 0.463 $ &$ 0.6714 $ &$ 0.6482 $ &$  3.20 $ &$  0.99 $ &$  1.51 $ &$ 3.68$ & $   319$ \\
$ 1500 $ &$ 0.05000 $ & $ 0.296 $ &$ 0.5489 $ &$ 0.5277 $ &$  3.18 $ &$  1.15 $ &$  1.51 $ &$ 3.70$ & $   319$ \\
$ 1500 $ &$ 0.08000 $ & $ 0.185 $ &$ 0.4925 $ &$ 0.4750 $ &$  3.23 $ &$  1.11 $ &$  1.53 $ &$ 3.74$ & $   319$ \\
$ 1500 $ &$ 0.13000 $ &$ 0.114 $ &$ 0.3524 $ &$ 0.3409 $ &$  4.47 $ &$  1.49 $ &$  1.56 $ &$ 4.96$ & $   319$ \\
$ 1500 $ &$ 0.18000 $ &$ 0.082 $ &$ 0.2950 $ &$ 0.2867 $ &$  4.44 $ &$  1.70 $ &$  1.57 $ &$ 5.01$ & $   319$ \\
\hline
\end{tabular} 
\end{center} 
\captcont
{continued.}
\end{table} 
\begin{table}[\tablepos] 
\begin{center} 
\footnotesize 
\begin{tabular}{|r|c|c|r|r|r|r|r|r|c|} 
\hline 
$Q^2$ & $x$ & $y$ & $\tilde{\sigma}_{\rm NC}$  & $F_{2}$ & 
$\delta_{\rm stat}$  & $\delta_{\rm unc}$ &   $\delta_{\rm cor}$ & $\delta_{\rm tot}$  & $\sqrt{s}$     \\ 
$(\rm GeV^2)$ &  &   &    &         & 
$(\%)$          & $(\%)$        & $(\%)$          & $(\%)$  & $(\rm GeV)$            \\
\hline 
$ 8000 $ &$ 0.08750 $ & $ 0.903 $ &$ 0.6314 $ & ---  & $  7.40 $ &$  2.49 $ &$  1.80 $ &$ 8.01$ & $   319$ \\
$ 8000 $ &$ 0.13000 $ &$ 0.608 $ &$ 0.5309 $ & ---  & $  5.55 $ &$  2.10 $ &$  1.52 $ &$ 6.12$ & $   319$ \\
$ 8000 $ &$ 0.18000 $ &$ 0.439 $ &$ 0.3795 $ &$ 0.2781 $ &$  6.23 $ &$  2.00 $ &$  1.50 $ &$ 6.71$ & $   319$ \\
$ 8000 $ &$ 0.25000 $ &$ 0.316 $ &$ 0.2627 $ &$ 0.2022 $ &$  7.12 $ &$  2.33 $ &$  1.57 $ &$ 7.65$ & $   319$ \\
$ 8000 $ &$ 0.40000 $ &$ 0.198 $ &$ 0.1231 $ &$ 0.1006 $ &$  9.69 $ &$  4.35 $ &$  2.09 $ &$10.83$ & $   319$ \\
$ 8000 $ &$ 0.65000 $ &$ 0.122 $ &$ 0.0123 $ &$ 0.0103 $ &$ 16.69 $ &$  6.54 $ &$  2.85 $ &$18.15$ & $   319$ \\
\hline
\end{tabular} 
\end{center} 
\captcont
{continued.}
\end{table} 

\renewcommand{\arraystretch}{1.22}

\begin{table}[\tablepos] 
\begin{center} 
\footnotesize 
\begin{tabular}{|r|c|c|r|r|r|r|r|r|c|} 
\hline 
$Q^2$ & $x$  & $y$ & $\tilde{\sigma}_{\rm NC}$  & $F_{2}$ & 
$\delta_{\rm stat}$  & $\delta_{\rm unc}$ &   $\delta_{\rm cor}$ & $\delta_{\rm tot}$  & $\sqrt{s}$     \\ 
 $(\rm GeV^2)$ &  &    &   &         & 
$(\%)$          & $(\%)$        & $(\%)$          & $(\%)$  & $(\rm GeV)$             \\
\hline 
$   60 $ &$ 0.00080 $ & $ 0.750 $ &$ 1.486  $ & ---  & $  0.87 $ &$  1.86 $ &$  1.21 $ &$ 2.39$ & $   319$ \\
\hline
$   90 $ &$ 0.00118 $ & $ 0.750 $ &$ 1.491  $ & ---  & $  0.71 $ &$  1.64 $ &$  1.17 $ &$ 2.14$ & $   319$ \\
\hline
$  100 $ &$ 0.00131 $ & $ 0.750 $ &$ 1.438  $ & ---  & $  1.61 $ &$  3.50 $ &$  1.49 $ &$ 4.13$ & $   319$ \\
$  100 $ &$ 0.00200 $ & $ 0.494 $ &$ 1.374  $ &$ 1.423  $ &$  1.41 $ &$  2.10 $ &$  1.46 $ &$ 2.92$ & $   319$ \\
\hline
$  120 $ &$ 0.00158 $ & $ 0.750 $ &$ 1.407  $ & ---  & $  0.77 $ &$  1.45 $ &$  1.15 $ &$ 2.00$ & $   319$ \\
$  120 $ &$ 0.00200 $ & $ 0.593 $ &$ 1.356  $ & ---  & $  0.59 $ &$  0.74 $ &$  1.35 $ &$ 1.65$ & $   319$ \\
\hline
\end{tabular} 
\end{center} 
\caption[RESULT] 
{ Combined HERA\,I+II NC $e^+p$ reduced cross sections for $P_e=0$.
Here $F_2$ is the dominant electromagnetic structure function and
$\delta_{\rm stat}$, $\delta_{\rm unc}$, $\delta_{\rm cor}$ and $\delta_{\rm tot}$ 
are the relative statistical, total uncorrelated, correlated systematic and total uncertainties, respectively.
The table continues on the next pages.}
\label{tab:nccombined_pos}
\end{table} 

\begin{table}[\tablepos] 
\begin{center} 
\footnotesize 
\begin{tabular}{|r|c|c|r|r|r|r|r|r|c|} 
\hline 
$Q^2$ & $x$  & $y$ & $\tilde{\sigma}_{\rm NC}$  & $F_{2}$ & 
$\delta_{\rm stat}$  & $\delta_{\rm unc}$ &   $\delta_{\rm cor}$ & $\delta_{\rm tot}$  & $\sqrt{s}$     \\ 
$(\rm GeV^2)$ &  &    &    &         & 
$(\%)$          & $(\%)$        & $(\%)$          & $(\%)$  & $(\rm GeV)$             \\
\hline 
$  250 $ &$ 0.00328 $ & $ 0.750 $ &$ 1.278  $ & ---  & $  1.24 $ &$  1.55 $ &$  1.12 $ &$ 2.27$ & $   319$ \\
$  250 $ &$ 0.00500 $ & $ 0.554 $ &$ 1.102  $ &$ 1.136  $ &$  2.31 $ &$  4.90 $ &$  1.31 $ &$ 5.57$ & $   301$ \\
$  250 $ &$ 0.00500 $ & $ 0.494 $ &$ 1.124  $ &$ 1.158  $ &$  0.75 $ &$  0.69 $ &$  1.12 $ &$ 1.52$ & $   319$ \\
$  250 $ &$ 0.00800 $ & $ 0.309 $ &$ 0.9611 $ &$ 0.9695 $ &$  0.68 $ &$  0.82 $ &$  1.10 $ &$ 1.53$ & $   319$ \\
$  250 $ &$ 0.01300 $ & $ 0.190 $ &$ 0.8127 $ &$ 0.8147 $ &$  0.74 $ &$  0.99 $ &$  1.13 $ &$ 1.67$ & $   319$ \\
$  250 $ &$ 0.02000 $ & $ 0.124 $ &$ 0.6982 $ &$ 0.6981 $ &$  0.74 $ &$  1.08 $ &$  1.12 $ &$ 1.73$ & $   319$ \\
\hline
\end{tabular} 
\end{center} 
\captcont{\label{tab:nccombined_pos_c}continued.}
\end{table} 
\begin{table}[\tablepos] 
\begin{center} 
\footnotesize 
\begin{tabular}{|r|c|c|r|r|r|r|r|r|c|} 
\hline 
$Q^2$ & $x$  & $y$ & $\tilde{\sigma}_{\rm NC}$  & $F_{2}$ & 
$\delta_{\rm stat}$  & $\delta_{\rm unc}$ &   $\delta_{\rm cor}$ & $\delta_{\rm tot}$  & $\sqrt{s}$     \\ 
$(\rm GeV^2)$ &  &    &    &         & 
$(\%)$          & $(\%)$      
   & $(\%)$          & $(\%)$  & $(\rm GeV)$             \\
\hline 
$  500 $ &$ 0.00656 $ & $ 0.750 $ &$ 1.050  $ & ---  & $  1.66 $ &$  1.55 $ &$  1.08 $ &$ 2.51$ & $   319$ \\
$  500 $ &$ 0.00800 $ & $ 0.675 $ &$ 1.022  $ & ---  & $  4.21 $ &$  5.00 $ &$  1.31 $ &$ 6.67$ & $   301$ \\
$  500 $ &$ 0.00800 $ & $ 0.618 $ &$ 0.9950 $ & ---  & $  1.73 $ &$  0.89 $ &$  1.12 $ &$ 2.24$ & $   319$ \\
$  500 $ &$ 0.01300 $ & $ 0.426 $ &$ 0.8868 $ &$ 0.8984 $ &$  3.34 $ &$  5.00 $ &$  1.25 $ &$ 6.14$ & $   301$ \\
$  500 $ &$ 0.01300 $ & $ 0.380 $ &$ 0.8717 $ &$ 0.8832 $ &$  1.22 $ &$  0.77 $ &$  1.09 $ &$ 1.81$ & $   319$ \\
$  500 $ &$ 0.02000 $ & $ 0.247 $ &$ 0.7423 $ &$ 0.7447 $ &$  1.15 $ &$  0.90 $ &$  1.11 $ &$ 1.83$ & $   319$ \\
\hline
\end{tabular} 
\end{center} 
\captcont{continued.}
\end{table} 

\begin{table}[\tablepos] 
\begin{center} 
\footnotesize 
\begin{tabular}{|r|c|c|r|r|r|r|r|r|c|} 
\hline 
$Q^2$ & $x$  & $y$ & $\tilde{\sigma}_{\rm NC}$  & $F_{2}$ & 
$\delta_{\rm stat}$  & $\delta_{\rm unc}$ &   $\delta_{\rm cor}$ & $\delta_{\rm tot}$  & $\sqrt{s}$     \\ 
$(\rm GeV^2)$ &  &    &    &         & 
$(\%)$          & $(\%)$        & $(\%)$          & $(\%)$  & $(\rm GeV)$             \\
\hline 
$ 1000 $ &$ 0.01300 $ & $ 0.750 $ &$ 0.8194 $ & ---  & $  2.33 $ &$  1.36 $ &$  1.53 $ &$ 3.11$ & $   319$ \\
$ 1000 $ &$ 0.02000 $ & $ 0.554 $ &$ 0.7428 $ &$ 0.7608 $ &$  5.44 $ &$  3.70 $ &$  1.28 $ &$ 6.70$ & $   301$ \\
$ 1000 $ &$ 0.02000 $ & $ 0.494 $ &$ 0.7340 $ &$ 0.7516 $ &$  1.92 $ &$  0.88 $ &$  1.09 $ &$ 2.38$ & $   319$ \\
$ 1000 $ &$ 0.03200 $ & $ 0.309 $ &$ 0.6236 $ &$ 0.6286 $ &$  1.81 $ &$  0.98 $ &$  1.09 $ &$ 2.33$ & $   319$ \\
$ 1000 $ &$ 0.05000 $ & $ 0.198 $ &$ 0.5322 $ &$ 0.5330 $ &$  1.88 $ &$  1.26 $ &$  1.09 $ &$ 2.51$ & $   319$ \\
$ 1000 $ &$ 0.08000 $ & $ 0.124 $ &$ 0.4424 $ &$ 0.4415 $ &$  2.03 $ &$  1.01 $ &$  1.09 $ &$ 2.52$ & $   319$ \\
\hline
\end{tabular} 
\end{center} 
\captcont{continued.}
\end{table} 

\begin{table}[\tablepos] 
\begin{center} 
\footnotesize 
\begin{tabular}{|r|c|c|r|r|r|r|r|r|c|} 
\hline 
$Q^2$ & $x$  & $y$ & $\tilde{\sigma}_{\rm NC}$  & $F_{2}$ & 
$\delta_{\rm stat}$  & $\delta_{\rm unc}$ &   $\delta_{\rm cor}$ & $\delta_{\rm tot}$  & $\sqrt{s}$     \\ 
 $(\rm GeV^2)$ &  &    &     &         & 
$(\%)$          & $(\%)$        & $(\%)$          & $(\%)$  & $(\rm GeV)$             \\
\hline 
$ 2000 $ &$ 0.02190 $ & $ 0.903 $ &$ 0.7034 $ & ---  & $  5.55 $ &$  2.40 $ &$  1.85 $ &$ 6.33$ & $   319$ \\
$ 2000 $ &$ 0.03200 $ & $ 0.675 $ &$ 0.6077 $ & ---  & $  9.05 $ &$  4.00 $ &$  1.30 $ &$ 9.98$ & $   301$ \\
$ 2000 $ &$ 0.03200 $ & $ 0.618 $ &$ 0.5687 $ & ---  & $  3.35 $ &$  1.33 $ &$  1.10 $ &$ 3.77$ & $   319$ \\
$ 2000 $ &$ 0.05000 $ & $ 0.443 $ &$ 0.5285 $ &$ 0.5406 $ &$  8.80 $ &$  4.00 $ &$  1.26 $ &$ 9.75$ & $   301$ \\
$ 2000 $ &$ 0.05000 $ & $ 0.395 $ &$ 0.5231 $ &$ 0.5345 $ &$  3.27 $ &$  1.07 $ &$  1.09 $ &$ 3.61$ & $   319$ \\
$ 2000 $ &$ 0.08000 $ & $ 0.247 $ &$ 0.4225 $ &$ 0.4267 $ &$  3.02 $ &$  1.17 $ &$  1.08 $ &$ 3.41$ & $   319$ \\
\hline
\end{tabular} 
\end{center} 
\captcont 
{continued.}
\end{table} 
\begin{table}[\tablepos] 
\begin{center} 
\footnotesize 
\begin{tabular}{|r|c|c|r|r|r|r|r|r|c|} 
\hline 
$Q^2$ & $x$  & $y$ & $\tilde{\sigma}_{\rm NC}$  & $F_{2}$ & 
$\delta_{\rm stat}$  & $\delta_{\rm unc}$ &   $\delta_{\rm cor}$ & $\delta_{\rm tot}$  & $\sqrt{s}$     \\ 
 $(\rm GeV^2)$ &  &    &     &         & 
$(\%)$          & $(\%)$        & $(\%)$          & $(\%)$  & $(\rm GeV)$             \\
\hline 
$12000 $ &$ 0.13000 $ &$ 0.912 $ &$ 0.2126 $ & ---  & $ 19.65 $ &$  5.03 $ &$  1.76 $ &$20.36$ & $   319$ \\
$12000 $ &$ 0.18000 $ &$ 0.750 $ &$ 0.1512 $ & ---  & $ 34.61 $ &$  3.80 $ &$  1.50 $ &$34.85$ & $   301$ \\
$12000 $ &$ 0.18000 $ &$ 0.675 $ &$ 0.2211 $ & ---  & $  9.76 $ &$  2.16 $ &$  1.24 $ &$10.07$ & $   319$ \\
$12000 $ &$ 0.25000 $ &$ 0.532 $ &$ 0.1239 $ &$ 0.1482 $ &$ 32.50 $ &$  6.10 $ &$  1.32 $ &$33.09$ & $   301$ \\
$12000 $ &$ 0.25000 $ &$ 0.474 $ &$ 0.1463 $ &$ 0.1744 $ &$ 10.52 $ &$  2.49 $ &$  1.11 $ &$10.87$ & $   319$ \\
$12000 $ &$ 0.40000 $ &$ 0.296 $ &$ 0.0859 $ &$ 0.0905 $ &$ 11.85 $ &$  4.07 $ &$  1.41 $ &$12.61$ & $   319$ \\
\hline
\end{tabular} 
\end{center} 
\captcont 
{continued.}
\end{table} 

\clearpage
\begin{table}[\tablepos]
\begin{center} 
\footnotesize 
\begin{tabular}{|r|c|c|l|r|r|r|r|r|c|} 
\hline 
 $Q^2$ & $x$ & $y$ & ${\rm d}^2\sigma_{\rm CC}/{\rm d}x{\rm d}Q^2$ & $\tilde{\sigma}_{\rm CC}$ & $\delta_{\rm stat}$  & $\delta_{\rm unc}$ &   $\delta_{\rm cor}$ & $\delta_{\rm tot}$ & $\sqrt{s}$      \\ 
$(\rm GeV^2)$ &  &   & $(\rm pb / \rm GeV^2)$    &    & $(\%)$          & $(\%)$        & $(\%)$          & $(\%)$   & $(\rm GeV)$           \\
\hline 
$     300  $ &$  0.008  $ &$  0.371  $ &$ 1.88 $ & $  1.95   $ &$ 31.4$ &$ 23.7$ &$ 13.2$ &$ 41.5$ & $   319$ \\
$     300  $ &$  0.013  $ &$  0.228  $ &$ 0.731 $ &$  1.23   $ &$ 12.9$ &$  9.6$ &$  7.5$ &$ 17.7$ & $   319$ \\
$     300  $ &$  0.032  $ &$  0.093  $ &$ 0.269 $ &$  1.12   $ &$ 10.8$ &$  4.5$ &$  4.6$ &$ 12.6$ & $   319$ \\
$     300  $ &$  0.080  $ &$  0.037  $ &$ 0.670 \cdot 10^{-1}$ & $  0.696  $ &$ 11.2$ &$  4.8$ &$  3.4$ &$ 12.6$ & $   319$ \\
\hline
$     500  $ &$  0.013  $ &$  0.380  $ &$ 0.646 $ &$  1.16   $ &$  8.5$ &$  6.3$ &$  6.6$ &$ 12.5$ & $   319$ \\
$     500  $ &$  0.032  $ &$  0.154  $ &$ 0.219 $ &$  0.963  $ &$  6.7$ &$  3.1$ &$  2.8$ &$  7.8$ & $   319$ \\
\hline
\end{tabular} 
\end{center} 
\caption[RESULT] 
{ Combined HERA\,I+II CC $e^-p$ cross sections for $P_e=0$. Here
$\delta_{\rm stat}$, $\delta_{\rm unc}$, $\delta_{\rm cor}$ and $\delta_{\rm tot}$ 
are the relative statistical, total uncorrelated, correlated systematic and total uncertainties, respectively.
}
\label{tab:cccombined_ele} 
\end{table}

\begin{table}[\tablepos] 
\begin{center} 
\footnotesize 
\begin{tabular}{|r|c|c|l|r|r|r|r|r|c|} 
\hline 
 $Q^2$ & $x$ & $y$ & ${\rm d}^2\sigma_{\rm CC}/{\rm d}x{\rm d}Q^2$ & $\tilde{\sigma}_{\rm CC}$ & $\delta_{\rm stat}$  & $\delta_{\rm unc}$ &   $\delta_{\rm cor}$ & $\delta_{\rm tot}$   & $\sqrt{s}$    \\ 
 $(\rm GeV^2)$ &  &   & $(\rm pb / \rm GeV^2)$    &    & $(\%)$          & $(\%)$        & $(\%)$          & $(\%)$   & $(\rm GeV)$          \\
\hline 
$     300  $ &$  0.008  $ &$  0.371  $ &$ 1.08 $ & $  1.12   $ &$ 28.5$ &$ 25.7$ &$ 13.9$ &$ 40.8$ & $   319$ \\
$     300  $ &$  0.013  $ &$  0.228  $ &$ 0.638 $ &$  1.08   $ &$ 11.1$ &$  6.2$ &$  6.6$ &$ 14.3$ & $   319$ \\
$     300  $ &$  0.032  $ &$  0.093  $ &$ 0.217 $ &$  0.901  $ &$  7.8$ &$  3.5$ &$  3.9$ &$  9.4$ & $   319$ \\
$     300  $ &$  0.080  $ &$  0.037  $ &$ 0.481 \cdot 10^{-1}$ & $  0.500  $ &$ 10.1$ &$  3.6$ &$  2.5$ &$ 11.0$ & $   319$ \\
\hline
$     500  $ &$  0.013  $ &$  0.380  $ &$ 0.550 $ &$  0.984  $ &$  7.5$ &$  4.5$ &$  5.3$ &$ 10.2$ & $   319$ \\
$     500  $ &$  0.032  $ &$  0.154  $ &$ 0.190 $ &$  0.839  $ &$  5.8$ &$  2.4$ &$  2.4$ &$  6.7$ & $   319$ \\
\hline
\end{tabular} 
\end{center} 
\caption[RESULT] 
{ Combined HERA\,I+II CC $e^+p$ cross sections for $P_e=0$. Here
$\delta_{\rm stat}$, $\delta_{\rm unc}$, $\delta_{\rm cor}$ and $\delta_{\rm tot}$ 
are the relative statistical, total uncorrelated, correlated systematic and total uncertainties, respectively.}
\label{tab:cccombined_pos} 
\end{table}

\clearpage
\begin{table}[htb] 
\begin{center} 
\footnotesize 
\begin{tabular}{|r|l|r|r|r|r|r|r|r|r|r|} 
\hline 
$Q^2$                 & ${\rm d}\sigma_{\rm NC}/{\rm d}Q^2$      & 
$\delta_{\rm stat}$      & $\delta_{\rm unc}$         &   $\delta_{\rm cor}$          & 
$\delta_{\rm tot}$       & $\delta_{\rm cor}^{E^+}$   & $\delta_{\rm cor}^{\theta^+}$ & $\delta_{\rm cor}^{h^+}$& 
$\delta_{\rm cor}^{N^+}$ & $\delta_{\rm cor}^{B^+}$ \\ 
$(\rm GeV^2)$ &$(\rm pb / \rm GeV^2)$ & 
$(\%)$ & $(\%)$  & $(\%)$ & $(\%)$ & $(\%)$ & 
$(\%)$ & $(\%)$  & $(\%)$ & $(\%)$ \\ 
\hline 
$200$ & $17.91$ & $0.42$ & $0.92$ & $0.62$ & $1.19$ & $-0.22$ & $-0.57$ & $0.00$ & $0.05$ & $-0.06$ \\ 
 
$250$ & $10.61$ & $0.43$ & $1.21$ & $0.67$ & $1.45$ & $0.39$ & $-0.54$ & $0.01$ & $0.05$ & $-0.04$ \\ 
 
$300$ & $6.800$ & $0.51$ & $1.24$ & $0.73$ & $1.53$ & $0.48$ & $-0.55$ & $0.01$ & $0.04$ & $-0.04$ \\ 
 
$400$ & $3.436$ & $0.60$ & $1.06$ & $0.62$ & $1.36$ & $0.46$ & $-0.41$ & $0.01$ & $0.04$ & $-0.03$ \\ 
 
$500$ & $2.038$ & $0.71$ & $1.08$ & $0.62$ & $1.44$ & $0.50$ & $-0.37$ & $0.01$ & $0.04$ & $-0.02$ \\ 
 
$650$ & $1.063$ & $0.86$ & $1.10$ & $0.63$ & $1.53$ & $0.51$ & $-0.35$ & $0.01$ & $0.05$ & $-0.02$ \\
\hline 
\end{tabular} 
\end{center} 
\caption[RESULT] 
{\label{tab:ncdq2_eleLH} The NC 
$e^-p$ cross section ${\rm d}\sigma_{\rm NC}/{\rm d}Q^2$ for lepton beam polarisation $P_e=-25.8$\% 
and $y<0.9$. 
The 
statistical $(\delta_{\rm stat})$, 
uncorrelated systematic $(\delta_{\rm unc})$, 
correlated systematic $(\delta_{\rm cor})$ 
and total $(\delta_{\rm tot})$ errors are provided. 
In addition the correlated systematic  
error contributions from a 
positive variation of one  
standard deviation of the 
electron energy error ($\delta_{\rm cor}^{E^+}$), of 
the polar electron angle error 
($\delta_{\rm cor}^{\theta^+}$), of the hadronic 
energy error ($\delta_{\rm cor}^{h^+}$), of the error 
due to noise subtraction ($\delta_{\rm cor}^{N^+}$) and 
of the error due to background subtraction 
($\delta_{\rm cor}^{B^+}$) are given. 
The normalisation and polarisation uncertainties are 
not included in the errors.} 
\end{table} 

\begin{table}[htb] 
\begin{center} 
\footnotesize 
\begin{tabular}{|r|l|r|r|r|r|r|r|r|r|r|} 
\hline 
$Q^2$                 & ${\rm d}\sigma_{\rm NC}/{\rm d}Q^2$      & 
$\delta_{\rm stat}$      & $\delta_{\rm unc}$         &   $\delta_{\rm cor}$          & 
$\delta_{\rm tot}$       & $\delta_{\rm cor}^{E^+}$   & $\delta_{\rm cor}^{\theta^+}$ & $\delta_{\rm cor}^{h^+}$& 
$\delta_{\rm cor}^{N^+}$ & $\delta_{\rm cor}^{B^+}$ \\ 
$(\rm GeV^2)$ &$(\rm pb / \rm GeV^2)$ & 
$(\%)$ & $(\%)$  & $(\%)$ & $(\%)$ & $(\%)$ & 
$(\%)$ & $(\%)$  & $(\%)$ & $(\%)$ \\ 
\hline 
$200$ & $17.78$ & $0.62$ & $0.91$ & $0.60$ & $1.25$ & $-0.19$ & $-0.56$ & $0.00$ & $0.05$ & $-0.06$ \\ 
 
$250$ & $10.32$ & $0.65$ & $1.22$ & $0.64$ & $1.52$ & $0.38$ & $-0.52$ & $0.00$ & $0.04$ & $-0.05$ \\ 
 
$300$ & $6.781$ & $0.76$ & $1.27$ & $0.77$ & $1.67$ & $0.53$ & $-0.57$ & $0.00$ & $0.05$ & $-0.03$ \\ 
 
$400$ & $3.343$ & $0.91$ & $1.05$ & $0.61$ & $1.52$ & $0.45$ & $-0.41$ & $0.01$ & $0.04$ & $-0.03$ \\ 
 
$500$ & $1.928$ & $1.08$ & $1.09$ & $0.68$ & $1.68$ & $0.53$ & $-0.42$ & $0.01$ & $0.04$ & $-0.02$ \\ 
 
$650$ & $1.030$ & $1.29$ & $1.11$ & $0.63$ & $1.82$ & $0.53$ & $-0.34$ & $0.01$ & $0.05$ & $-0.02$ \\ 
 
$800$ & $0.6080$ & $1.59$ & $1.09$ & $0.55$ & $2.00$ & $0.46$ & $-0.30$ & $0.01$ & $0.05$ & $-0.02$ \\ 
 
$1000$ & $0.3480$ & $1.88$ & $1.17$ & $0.60$ & $2.29$ & $0.48$ & $-0.31$ & $0.01$ & $0.05$ & $-0.18$ \\ 
 
$1200$ & $0.2160$ & $2.18$ & $1.29$ & $0.59$ & $2.60$ & $0.50$ & $-0.23$ & $0.01$ & $0.04$ & $-0.20$ \\ 
 
$1500$ & $0.1210$ & $2.72$ & $1.35$ & $0.54$ & $3.08$ & $0.48$ & $-0.23$ & $0.01$ & $0.05$ & $-0.10$ \\ 
 
$2000$ & $0.5700 \cdot 10^{-1}$ & $3.21$ & $1.45$ & $0.57$ & $3.57$ & $0.51$ & $-0.18$ & $0.01$ & $0.05$ & $-0.18$ \\ 
 
$3000$ & $0.2020 \cdot 10^{-1}$ & $2.99$ & $1.58$ & $0.60$ & $3.43$ & $0.54$ & $-0.17$ & $0.01$ & $0.05$ & $-0.18$ \\ 
 
$5000$ & $0.5190 \cdot 10^{-2}$ & $4.10$ & $1.72$ & $0.42$ & $4.47$ & $0.37$ & $-0.11$ & $0.01$ & $0.05$ & $-0.16$ \\ 
 
$8000$ & $0.1490 \cdot 10^{-2}$ & $5.83$ & $2.33$ & $0.63$ & $6.31$ & $0.56$ & $-0.09$ & $0.01$ & $0.06$ & $-0.27$ \\ 
 
$12000$ & $0.4650 \cdot 10^{-3}$ & $9.30$ & $2.90$ & $0.79$ & $9.77$ & $0.76$ & $-0.03$ & $0.02$ & $0.05$ & $-0.21$ \\ 
 
$20000$ & $0.7660 \cdot 10^{-4}$ & $17.86$ & $3.82$ & $0.81$ & $18.28$ & $0.80$ & $0.11$ & $0.02$ & $0.05$ & $-0.04$ \\ 
 
$30000$ & $0.2200 \cdot 10^{-4}$ & $34.74$ & $6.25$ & $2.43$ & $35.38$ & $1.67$ & $0.07$ & $0.01$ & $0.04$ & $-1.77$ \\ 
 
\hline 
\end{tabular} 
\end{center} 
\caption[RESULT] 
{\label{tab:ncdq2_eleRH} The NC 
$e^-p$ cross section ${\rm d}\sigma_{\rm NC}/{\rm d}Q^2$ for lepton beam polarisation $P_e=+36.0$\% 
and $y<0.9$. 
The 
statistical $(\delta_{\rm stat})$, 
uncorrelated systematic $(\delta_{\rm unc})$, 
correlated systematic $(\delta_{\rm cor})$ 
and total $(\delta_{\rm tot})$ errors are provided. 
In addition the correlated systematic  
error contributions from a 
positive variation of one  
standard deviation of the 
electron energy error ($\delta_{\rm cor}^{E^+}$), of 
the polar electron angle error 
($\delta_{\rm cor}^{\theta^+}$), of the hadronic 
energy error ($\delta_{\rm cor}^{h^+}$), of the error 
due to noise subtraction ($\delta_{\rm cor}^{N^+}$) and 
of the error due to background subtraction 
($\delta_{\rm cor}^{B^+}$) are given. 
The normalisation and polarisation uncertainties are 
not included in the errors.} 
\end{table} 

\begin{table}[htb] 
\begin{center} 
\footnotesize 
\begin{tabular}{|r|l|r|r|r|r|r|r|r|r|r|} 
\hline 
$Q^2$                 & ${\rm d}\sigma_{\rm NC}/{\rm d}Q^2$      & 
$\delta_{\rm stat}$      & $\delta_{\rm unc}$         &   $\delta_{\rm cor}$          & 
$\delta_{\rm tot}$       & $\delta_{\rm cor}^{E^+}$   & $\delta_{\rm cor}^{\theta^+}$ & $\delta_{\rm cor}^{h^+}$& 
$\delta_{\rm cor}^{N^+}$ & $\delta_{\rm cor}^{B^+}$ \\ 
$(\rm GeV^2)$ &$(\rm pb / \rm GeV^2)$ & 
$(\%)$ & $(\%)$  & $(\%)$ & $(\%)$ & $(\%)$ & 
$(\%)$ & $(\%)$  & $(\%)$ & $(\%)$ \\ 
\hline 
$200$ & $18.16$ & $0.47$ & $0.80$ & $0.65$ & $1.14$ & $-0.23$ & $-0.61$ & $0.01$ & $0.05$ & $-0.06$ \\ 
 
$250$ & $10.61$ & $0.49$ & $1.12$ & $0.66$ & $1.39$ & $0.40$ & $-0.52$ & $0.01$ & $0.04$ & $-0.04$ \\ 
 
$300$ & $6.897$ & $0.57$ & $1.15$ & $0.72$ & $1.47$ & $0.47$ & $-0.54$ & $0.01$ & $0.04$ & $-0.04$ \\ 
 
$400$ & $3.419$ & $0.67$ & $0.98$ & $0.63$ & $1.35$ & $0.46$ & $-0.43$ & $0.00$ & $0.04$ & $-0.03$ \\ 
 
$500$ & $1.995$ & $0.81$ & $0.98$ & $0.64$ & $1.43$ & $0.50$ & $-0.40$ & $0.01$ & $0.04$ & $-0.03$ \\ 
 
$650$ & $1.029$ & $0.98$ & $1.02$ & $0.68$ & $1.57$ & $0.55$ & $-0.40$ & $0.01$ & $0.04$ & $-0.02$ \\ 
 
$800$ & $0.6030$ & $1.19$ & $1.06$ & $0.66$ & $1.73$ & $0.55$ & $-0.36$ & $0.01$ & $0.05$ & $-0.02$ \\ 
 
$1000$ & $0.3390$ & $1.41$ & $1.03$ & $0.54$ & $1.83$ & $0.42$ & $-0.27$ & $0.01$ & $0.05$ & $-0.19$ \\ 
 
$1200$ & $0.2130$ & $1.66$ & $1.21$ & $0.63$ & $2.14$ & $0.52$ & $-0.28$ & $0.01$ & $0.05$ & $-0.18$ \\ 
 
$1500$ & $0.1190$ & $2.06$ & $1.39$ & $0.65$ & $2.57$ & $0.59$ & $-0.20$ & $0.01$ & $0.05$ & $-0.17$ \\ 
 
$2000$ & $0.5470 \cdot 10^{-1}$ & $2.45$ & $1.39$ & $0.57$ & $2.87$ & $0.49$ & $-0.23$ & $0.01$ & $0.04$ & $-0.17$ \\ 
 
$3000$ & $0.1830 \cdot 10^{-1}$ & $2.34$ & $1.60$ & $0.68$ & $2.92$ & $0.64$ & $-0.17$ & $0.01$ & $0.05$ & $-0.18$ \\ 
 
$5000$ & $0.4070 \cdot 10^{-2}$ & $3.27$ & $1.75$ & $0.55$ & $3.75$ & $0.46$ & $-0.12$ & $0.01$ & $0.04$ & $-0.28$ \\ 
 
$8000$ & $0.9380 \cdot 10^{-3}$ & $5.47$ & $2.49$ & $0.91$ & $6.08$ & $0.76$ & $-0.11$ & $0.02$ & $0.03$ & $-0.49$ \\ 
 
$12000$ & $0.2170 \cdot 10^{-3}$ & $10.30$ & $3.20$ & $0.91$ & $10.83$ & $0.76$ & $-0.04$ & $0.01$ & $0.05$ & $-0.49$ \\ 
 
$20000$ & $0.2920 \cdot 10^{-4}$ & $21.41$ & $4.38$ & $1.15$ & $21.89$ & $1.14$ & $0.07$ & $0.01$ & $0.05$ & $-0.13$ \\ 
 
\hline 
\end{tabular} 
\end{center} 
\caption[RESULT] 
{\label{tab:ncdq2_posLH} The NC 
$e^+p$ cross section ${\rm d}\sigma_{\rm NC}/{\rm d}Q^2$ for lepton beam polarisation $P_e=-37.0$\% 
and $y<0.9$. 
The 
statistical $(\delta_{\rm stat})$, 
uncorrelated systematic $(\delta_{\rm unc})$, 
correlated systematic $(\delta_{\rm cor})$ 
and total $(\delta_{\rm tot})$ errors are provided. 
In addition the correlated systematic  
error contributions from a 
positive variation of one  
standard deviation of the 
electron energy error ($\delta_{\rm cor}^{E^+}$), of 
the polar electron angle error 
($\delta_{\rm cor}^{\theta^+}$), of the hadronic 
energy error ($\delta_{\rm cor}^{h^+}$), of the error 
due to noise subtraction ($\delta_{\rm cor}^{N^+}$) and 
of the error due to background subtraction 
($\delta_{\rm cor}^{B^+}$) are given. 
The normalisation and polarisation uncertainties are 
not included in the errors.} 
\end{table} 

\begin{table}[htb] 
\begin{center} 
\footnotesize 
\begin{tabular}{|r|l|r|r|r|r|r|r|r|r|r|} 
\hline 
$Q^2$                 & ${\rm d}\sigma_{\rm NC}/{\rm d}Q^2$      & 
$\delta_{\rm stat}$      & $\delta_{\rm unc}$         &   $\delta_{\rm cor}$          & 
$\delta_{\rm tot}$       & $\delta_{\rm cor}^{E^+}$   & $\delta_{\rm cor}^{\theta^+}$ & $\delta_{\rm cor}^{h^+}$& 
$\delta_{\rm cor}^{N^+}$ & $\delta_{\rm cor}^{B^+}$ \\ 
$(\rm GeV^2)$ &$(\rm pb / \rm GeV^2)$ & 
$(\%)$ & $(\%)$  & $(\%)$ & $(\%)$ & $(\%)$ & 
$(\%)$ & $(\%)$  & $(\%)$ & $(\%)$ \\ 
\hline 
$200$ & $17.84$ & $0.42$ & $0.80$ & $0.63$ & $1.10$ & $-0.23$ & $-0.57$ & $0.01$ & $0.05$ & $-0.07$ \\ 
 
$250$ & $10.49$ & $0.44$ & $1.14$ & $0.69$ & $1.40$ & $0.42$ & $-0.55$ & $0.00$ & $0.04$ & $-0.05$ \\ 
 
$300$ & $6.814$ & $0.52$ & $1.15$ & $0.70$ & $1.44$ & $0.46$ & $-0.53$ & $0.00$ & $0.04$ & $-0.04$ \\ 
 
$400$ & $3.392$ & $0.61$ & $0.96$ & $0.61$ & $1.29$ & $0.45$ & $-0.41$ & $0.00$ & $0.04$ & $-0.03$ \\ 
 
$500$ & $2.004$ & $0.72$ & $0.98$ & $0.65$ & $1.38$ & $0.51$ & $-0.40$ & $0.01$ & $0.04$ & $-0.03$ \\ 
 
$650$ & $1.059$ & $0.87$ & $1.03$ & $0.66$ & $1.50$ & $0.54$ & $-0.38$ & $0.01$ & $0.05$ & $-0.03$ \\ 
 
$800$ & $0.6260$ & $1.04$ & $1.03$ & $0.63$ & $1.60$ & $0.53$ & $-0.34$ & $0.01$ & $0.06$ & $-0.03$ \\ 
 
$1000$ & $0.3490$ & $1.26$ & $1.02$ & $0.54$ & $1.71$ & $0.42$ & $-0.28$ & $0.01$ & $0.06$ & $-0.21$ \\ 
 
$1200$ & $0.2180$ & $1.49$ & $1.21$ & $0.57$ & $2.00$ & $0.48$ & $-0.26$ & $0.01$ & $0.05$ & $-0.17$ \\ 
 
$1500$ & $0.1220$ & $1.75$ & $1.28$ & $0.58$ & $2.25$ & $0.52$ & $-0.22$ & $0.01$ & $0.06$ & $-0.14$ \\ 
 
$2000$ & $0.5850 \cdot 10^{-1}$ & $2.15$ & $1.43$ & $0.64$ & $2.66$ & $0.57$ & $-0.23$ & $0.01$ & $0.05$ & $-0.16$ \\ 
 
$3000$ & $0.2020 \cdot 10^{-1}$ & $2.06$ & $1.54$ & $0.59$ & $2.64$ & $0.54$ & $-0.13$ & $0.01$ & $0.04$ & $-0.17$ \\ 
 
$5000$ & $0.4920 \cdot 10^{-2}$ & $2.68$ & $1.76$ & $0.51$ & $3.25$ & $0.44$ & $-0.15$ & $0.01$ & $0.04$ & $-0.20$ \\ 
 
$8000$ & $0.1140 \cdot 10^{-2}$ & $4.75$ & $2.39$ & $0.74$ & $5.37$ & $0.68$ & $-0.04$ & $0.01$ & $0.04$ & $-0.28$ \\ 
 
$12000$ & $0.2750 \cdot 10^{-3}$ & $8.24$ & $2.91$ & $0.82$ & $8.77$ & $0.74$ & $-0.13$ & $0.01$ & $0.05$ & $-0.33$ \\ 
 
$20000$ & $0.4080 \cdot 10^{-4}$ & $16.20$ & $4.65$ & $1.14$ & $16.90$ & $1.08$ & $0.20$ & $0.03$ & $0.06$ & $-0.29$ \\ 
 
$30000$ & $0.6680 \cdot 10^{-5}$ & $41.62$ & $6.22$ & $1.46$ & $42.11$ & $1.45$ & $0.05$ & $0.01$ & $0.04$ & $-0.15$ \\ 
 
\hline 
\end{tabular} 
\end{center} 
\caption[RESULT] 
{\label{tab:ncdq2_posRH} The NC 
$e^+p$ cross section ${\rm d}\sigma_{\rm NC}/{\rm d}Q^2$ for lepton beam polarisation $P_e=+32.5$\% 
and $y<0.9$. 
The 
statistical $(\delta_{\rm stat})$, 
uncorrelated systematic $(\delta_{\rm unc})$, 
correlated systematic $(\delta_{\rm cor})$ 
and total $(\delta_{\rm tot})$ errors are provided. 
In addition the correlated systematic  
error contributions from a 
positive variation of one  
standard deviation of the 
electron energy error ($\delta_{\rm cor}^{E^+}$), of 
the polar electron angle error 
($\delta_{\rm cor}^{\theta^+}$), of the hadronic 
energy error ($\delta_{\rm cor}^{h^+}$), of the error 
due to noise subtraction ($\delta_{\rm cor}^{N^+}$) and 
of the error due to background subtraction 
($\delta_{\rm cor}^{B^+}$) are given. 
The normalisation and polarisation uncertainties are 
not included in the errors.} 
\end{table}

\clearpage
\newpage
\begin{table}[htb] 
\begin{center} 
\footnotesize 
\renewcommand{\arraystretch}{1.22}
\begin{tabular}{|r|l|l|r|r|r|r|r|r|r|r|} 
\hline 
$Q^2$                 & ${\rm d}\sigma_{\rm CC}/{\rm d}Q^2$      &   $k_{\rm cor}$           & 
$\delta_{\rm stat}$      & $\delta_{\rm unc}$         &   $\delta_{\rm cor}$          & 
$\delta_{\rm tot}$       & $\delta_{\rm cor}^{V^+}$   &   $\delta_{\rm cor}^{h^+}$& 
$\delta_{\rm cor}^{N^+}$ & $\delta_{\rm cor}^{B^+}$ \\ 
$(\rm GeV^2)$ &$(\rm pb / \rm GeV^2)$ &   & 
$(\%)$ & $(\%)$  & $(\%)$ & $(\%)$ & $(\%)$ & 
$(\%)$ & $(\%)$  & $(\%)$ \\ 
\hline 
$300$ & $0.426 \cdot 10^{-1}$ & $1.493$ & $8.7$ & $7.5$ & $6.4$ & $13.1$ & $5.7$ & $-0.5$ & $-1.3$ & $-2.2$ \\ 
 
$500$ & $0.282 \cdot 10^{-1}$ & $1.246$ & $4.9$ & $4.4$ & $4.0$ & $7.8$ & $3.7$ & $-0.6$ & $-0.9$ & $-0.3$ \\ 
 
$1000$ & $0.203 \cdot 10^{-1}$ & $1.071$ & $3.8$ & $3.2$ & $2.2$ & $5.5$ & $2.1$ & $-0.4$ & $-0.5$ & $-0.1$ \\ 
 
$2000$ & $0.118 \cdot 10^{-1}$ & $1.024$ & $3.3$ & $2.5$ & $1.0$ & $4.3$ & $0.9$ & $-0.2$ & $-0.2$ & $-0.0$ \\ 
 
$3000$ & $0.713 \cdot 10^{-2}$ & $1.027$ & $3.4$ & $2.1$ & $0.6$ & $4.1$ & $0.2$ & $-0.0$ & $0.3$ & $-0.1$ \\ 
 
$5000$ & $0.374 \cdot 10^{-2}$ & $1.033$ & $3.7$ & $2.3$ & $0.7$ & $4.4$ & $-0.0$ & $0.1$ & $0.4$ & $-0.0$ \\ 
 
$8000$ & $0.168 \cdot 10^{-2}$ & $1.049$ & $4.7$ & $3.5$ & $1.6$ & $6.2$ & $-0.0$ & $0.8$ & $1.2$ & $0.0$ \\ 
 
$15000$ & $0.425 \cdot 10^{-3}$ & $1.082$ & $6.5$ & $5.9$ & $2.7$ & $9.2$ & $0.0$ & $1.6$ & $1.9$ & $0.0$ \\ 
 
$30000$ & $0.388 \cdot 10^{-4}$ & $1.195$ & $14.5$ & $11.4$ & $5.3$ & $19.2$ & $0.0$ & $3.3$ & $3.3$ & $0.0$ \\ 
 
\hline 
\end{tabular} 
\end{center} 
\caption[RESULT] 
{\label{tab:ccdq2_eleLH} The CC 
$e^-p$ cross section ${\rm d}\sigma_{\rm CC}/{\rm d}Q^2$ for lepton beam polarisation $P_e=-25.8$\% 
and $y<0.9$ after 
correction ($k_{\rm cor}$) according to the  
Standard Model expectation for 
the kinematic cuts $P_{T,h}>12$~GeV 
and $0.03<y<0.85$. The 
statistical $(\delta_{\rm stat})$, 
uncorrelated systematic $(\delta_{\rm unc})$, 
correlated systematic $(\delta_{\rm cor})$ 
and total $(\delta_{\rm tot})$ errors are provided. 
In addition the correlated systematic  
error contributions from a 
positive variation of one  
standard deviation of the 
cuts against photoproduction ($\delta_{\rm cor}^{V^+}$), of 
the hadronic 
energy error ($\delta_{\rm cor}^{h^+}$), of the error 
due to noise subtraction ($\delta_{\rm cor}^{N^+}$) and 
of the error due to background subtraction 
($\delta_{\rm cor}^{B^+}$) are given. 
The luminosity and 
polarisation uncertainties are not included in the errors.} 
\end{table} 

\begin{table}[htb] 
\begin{center} 
\footnotesize 
\renewcommand{\arraystretch}{1.22}
\begin{tabular}{|r|l|l|r|r|r|r|r|r|r|r|} 
\hline 
$Q^2$                 & ${\rm d}\sigma_{\rm CC}/{\rm d}Q^2$      &   $k_{\rm cor}$           & 
$\delta_{\rm stat}$      & $\delta_{\rm unc}$         &   $\delta_{\rm cor}$          & 
$\delta_{\rm tot}$       & $\delta_{\rm cor}^{V^+}$   &   $\delta_{\rm cor}^{h^+}$& 
$\delta_{\rm cor}^{N^+}$ & $\delta_{\rm cor}^{B^+}$ \\ 
$(\rm GeV^2)$ &$(\rm pb / \rm GeV^2)$ &   & 
$(\%)$ & $(\%)$  & $(\%)$ & $(\%)$ & $(\%)$ & 
$(\%)$ & $(\%)$  & $(\%)$ \\ 
\hline 
$300$ & $0.207 \cdot 10^{-1}$ & $1.493$ & $15.1$ & $7.6$ & $6.4$ & $18.1$ & $5.7$ & $-0.5$ & $-0.8$ & $-1.9$ \\ 
 
$500$ & $0.159 \cdot 10^{-1}$ & $1.246$ & $10.0$ & $4.6$ & $4.1$ & $11.7$ & $3.8$ & $-0.6$ & $-1.0$ & $-0.8$ \\ 
 
$1000$ & $0.110 \cdot 10^{-1}$ & $1.071$ & $7.6$ & $3.0$ & $2.1$ & $8.5$ & $2.0$ & $-0.3$ & $-0.1$ & $-0.2$ \\ 
 
$2000$ & $0.528 \cdot 10^{-2}$ & $1.024$ & $7.4$ & $2.5$ & $1.2$ & $7.9$ & $0.8$ & $-0.4$ & $-0.5$ & $-0.2$ \\ 
 
$3000$ & $0.366 \cdot 10^{-2}$ & $1.027$ & $7.0$ & $2.2$ & $0.7$ & $7.4$ & $0.2$ & $-0.0$ & $0.3$ & $-0.1$ \\ 
 
$5000$ & $0.190 \cdot 10^{-2}$ & $1.033$ & $7.5$ & $2.4$ & $0.9$ & $8.0$ & $-0.0$ & $0.3$ & $0.6$ & $-0.0$ \\ 
 
$8000$ & $0.838 \cdot 10^{-3}$ & $1.049$ & $9.8$ & $3.3$ & $1.5$ & $10.5$ & $-0.0$ & $0.8$ & $1.0$ & $0.0$ \\ 
 
$15000$ & $0.215 \cdot 10^{-3}$ & $1.082$ & $14.0$ & $5.8$ & $2.7$ & $15.5$ & $-0.0$ & $1.5$ & $2.0$ & $0.0$ \\ 
 
$30000$ & $0.188 \cdot 10^{-4}$ & $1.195$ & $28.9$ & $10.8$ & $4.8$ & $31.3$ & $0.0$ & $3.0$ & $2.8$ & $0.0$ \\ 
 
\hline 
\end{tabular} 
\end{center} 
\caption[RESULT] 
{\label{tab:ccdq2_eleRH} The CC 
$e^-p$ cross section ${\rm d}\sigma_{\rm CC}/{\rm d}Q^2$ for lepton beam polarisation $P_e=+36.0$\% 
and $y<0.9$ after 
correction ($k_{\rm cor}$) according to the  
Standard Model expectation for 
the kinematic cuts $P_{T,h}>12$~GeV 
and $0.03<y<0.85$. The 
statistical $(\delta_{\rm stat})$, 
uncorrelated systematic $(\delta_{\rm unc})$, 
correlated systematic $(\delta_{\rm cor})$ 
and total $(\delta_{\rm tot})$ errors are provided. 
In addition the correlated systematic  
error contributions from a 
positive variation of one  
standard deviation of the 
cuts against photoproduction ($\delta_{\rm cor}^{V^+}$), of 
the hadronic 
energy error ($\delta_{\rm cor}^{h^+}$), of the error 
due to noise subtraction ($\delta_{\rm cor}^{N^+}$) and 
of the error due to background subtraction 
($\delta_{\rm cor}^{B^+}$) are given. 
The luminosity and 
polarisation uncertainties are not included in the errors.} 
\end{table} 

\begin{table}[htb] 
\begin{center} 
\footnotesize 
\renewcommand{\arraystretch}{1.22}
\begin{tabular}{|r|l|l|r|r|r|r|r|r|r|r|} 
\hline 
$Q^2$                 & ${\rm d}\sigma_{\rm CC}/{\rm d}Q^2$      &   $k_{\rm cor}$           & 
$\delta_{\rm stat}$      & $\delta_{\rm unc}$         &   $\delta_{\rm cor}$          & 
$\delta_{\rm tot}$       & $\delta_{\rm cor}^{V^+}$   &   $\delta_{\rm cor}^{h^+}$& 
$\delta_{\rm cor}^{N^+}$ & $\delta_{\rm cor}^{B^+}$ \\ 
$(\rm GeV^2)$ &$(\rm pb / \rm GeV^2)$ &   & 
$(\%)$ & $(\%)$  & $(\%)$ & $(\%)$ & $(\%)$ & 
$(\%)$ & $(\%)$  & $(\%)$ \\ 
\hline 
$300$ & $0.157 \cdot 10^{-1}$ & $1.408$ & $14.2$ & $8.0$ & $8.0$ & $18.2$ & $6.2$ & $-0.6$ & $-0.8$ & $-4.4$ \\ 
 
$500$ & $0.124 \cdot 10^{-1}$ & $1.181$ & $8.9$ & $4.3$ & $4.7$ & $11.0$ & $4.3$ & $-0.4$ & $0.2$ & $-1.5$ \\ 
 
$1000$ & $0.761 \cdot 10^{-2}$ & $1.043$ & $7.1$ & $3.0$ & $2.4$ & $8.1$ & $2.3$ & $-0.2$ & $-0.1$ & $-0.4$ \\ 
 
$2000$ & $0.327 \cdot 10^{-2}$ & $1.026$ & $7.0$ & $2.4$ & $1.2$ & $7.5$ & $1.0$ & $0.0$ & $0.3$ & $-0.1$ \\ 
 
$3000$ & $0.177 \cdot 10^{-2}$ & $1.030$ & $7.8$ & $2.6$ & $1.1$ & $8.3$ & $0.3$ & $0.4$ & $0.8$ & $-0.0$ \\ 
 
$5000$ & $0.600 \cdot 10^{-3}$ & $1.035$ & $10.6$ & $3.1$ & $1.6$ & $11.3$ & $-0.1$ & $0.6$ & $1.3$ & $-0.0$ \\ 
 
$8000$ & $0.184 \cdot 10^{-3}$ & $1.049$ & $15.8$ & $5.7$ & $2.7$ & $17.3$ & $-0.1$ & $1.4$ & $1.9$ & $0.0$ \\ 
 
$15000$ & $0.198 \cdot 10^{-4}$ & $1.077$ & $30.7$ & $7.0$ & $3.9$ & $31.9$ & $-0.0$ & $2.4$ & $2.1$ & $0.0$ \\ 
 
\hline 
\end{tabular} 
\end{center} 
\caption[RESULT] 
{\label{tab:ccdq2_posLH} The CC 
$e^+p$ cross section ${\rm d}\sigma_{\rm CC}/{\rm d}Q^2$ for lepton beam polarisation $P_e=-37.0$\% 
and $y<0.9$ after 
correction ($k_{\rm cor}$) according to the  
Standard Model expectation for 
the kinematic cuts $P_{T,h}>12$~GeV 
and $0.03<y<0.85$. The 
statistical $(\delta_{\rm stat})$, 
uncorrelated systematic $(\delta_{\rm unc})$, 
correlated systematic $(\delta_{\rm cor})$ 
and total $(\delta_{\rm tot})$ errors are provided. 
In addition the correlated systematic  
error contributions from a 
positive variation of one  
standard deviation of the 
cuts against photoproduction ($\delta_{\rm cor}^{V^+}$), of 
the hadronic 
energy error ($\delta_{\rm cor}^{h^+}$), of the error 
due to noise subtraction ($\delta_{\rm cor}^{N^+}$) and 
of the error due to background subtraction 
($\delta_{\rm cor}^{B^+}$) are given. 
The luminosity and 
polarisation uncertainties are not included in the errors.} 
\end{table} 

\begin{table}[htb] 
\begin{center} 
\footnotesize 
\renewcommand{\arraystretch}{1.22}
\begin{tabular}{|r|l|l|r|r|r|r|r|r|r|r|} 
\hline 
$Q^2$                 & ${\rm d}\sigma_{\rm CC}/{\rm d}Q^2$      &   $k_{\rm cor}$           & 
$\delta_{\rm stat}$      & $\delta_{\rm unc}$         &   $\delta_{\rm cor}$          & 
$\delta_{\rm tot}$       & $\delta_{\rm cor}^{V^+}$   &   $\delta_{\rm cor}^{h^+}$& 
$\delta_{\rm cor}^{N^+}$ & $\delta_{\rm cor}^{B^+}$ \\ 
$(\rm GeV^2)$ &$(\rm pb / \rm GeV^2)$ &   & 
$(\%)$ & $(\%)$  & $(\%)$ & $(\%)$ & $(\%)$ & 
$(\%)$ & $(\%)$  & $(\%)$ \\ 
\hline 
$300$ & $0.294 \cdot 10^{-1}$ & $1.408$ & $8.8$ & $8.0$ & $7.1$ & $13.8$ & $6.5$ & $-0.4$ & $-0.5$ & $-2.5$ \\ 
 
$500$ & $0.259 \cdot 10^{-1}$ & $1.181$ & $5.3$ & $4.4$ & $4.4$ & $8.2$ & $4.2$ & $-0.4$ & $-0.5$ & $-0.4$ \\ 
 
$1000$ & $0.145 \cdot 10^{-1}$ & $1.043$ & $4.5$ & $3.0$ & $2.4$ & $5.9$ & $2.3$ & $-0.4$ & $-0.0$ & $-0.1$ \\ 
 
$2000$ & $0.707 \cdot 10^{-2}$ & $1.026$ & $4.2$ & $2.4$ & $1.2$ & $5.0$ & $1.0$ & $0.1$ & $0.4$ & $-0.1$ \\ 
 
$3000$ & $0.382 \cdot 10^{-2}$ & $1.030$ & $4.7$ & $2.5$ & $1.1$ & $5.5$ & $0.3$ & $0.4$ & $0.8$ & $0.0$ \\ 
 
$5000$ & $0.136 \cdot 10^{-2}$ & $1.035$ & $6.2$ & $3.8$ & $1.9$ & $7.7$ & $-0.0$ & $0.8$ & $1.5$ & $-0.0$ \\ 
 
$8000$ & $0.438 \cdot 10^{-3}$ & $1.049$ & $9.1$ & $5.1$ & $2.5$ & $11.1$ & $-0.1$ & $1.4$ & $1.7$ & $0.0$ \\ 
 
$15000$ & $0.544 \cdot 10^{-4}$ & $1.077$ & $16.3$ & $7.3$ & $3.6$ & $18.5$ & $0.1$ & $1.8$ & $2.2$ & $0.0$ \\ 
 
\hline 
\end{tabular} 
\end{center} 
\caption[RESULT] 
{\label{tab:ccdq2_posRH} The CC 
$e^+p$ cross section ${\rm d}\sigma_{\rm CC}/{\rm d}Q^2$ for lepton beam polarisation $P_e=+32.5$\% 
and $y<0.9$ after 
correction ($k_{\rm cor}$) according to the  
Standard Model expectation for 
the kinematic cuts $P_{T,h}>12$~GeV 
and $0.03<y<0.85$. The 
statistical $(\delta_{\rm stat})$, 
uncorrelated systematic $(\delta_{\rm unc})$, 
correlated systematic $(\delta_{\rm cor})$ 
and total $(\delta_{\rm tot})$ errors are provided. 
In addition the correlated systematic  
error contributions from a 
positive variation of one  
standard deviation of the 
cuts against photoproduction ($\delta_{\rm cor}^{V^+}$), of 
the hadronic 
energy error ($\delta_{\rm cor}^{h^+}$), of the error 
due to noise subtraction ($\delta_{\rm cor}^{N^+}$) and 
of the error due to background subtraction 
($\delta_{\rm cor}^{B^+}$) are given. 
The luminosity and 
polarisation uncertainties are not included in the errors.} 
\end{table}

\clearpage
\begin{table}[htb] 
\begin{center} 
\footnotesize 
\begin{tabular}{|r|l|r|r|r|r|r|r|r|r|r|} 
\hline 
$Q^2$                 & ${\rm d}\sigma_{\rm NC}/{\rm d}Q^2$      & 
$\delta_{\rm stat}$      & $\delta_{\rm unc}$         &   $\delta_{\rm cor}$          & 
$\delta_{\rm tot}$       & $\delta_{\rm cor}^{E^+}$   & $\delta_{\rm cor}^{\theta^+}$ & $\delta_{\rm cor}^{h^+}$& 
$\delta_{\rm cor}^{N^+}$ & $\delta_{\rm cor}^{B^+}$ \\ 
$(\rm GeV^2)$ &$(\rm pb / \rm GeV^2)$ & 
$(\%)$ & $(\%)$  & $(\%)$ & $(\%)$ & $(\%)$ & 
$(\%)$ & $(\%)$  & $(\%)$ & $(\%)$ \\ 
\hline 
$200$ & $17.85$ & $0.35$ & $0.91$ & $0.61$ & $1.15$ & $-0.21$ & $-0.57$ & $0.00$ & $0.05$ & $-0.06$ \\ 
 
$250$ & $10.50$ & $0.36$ & $1.21$ & $0.66$ & $1.42$ & $0.39$ & $-0.53$ & $0.00$ & $0.04$ & $-0.04$ \\ 
 
$300$ & $6.785$ & $0.42$ & $1.24$ & $0.75$ & $1.51$ & $0.50$ & $-0.55$ & $0.01$ & $0.04$ & $-0.04$ \\ 
 
$400$ & $3.401$ & $0.50$ & $1.05$ & $0.61$ & $1.31$ & $0.45$ & $-0.41$ & $0.01$ & $0.04$ & $-0.03$ \\ 
 
$500$ & $1.999$ & $0.59$ & $1.08$ & $0.64$ & $1.39$ & $0.51$ & $-0.39$ & $0.01$ & $0.04$ & $-0.02$ \\ 
 
$650$ & $1.050$ & $0.72$ & $1.09$ & $0.63$ & $1.45$ & $0.52$ & $-0.35$ & $0.01$ & $0.05$ & $-0.02$ \\ 
 
$800$ & $0.6210$ & $0.87$ & $1.07$ & $0.57$ & $1.49$ & $0.48$ & $-0.31$ & $0.01$ & $0.05$ & $-0.02$ \\ 
 
$1000$ & $0.3580$ & $1.02$ & $1.13$ & $0.60$ & $1.63$ & $0.47$ & $-0.31$ & $0.01$ & $0.05$ & $-0.19$ \\ 
 
$1200$ & $0.2240$ & $1.21$ & $1.27$ & $0.55$ & $1.84$ & $0.47$ & $-0.24$ & $0.01$ & $0.04$ & $-0.16$ \\ 
 
$1500$ & $0.1280$ & $1.45$ & $1.34$ & $0.58$ & $2.06$ & $0.51$ & $-0.21$ & $0.01$ & $0.05$ & $-0.16$ \\ 
 
$2000$ & $0.5930 \cdot 10^{-1}$ & $1.75$ & $1.43$ & $0.59$ & $2.34$ & $0.52$ & $-0.22$ & $0.01$ & $0.05$ & $-0.19$ \\ 
 
$3000$ & $0.2120 \cdot 10^{-1}$ & $1.64$ & $1.51$ & $0.54$ & $2.29$ & $0.50$ & $-0.16$ & $0.01$ & $0.04$ & $-0.14$ \\ 
 
$5000$ & $0.5610 \cdot 10^{-2}$ & $2.11$ & $1.71$ & $0.46$ & $2.76$ & $0.41$ & $-0.14$ & $0.01$ & $0.04$ & $-0.15$ \\ 
 
$8000$ & $0.1520 \cdot 10^{-2}$ & $3.19$ & $2.30$ & $0.61$ & $3.98$ & $0.55$ & $-0.07$ & $0.01$ & $0.04$ & $-0.24$ \\ 
 
$12000$ & $0.5190 \cdot 10^{-3}$ & $4.88$ & $2.71$ & $0.72$ & $5.63$ & $0.67$ & $0.00$ & $0.02$ & $0.06$ & $-0.24$ \\ 
 
$20000$ & $0.1030 \cdot 10^{-3}$ & $8.53$ & $3.86$ & $1.02$ & $9.42$ & $0.89$ & $0.11$ & $0.02$ & $0.04$ & $-0.49$ \\ 
 
$30000$ & $0.1920 \cdot 10^{-4}$ & $20.43$ & $6.20$ & $2.10$ & $21.46$ & $1.58$ & $0.25$ & $0.02$ & $0.08$ & $-1.36$ \\ 
 
$50000$ & $0.1770 \cdot 10^{-5}$ & $57.80$ & $9.87$ & $2.73$ & $58.70$ & $2.66$ & $0.60$ & $0.00$ & $0.00$ & $0.00$ \\ 
 
\hline 
\end{tabular} 
\end{center} 
\caption[RESULT] 
{\label{tab:ncdq2_eleP0} The NC 
$e^-p$ cross section ${\rm d}\sigma_{\rm NC}/{\rm d}Q^2$ for $P_e=0$ 
and $y<0.9$. 
The 
statistical $(\delta_{\rm stat})$, 
uncorrelated systematic $(\delta_{\rm unc})$, 
correlated systematic $(\delta_{\rm cor})$ 
and total $(\delta_{\rm tot})$ errors are provided. 
In addition the correlated systematic  
error contributions from a 
positive variation of one  
standard deviation of the 
electron energy error ($\delta_{\rm cor}^{E^+}$), of 
the polar electron angle error 
($\delta_{\rm cor}^{\theta^+}$), of the hadronic 
energy error ($\delta_{\rm cor}^{h^+}$), of the error 
due to noise subtraction ($\delta_{\rm cor}^{N^+}$) and 
of the error due to background subtraction 
($\delta_{\rm cor}^{B^+}$) are given. 
The normalisation and polarisation uncertainties are 
not included in the errors.} 
\end{table} 

\begin{table}[htb] 
\begin{center} 
\footnotesize 
\begin{tabular}{|r|l|r|r|r|r|r|r|r|r|r|} 
\hline 
$Q^2$                 & ${\rm d}\sigma_{\rm NC}/{\rm d}Q^2$      & 
$\delta_{\rm stat}$      & $\delta_{\rm unc}$         &   $\delta_{\rm cor}$          & 
$\delta_{\rm tot}$       & $\delta_{\rm cor}^{E^+}$   & $\delta_{\rm cor}^{\theta^+}$ & $\delta_{\rm cor}^{h^+}$& 
$\delta_{\rm cor}^{N^+}$ & $\delta_{\rm cor}^{B^+}$ \\ 
$(\rm GeV^2)$ &$(\rm pb / \rm GeV^2)$ & 
$(\%)$ & $(\%)$  & $(\%)$ & $(\%)$ & $(\%)$ & 
$(\%)$ & $(\%)$  & $(\%)$ & $(\%)$ \\ 
\hline 
$200$ & $17.98$ & $0.31$ & $0.80$ & $0.64$ & $1.07$ & $-0.23$ & $-0.59$ & $0.01$ & $0.05$ & $-0.07$ \\ 
 
$250$ & $10.54$ & $0.33$ & $1.13$ & $0.68$ & $1.35$ & $0.41$ & $-0.54$ & $0.00$ & $0.04$ & $-0.05$ \\ 
 
$300$ & $6.848$ & $0.38$ & $1.14$ & $0.71$ & $1.40$ & $0.47$ & $-0.53$ & $0.00$ & $0.04$ & $-0.04$ \\ 
 
$400$ & $3.403$ & $0.45$ & $0.96$ & $0.62$ & $1.23$ & $0.46$ & $-0.42$ & $0.00$ & $0.04$ & $-0.03$ \\ 
 
$500$ & $1.999$ & $0.54$ & $0.97$ & $0.65$ & $1.28$ & $0.51$ & $-0.40$ & $0.01$ & $0.04$ & $-0.03$ \\ 
 
$650$ & $1.045$ & $0.65$ & $1.01$ & $0.67$ & $1.38$ & $0.54$ & $-0.39$ & $0.01$ & $0.04$ & $-0.02$ \\ 
 
$800$ & $0.6150$ & $0.79$ & $1.02$ & $0.65$ & $1.44$ & $0.54$ & $-0.35$ & $0.01$ & $0.05$ & $-0.03$ \\ 
 
$1000$ & $0.3440$ & $0.94$ & $1.00$ & $0.54$ & $1.47$ & $0.42$ & $-0.27$ & $0.01$ & $0.05$ & $-0.20$ \\ 
 
$1200$ & $0.2150$ & $1.11$ & $1.19$ & $0.59$ & $1.73$ & $0.50$ & $-0.27$ & $0.01$ & $0.05$ & $-0.17$ \\ 
 
$1500$ & $0.1200$ & $1.34$ & $1.31$ & $0.61$ & $1.97$ & $0.55$ & $-0.22$ & $0.01$ & $0.05$ & $-0.15$ \\ 
 
$2000$ & $0.5670 \cdot 10^{-1}$ & $1.62$ & $1.38$ & $0.61$ & $2.21$ & $0.54$ & $-0.23$ & $0.01$ & $0.05$ & $-0.17$ \\ 
 
$3000$ & $0.1930 \cdot 10^{-1}$ & $1.55$ & $1.55$ & $0.63$ & $2.28$ & $0.58$ & $-0.15$ & $0.01$ & $0.05$ & $-0.17$ \\ 
 
$5000$ & $0.4520 \cdot 10^{-2}$ & $2.07$ & $1.73$ & $0.52$ & $2.75$ & $0.45$ & $-0.14$ & $0.01$ & $0.04$ & $-0.23$ \\ 
 
$8000$ & $0.1050 \cdot 10^{-2}$ & $3.60$ & $2.37$ & $0.81$ & $4.38$ & $0.71$ & $-0.07$ & $0.01$ & $0.03$ & $-0.37$ \\ 
 
$12000$ & $0.2480 \cdot 10^{-3}$ & $6.43$ & $2.93$ & $0.85$ & $7.12$ & $0.75$ & $-0.09$ & $0.01$ & $0.05$ & $-0.40$ \\ 
 
$20000$ & $0.3540 \cdot 10^{-4}$ & $12.92$ & $4.48$ & $1.14$ & $13.72$ & $1.10$ & $0.15$ & $0.02$ & $0.06$ & $-0.23$ \\ 
 
$30000$ & $0.4190 \cdot 10^{-5}$ & $38.94$ & $5.92$ & $1.47$ & $39.41$ & $1.44$ & $0.21$ & $0.01$ & $0.04$ & $-0.16$ \\ 
 
\hline 
\end{tabular} 
\end{center} 
\caption[RESULT] 
{\label{tab:ncdq2_posP0} The NC 
$e^+p$ cross section ${\rm d}\sigma_{\rm NC}/{\rm d}Q^2$ for $P_e=0$ 
and $y<0.9$. 
The 
statistical $(\delta_{\rm stat})$, 
uncorrelated systematic $(\delta_{\rm unc})$, 
correlated systematic $(\delta_{\rm cor})$ 
and total $(\delta_{\rm tot})$ errors are provided. 
In addition the correlated systematic  
error contributions from a 
positive variation of one  
standard deviation of the 
electron energy error ($\delta_{\rm cor}^{E^+}$), of 
the polar electron angle error 
($\delta_{\rm cor}^{\theta^+}$), of the hadronic 
energy error ($\delta_{\rm cor}^{h^+}$), of the error 
due to noise subtraction ($\delta_{\rm cor}^{N^+}$) and 
of the error due to background subtraction 
($\delta_{\rm cor}^{B^+}$) are given. 
The normalisation and polarisation uncertainties are 
not included in the errors.} 
\end{table} 

\begin{table}[htb] 
\begin{center} 
\renewcommand{\arraystretch}{1.22}
\footnotesize 
\begin{tabular}{|r|l|l|r|r|r|r|r|r|r|r|} 
\hline 
$Q^2$                 & ${\rm d}\sigma_{\rm CC}/{\rm d}Q^2$      &   $k_{\rm cor}$           & 
$\delta_{\rm stat}$      & $\delta_{\rm unc}$         &   $\delta_{\rm cor}$          & 
$\delta_{\rm tot}$       & $\delta_{\rm cor}^{V^+}$   &   $\delta_{\rm cor}^{h^+}$& 
$\delta_{\rm cor}^{N^+}$ & $\delta_{\rm cor}^{B^+}$ \\ 
$(\rm GeV^2)$ &$(\rm pb / \rm GeV^2)$ &   & 
$(\%)$ & $(\%)$  & $(\%)$ & $(\%)$ & $(\%)$ & 
$(\%)$ & $(\%)$  & $(\%)$ \\ 
\hline 
$300$ & $0.336 \cdot 10^{-1}$ & $1.493$ & $7.6$ & $7.5$ & $6.4$ & $12.4$ & $5.7$ & $-0.5$ & $-1.2$ & $-2.2$ \\ 
 
$500$ & $0.229 \cdot 10^{-1}$ & $1.246$ & $4.4$ & $4.4$ & $4.0$ & $7.4$ & $3.7$ & $-0.6$ & $-0.9$ & $-0.4$ \\ 
 
$1000$ & $0.163 \cdot 10^{-1}$ & $1.071$ & $3.4$ & $3.2$ & $2.2$ & $5.2$ & $2.0$ & $-0.4$ & $-0.4$ & $-0.1$ \\ 
 
$2000$ & $0.917 \cdot 10^{-2}$ & $1.024$ & $3.0$ & $2.5$ & $1.0$ & $4.1$ & $0.9$ & $-0.2$ & $-0.2$ & $-0.0$ \\ 
 
$3000$ & $0.568 \cdot 10^{-2}$ & $1.027$ & $3.0$ & $2.1$ & $0.6$ & $3.8$ & $0.2$ & $-0.0$ & $0.3$ & $-0.1$ \\ 
 
$5000$ & $0.297 \cdot 10^{-2}$ & $1.033$ & $3.3$ & $2.3$ & $0.7$ & $4.2$ & $-0.0$ & $0.2$ & $0.4$ & $-0.0$ \\ 
 
$8000$ & $0.133 \cdot 10^{-2}$ & $1.049$ & $4.2$ & $3.5$ & $1.5$ & $5.8$ & $-0.0$ & $0.8$ & $1.1$ & $0.0$ \\ 
 
$15000$ & $0.337 \cdot 10^{-3}$ & $1.082$ & $5.9$ & $5.9$ & $2.6$ & $8.8$ & $-0.0$ & $1.6$ & $1.9$ & $0.0$ \\ 
 
$30000$ & $0.305 \cdot 10^{-4}$ & $1.195$ & $12.9$ & $11.3$ & $5.0$ & $17.9$ & $0.0$ & $3.2$ & $3.2$ & $0.0$ \\ 
 
\hline 
\end{tabular} 
\end{center} 
\caption[RESULT] 
{\label{tab:ccdq2_eleP0} The CC 
$e^-p$ cross section ${\rm d}\sigma_{\rm CC}/{\rm d}Q^2$ for $P_e=0$ 
and $y<0.9$ after 
correction ($k_{\rm cor}$) according to the  
Standard Model expectation for 
the kinematic cuts $P_{T,h}>12$~GeV 
and $0.03<y<0.85$. The 
statistical $(\delta_{\rm stat})$, 
uncorrelated systematic $(\delta_{\rm unc})$, 
correlated systematic $(\delta_{\rm cor})$ 
and total $(\delta_{\rm tot})$ errors are provided. 
In addition the correlated systematic  
error contributions from a 
positive variation of one  
standard deviation of the 
cuts against photoproduction ($\delta_{\rm cor}^{V^+}$), of 
the hadronic 
energy error ($\delta_{\rm cor}^{h^+}$), of the error 
due to noise subtraction ($\delta_{\rm cor}^{N^+}$) and 
of the error due to background subtraction 
($\delta_{\rm cor}^{B^+}$) are given. 
The luminosity and 
polarisation uncertainties are not included in the errors.} 
\end{table} 

\begin{table}[htb] 
\begin{center} 
\renewcommand{\arraystretch}{1.22}
\footnotesize 
\begin{tabular}{|r|l|l|r|r|r|r|r|r|r|r|} 
\hline 
$Q^2$                 & ${\rm d}\sigma_{\rm CC}/{\rm d}Q^2$      &   $k_{\rm cor}$           & 
$\delta_{\rm stat}$      & $\delta_{\rm unc}$         &   $\delta_{\rm cor}$          & 
$\delta_{\rm tot}$       & $\delta_{\rm cor}^{V^+}$   &   $\delta_{\rm cor}^{h^+}$& 
$\delta_{\rm cor}^{N^+}$ & $\delta_{\rm cor}^{B^+}$ \\ 
$(\rm GeV^2)$ &$(\rm pb / \rm GeV^2)$ &   & 
$(\%)$ & $(\%)$  & $(\%)$ & $(\%)$ & $(\%)$ & 
$(\%)$ & $(\%)$  & $(\%)$ \\ 
\hline 
$300$ & $0.229 \cdot 10^{-1}$ & $1.408$ & $7.5$ & $8.0$ & $7.3$ & $13.1$ & $6.4$ & $-0.4$ & $-0.6$ & $-3.0$ \\ 
 
$500$ & $0.196 \cdot 10^{-1}$ & $1.181$ & $4.6$ & $4.3$ & $4.4$ & $7.7$ & $4.3$ & $-0.4$ & $-0.4$ & $-0.7$ \\ 
 
$1000$ & $0.113 \cdot 10^{-1}$ & $1.043$ & $3.8$ & $3.0$ & $2.3$ & $5.4$ & $2.3$ & $-0.3$ & $-0.1$ & $-0.2$ \\ 
 
$2000$ & $0.529 \cdot 10^{-2}$ & $1.026$ & $3.6$ & $2.4$ & $1.2$ & $4.5$ & $1.0$ & $0.0$ & $0.4$ & $-0.1$ \\ 
 
$3000$ & $0.287 \cdot 10^{-2}$ & $1.030$ & $4.0$ & $2.6$ & $1.0$ & $4.9$ & $0.3$ & $0.4$ & $0.8$ & $-0.0$ \\ 
 
$5000$ & $0.101 \cdot 10^{-2}$ & $1.035$ & $5.3$ & $3.6$ & $1.7$ & $6.9$ & $-0.0$ & $0.7$ & $1.4$ & $-0.0$ \\ 
 
$8000$ & $0.320 \cdot 10^{-3}$ & $1.049$ & $7.9$ & $5.2$ & $2.4$ & $10.2$ & $-0.1$ & $1.4$ & $1.8$ & $0.0$ \\ 
 
$15000$ & $0.384 \cdot 10^{-4}$ & $1.077$ & $14.4$ & $7.2$ & $3.4$ & $16.7$ & $-0.1$ & $2.0$ & $2.2$ & $0.0$ \\ 
 
\hline 
\end{tabular} 
\end{center} 
\caption[RESULT] 
{\label{tab:ccdq2_posP0} The CC 
$e^+p$ cross section ${\rm d}\sigma_{\rm CC}/{\rm d}Q^2$ for $P_e=0$ 
and $y<0.9$ after 
correction ($k_{\rm cor}$) according to the  
Standard Model expectation for 
the kinematic cuts $P_{T,h}>12$~GeV 
and $0.03<y<0.85$. The 
statistical $(\delta_{\rm stat})$, 
uncorrelated systematic $(\delta_{\rm unc})$, 
correlated systematic $(\delta_{\rm cor})$ 
and total $(\delta_{\rm tot})$ errors are provided. 
In addition the correlated systematic  
error contributions from a 
positive variation of one  
standard deviation of the 
cuts against photoproduction ($\delta_{\rm cor}^{V^+}$), of 
the hadronic 
energy error ($\delta_{\rm cor}^{h^+}$), of the error 
due to noise subtraction ($\delta_{\rm cor}^{N^+}$) and 
of the error due to background subtraction 
($\delta_{\rm cor}^{B^+}$) are given. 
The luminosity and 
polarisation uncertainties are not included in the errors.} 
\end{table}

\clearpage
\begin{table}[\tablepos] 
\begin{center} 
\footnotesize
\renewcommand{\arraystretch}{1.22}
\begin{tabular}{|r|l|r|r|r|r|} 
\hline 
$Q^2$ & ${\rm d}\sigma_{\rm NC}/{\rm d}Q^2$ & $\delta_{\rm stat}$  & $\delta_{\rm unc}$ & 
$\delta_{\rm cor}$& $\delta_{\rm tot}$       \\ 
  $(\rm GeV^2)$ & $(\rm pb/GeV^2)$ & (\%) & (\%) & (\%) & (\%) \\
\hline 
$   200 $ &$ 18.21 $ & $  0.34 $ &$  0.87 $ &$  1.70 $ &$ 1.94$  \\
$   250 $ &$ 10.72 $ & $  0.35 $ &$  1.13 $ &$  1.69 $ &$ 2.06$  \\
$   300 $ &$ 6.958 $ & $  0.40 $ &$  1.17 $ &$  1.71 $ &$ 2.11$  \\
$   400 $ &$ 3.485 $ & $  0.48 $ &$  0.98 $ &$  1.67 $ &$ 2.00$  \\
$   500 $ &$ 2.047 $ & $  0.56 $ &$  1.03 $ &$  1.71 $ &$ 2.07$  \\
$   650 $ &$ 1.077 $ & $  0.69 $ &$  1.04 $ &$  1.69 $ &$ 2.10$  \\
$   800 $ &$ 0.6402 $ & $  0.83 $ &$  1.02 $ &$  1.67 $ &$ 2.13$  \\
$  1000 $ &$ 0.3670 $ & $  0.97 $ &$  1.06 $ &$  1.68 $ &$ 2.21$  \\
$  1200 $ &$ 0.2310 $ & $  1.15 $ &$  1.18 $ &$  1.66 $ &$ 2.34$  \\
$  1500 $ &$ 0.1314 $ & $  1.38 $ &$  1.24 $ &$  1.67 $ &$ 2.49$  \\
$  2000 $ &$ 0.6081 \cdot 10^{-1}$ & $  1.66 $ &$  1.33 $ &$  1.69 $ &$ 2.72$  \\
$  3000 $ &$ 0.2201 \cdot 10^{-1}$ & $  1.55 $ &$  1.38 $ &$  1.66 $ &$ 2.66$  \\
$  5000 $ &$ 0.5857 \cdot 10^{-2}$ & $  1.98 $ &$  1.57 $ &$  1.65 $ &$ 3.01$  \\
$  8000 $ &$ 0.1596 \cdot 10^{-2}$ & $  2.98 $ &$  2.14 $ &$  1.69 $ &$ 4.04$  \\
$ 12000 $ &$ 0.5190 \cdot 10^{-3}$ & $  4.67 $ &$  2.53 $ &$  1.72 $ &$ 5.58$  \\
$ 20000 $ &$ 0.1049 \cdot 10^{-3}$ & $  8.12 $ &$  3.63 $ &$  1.83 $ &$ 9.08$  \\
$ 30000 $ &$ 0.1995 \cdot 10^{-4}$ & $ 19.00 $ &$  5.99 $ &$  2.41 $ &$20.06$  \\
$ 50000 $ &$ 0.1839 \cdot 10^{-5}$ & $ 56.69 $ &$  9.87 $ &$  2.90 $ &$57.62$  \\

\hline
\end{tabular} 
\end{center} 
\caption[RESULT] 
{ Combined HERA\,I+II NC $e^-p$ cross sections ${\rm d}\sigma_{\rm NC}/{\rm d}Q^2$
for $P_e=0$ and $y<0.9$. Here
$\delta_{\rm stat}$, $\delta_{\rm unc}$, $\delta_{\rm cor}$ and $\delta_{\rm tot}$ 
are the relative statistical, total uncorrelated, correlated systematic and total uncertainties, respectively.}
\label{tab:dq2_comb_em} 
\end{table} 

\begin{table}[\tablepos] 
\begin{center} 
\footnotesize
\renewcommand{\arraystretch}{1.22}
\begin{tabular}{|r|l|r|r|r|r|} 
\hline 
$Q^2$ & ${\rm d}\sigma_{\rm NC}/{\rm d}Q^2$ & $\delta_{\rm stat}$  & $\delta_{\rm unc}$ & 
$\delta_{\rm cor}$& $\delta_{\rm tot}$       \\ 
  $(\rm GeV^2)$ & $(\rm pb/GeV^2)$ & (\%) & (\%) & (\%) & (\%) \\
\hline 
$   200 $ &$ 18.30 $ & $  0.25 $ &$  0.68 $ &$  1.22 $ &$ 1.42$  \\
$   250 $ &$ 10.77 $ & $  0.27 $ &$  0.92 $ &$  1.20 $ &$ 1.54$  \\
$   300 $ &$ 7.008 $ & $  0.30 $ &$  0.95 $ &$  1.22 $ &$ 1.57$  \\
$   400 $ &$ 3.495 $ & $  0.37 $ &$  0.79 $ &$  1.20 $ &$ 1.48$  \\
$   500 $ &$ 2.042 $ & $  0.43 $ &$  0.82 $ &$  1.21 $ &$ 1.53$  \\
$   650 $ &$ 1.067 $ & $  0.52 $ &$  0.83 $ &$  1.23 $ &$ 1.57$  \\
$   800 $ &$ 0.6323 $ & $  0.64 $ &$  0.85 $ &$  1.22 $ &$ 1.61$  \\
$  1000 $ &$ 0.3578 $ & $  0.75 $ &$  0.85 $ &$  1.21 $ &$ 1.65$  \\
$  1200 $ &$ 0.2230 $ & $  0.89 $ &$  0.95 $ &$  1.22 $ &$ 1.78$  \\
$  1500 $ &$ 0.1229 $ & $  1.08 $ &$  1.01 $ &$  1.20 $ &$ 1.90$  \\
$  2000 $ &$ 0.5794 \cdot 10^{-1}$ & $  1.29 $ &$  1.11 $ &$  1.23 $ &$ 2.10$  \\
$  3000 $ &$ 0.1979 \cdot 10^{-1}$ & $  1.24 $ &$  1.21 $ &$  1.22 $ &$ 2.12$  \\
$  5000 $ &$ 0.4579 \cdot 10^{-2}$ & $  1.69 $ &$  1.42 $ &$  1.19 $ &$ 2.51$  \\
$  8000 $ &$ 0.1109 \cdot 10^{-2}$ & $  2.84 $ &$  1.99 $ &$  1.31 $ &$ 3.71$  \\
$ 12000 $ &$ 0.2387 \cdot 10^{-3}$ & $  5.37 $ &$  2.57 $ &$  1.29 $ &$ 6.09$  \\
$ 20000 $ &$ 0.3529 \cdot 10^{-4}$ & $ 10.62 $ &$  4.02 $ &$  1.46 $ &$11.45$  \\
$ 30000 $ &$ 0.6517 \cdot 10^{-5}$ & $ 25.70 $ &$  6.07 $ &$  1.84 $ &$26.47$  \\

\hline
\end{tabular} 
\end{center} 
\caption[RESULT] 
{ Combined HERA\,I+II NC $e^+p$ cross sections ${\rm d}\sigma_{\rm NC}/{\rm d}Q^2$
for $P_e=0$ and $y<0.9$. Here
$\delta_{\rm stat}$, $\delta_{\rm unc}$, $\delta_{\rm cor}$ and $\delta_{\rm tot}$ 
are the relative statistical, total uncorrelated, correlated systematic and total uncertainties, respectively.}
\label{tab:dq2_comb_ep} 
\end{table} 

\begin{table}[\tablepos] 
\begin{center} 
\footnotesize
\renewcommand{\arraystretch}{1.22}
\begin{tabular}{|r|c|r|r|r|r|} 
\hline 
$Q^2$ & ${\rm d}\sigma_{\rm CC}/{\rm d}Q^2$ & $\delta_{\rm stat}$  & $\delta_{\rm unc}$ & 
$\delta_{\rm cor}$& $\delta_{\rm tot}$       \\ 
  $(\rm GeV^2)$ & $(\rm pb/GeV^2)$ & (\%) & (\%) & (\%) & (\%) \\
\hline 
$   300 $ &$ 0.355 \cdot 10^{-1}$ & $  7.1 $ &$  6.2 $ &$  4.9 $ &$ 10.6 $  \\
$   500 $ &$ 0.245 \cdot 10^{-1}$ & $  4.2 $ &$  3.9 $ &$  3.3 $ &$  6.5 $  \\
$  1000 $ &$ 0.166 \cdot 10^{-1}$ & $  3.2 $ &$  2.8 $ &$  2.2 $ &$  4.8 $  \\
$  2000 $ &$ 0.951 \cdot 10^{-2}$ & $  2.8 $ &$  2.3 $ &$  1.8 $ &$  4.0 $  \\
$  3000 $ &$ 0.593 \cdot 10^{-2}$ & $  2.8 $ &$  2.0 $ &$  1.6 $ &$  3.8 $  \\
$  5000 $ &$ 0.301 \cdot 10^{-2}$ & $  3.1 $ &$  2.2 $ &$  1.7 $ &$  4.2 $  \\
$  8000 $ &$ 0.136 \cdot 10^{-2}$ & $  3.9 $ &$  3.3 $ &$  2.0 $ &$  5.5 $  \\
$ 15000 $ &$ 0.341 \cdot 10^{-3}$ & $  5.5 $ &$  5.5 $ &$  2.7 $ &$  8.3 $  \\
$ 30000 $ &$ 0.291 \cdot 10^{-4}$ & $ 12.6 $ &$ 10.3 $ &$  4.3 $ &$ 16.8 $  \\

\hline
\end{tabular} 
\end{center} 
\caption[RESULT] 
{ Combined HERA\,I+II CC $e^-p$ cross sections ${\rm d}\sigma_{\rm CC}/{\rm d}Q^2$
for $P_e=0$ and $y<0.9$. Here
$\delta_{\rm stat}$, $\delta_{\rm unc}$, $\delta_{\rm cor}$ and $\delta_{\rm tot}$ 
are the relative statistical, total uncorrelated, correlated systematic and total uncertainties, respectively.}
\label{tab:dq2cc_comb_em} 
\end{table} 

\begin{table}[\tablepos] 
\begin{center} 
\footnotesize
\renewcommand{\arraystretch}{1.22}
\begin{tabular}{|r|c|r|r|r|r|} 
\hline 
$Q^2$ & ${\rm d}\sigma_{\rm CC}/{\rm d}Q^2$ & $\delta_{\rm stat}$  & $\delta_{\rm unc}$ & 
$\delta_{\rm cor}$& $\delta_{\rm tot}$       \\ 
  $(\rm GeV^2)$ & $(\rm pb/GeV^2)$ & (\%) & (\%) & (\%) & (\%) \\
\hline 
$   300 $ &$ 0.266 \cdot 10^{-1}$ & $  5.5 $ &$  4.8 $ &$  4.0 $ &$ 8.3$  \\
$   500 $ &$ 0.200 \cdot 10^{-1}$ & $  3.6 $ &$  3.0 $ &$  2.8 $ &$ 5.4$  \\
$  1000 $ &$ 0.113 \cdot 10^{-1}$ & $  2.9 $ &$  2.3 $ &$  1.9 $ &$ 4.2$  \\
$  2000 $ &$ 0.535 \cdot 10^{-2}$ & $  2.8 $ &$  2.0 $ &$  1.4 $ &$ 3.7$  \\
$  3000 $ &$ 0.295 \cdot 10^{-2}$ & $  3.0 $ &$  2.1 $ &$  1.4 $ &$ 3.9$  \\
$  5000 $ &$ 0.104 \cdot 10^{-2}$ & $  4.1 $ &$  2.8 $ &$  1.8 $ &$ 5.3$  \\
$  8000 $ &$ 0.322 \cdot 10^{-3}$ & $  6.3 $ &$  4.4 $ &$  2.5 $ &$ 8.1$  \\
$ 15000 $ &$ 0.424 \cdot 10^{-4}$ & $ 11.4 $ &$  6.5 $ &$  2.9 $ &$13.4$  \\

\hline
\end{tabular} 
\end{center} 
\caption[RESULT] 
{ Combined HERA\,I+II CC $e^+p$ cross sections ${\rm d}\sigma_{\rm CC}/{\rm d}Q^2$
for $P_e=0$ and $y<0.9$. Here
$\delta_{\rm stat}$, $\delta_{\rm unc}$, $\delta_{\rm cor}$ and $\delta_{\rm tot}$ 
are the relative statistical, total uncorrelated, correlated systematic and total uncertainties, respectively.}
\label{tab:dq2cc_comb_ep} 
\end{table} 
\renewcommand{\arraystretch}{1.22}

\begin{table}[\tablepos] 
\begin{center} 
\footnotesize
\renewcommand{\arraystretch}{1.22}
\begin{tabular}{|r|c|r|r|r|r|r|} 
\hline 
$Q^2$ & $x$ & $F^{\gamma Z}_2$ & $\Delta_{\rm stat}$ & $\Delta_{\rm unc}$ & $\Delta_{\rm cor}$ & $\Delta_{\rm tot}$ \\
 $({\rm GeV}^2)$ & & & & & & \\
\hline
$   200 $ &$  0.020 $ &$ -0.412 $ &$ 0.908 $ &$ 0.682 $ &$ 0.946 $ &$ 1.478 $ \\
$   200 $ &$  0.032 $ &$  0.647 $ &$ 0.873 $ &$ 0.753 $ &$ 0.799 $ &$ 1.403 $ \\
$   200 $ &$  0.050 $ &$ -1.814 $ &$ 0.922 $ &$ 0.871 $ &$ 0.724 $ &$ 1.460 $ \\
$   200 $ &$  0.080 $ &$  0.547 $ &$ 0.807 $ &$ 0.861 $ &$ 0.600 $ &$ 1.323 $ \\
$   200 $ &$  0.130 $ &$ -0.836 $ &$ 0.766 $ &$ 0.742 $ &$ 0.497 $ &$ 1.177 $ \\
$   200 $ &$  0.180 $ &$ -1.246 $ &$ 0.909 $ &$ 0.819 $ &$ 0.458 $ &$ 1.306 $ \\
\hline

\hline
\end{tabular} 
\end{center} 
\caption[RESULT] 
{ The measured structure function $F^{\gamma Z}_2$ determined
using the polarised HERA\,II data set. The absolute statistical,
uncorrelated, correlated and total uncertainties $\Delta_{\rm stat}$,
$\Delta_{\rm unc}$, $\Delta_{\rm cor}$ and $\Delta_{\rm tot}$ are also
given. The luminosity uncertainty of 2.3\% is not included in the errors.
Table continues on next pages.}
\label{tab:f2gZ} 
\end{table} 

\begin{table}[\tablepos] 
\begin{center} 
\footnotesize
\renewcommand{\arraystretch}{1.22}
\begin{tabular}{|r|c|r|r|r|r|r|} 
\hline 
$Q^2$ & $x$ & $F^{\gamma Z}_2$ & $\Delta_{\rm stat}$ & $\Delta_{\rm unc}$ & $\Delta_{\rm cor}$ & $\Delta_{\rm tot}$ \\
 $({\rm GeV}^2)$ & & & & & & \\
\hline
$   650 $ &$  0.020 $ &$  0.408 $ &$ 0.563 $ &$ 0.242 $ &$ 0.330 $ &$ 0.696 $ \\
$   650 $ &$  0.032 $ &$  0.730 $ &$ 0.501 $ &$ 0.254 $ &$ 0.287 $ &$ 0.630 $ \\
$   650 $ &$  0.050 $ &$  0.027 $ &$ 0.439 $ &$ 0.184 $ &$ 0.235 $ &$ 0.530 $ \\
$   650 $ &$  0.080 $ &$  0.905 $ &$ 0.390 $ &$ 0.147 $ &$ 0.190 $ &$ 0.458 $ \\
$   650 $ &$  0.130 $ &$  0.848 $ &$ 0.389 $ &$ 0.159 $ &$ 0.165 $ &$ 0.452 $ \\
$   650 $ &$  0.180 $ &$  0.102 $ &$ 0.361 $ &$ 0.168 $ &$ 0.164 $ &$ 0.430 $ \\
\hline

\hline
\end{tabular} 
\end{center} 
\captcont{\label{tab:f2gZ_c}continued.}
\end{table} 

\begin{table}[\tablepos] 
\begin{center} 
\footnotesize
\renewcommand{\arraystretch}{1.22}
\begin{tabular}{|r|c|r|r|r|r|r|} 
\hline 
$Q^2$ & $x$ & $F^{\gamma Z}_2$ & $\Delta_{\rm stat}$ & $\Delta_{\rm unc}$ & $\Delta_{\rm cor}$ & $\Delta_{\rm tot}$ \\
 $({\rm GeV}^2)$ & & & & & & \\
\hline
$  2000 $ &$  0.022 $ &$  1.246 $ &$ 0.709 $ &$ 0.227 $ &$ 0.142 $ &$ 0.758 $ \\
$  2000 $ &$  0.032 $ &$  0.071 $ &$ 0.397 $ &$ 0.100 $ &$ 0.100 $ &$ 0.422 $ \\
$  2000 $ &$  0.050 $ &$  0.467 $ &$ 0.342 $ &$ 0.072 $ &$ 0.089 $ &$ 0.361 $ \\
$  2000 $ &$  0.080 $ &$ -0.081 $ &$ 0.286 $ &$ 0.068 $ &$ 0.073 $ &$ 0.302 $ \\
$  2000 $ &$  0.130 $ &$  0.683 $ &$ 0.280 $ &$ 0.065 $ &$ 0.059 $ &$ 0.293 $ \\
$  2000 $ &$  0.180 $ &$  0.161 $ &$ 0.252 $ &$ 0.065 $ &$ 0.050 $ &$ 0.265 $ \\
\hline
\end{tabular} 
\end{center} 
\captcont{continued.} 
\end{table}


\begin{table}[\tablepos] 
\begin{center} 
\footnotesize
\renewcommand{\arraystretch}{1.22}
\begin{tabular}{|c|c|c|c|c|c|} 
\hline 
$x$ & $F^{\gamma Z}_2$ & $\Delta_{\rm stat}$ & $\Delta_{\rm unc}$ & $\Delta_{\rm cor}$ & $\Delta_{\rm tot}$ \\\hline
$  0.020 $ &$  0.751 $ &$ 0.188 $ &$ 0.080 $ &$ 0.326 $ &$ 0.385 $ \\
$  0.032 $ &$  0.546 $ &$ 0.139 $ &$ 0.053 $ &$ 0.213 $ &$ 0.259 $ \\
$  0.050 $ &$  0.299 $ &$ 0.104 $ &$ 0.036 $ &$ 0.139 $ &$ 0.177 $ \\
$  0.080 $ &$  0.398 $ &$ 0.081 $ &$ 0.025 $ &$ 0.098 $ &$ 0.129 $ \\
$  0.130 $ &$  0.286 $ &$ 0.074 $ &$ 0.023 $ &$ 0.074 $ &$ 0.107 $ \\
$  0.180 $ &$  0.264 $ &$ 0.065 $ &$ 0.020 $ &$ 0.055 $ &$ 0.087 $ \\
$  0.250 $ &$  0.051 $ &$ 0.056 $ &$ 0.015 $ &$ 0.032 $ &$ 0.067 $ \\
$  0.400 $ &$  0.068 $ &$ 0.035 $ &$ 0.013 $ &$ 0.017 $ &$ 0.041 $ \\
$  0.650 $ &$  0.019 $ &$ 0.008 $ &$ 0.003 $ &$ 0.002 $ &$ 0.009 $ \\

\hline
\end{tabular} 
\end{center} 
\caption[RESULT] 
{ Averaged structure function $F^{\gamma Z}_2$ for
$Q^2=1\,500\,{\rm GeV}^2$ determined using the polarised HERA\,II data
set. The absolute statistical, uncorrelated, correlated and total
uncertainties $\Delta_{\rm stat}$, $\Delta_{\rm unc}$, $\Delta_{\rm
cor}$ and $\Delta_{\rm tot}$ are also given. The luminosity uncertainty of 
2.3\% is not included in the errors.}
\label{tab:f2gZ_1500} 
\end{table}

\begin{table}[\tablepos] 
\begin{center} 
\footnotesize
\renewcommand{\arraystretch}{1.22}
\footnotesize
\begin{tabular}{|r|c|r|c|c|c|c|} 
\hline 
$Q^2$ & $x$ & $xF^{\gamma Z}_3$ & $\Delta_{\rm stat}$  & $\Delta_{\rm unc}$ & 
$\Delta_{\rm cor}$& $\Delta_{\rm tot}$       \\ 
  $(\rm GeV^2)$ &  &  & & &  & \\
\hline 
$ 1200$ &$  0.013 $ &$  0.256 $ &$ 0.301 $ &$ 0.158 $ &$ 0.086 $ &$0.350$  \\
$ 1200$ &$  0.020 $ &$  0.410 $ &$ 0.211 $ &$ 0.098 $ &$ 0.098 $ &$0.253$  \\
$ 1200$ &$  0.032 $ &$  0.726 $ &$ 0.277 $ &$ 0.114 $ &$ 0.140 $ &$0.331$  \\
$ 1200$ &$  0.050 $ &$  0.666 $ &$ 0.373 $ &$ 0.190 $ &$ 0.184 $ &$0.457$  \\
$ 1200$ &$  0.080 $ &$  0.339 $ &$ 0.560 $ &$ 0.274 $ &$ 0.259 $ &$0.676$  \\
$ 1200$ &$  0.130 $ &$ -0.744 $ &$ 0.961 $ &$ 0.507 $ &$ 0.349 $ &$1.141$  \\
\hline

\hline
\end{tabular} 
\end{center} 
\caption[RESULT] 
{ The measured structure function $xF^{\gamma Z}_3$ determined
using the complete HERA\,I+II data set. The absolute statistical,
uncorrelated, correlated and total uncertainties $\Delta_{\rm stat}$,
$\Delta_{\rm unc}$, $\Delta_{\rm cor}$ and $\Delta_{\rm tot}$ are also
given.}
\label{tab:xf3gz} 
\end{table} 
\begin{table}[\tablepos] 
\begin{center} 
\renewcommand{\arraystretch}{1.22}
\footnotesize
\begin{tabular}{|r|c|r|c|c|c|c|} 
\hline 
$Q^2$ & $x$ & $xF^{\gamma Z}_3$ & $\Delta_{\rm stat}$  & $\Delta_{\rm unc}$ & 
$\Delta_{\rm cor}$& $\Delta_{\rm tot}$       \\ 
  $(\rm GeV^2)$ &  &  & & &  & \\
\hline 
$ 5000$ &$  0.055 $ &$  0.297 $ &$ 0.073 $ &$ 0.033 $ &$ 0.019 $ &$0.082$  \\
$ 5000$ &$  0.080 $ &$  0.308 $ &$ 0.058 $ &$ 0.026 $ &$ 0.020 $ &$0.067$  \\
$ 5000$ &$  0.130 $ &$  0.516 $ &$ 0.096 $ &$ 0.045 $ &$ 0.028 $ &$0.109$  \\
$ 5000$ &$  0.180 $ &$  0.578 $ &$ 0.123 $ &$ 0.055 $ &$ 0.033 $ &$0.139$  \\
$ 5000$ &$  0.250 $ &$  0.132 $ &$ 0.167 $ &$ 0.081 $ &$ 0.034 $ &$0.188$  \\
$ 5000$ &$  0.400 $ &$  0.023 $ &$ 0.151 $ &$ 0.078 $ &$ 0.027 $ &$0.172$  \\
\hline
\end{tabular} 
\end{center}
\captcont{continued.} 
\end{table}

\begin{table}[\tablepos] 
\begin{center} 
\footnotesize
\renewcommand{\arraystretch}{1.22}
\begin{tabular}{|c|c|c|c|c|c|} 
\hline 
 $x$ & $xF^{\gamma Z}_3$ & $\Delta_{\rm stat}$  & $\Delta_{\rm unc}$ & 
$\Delta_{\rm cor}$& $\Delta_{\rm tot}$       \\ 
\hline 
      $ 0.013 $&$ 0.258 $&$ 0.304 $&$ 0.159 $&$ 0.086 $&$ 0.354$\\
      $ 0.020 $&$ 0.470 $&$ 0.135 $&$ 0.071 $&$ 0.080 $&$ 0.172$\\
      $ 0.032 $&$ 0.393 $&$ 0.070 $&$ 0.034 $&$ 0.048 $&$ 0.091$\\
      $ 0.050 $&$ 0.266 $&$ 0.073 $&$ 0.033 $&$ 0.050 $&$ 0.095$\\
      $ 0.080 $&$ 0.327 $&$ 0.051 $&$ 0.023 $&$ 0.031 $&$ 0.064$\\
      $ 0.130 $&$ 0.489 $&$ 0.047 $&$ 0.020 $&$ 0.022 $&$ 0.055$\\
      $ 0.180 $&$ 0.360 $&$ 0.046 $&$ 0.016 $&$ 0.018 $&$ 0.052$\\
      $ 0.250 $&$ 0.330 $&$ 0.040 $&$ 0.012 $&$ 0.012 $&$ 0.044$\\
      $ 0.400 $&$ 0.227 $&$ 0.035 $&$ 0.013 $&$ 0.008 $&$ 0.038$\\
      $ 0.650 $&$ 0.018 $&$ 0.015 $&$ 0.006 $&$ 0.002 $&$ 0.016$\\

\hline
\end{tabular} 
\end{center} 
\caption[RESULT] 
{ Averaged structure function $xF^{\gamma Z}_3$ for
$Q^2=1\,500\,{\rm GeV}^2$ determined using the complete HERA\,I+II data
set. The absolute statistical, uncorrelated, correlated and total
uncertainties $\Delta_{\rm stat}$, $\Delta_{\rm unc}$, $\Delta_{\rm
cor}$ and $\Delta_{\rm tot}$ are also given.}
\label{tab:xf3gz_1500} 
\end{table} 

\clearpage
\newpage

\begin{figure}[\tablepos]
\begin{center}
\includegraphics[width=\columnwidth]{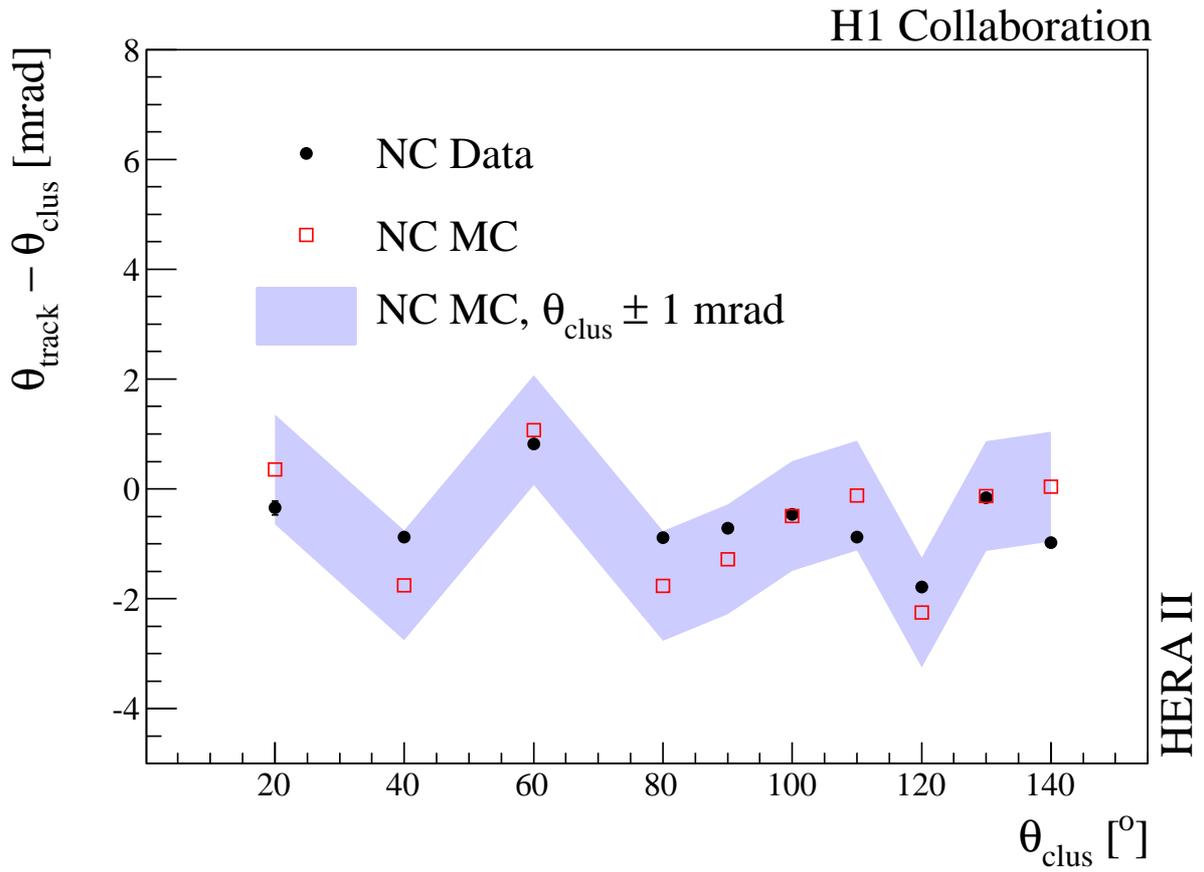}
\end{center}
\caption{ The mean value of $\Delta\theta=\theta_{\rm
  track}-\theta_{\rm clus}$ as a function of the polar angle of the
  cluster $\theta_{\rm clus}$ after alignment for data (solid points) 
  covering the complete HERA\,II data set and simulation (open squares). 
  The shaded band corresponds to a $\pm1$~{\rm mrad} uncertainty.}
\label{fig:align}
\end{figure}

\begin{figure}[\tablepos]
\begin{center}
\subfigure[\label{fig:em-calib-eda}]{\includegraphics[width=\columnwidth]{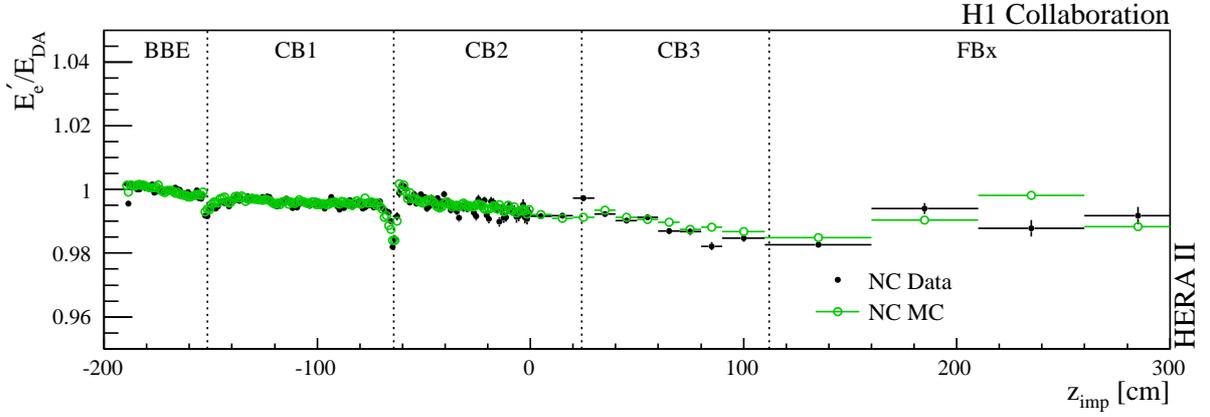} }
\subfigure[\label{fig:em-calib-eop}]   {\includegraphics[width=0.45\columnwidth]{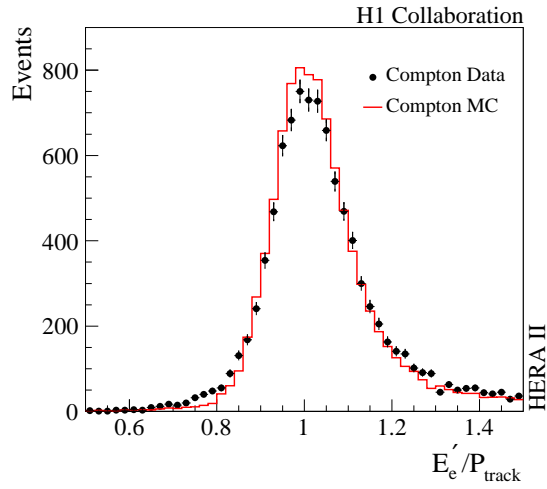} }
\end{center}
\caption{ (a) The mean value of $E_e^{\prime}/E_{DA}$ as a function of 
  $z_{\rm imp}$ for data (solid points) and simulation (open circles) based on 
  a NC DIS sample for the complete HERA\,II data sample, where BBE, CB1, CB2, 
  CB3 and FBx stand for Backward Barrel Electromagnetic, Central Barrel and 
  Forward Barrel wheels. (b) The distribution of $E_{e}^{\prime}/P_{\rm track}$ 
  with data (solid points) and simulation (histogram), based on a QED Compton 
  selection in the energy range between $3-8$\,GeV for the complete HERA\,II 
  data set.}
\label{fig:em-calib}
\end{figure}

\begin{figure}[\tablepos]
\begin{center}
\subfigure[\label{fig:hadscale-a}]{\includegraphics[width=0.7\columnwidth]{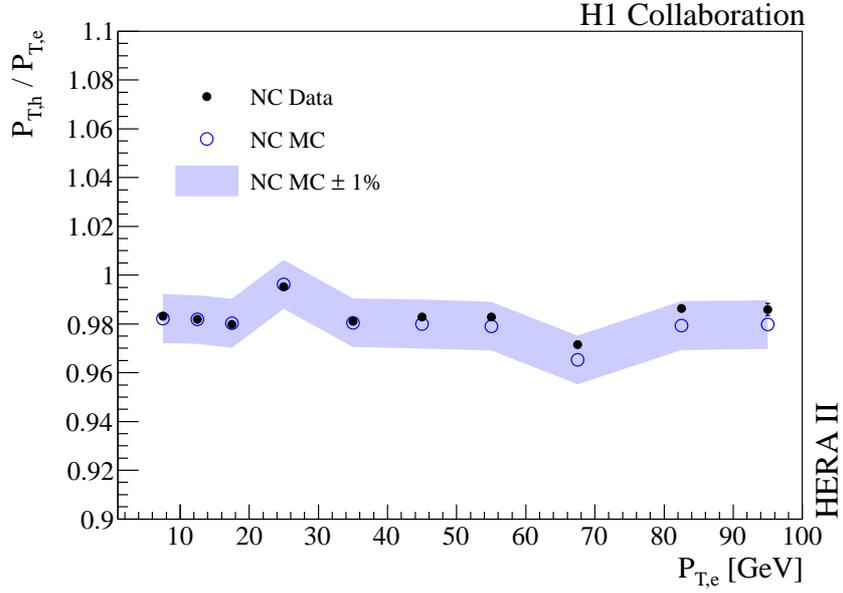} }
\subfigure[\label{fig:hadscale-b}]{\includegraphics[width=0.7\columnwidth]{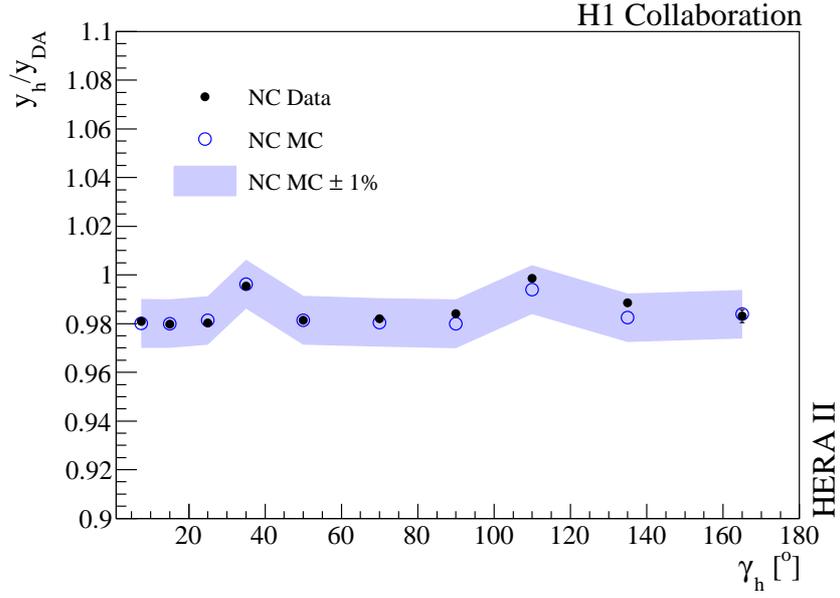} }
\end{center}
\caption{ (a) Mean values of $P_{T,h}/P_{T,e}$ as a function of
  $P_{T,e}$ and (b) $y_h/y_{\rm DA}$ as a function of $\gamma_h$ for
  neutral current data (solid points) and simulation (open circles) for
  the complete HERA\,II data set. The shaded bands correspond to a $\pm 1\%$
  variation around the simulation.}
\label{fig:hadscale}
\end{figure}

\begin{figure}[\tablepos]
\begin{center}
\subfigure[\label{fig:nc-control-ele}]{ \includegraphics[width=\columnwidth]{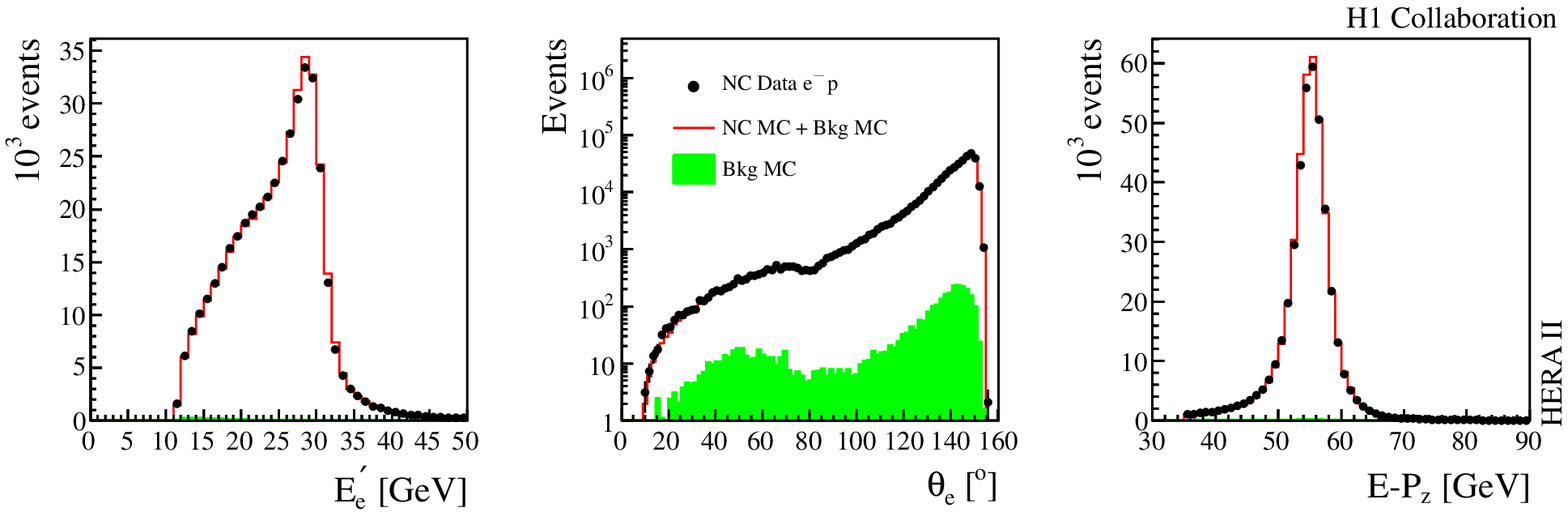} }
\subfigure[\label{fig:nc-control-pos}]{ \includegraphics[width=\columnwidth]{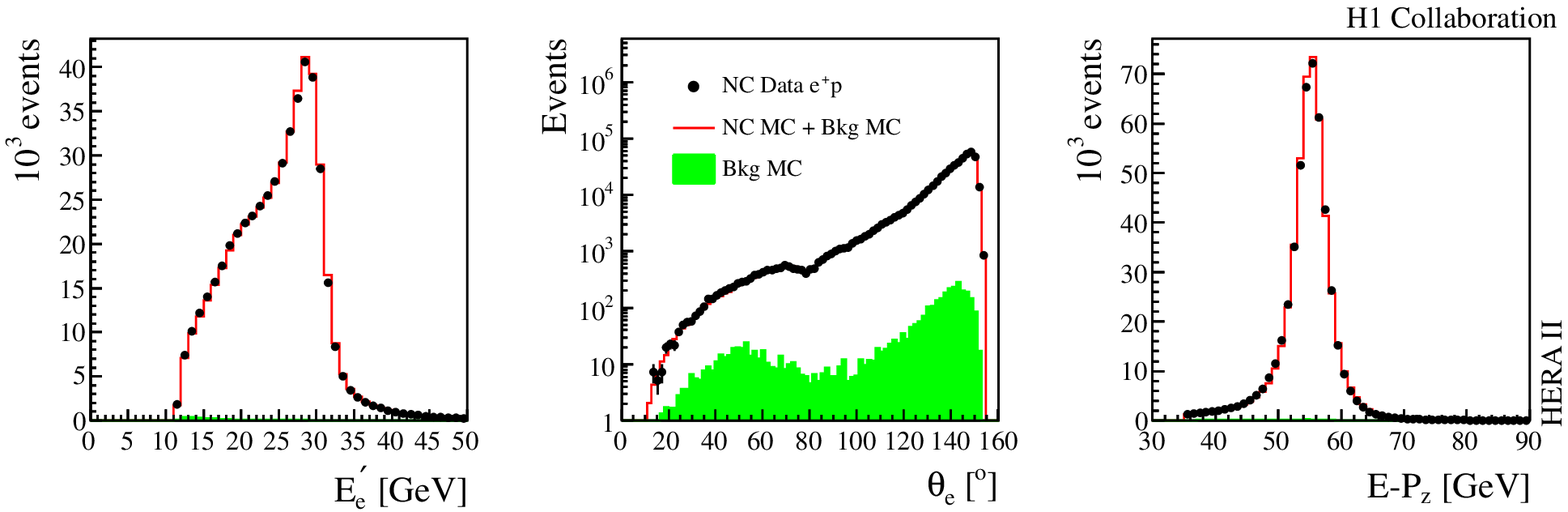} }
\end{center}
\caption{ Distributions of $E^\prime_e$, $\theta_e$ and $E-P_z$
  for (a) $e^-p$ and (b) $e^+p$ NC data (solid points) and simulation
  (histograms). The estimated background contribution is shown as the
  shaded histogram.}
\label{fig:nc-control}
\end{figure}

\begin{figure}[\tablepos]
\begin{center}
\subfigure[\label{fig:jet-control-ele}]{ \includegraphics[width=\columnwidth]{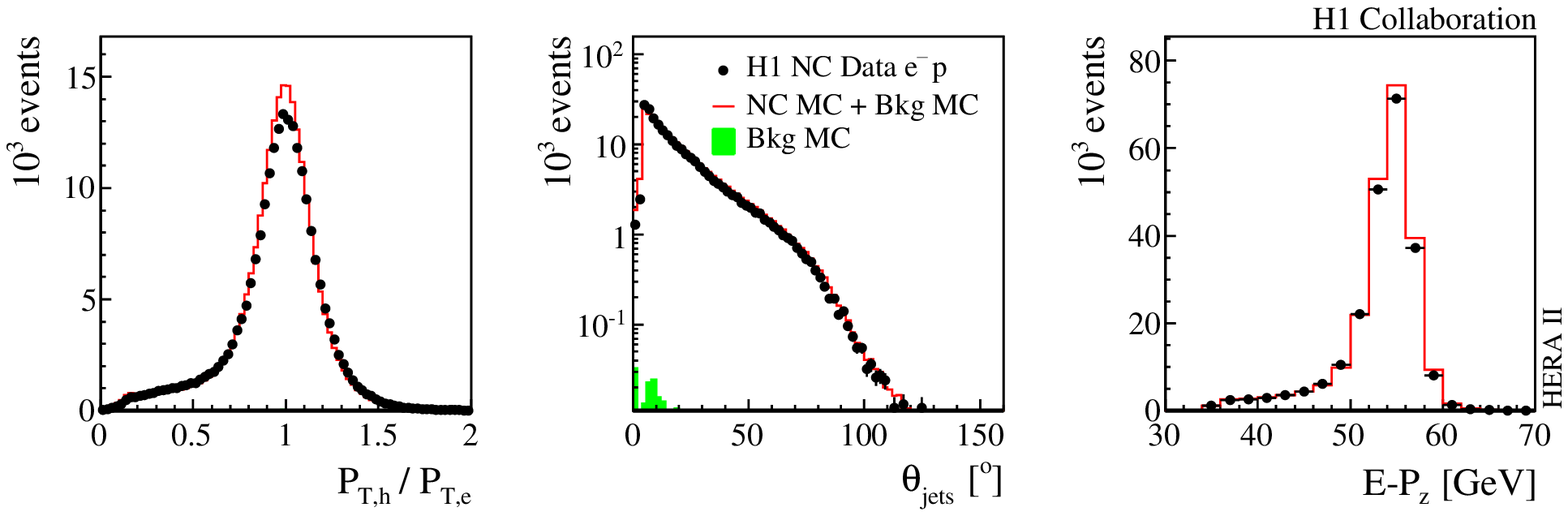} }
\subfigure[\label{fig:jet-control-pos}]{ \includegraphics[width=\columnwidth]{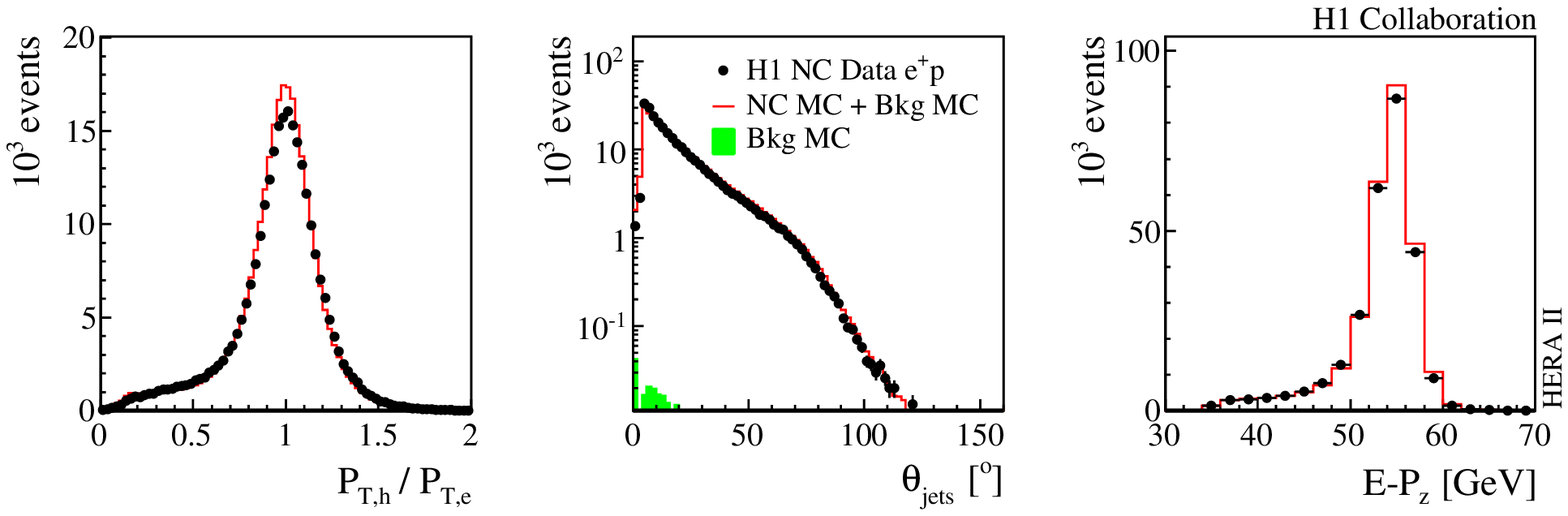} }
\end{center}
\caption{ Distributions of $P_{T,h}/P_{T,e}$, $\theta_{\rm jets}$ and
  $E-P_z$ in the region $y<0.19$ for (a) $e^-p$ and (b) $e^+p$ data
  (solid points) and for simulation (histograms). The estimated background 
  contribution is shown as the shaded histograms. The quantities $P_{T,h}$, 
  $\theta_{\rm jets}$ and the hadronic contribution to $E-P_z$ are calculated 
  using energies associated within jets (see text).}
\label{fig:jet-control}
\end{figure}

\begin{figure}[\tablepos]
\begin{center}
\subfigure[\label{fig:hiy-control-ele}]{ \includegraphics[width=0.33\columnwidth]{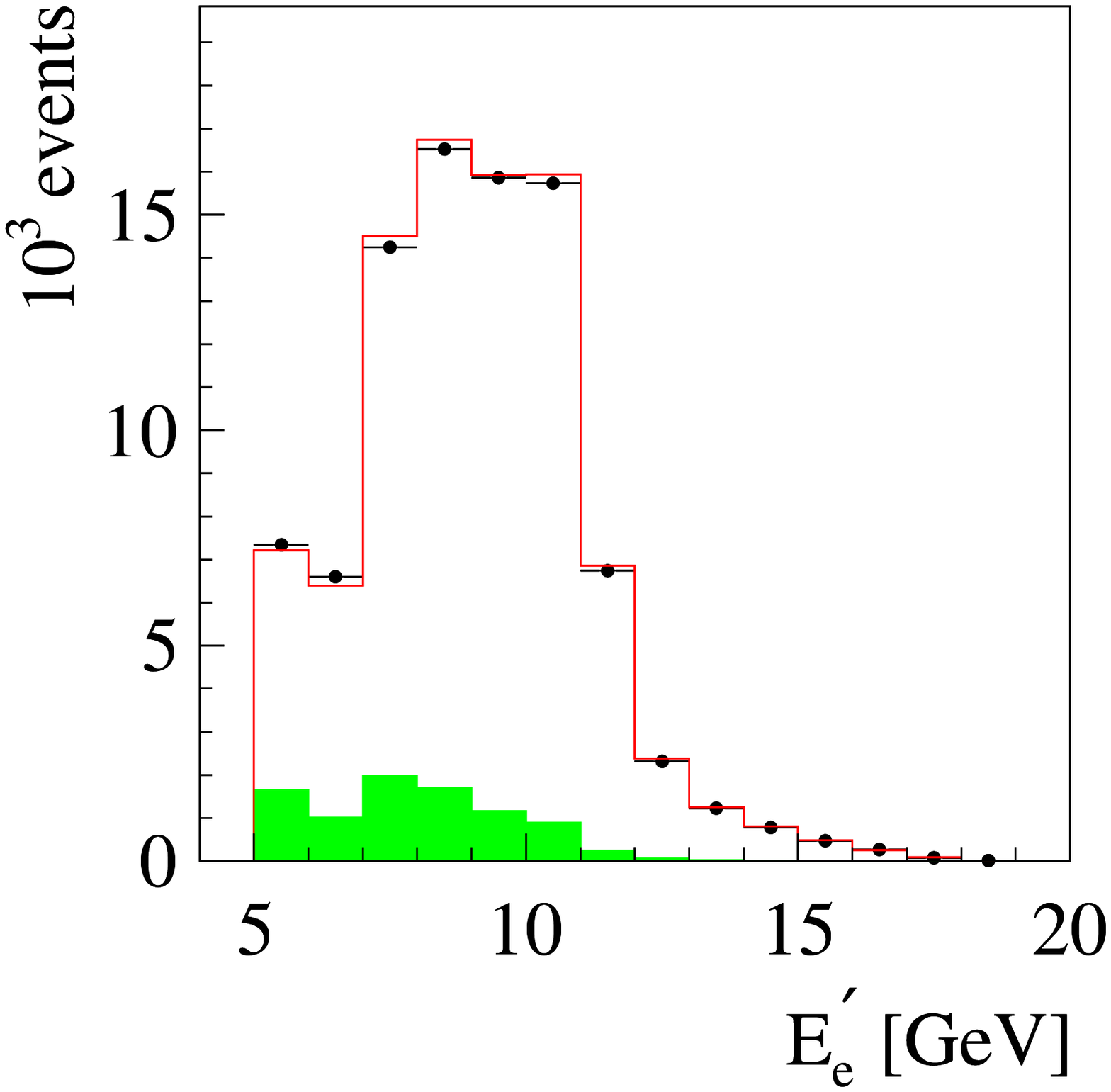}
\includegraphics[width=0.33\columnwidth]{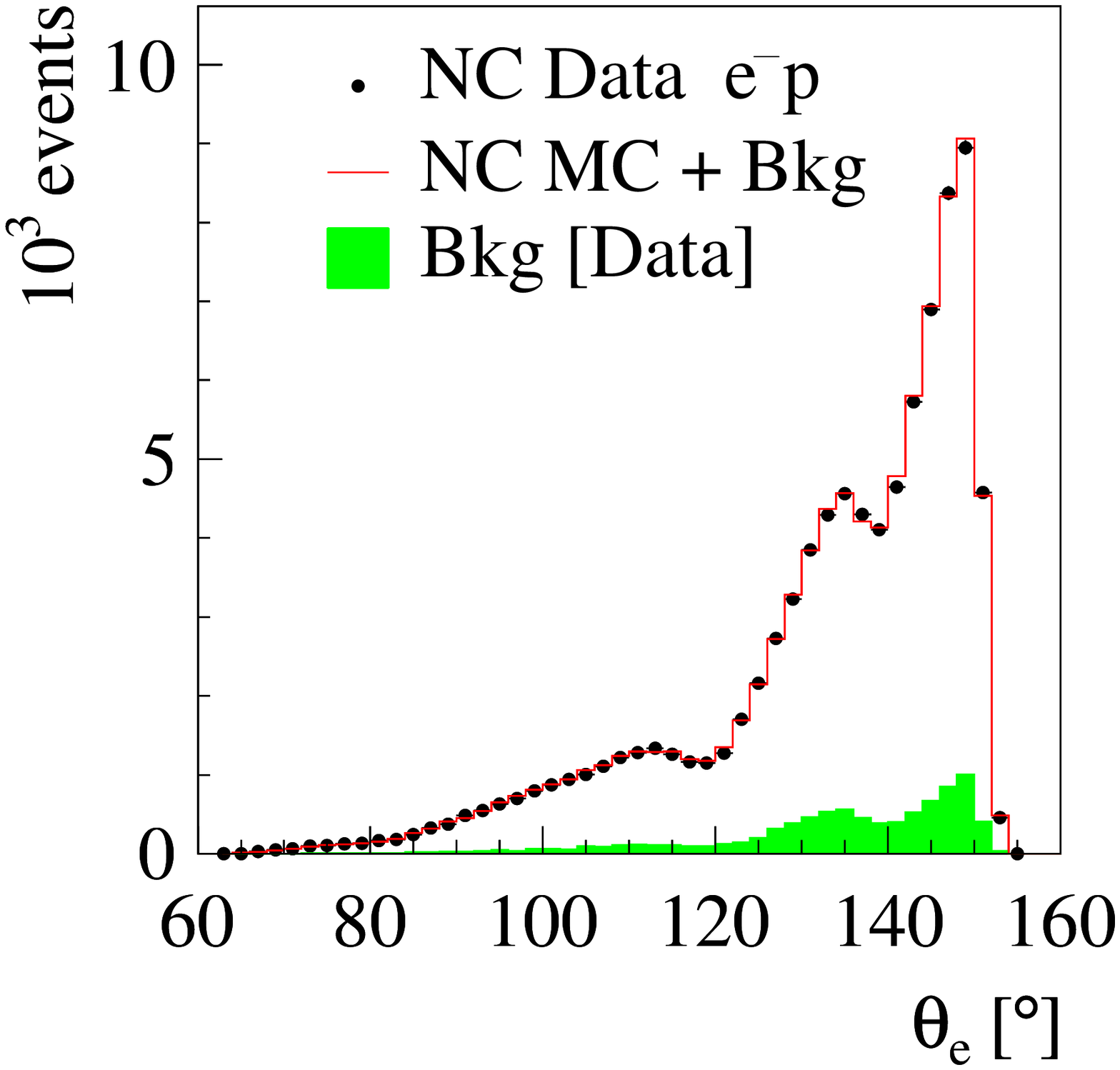}
\includegraphics[width=0.33\columnwidth]{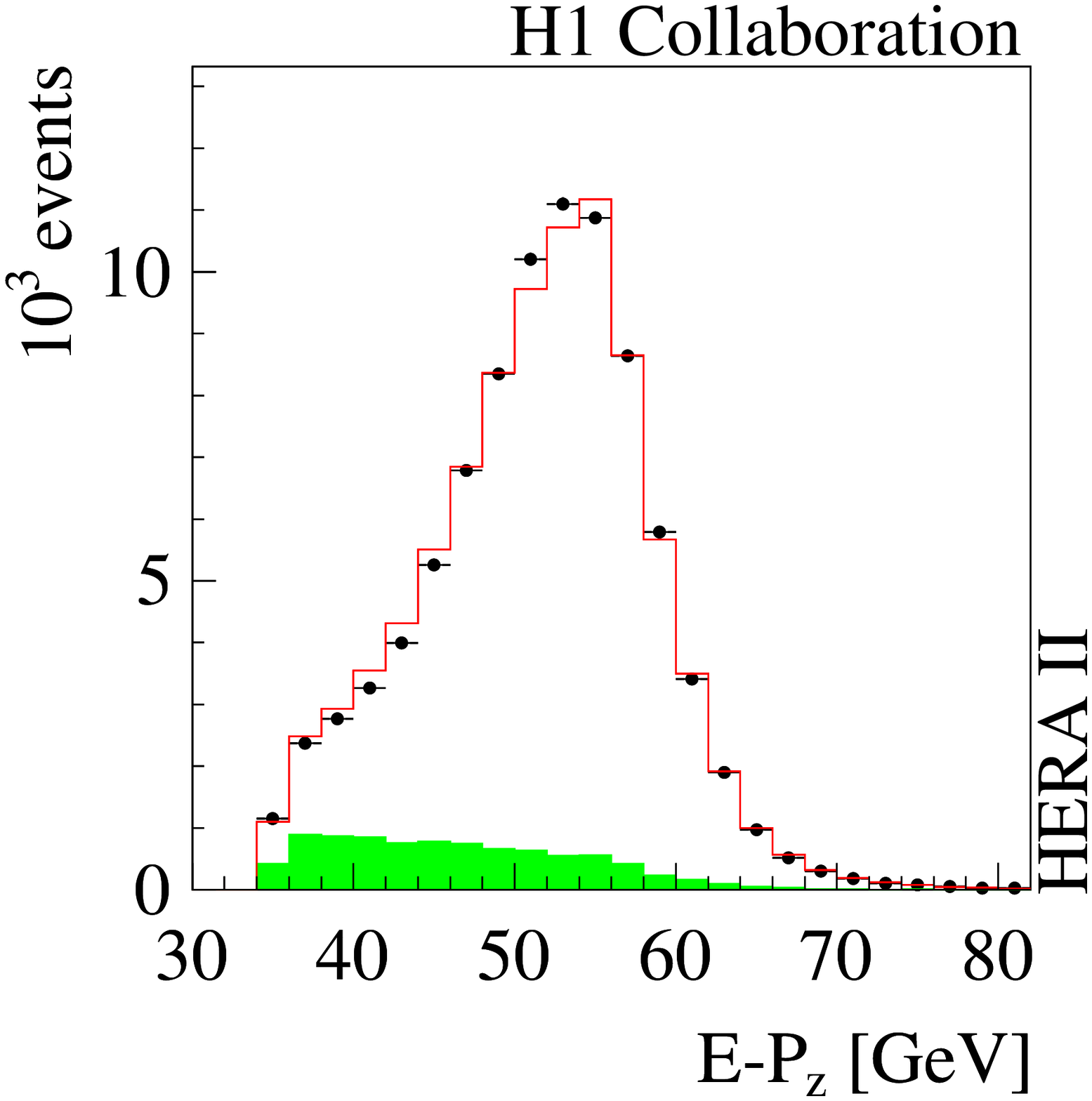} }
\subfigure[\label{fig:hiy-control-pos}]{ \includegraphics[width=0.33\columnwidth]{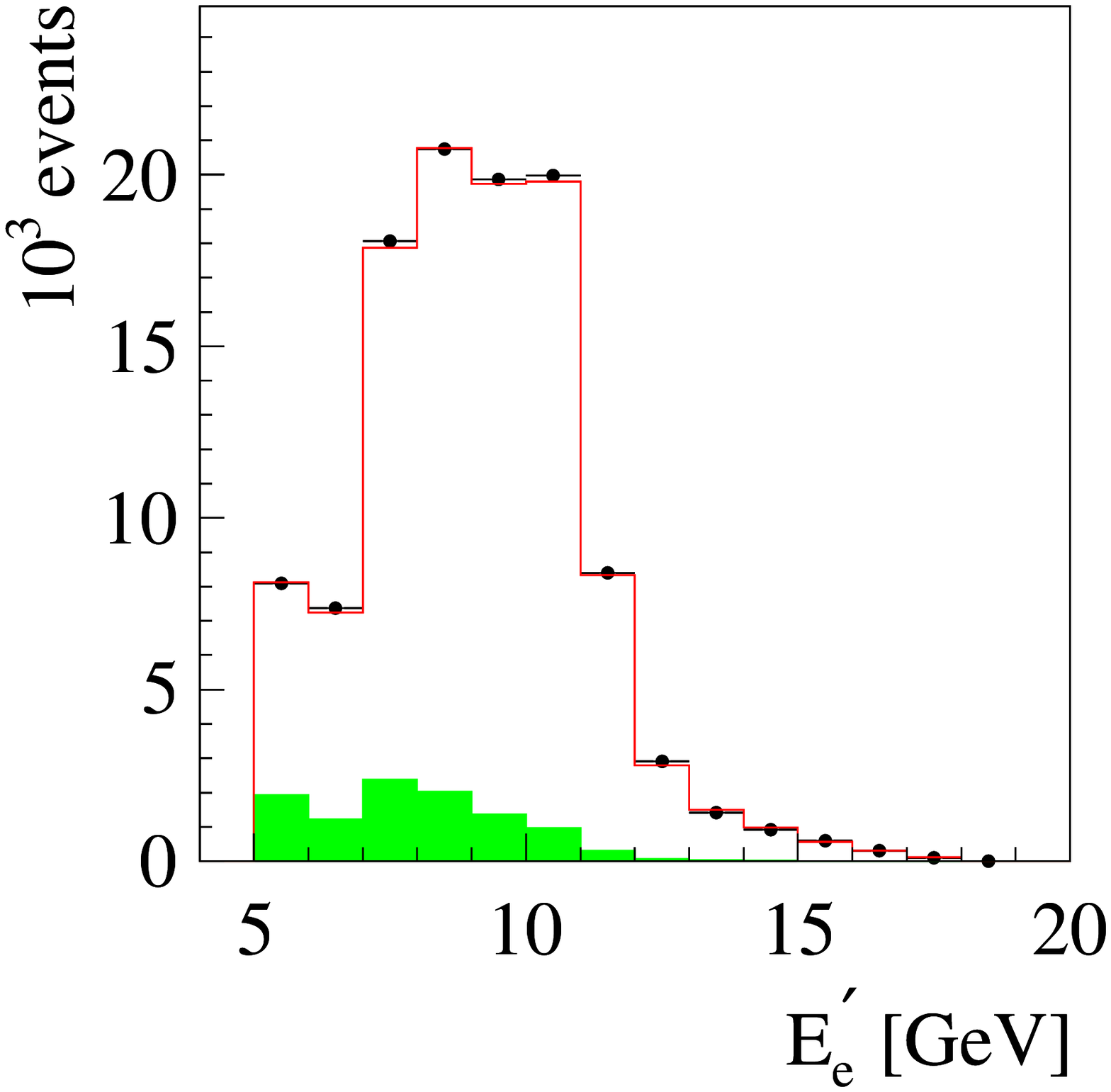}
\includegraphics[width=0.33\columnwidth]{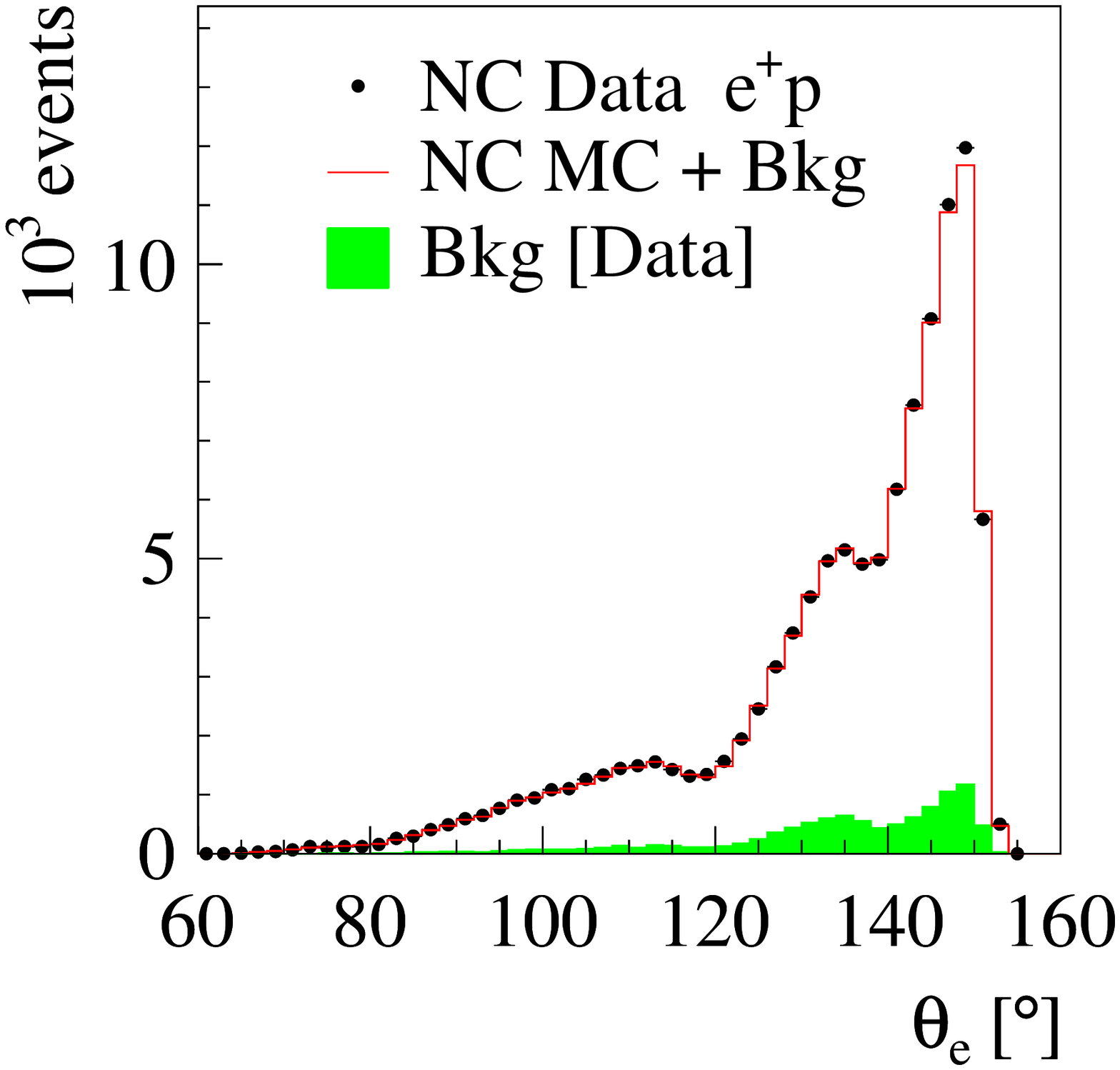}
\includegraphics[width=0.33\columnwidth]{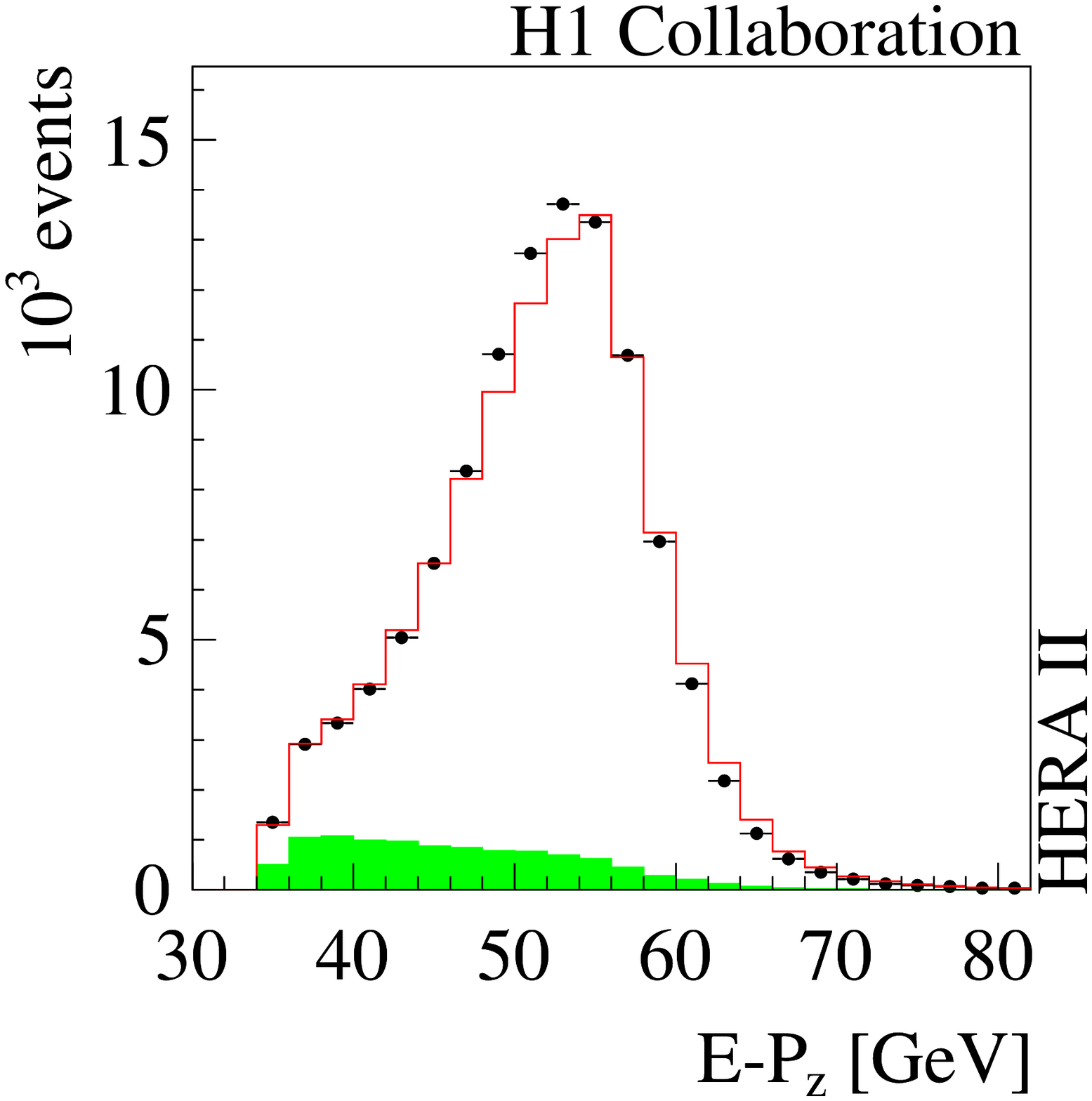} }
\end{center}
\caption{ Distributions of $E_e^\prime$, $\theta_e$ and $E-P_z$ for
  (a) $e^-p$ and (b) $e^+p$ NC {\em high $y$} analysis data. The data is shown
  as solid points and the histogram represent the MC simulation and the
  background obtained from data. The background corresponds to the wrongly 
  charged lepton candidates in the data corrected for the charge asymmetry 
  and is shown as the shaded histogram.}
\label{fig:hiy-control}
\end{figure}

\begin{figure}[\tablepos]
\begin{center}
\subfigure[\label{fig:cc-control-ele}]{ \includegraphics[width=\columnwidth]{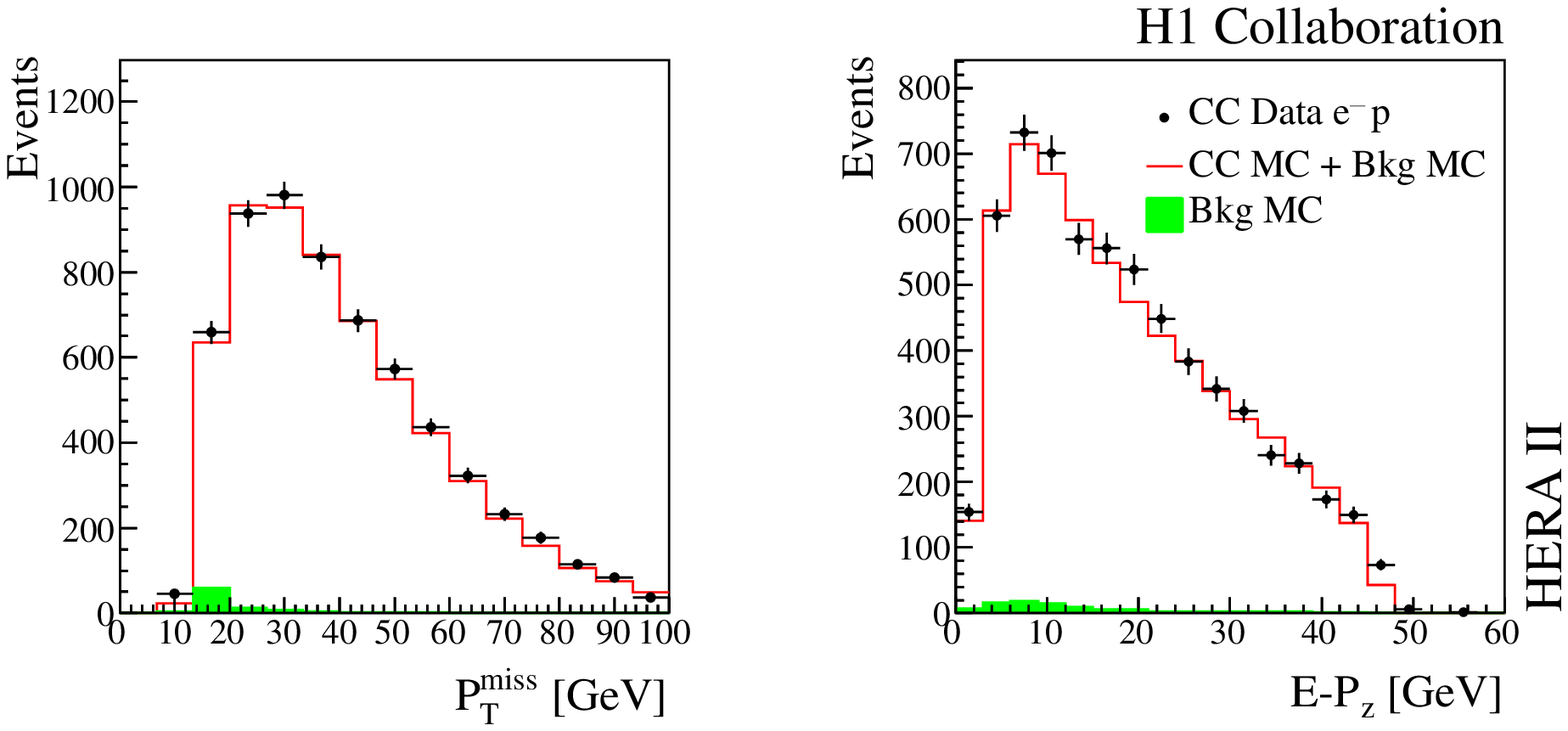} }\vspace*{0.5cm}
\subfigure[\label{fig:cc-control-pos}]{ \includegraphics[width=\columnwidth]{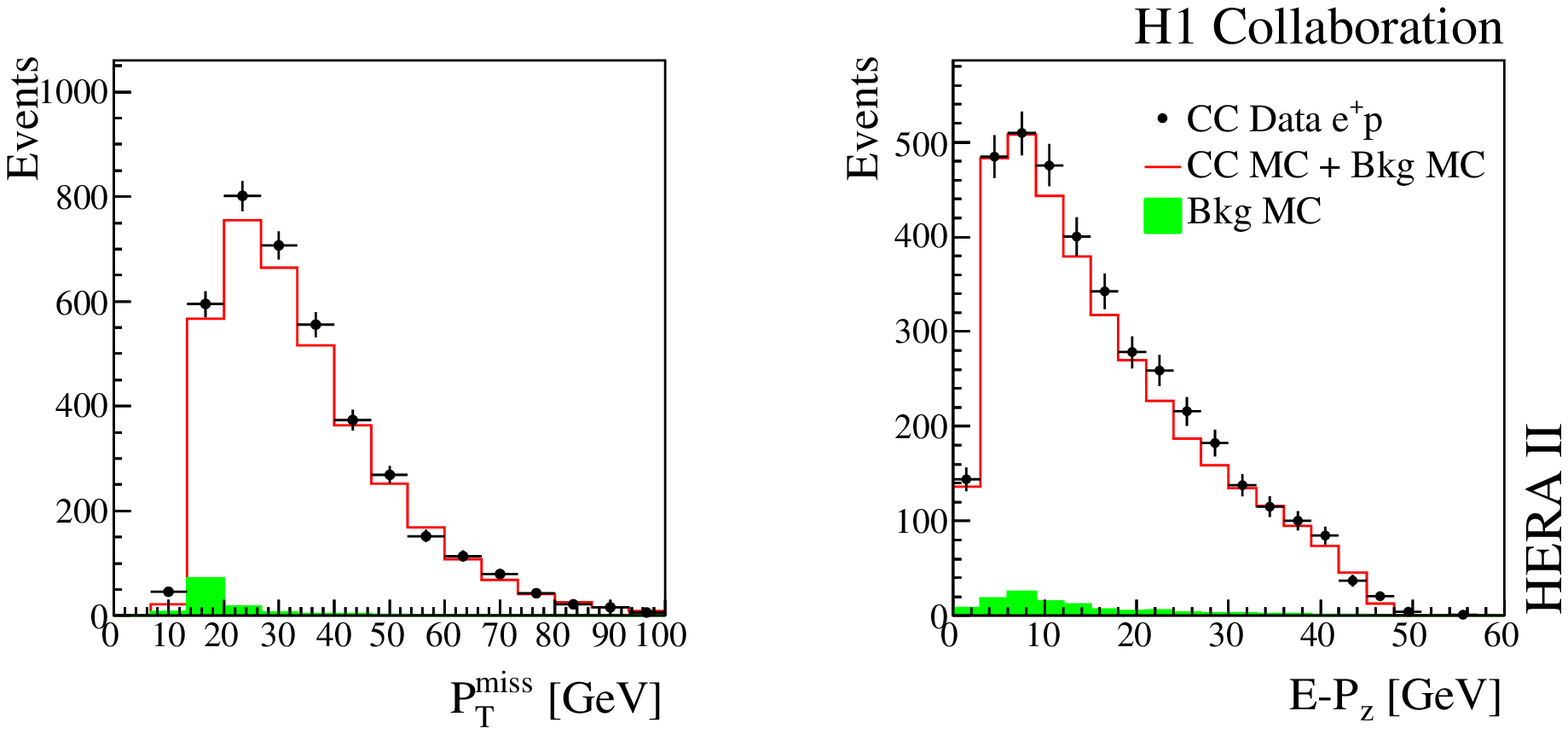} }
\end{center}
\caption{ Distributions of $P^{\rm miss}_T$ and $E-P_z$ for (a) $e^-p$ and 
  (b) $e^+p$ CC data (solid points) and simulation (histograms). 
  The estimated background contribution is shown as the shaded histograms.}
\label{fig:cc-control}
\end{figure}

\clearpage
\begin{figure}[\tablepos]
\begin{center}
\includegraphics[width=\columnwidth]{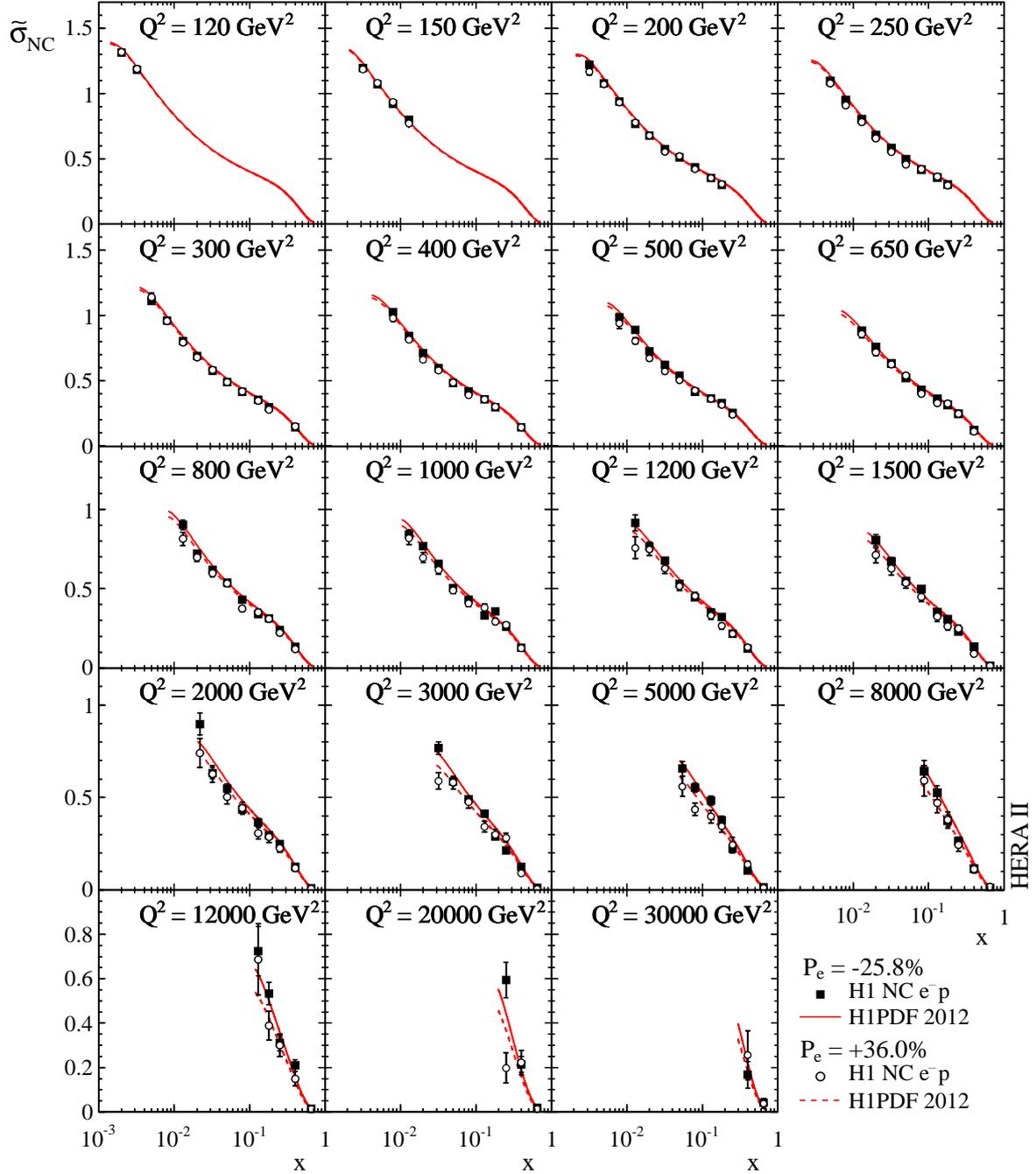}
\end{center}
\caption{ NC reduced cross sections $\tilde{\sigma}_{NC}$ for
  $e^-p$ $L$ (solid squares) and $R$ (open circles) data sets shown for
  various fixed $Q^2$ as a function of $x$. The measurement at 
  $Q^2=50\,000\,{\rm GeV}^2$ is not shown. The inner and outer 
  error bars represent the statistical and total errors, respectively.
  The luminosity and polarisation uncertainties are not included in the error 
  bars. The curves show the corresponding expectations from H1PDF\,2012.
}
\label{fig:ncdxdq2_ele} 
\end{figure}

\begin{figure}[\tablepos]
\begin{center}
\includegraphics[width=\columnwidth]{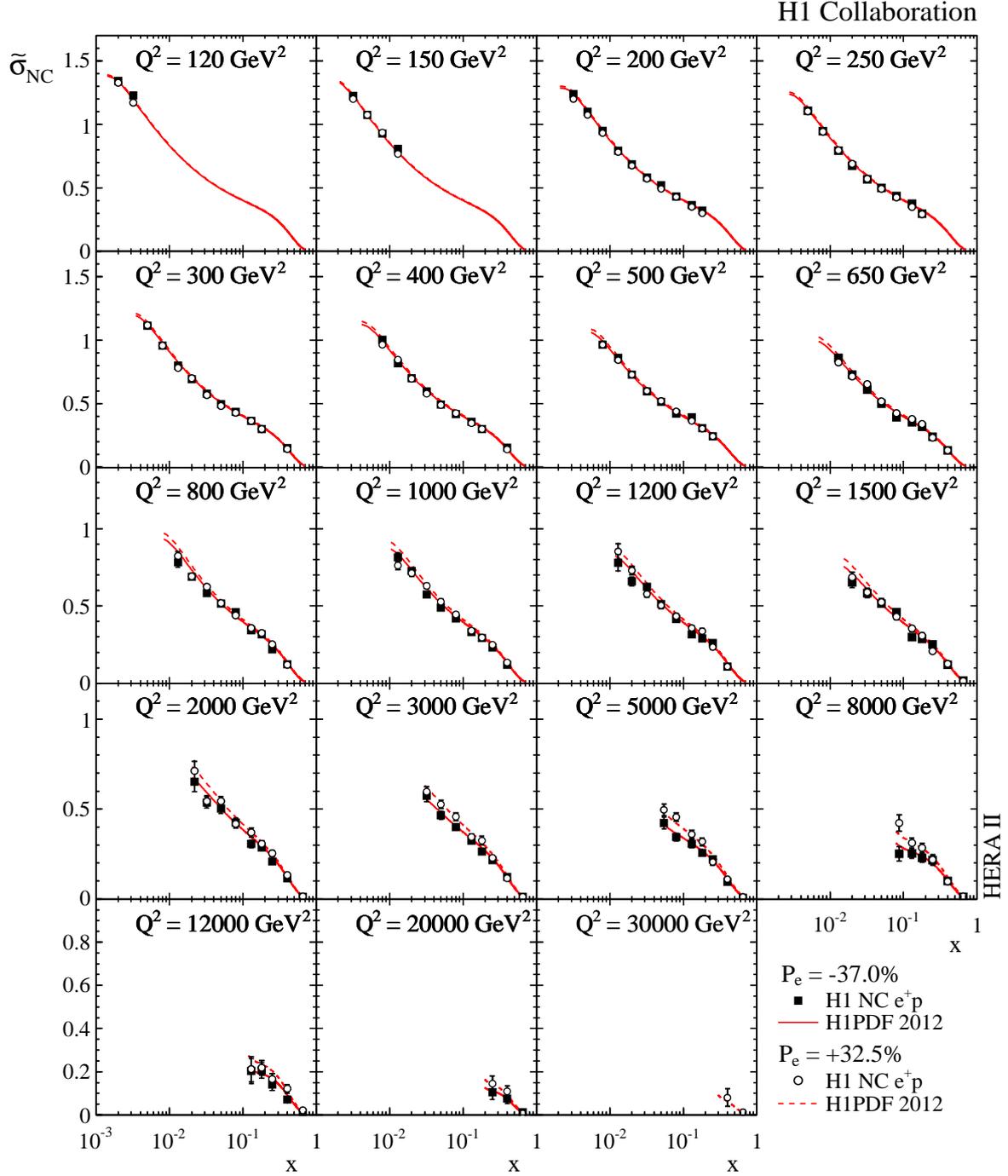}
\end{center}
\caption{ NC reduced cross sections $\tilde{\sigma}_{NC}$ for
  $e^+p$ $L$ (solid squares) and $R$ (open circles) data sets shown for
  various fixed $Q^2$ as a function of $x$. The inner and outer 
  error bars represent the statistical and total errors, respectively.
  The luminosity and polarisation uncertainties are not included in the error 
  bars. The curves show the corresponding expectations from H1PDF\,2012.
}
\label{fig:ncdxdq2_pos} 
\end{figure}

\begin{figure}[\tablepos]
\begin{center}
\includegraphics[width=\columnwidth]{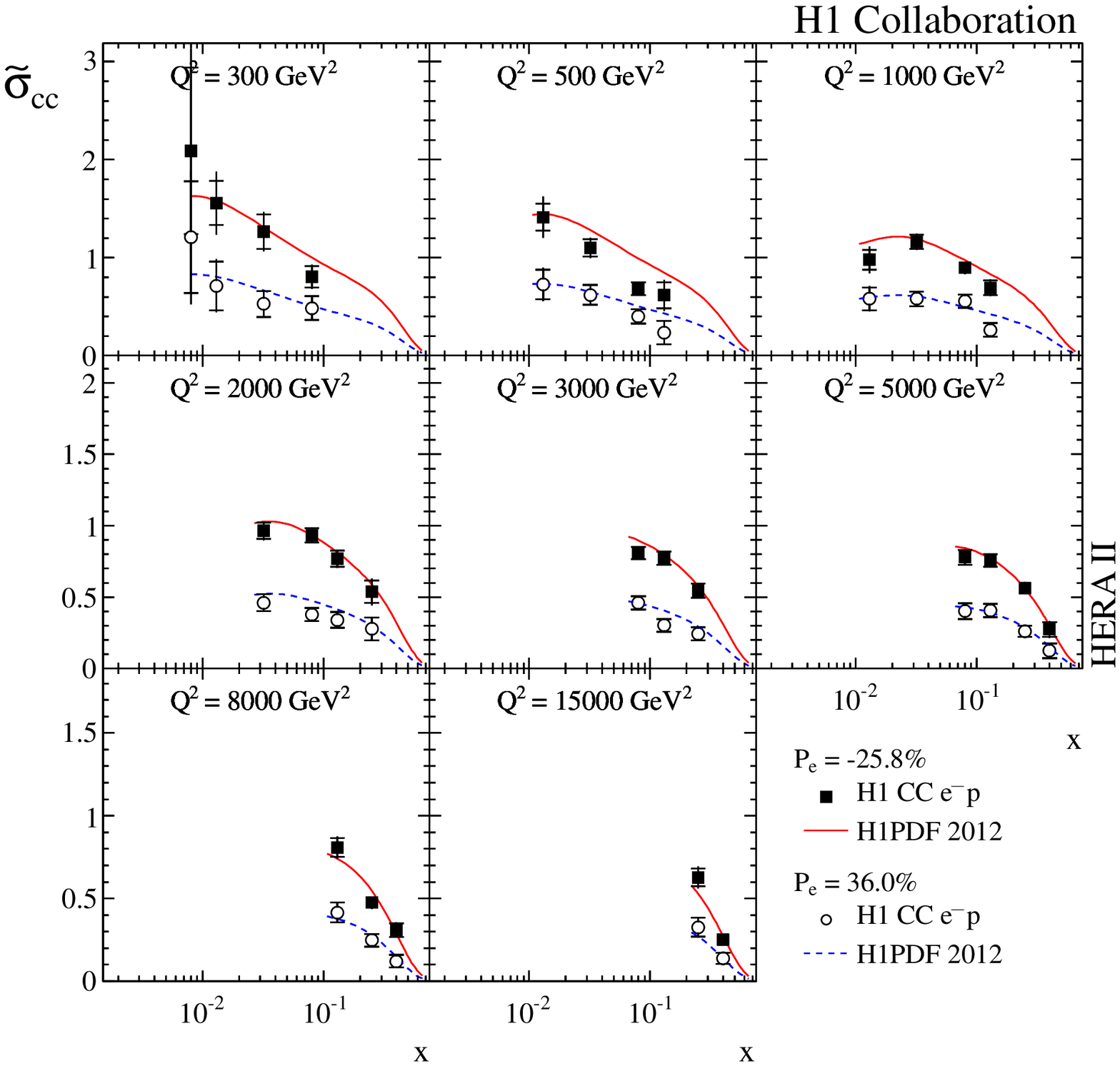}
\end{center}
\caption{ CC reduced cross sections $\tilde{\sigma}_{CC}$ for
  $e^-p$ $L$ (solid squares) and $R$ (open circles) handed data sets
  shown for various fixed $Q^2$ as a function of $x$. The inner and outer 
  error bars represent the statistical and total errors, respectively.
  The luminosity and polarisation uncertainties are not included in the error 
  bars. The curves show the corresponding expectations from H1PDF\,2012.
}
\label{fig:ccdxdq2_ele} 
\end{figure}

\begin{figure}[\tablepos]
\begin{center}
\includegraphics[width=\columnwidth]{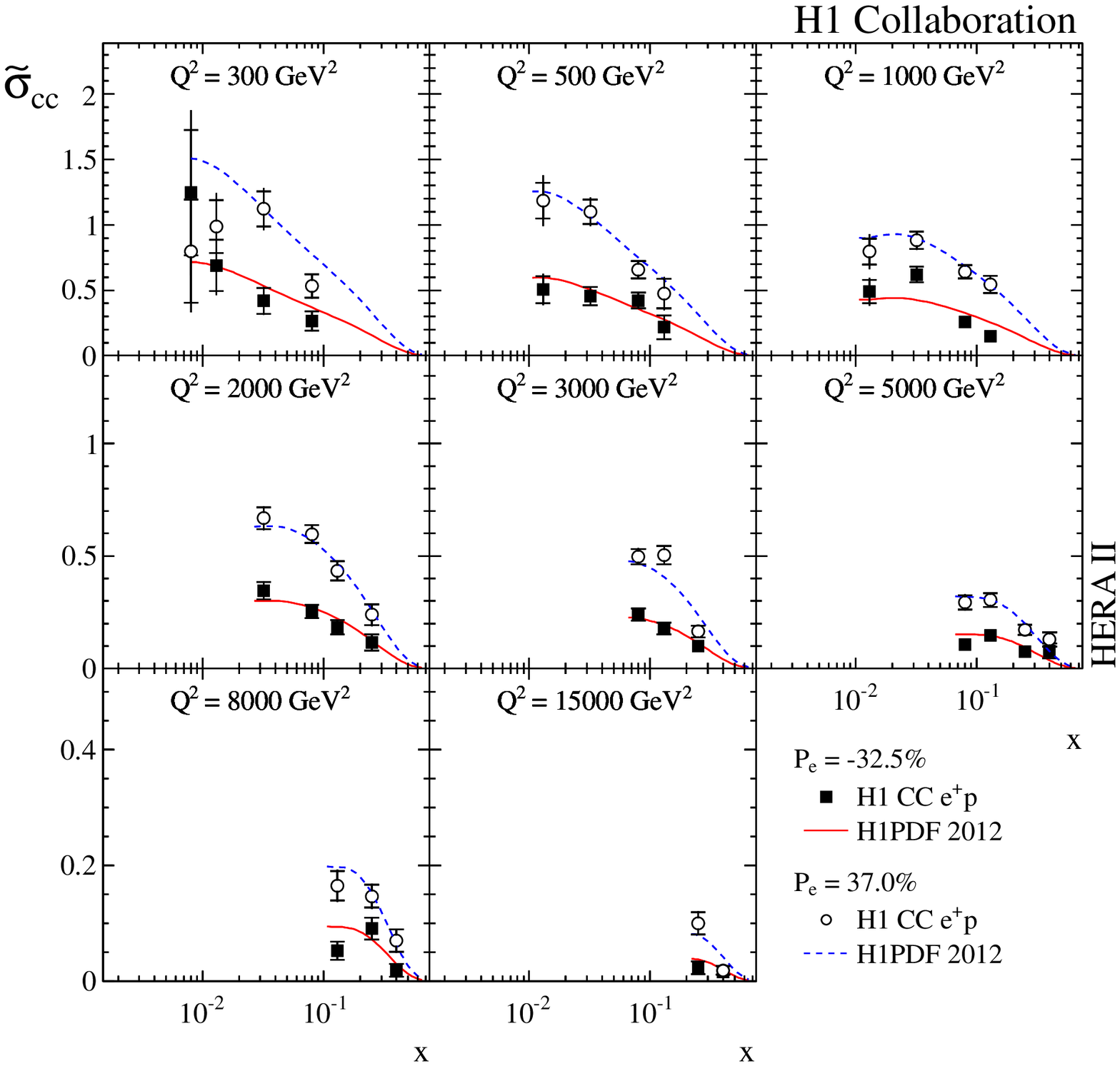}
\end{center}
\caption{ CC reduced cross sections $\tilde{\sigma}_{CC}$ for
  $e^+p$ $L$ (solid squares) and $R$ (open circles) handed data sets
  shown for various fixed $Q^2$ as a function of $x$. The inner and outer 
  error bars represent the statistical and total errors, respectively.
  The luminosity and polarisation uncertainties are not included in the error 
  bars. The curves show the corresponding expectations from H1PDF\,2012.
}
\label{fig:ccdxdq2_pos} 
\end{figure}

\begin{figure}[\tablepos]
\begin{center}
\includegraphics[width=\columnwidth]{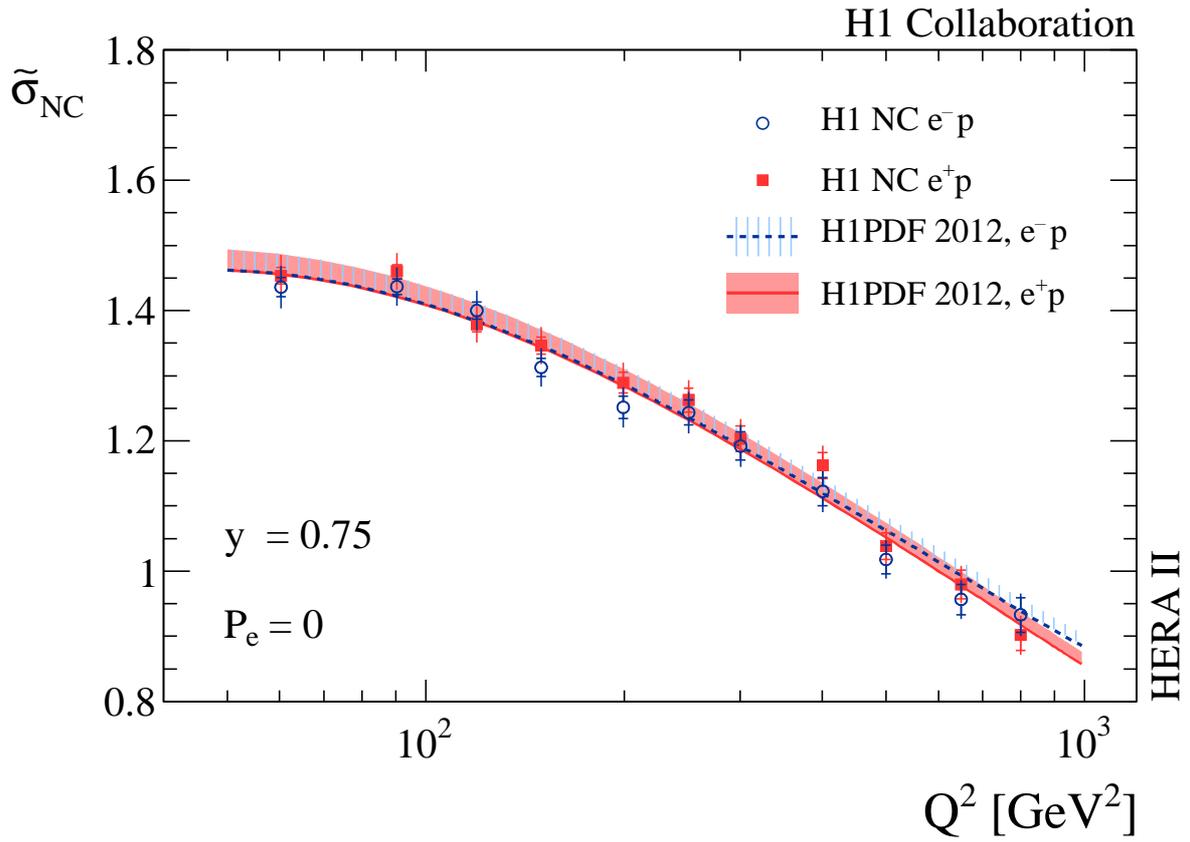}
\end{center}
\caption{ NC high $y$ reduced cross sections
  $\tilde{\sigma}_{NC}$ for $e^-p$ (open circles) and $e^+p$ (solid squares) 
  data shown as a function of $Q^2$. The inner and outer error bars represent 
  the statistical and total errors, respectively.
  The luminosity and polarisation uncertainties are not included in the error 
  bars. The error bands show the total uncertainty of the H1PDF\,2012 fit.}
\label{fig:nchighy} 
\end{figure}

\clearpage
\begin{figure}[\tablepos]
\begin{center}
\includegraphics[width=\columnwidth]{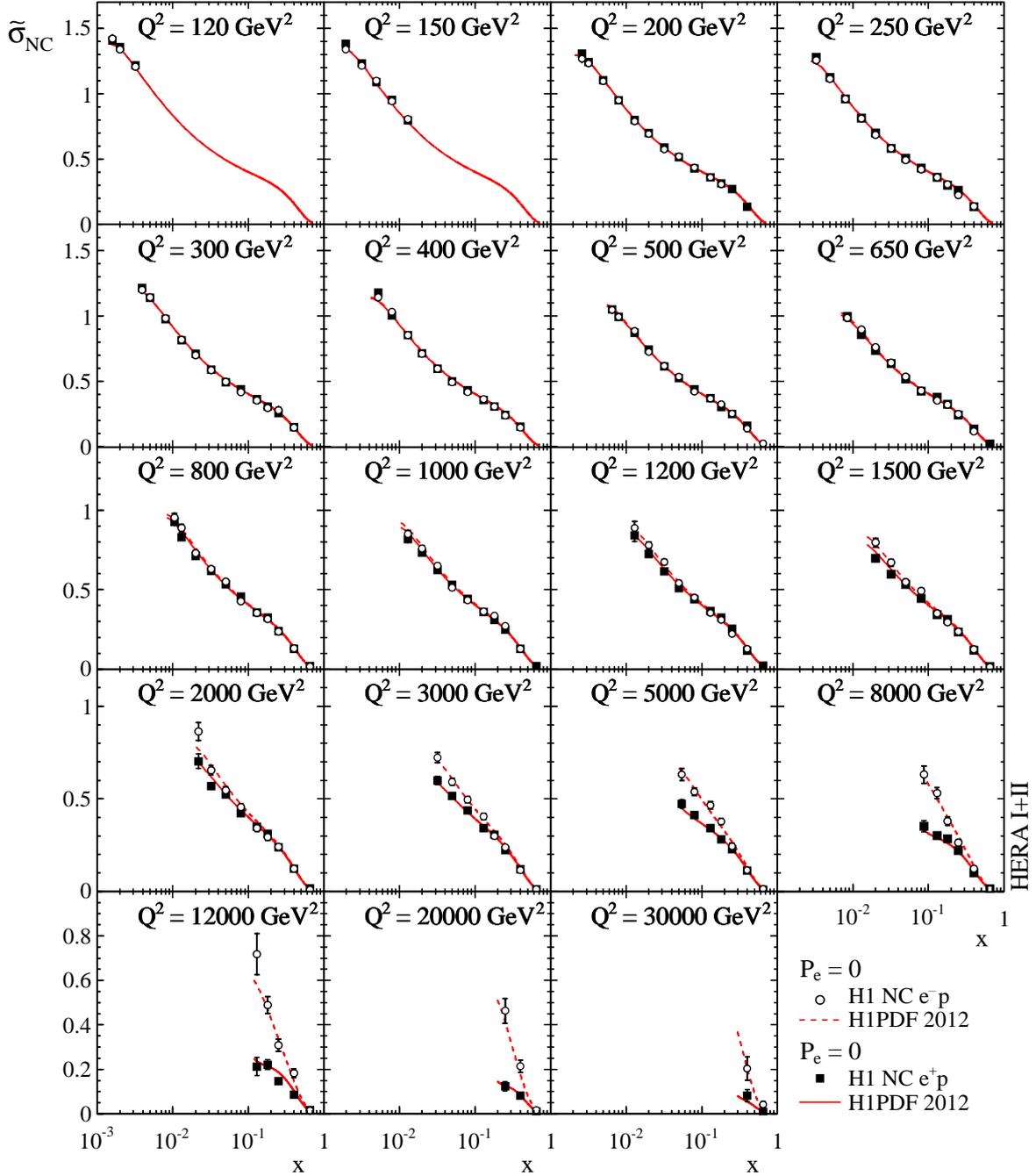}
\end{center}
\caption{ Combined HERA\,I+II unpolarised NC reduced cross sections
  $\tilde{\sigma}_{NC}$ for $e^-p$ (open circles) and $e^+p$ (solid squares) 
  data shown for various fixed $Q^2$ as a function of $x$. Only the measurements at
$\sqrt{s}=319~\rm{GeV}$ in the range $120\le Q^2\le 30\,000\,\rm{GeV}^2$ are shown.
The inner and outer 
  error bars represent the statistical and total errors, respectively.
  The curves show the corresponding expectations from H1PDF\,2012.
}
\label{fig:nccombined} 
\end{figure}

\begin{figure}[\tablepos]
\begin{center}
\includegraphics[width=\columnwidth]{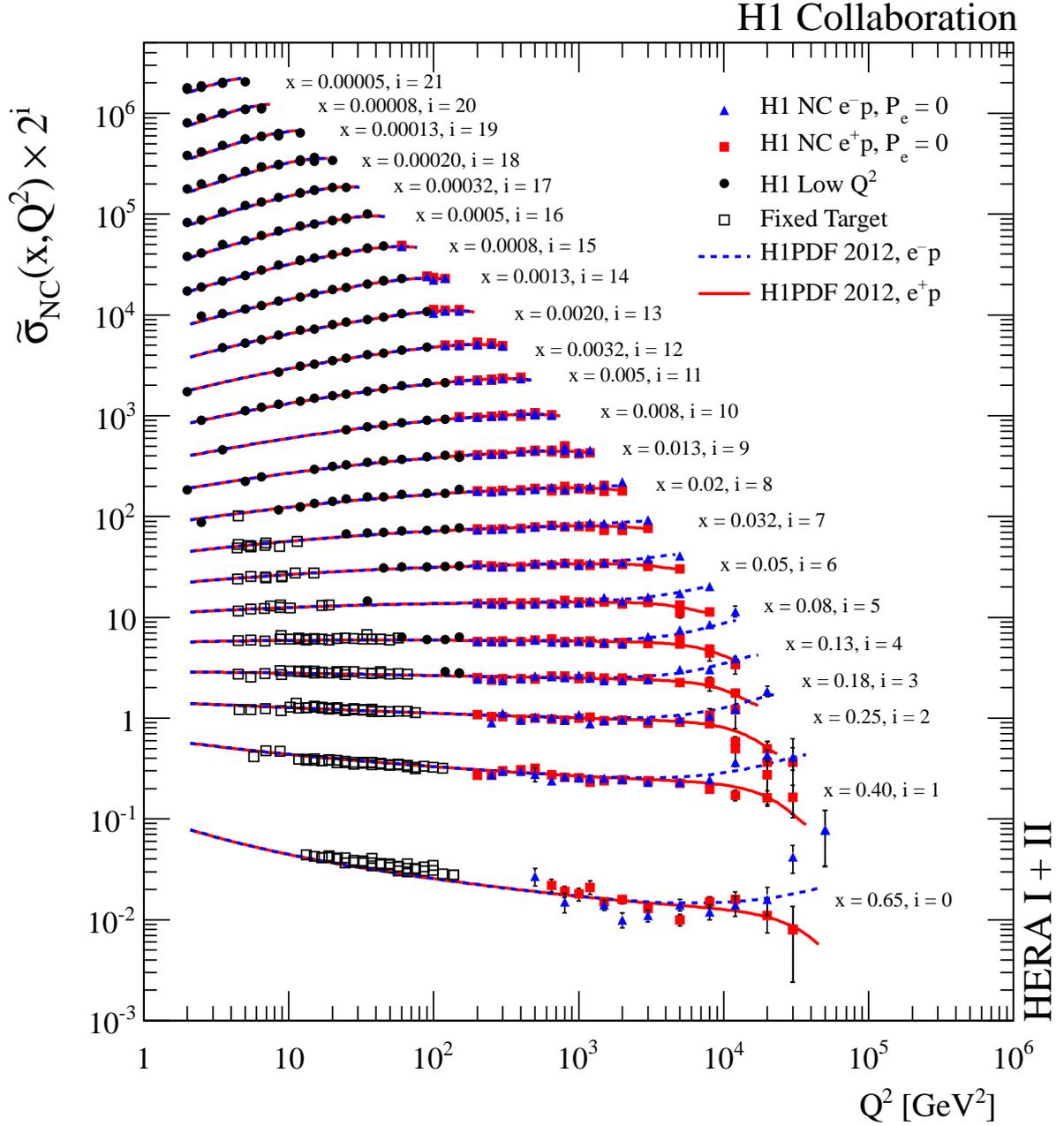}
\end{center}
\caption{ Combined HERA\,I+II unpolarised NC reduced cross
  sections $\tilde{\sigma}_{NC}$ for $e^-p$ (solid triangles), $e^+p$ 
  (solid squares) and low $Q^2$ (solid points) data shown for
  various fixed $x$ as a function of $Q^2$. The inner and outer 
  error bars represent the statistical and total errors, respectively.
  The curves show the corresponding expectations from H1PDF\,2012.
  Also shown in open squares are the fixed target data from BCDMS~\cite{bcdms}.
}
\label{fig:nccombined_scaling} 
\end{figure}

\begin{figure}[\tablepos]
\includegraphics[width=\columnwidth]{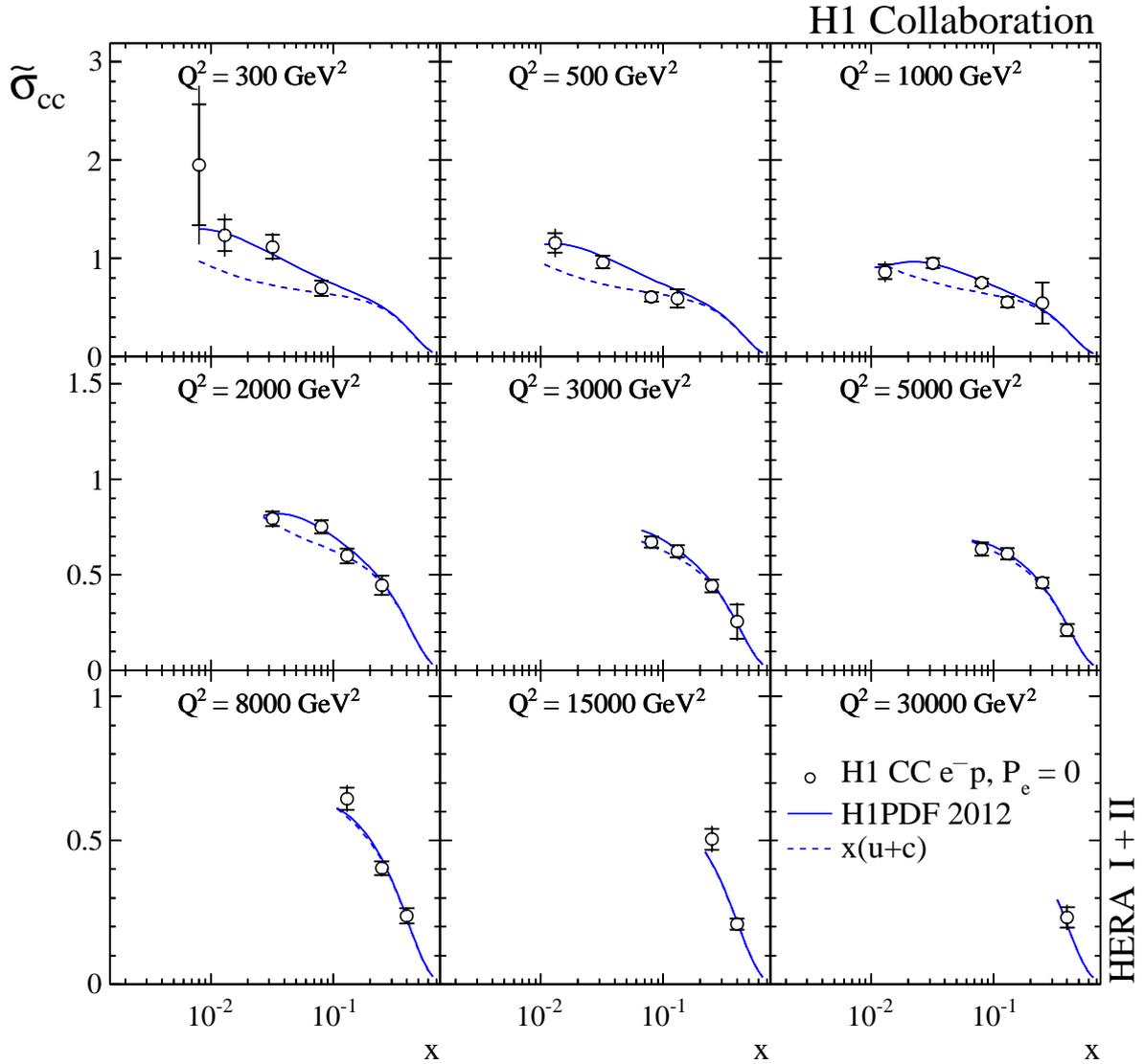}
\caption{ Combined HERA\,I+II unpolarised CC reduced cross sections
  $\tilde{\sigma}_{CC}$ for $e^-p$ data shown for various fixed $Q^2$ as a
  function of $x$ in comparison with the expectation from H1PDF\,2012. 
  The inner and outer error bars represent the statistical and total errors,
  respectively. 
  The dominant contribution $x(u+c)$ is also shown.}
\label{fig:cccombined_ele} 
\end{figure}

\begin{figure}[\tablepos]
\includegraphics[width=\columnwidth]{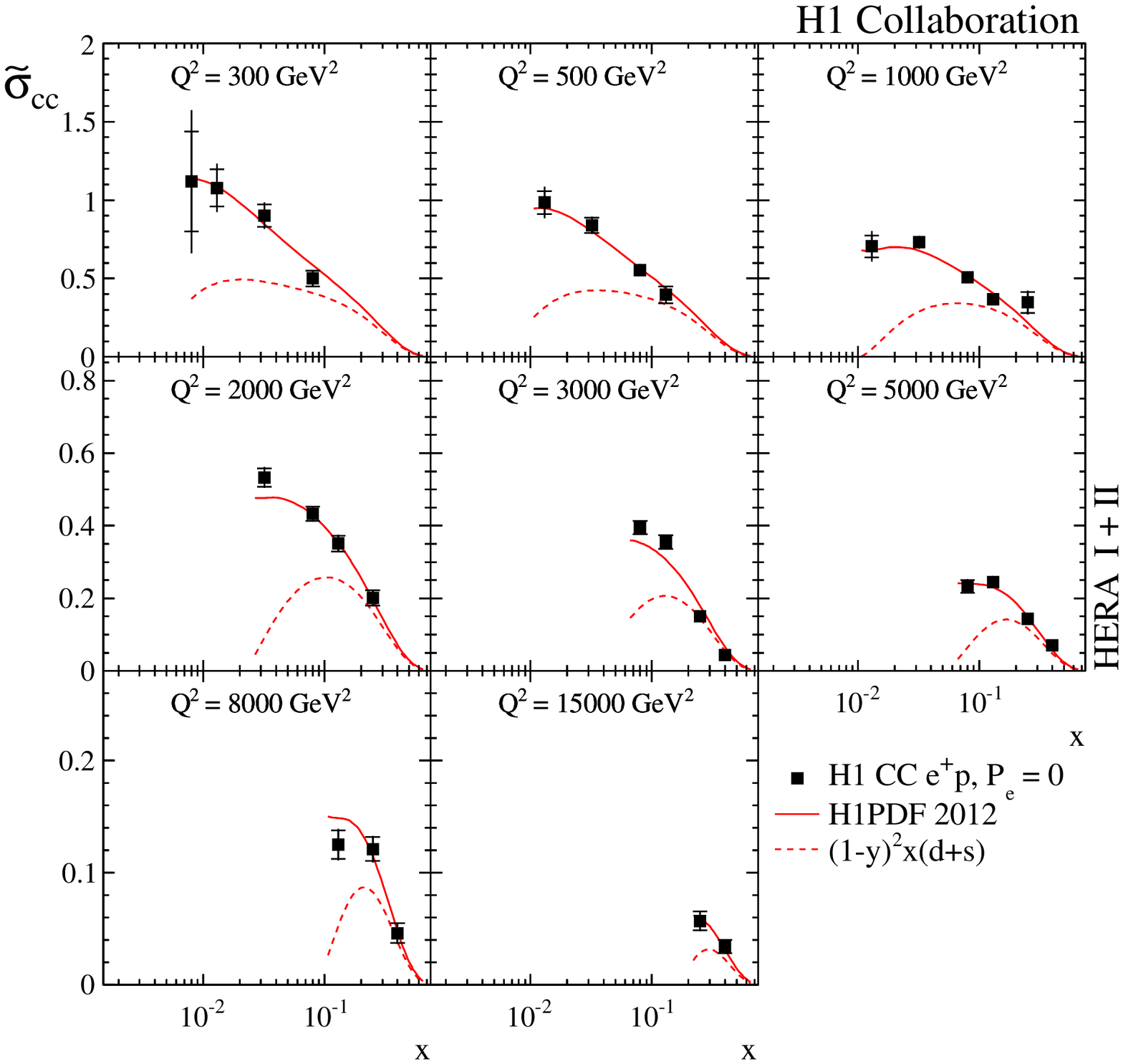}
\caption{ Combined HERA\,I+II unpolarised CC reduced cross
  sections $\tilde{\sigma}_{CC}$ for $e^+p$ data shown for various fixed $Q^2$
  as a function of $x$ in comparison with the expectation from H1PDF\,2012. 
  The inner and outer error bars represent the statistical and total errors,
  respectively. 
  The contribution $(1-y)^2x(d+s)$ is also shown.}
\label{fig:cccombined_pos} 
\end{figure}

\begin{figure}[\tablepos]
\center
\includegraphics[width=0.475\columnwidth]{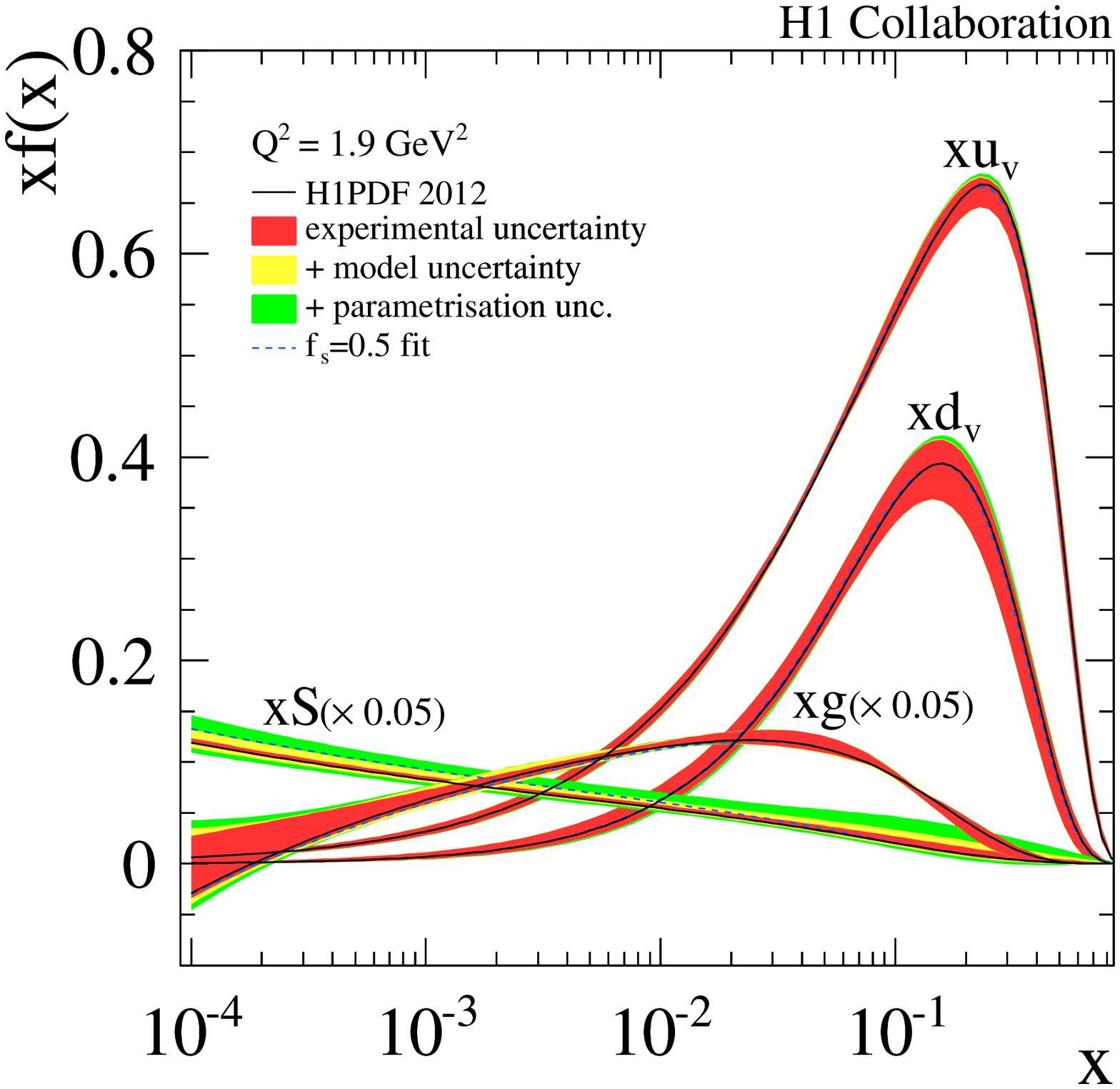}
\hspace{2mm}
\includegraphics[width=0.475\columnwidth]{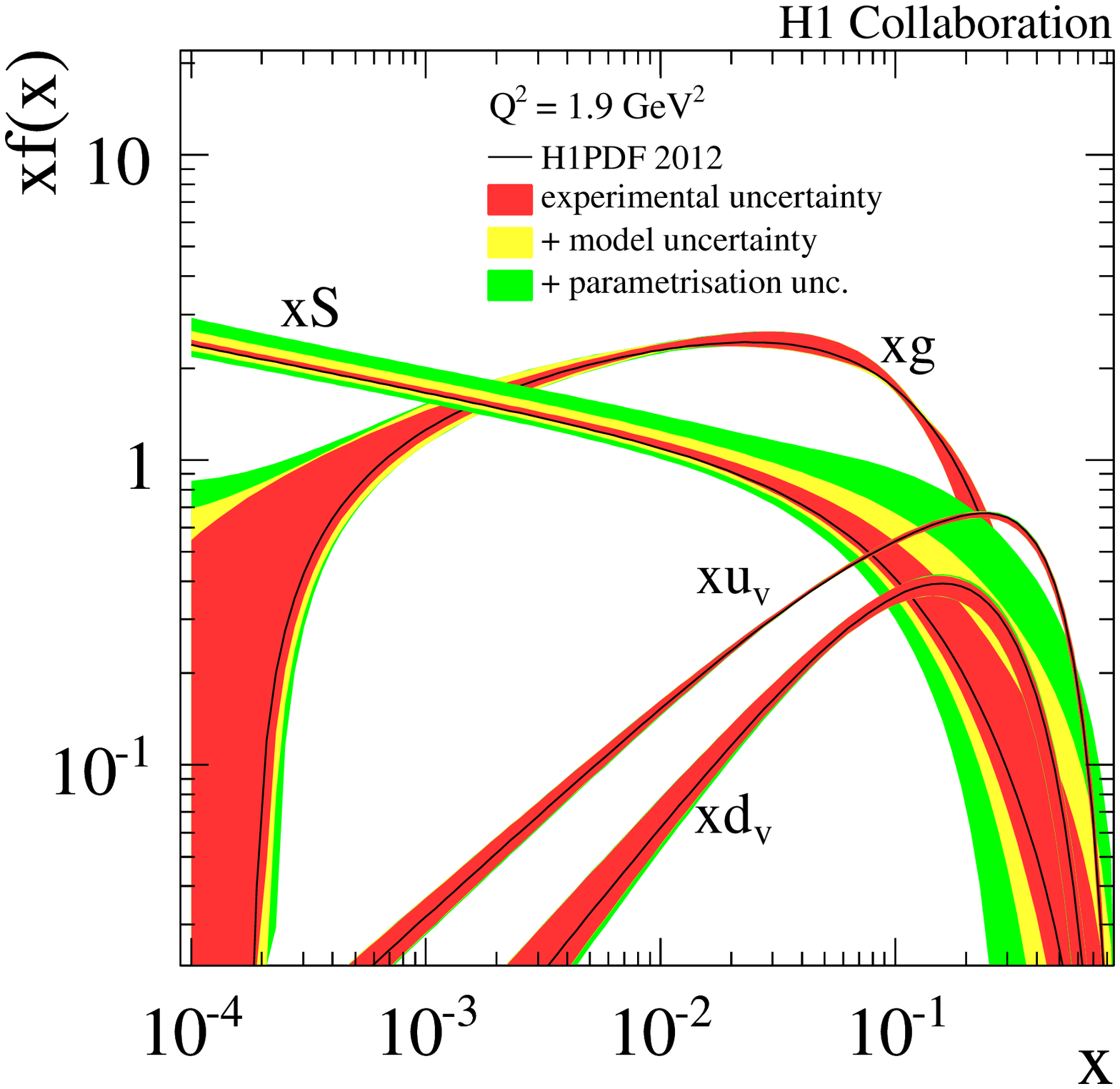}
\caption{ Parton distribution functions of H1PDF\,2012 at the starting 
  scale $Q^2=1.9\,{\rm GeV}^2$. The gluon and sea distributions in the linear 
  scale plot (left) are scaled by a factor $0.05$. The PDFs with $f_s=0.5$ 
  are also shown. The uncertainties include the experimental uncertainties 
  (inner), the model uncertainties (middle) and the parametrisation variation 
  (outer). All uncertainties are added in quadrature.}
\label{fig:pdfa} 
\end{figure}

\begin{figure}[\tablepos]
\center
\includegraphics[width=0.475\columnwidth]{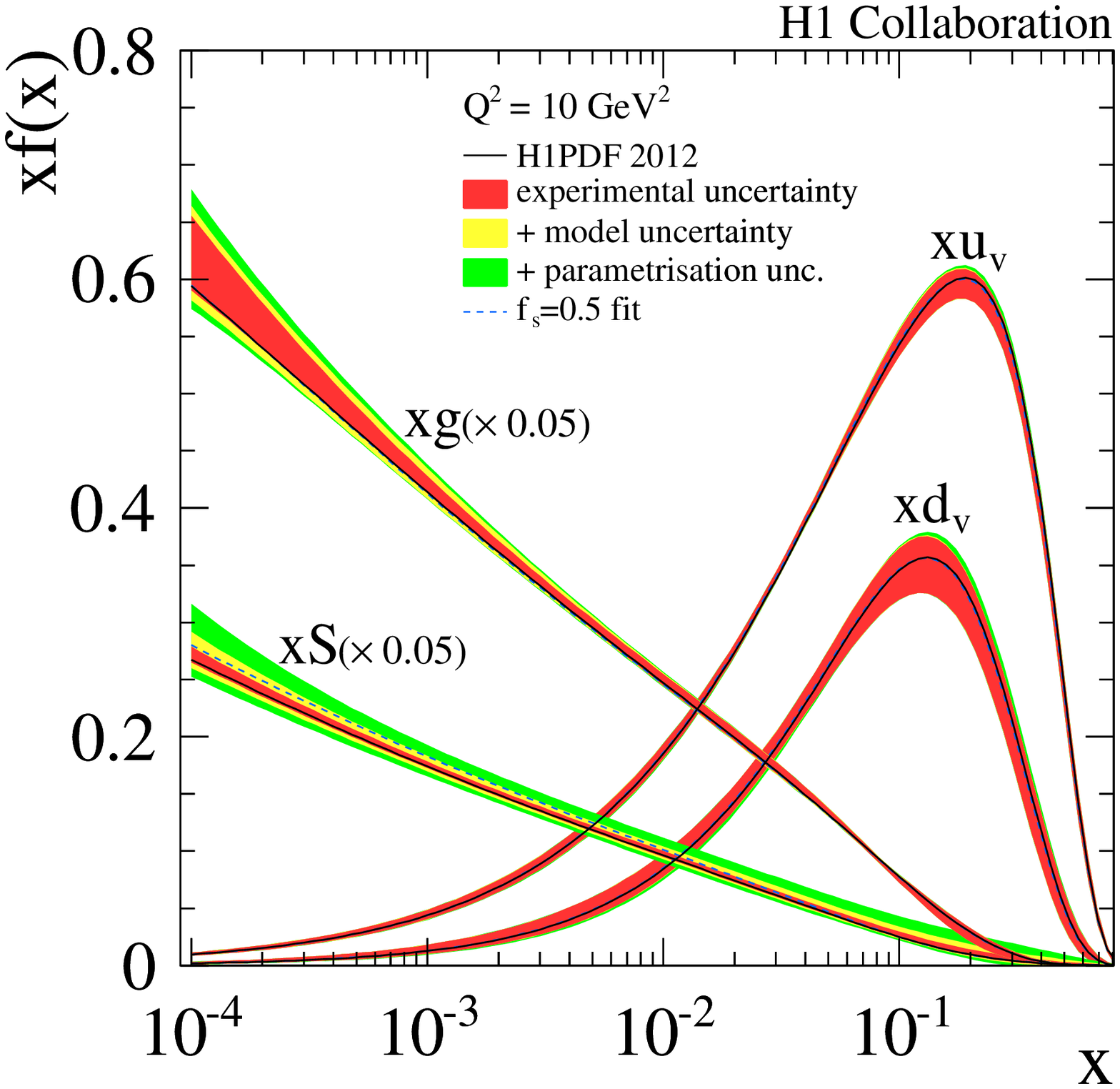}
\hspace{2mm}
\includegraphics[width=0.475\columnwidth]{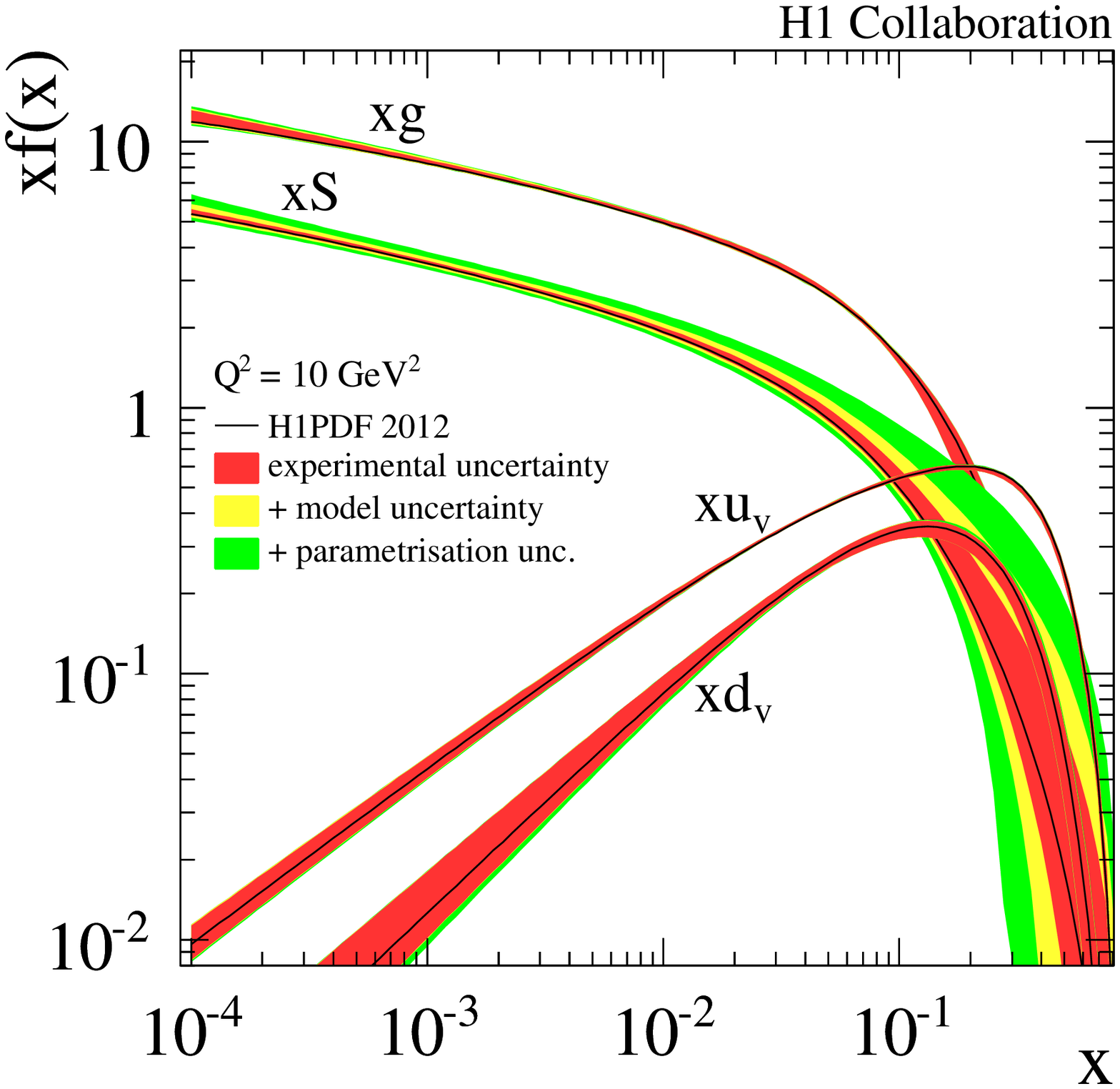}
\caption{ Parton distribution functions of H1PDF\,2012 at the evolved scale 
 of $10\,{\rm GeV}^2$. The gluon and sea distributions in the linear scale plot
 (left) are scaled by a factor $0.05$. The PDFs with $f_s=0.5$ are also shown. 
 The uncertainties include the experimental uncertainties (inner), the model 
 uncertainties (middle) and the parametrisation variation (outer). All 
 uncertainties are added in quadrature.}
\label{fig:pdfb} 
\end{figure}

\begin{figure}[\tablepos]
\center
\includegraphics[width=0.475\columnwidth]{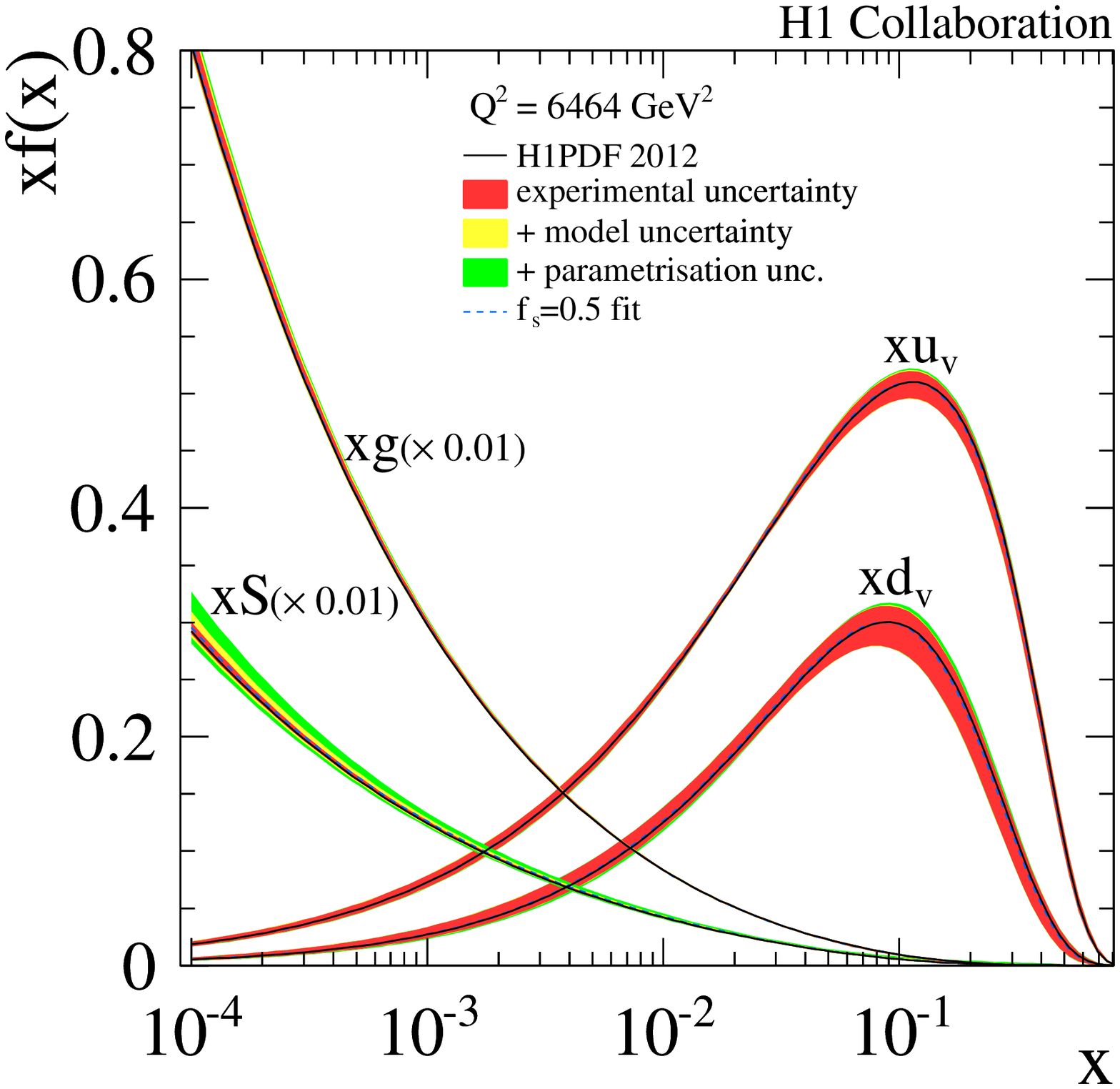}
\hspace{2mm}
\includegraphics[width=0.475\columnwidth]{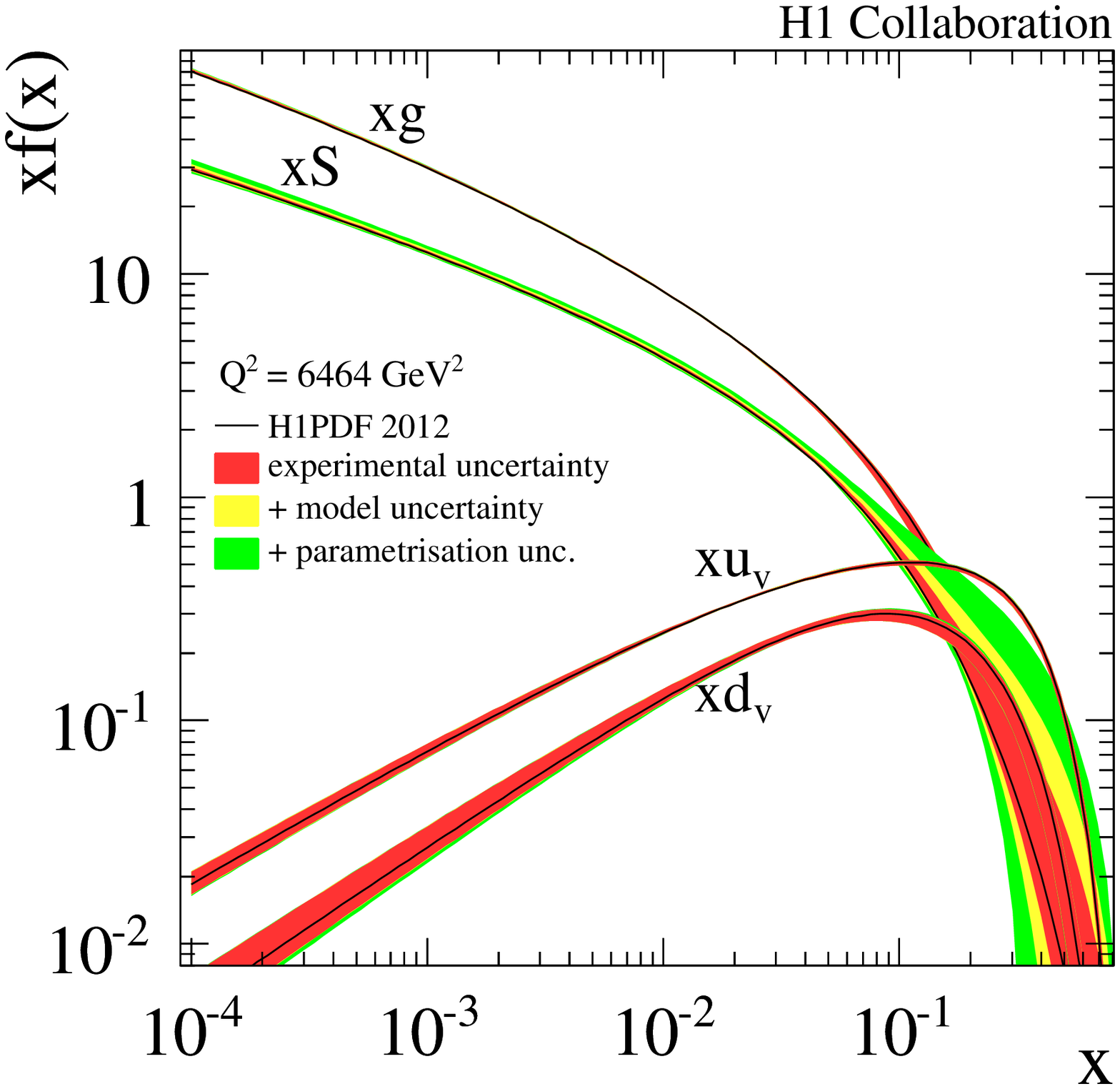}
\caption{ Parton distribution functions of H1PDF\,2012 at the evolved scale 
 of $M_W^2$. The gluon and sea distributions in the linear scale plot (left) 
 are scaled by a factor $0.01$. The PDFs with $f_s=0.5$ are also shown. 
 The uncertainties include the experimental uncertainties (inner), the model 
 uncertainties (middle) and the parametrisation variation (outer). 
 All uncertainties are added in quadrature.}
\label{fig:pdfc} 
\end{figure}

\begin{figure}[\tablepos]
\center
\includegraphics[width=\columnwidth]{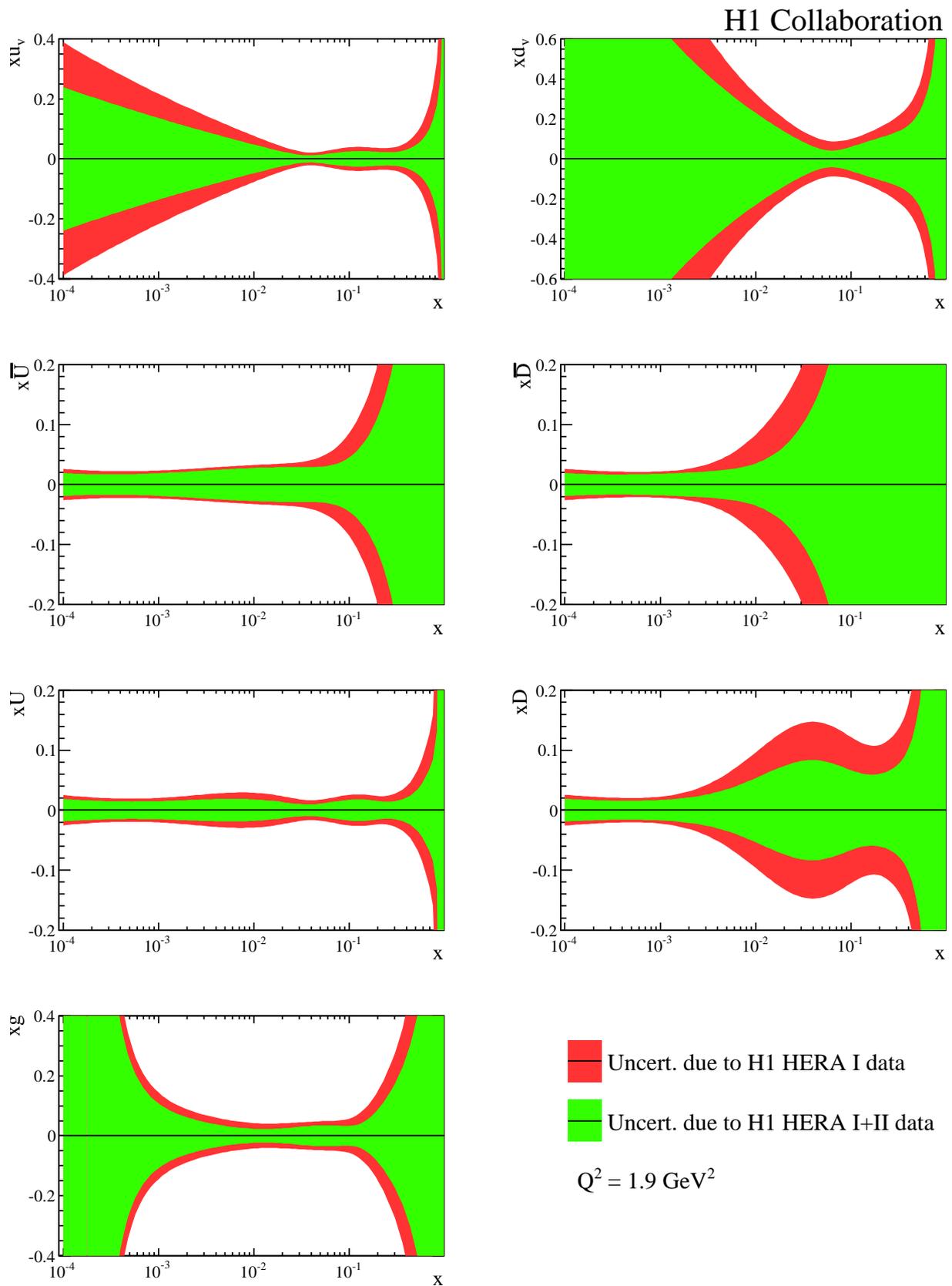}
\caption{ Comparison of relative experimental uncertainties of the PDFs 
extracted from HERA\,I (outer) vs HERA\,I+II (inner) data sets under the same 
fit conditions to better assess the effect of the new high $Q^2$ measurements. 
}
\label{fig:hera1vs2} 
\end{figure}

\begin{figure}[\tablepos]
\begin{center}
\subfigure[\label{fig:ncdq2-ele}]{ \includegraphics[width=0.49\columnwidth]{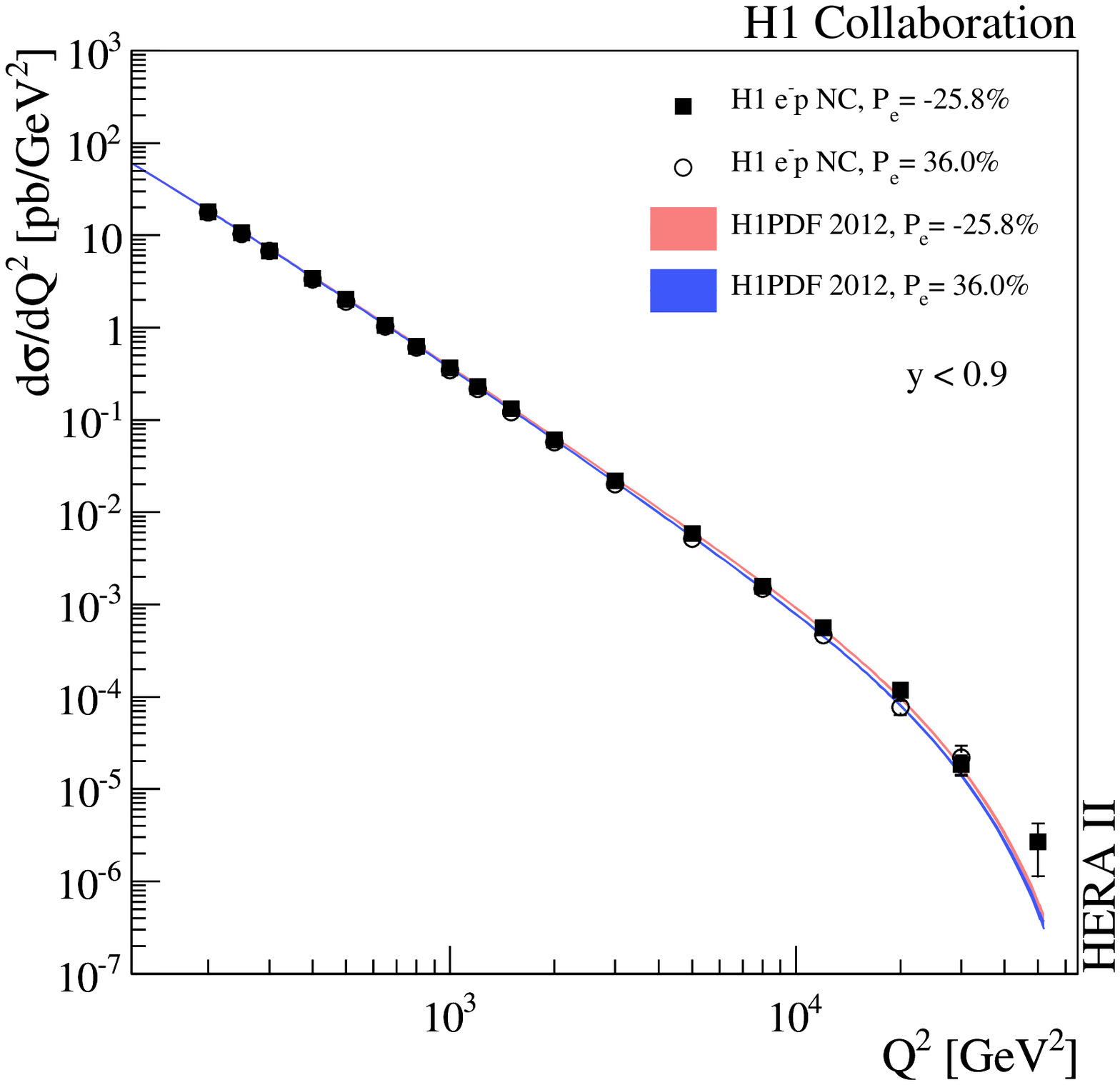}}
\subfigure[\label{fig:ncdq2-pos}]{ \includegraphics[width=0.49\columnwidth]{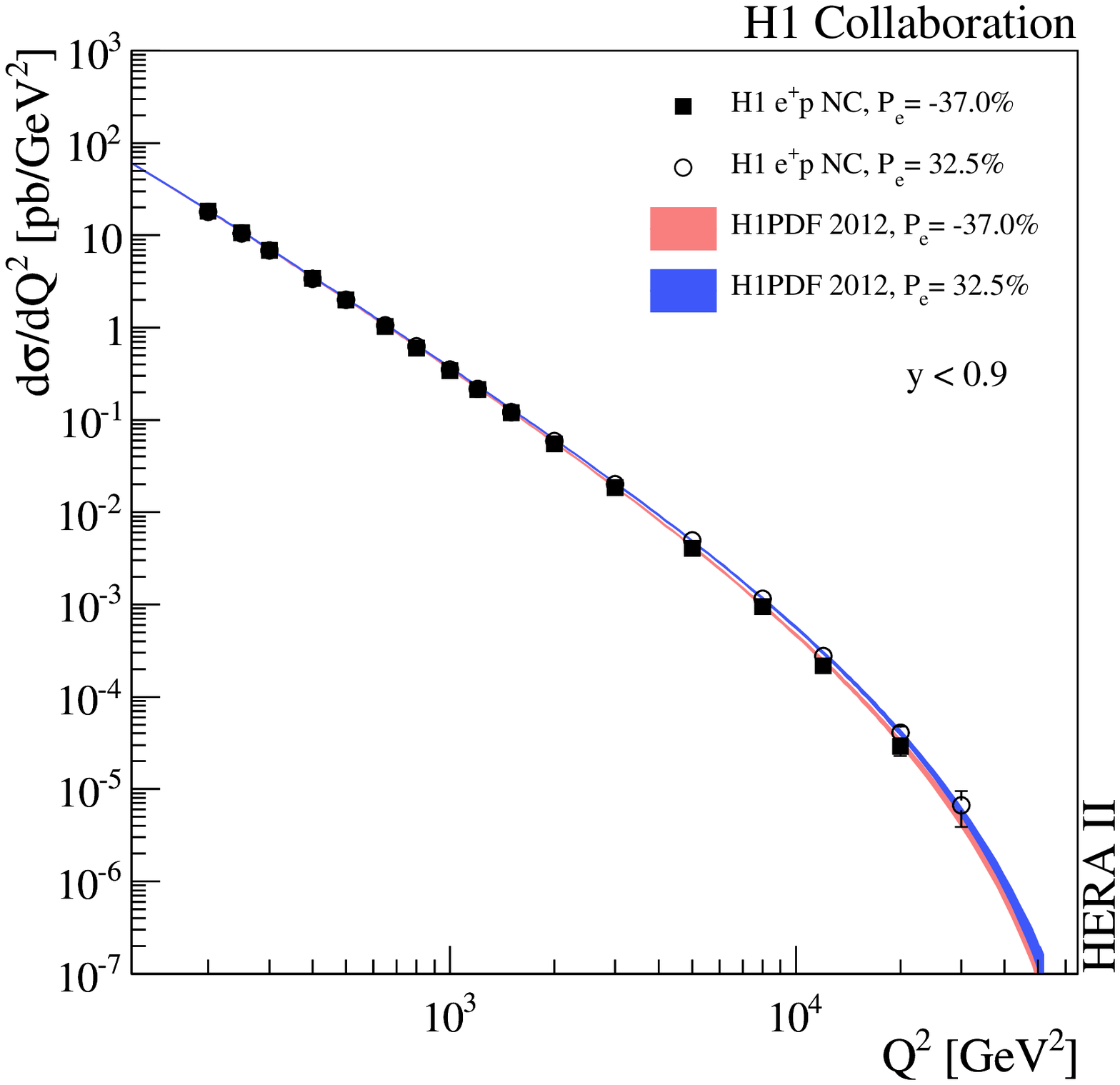}}
\subfigure[\label{fig:ncdq2ratio-ele}]{ \includegraphics[width=0.49\columnwidth]{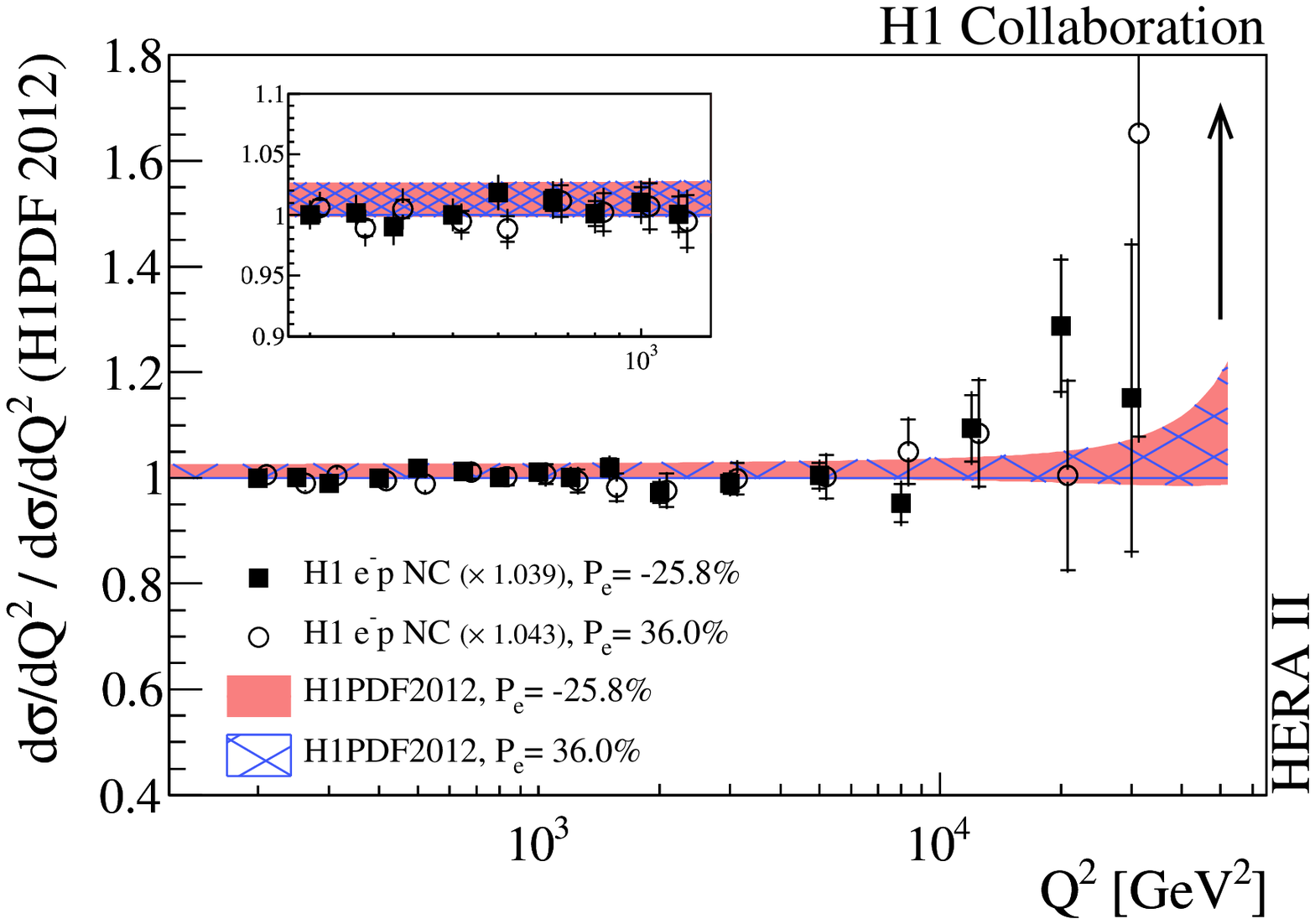}}
\subfigure[\label{fig:ncdq2ratio-pos}]{ \includegraphics[width=0.49\columnwidth]{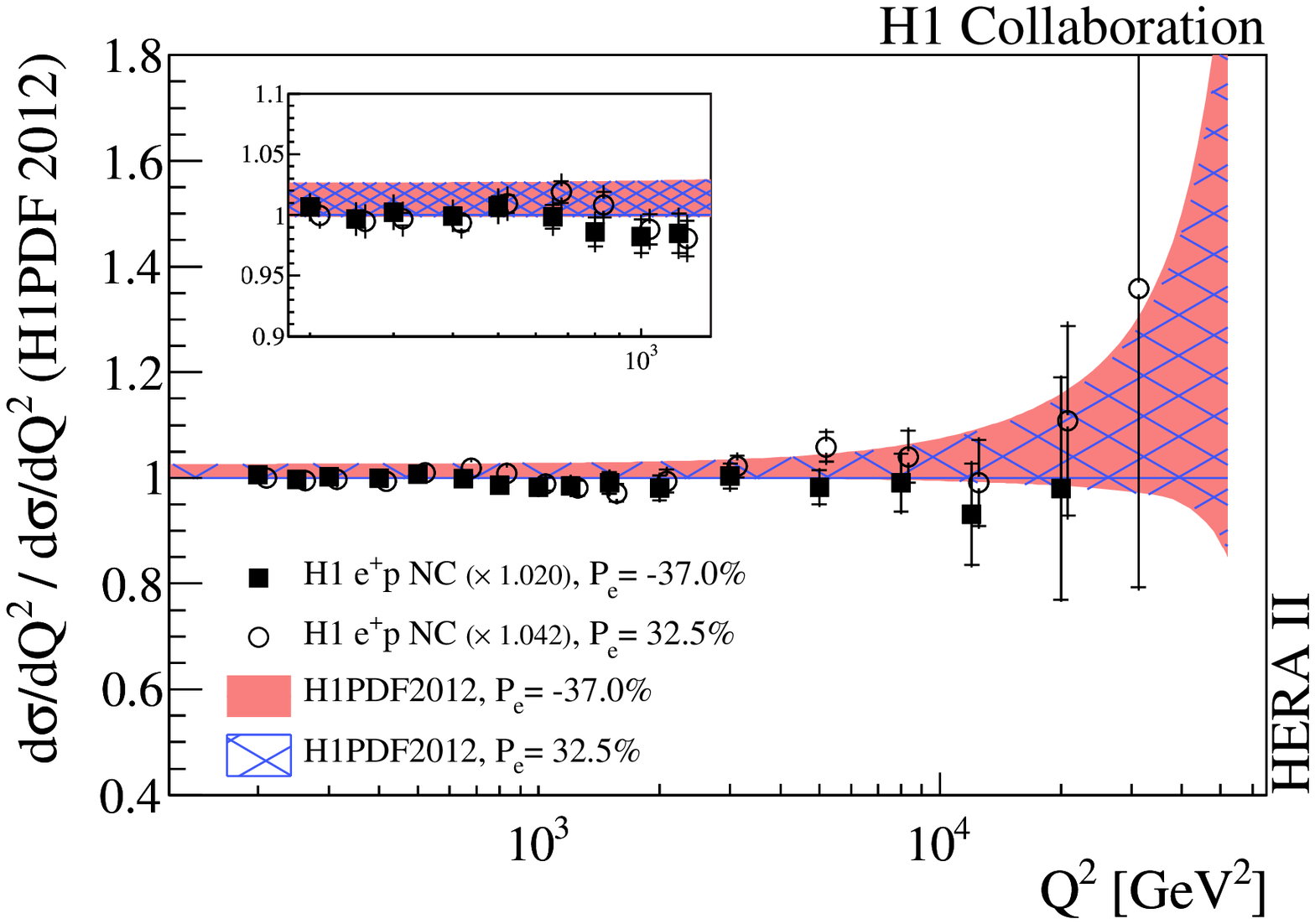}}
\end{center}
\caption{ $Q^2$ dependence of the NC cross sections 
  ${\rm d}\sigma/{\rm d}Q^2$ for the $e^-p$ (a) and $e^+p$ (b) $L$ and $R$ data 
  sets. The ratios of the $L$ and $R$ cross sections to the corresponding
  Standard Model expectations are shown for the $e^-p$ (c) and $e^+p$
  (d) data, where the normalisation shifts as determined from the QCD fit are 
  applied to the data (see Table~\ref{tab:fitresult}). The inner and outer 
  error bars represent the statistical and total errors, respectively.
  The luminosity and polarisation uncertainties are not included in the error 
  bars.}
\label{fig:ncdq2}

\end{figure}

\begin{figure}[\tablepos]
\begin{center}
\subfigure[\label{fig:ccdq2-ele}]{ \includegraphics[width=0.49\columnwidth]{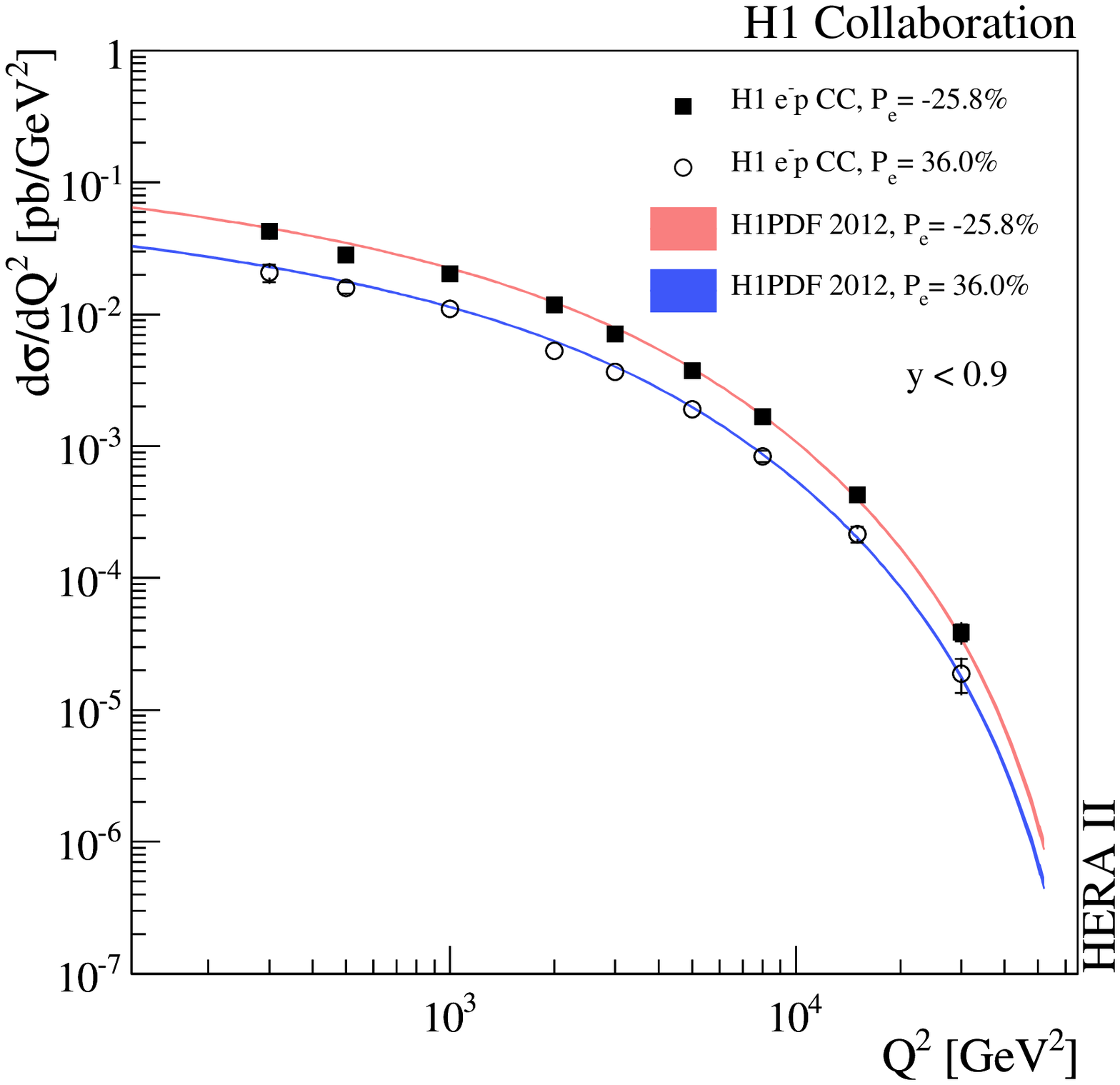}}
\subfigure[\label{fig:ccdq2-pos}]{ \includegraphics[width=0.49\columnwidth]{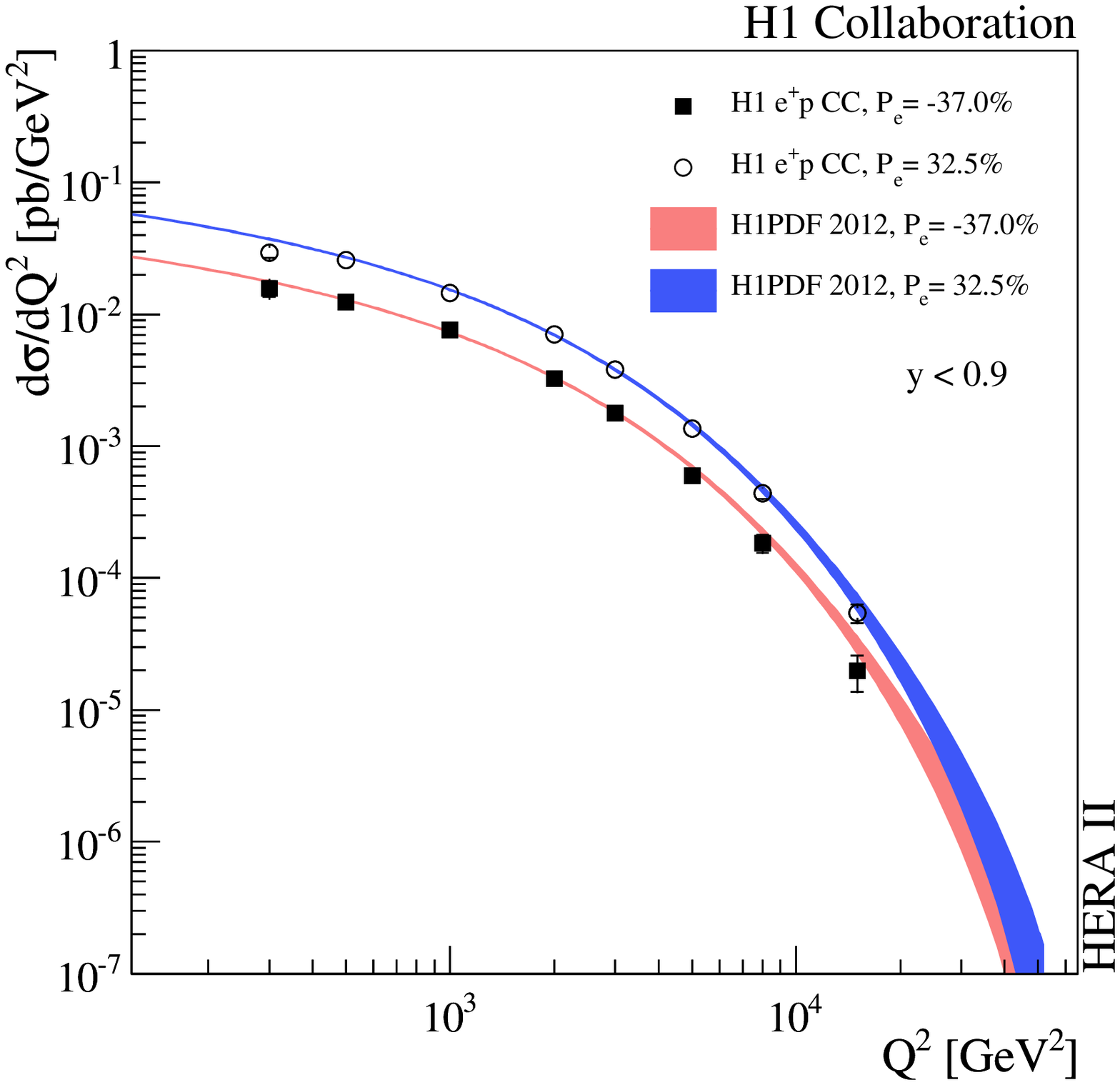}}
\subfigure[\label{fig:ccdq2ratio-ele}]{ \includegraphics[width=0.49\columnwidth]{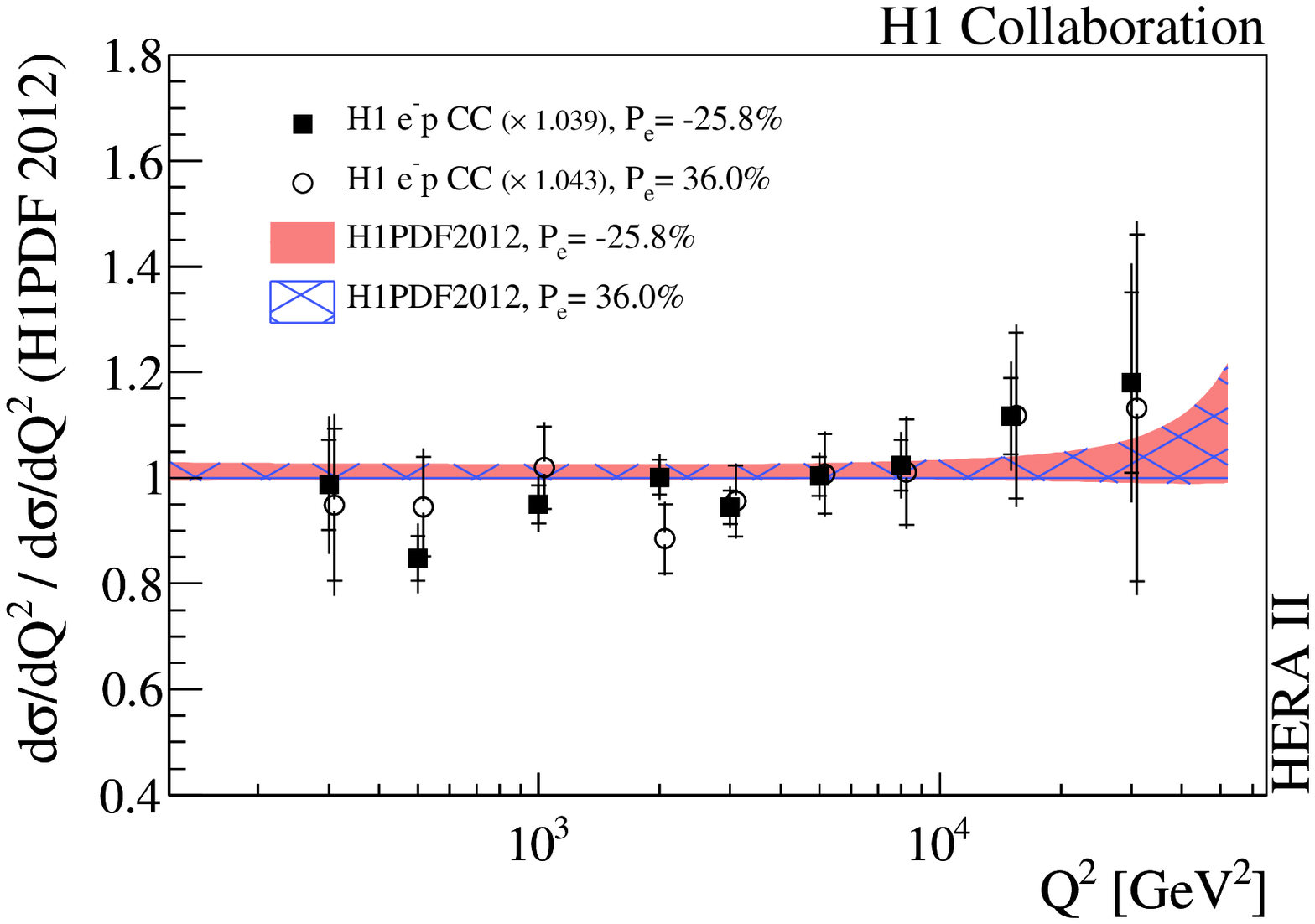}}
\subfigure[\label{fig:ccdq2ratio-pos}]{ \includegraphics[width=0.49\columnwidth]{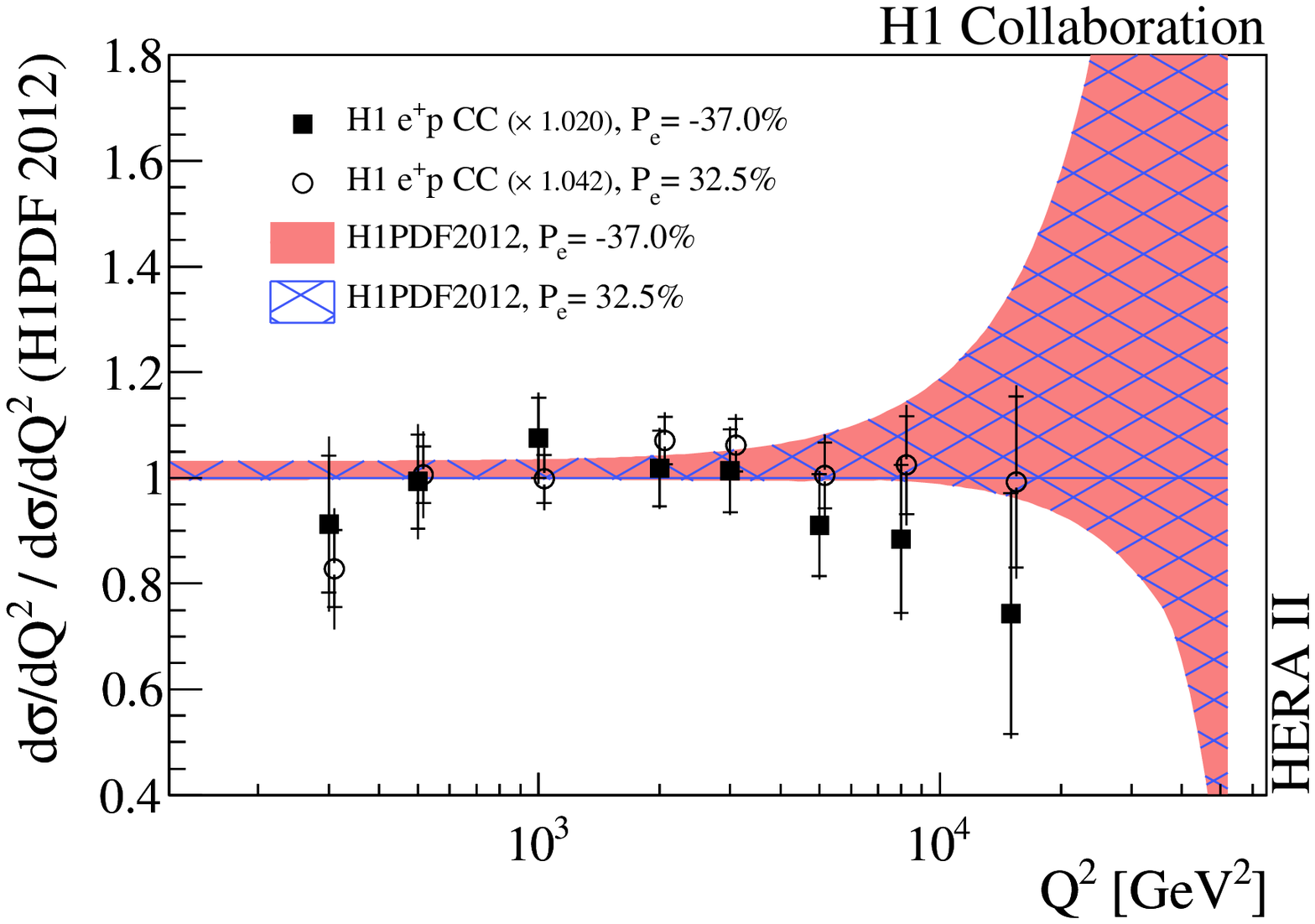}}
\end{center}
\caption{ $Q^2$ dependence of the CC cross sections 
  ${\rm d}\sigma/{\rm d}Q^2$ for the $e^-p$ (a) and $e^+p$ (b) $L$ and $R$ data 
  sets. The ratios of the $L$ and $R$ cross sections to the corresponding
  Standard Model expectations are shown for the $e^-p$ (c) and $e^+p$
  (d) data, where the normalisation shifts as determined from the QCD fit are 
  applied to the data (see Table~\ref{tab:fitresult}). The inner and outer 
  error bars represent the statistical and total errors, respectively.
  The luminosity and polarisation uncertainties are not included in the error 
  bars.}
\label{fig:ccdq2}
\end{figure}

\begin{figure}[\tablepos]
\begin{center}
\includegraphics[width=\columnwidth]{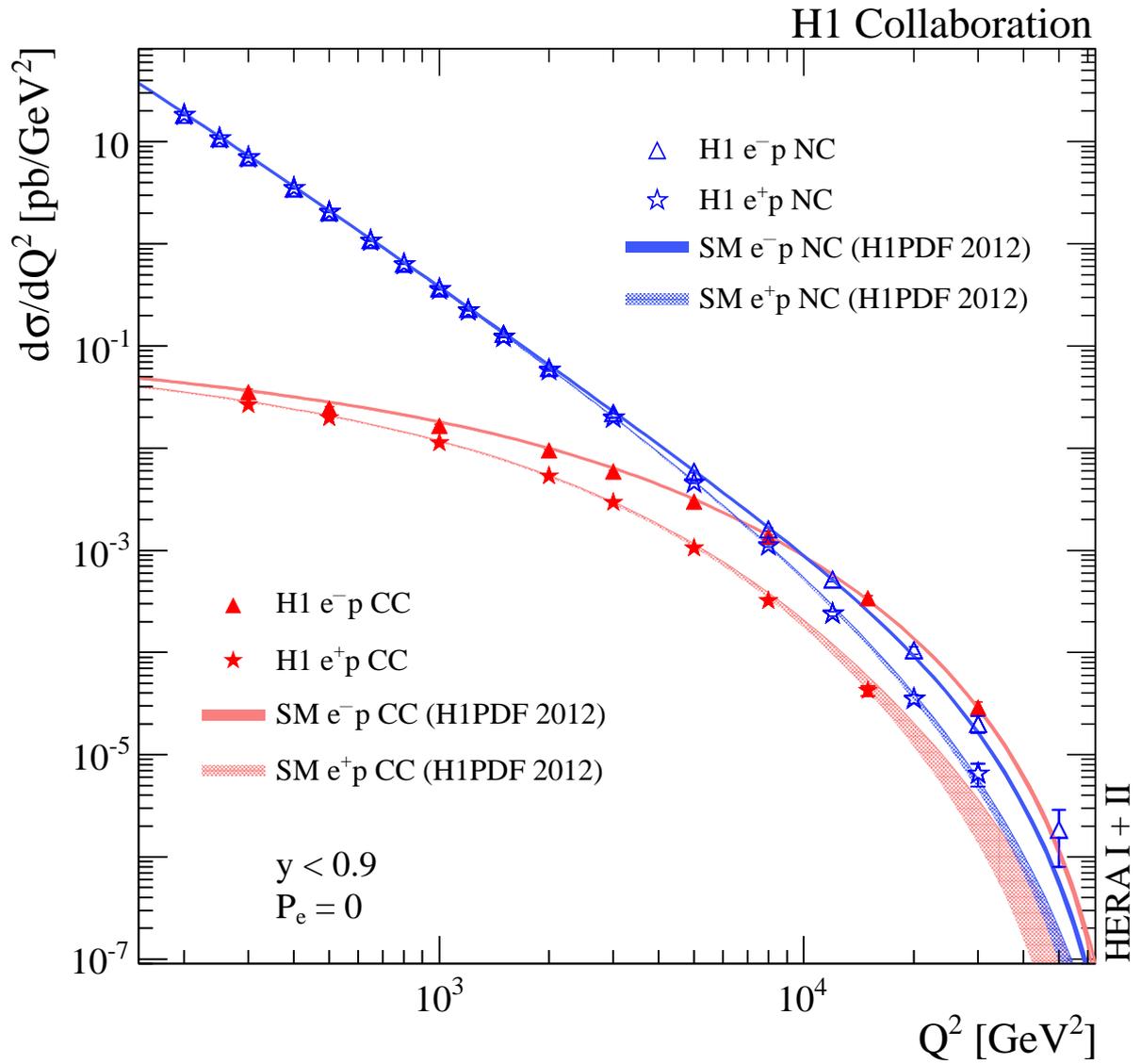}
\end{center}
\caption{ $Q^2$ dependence of the NC and CC cross sections
  $d\sigma/dQ^2$ for the combined HERA\,I+II unpolarised
  $e^-p$ and $e^+p$ data. The inner and outer error bars represent the 
  statistical and total errors, respectively.}
\label{fig:ncccdq2}
\end{figure}

\begin{figure}[\tablepos]
\begin{center}
\includegraphics[width=0.85\columnwidth]{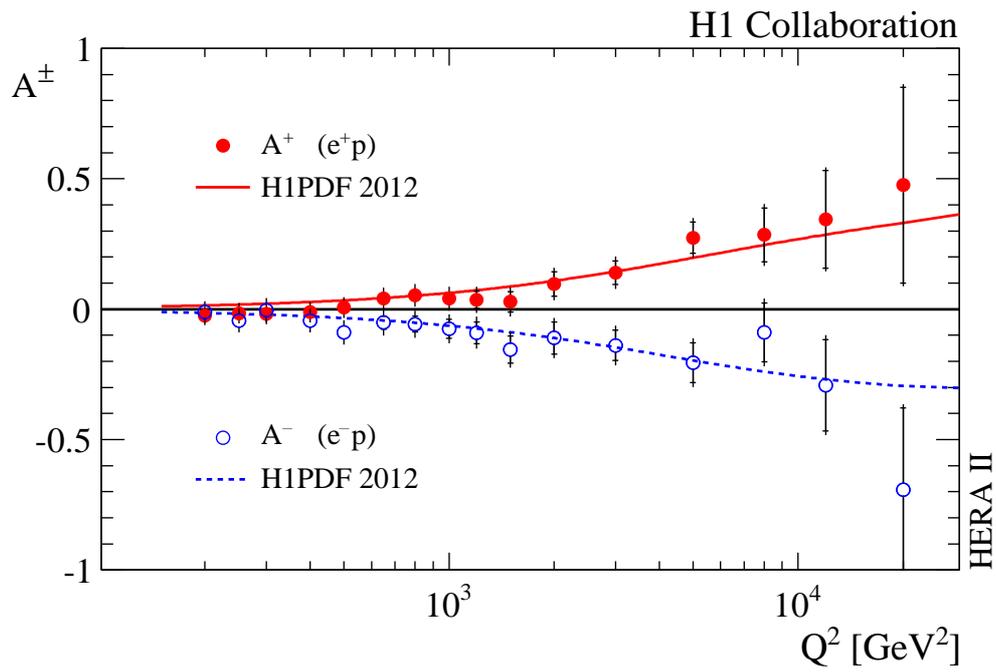}
\end{center}
\caption{ $Q^2$ dependence of the polarisation asymmetry $A^\pm$,
  for $e^+p$ (solid points) and $e^-p$ (open circles). The data are
  compared to the Standard Model expectation. The inner error bars
  represent the statistical uncertainties and the outer error bars
  represent the total errors. The normalisation uncertainty is not
  included in the error bars.}
\label{fig:pol_asy} 
\end{figure}

\begin{figure}[\tablepos]
\begin{center}
\includegraphics[width=\columnwidth]{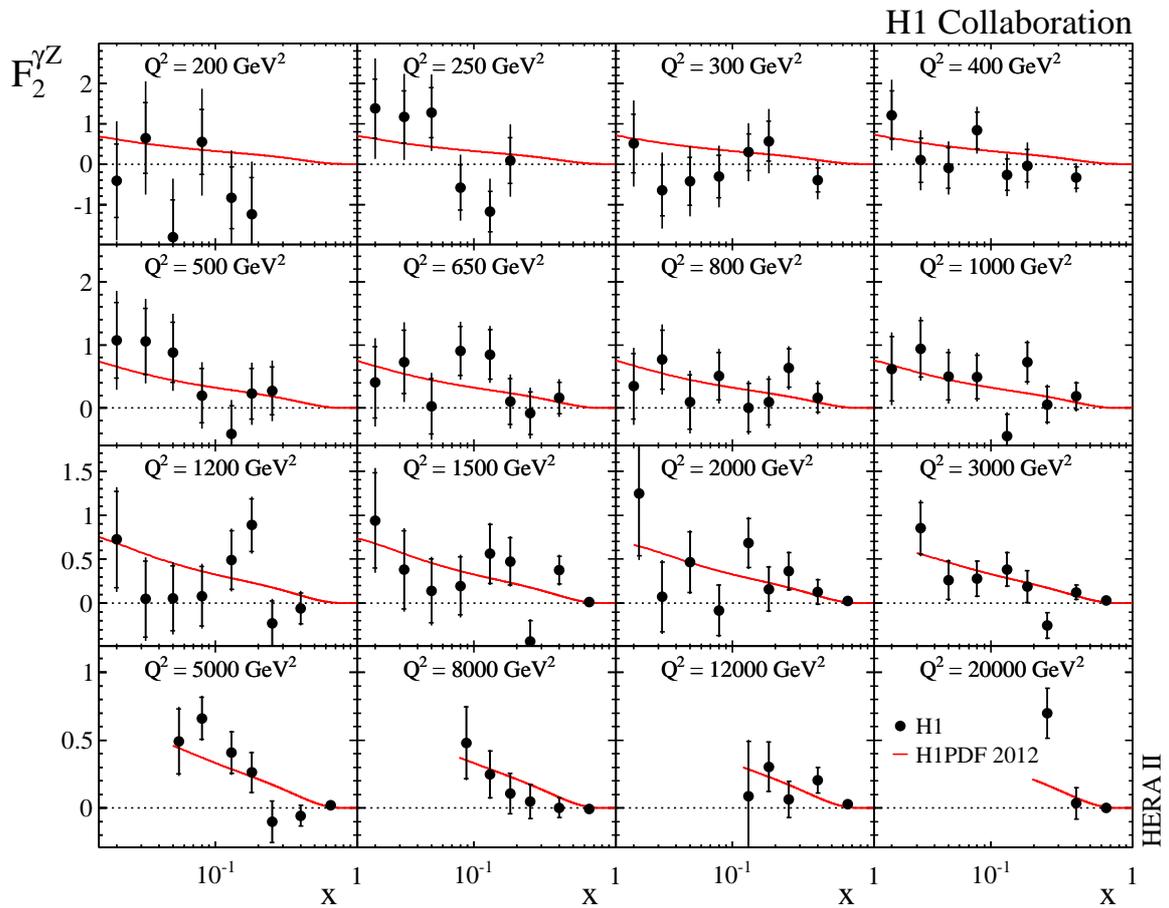}
\end{center}
\caption{ Structure function $F_2^{\gamma Z}$ for data (solid
  points) and the expectation from H1PDF\,2012 (solid curve). The 
  measurement at $Q^2=30\,000\,{\rm GeV}^2$ is now shown. The inner
  error bars represent the statistical uncertainties and the full
  error bar corresponds to the total measurement uncertainty.
}
\label{fig:f2gZ} 
\end{figure}

\begin{figure}[\tablepos]
\begin{center}
\includegraphics[width=0.95\columnwidth]{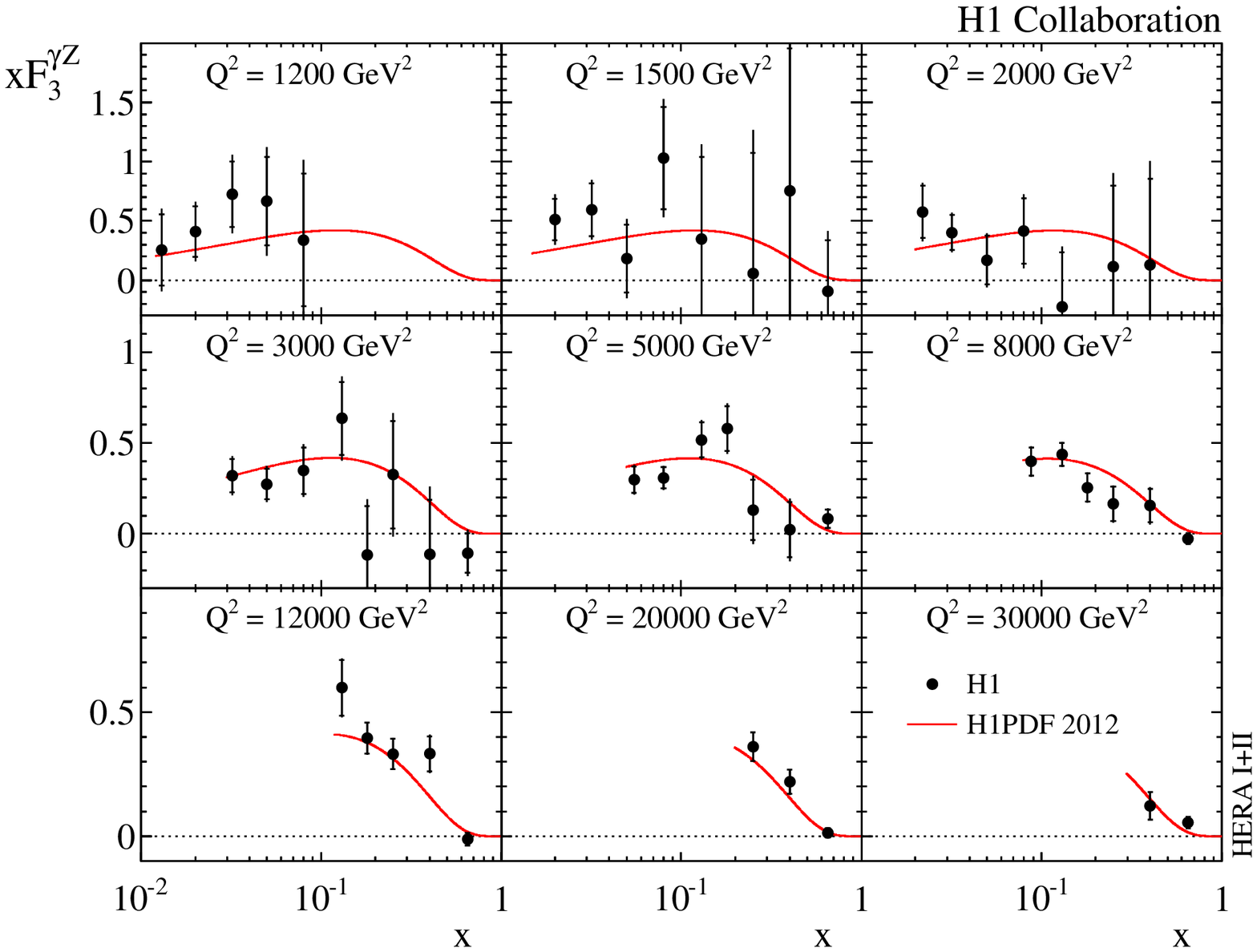}
\end{center}
\caption{ Structure function $xF_3^{\gamma Z}$ for the combined HERA\,I+II 
  data (solid points) and the expectation from H1PDF\,2012 (solid curve). 
  The inner error bars represent the statistical uncertainties and the full 
  error bar corresponds to the total measurement uncertainty.}
\label{fig:xf3} 
\end{figure}

\begin{figure}[\tablepos]
\begin{center}
\includegraphics[width=\columnwidth]{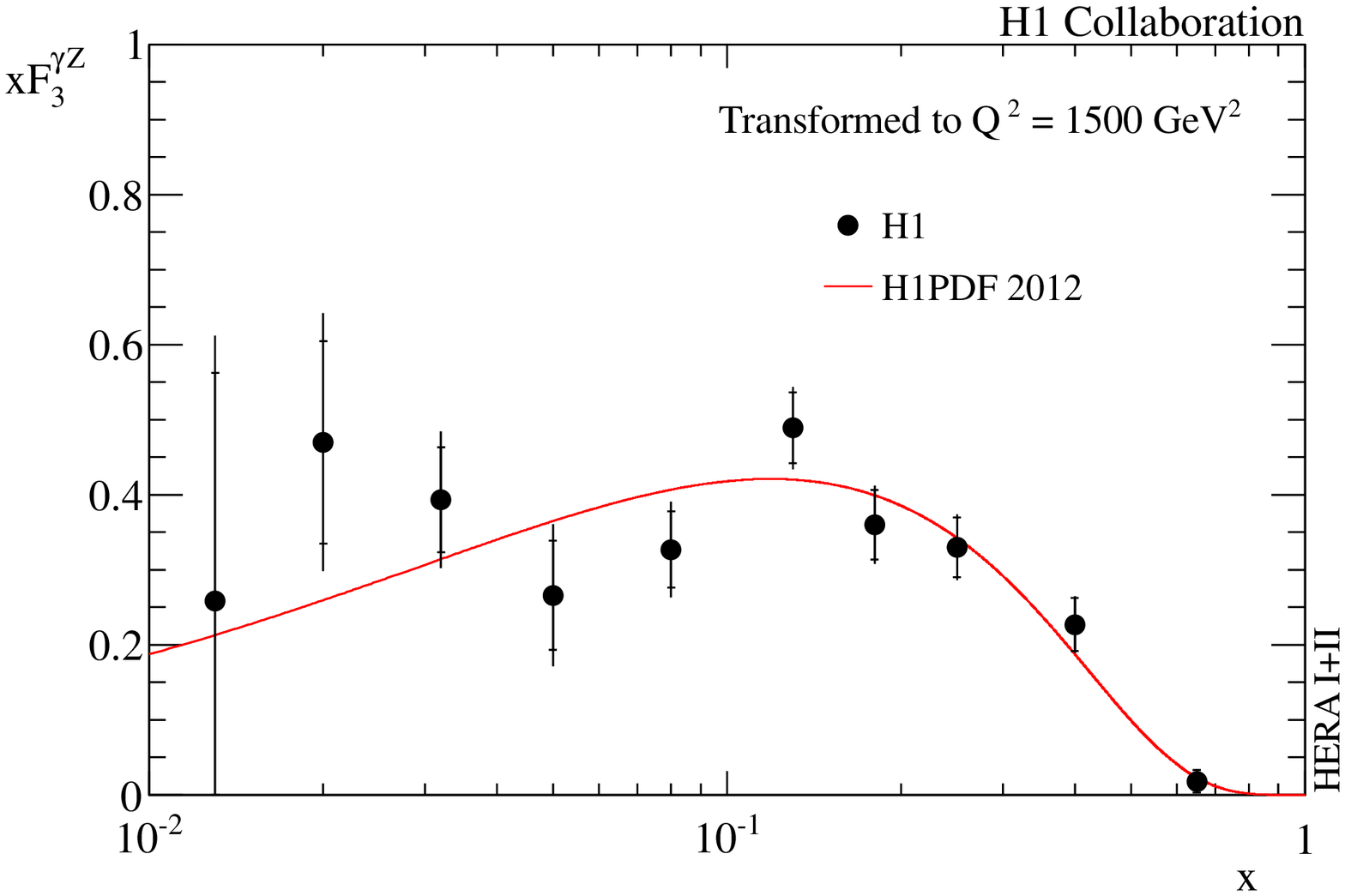}
\end{center}
\caption{ Structure function $xF_3^{\gamma Z}$ transformed to 
  $Q^2=1\,500$\,GeV$^2$ for data (solid points) and the expectation
  from H1PDF\,2012 (solid curve). The inner error bars represent the
  statistical uncertainties and the full error bar corresponds to the
  total measurement uncertainty.}
\label{fig:xf3gZ_1500} 
\end{figure}

\begin{figure}[\tablepos]
\begin{center}
\includegraphics[width=\columnwidth]{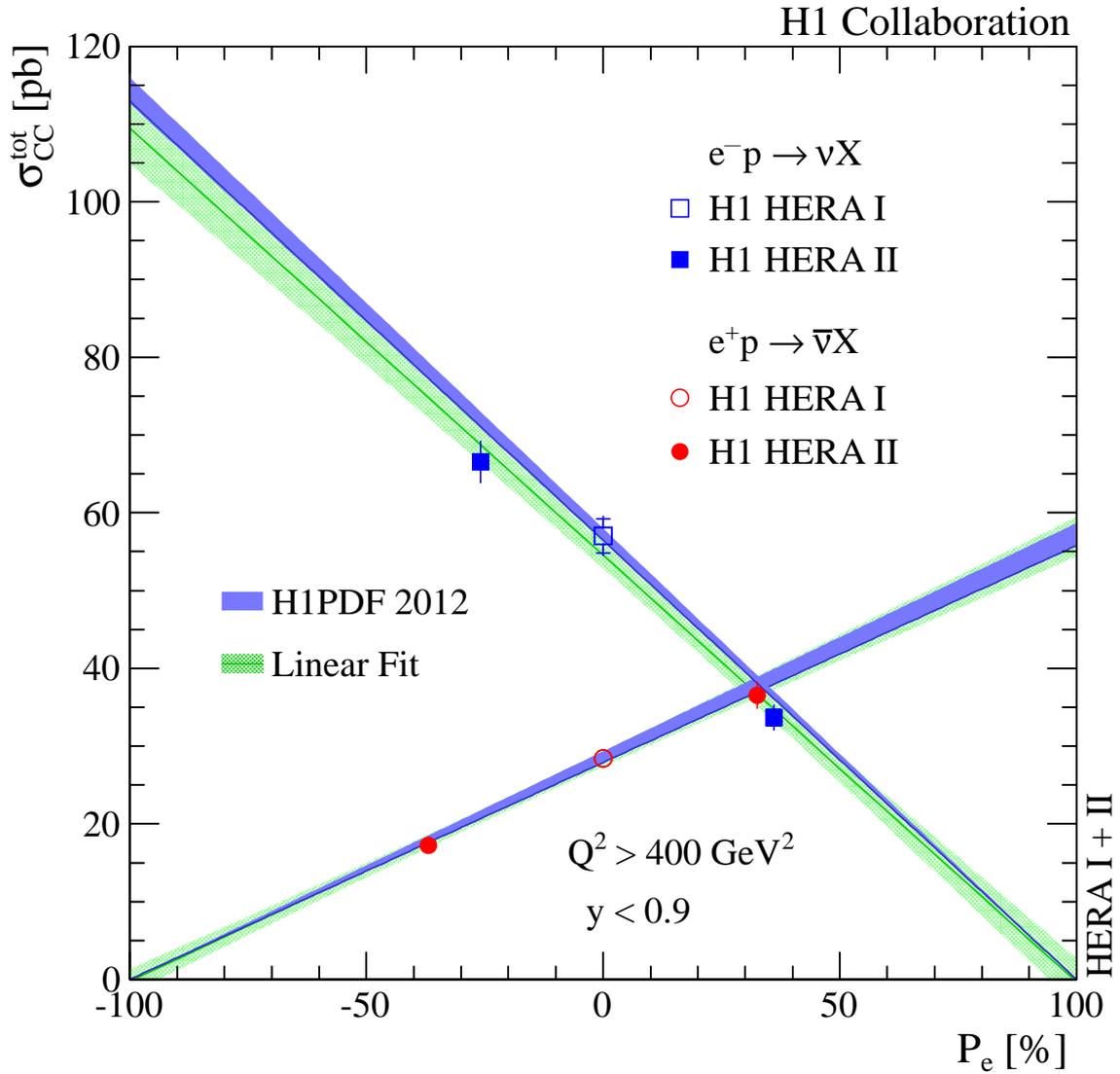}
\end{center}
\caption{ Dependence of the $e^\pm p$ CC cross sections on the
  longitudinal lepton beam polarisation $P_e$. The inner and outer error bars
  represent the statistical and total errors respectively. The
  uncertainties on the polarisation measurement are shown with
  horizontal error bars which are mostly smaller than the symbol
  size. The data are compared to the Standard Model expectation 
  based on the H1PDF\,2012 parametrisation (dark
  shaded band). The light shaded band corresponds to the resulting
  one standard deviation contour of a linear fit to the data shown as 
  the central line.}
\label{fig:cctot} 
\end{figure}

\end{document}